\documentclass[12pt]{article}\usepackage{etex}
\usepackage[textheight=9in, textwidth=6.5in, letterpaper]{geometry}
\usepackage{slashed}
\usepackage[nottoc,numbib]{tocbibind}
\usepackage{epsfig}
\usepackage{comment}
\usepackage{cancel}
\usepackage{bbm}
\usepackage{array}
\usepackage{amssymb,amsfonts,amsmath}
\usepackage{bigints}
\usepackage{booktabs}
\usepackage{color}
\usepackage{dsfont}
\usepackage{float}
\usepackage{framed}
\usepackage{graphicx}
\usepackage{indentfirst}
\usepackage{mathrsfs}
\usepackage{multirow}
\usepackage{subdepth}
\usepackage{subfig}
\usepackage{titlesec}
\usepackage[dotinlabels]{titletoc}
\usepackage{wrapfig}
\usepackage[all]{xy}
\usepackage{xcolor}
\usepackage{young}
\usepackage[vcentermath]{youngtab}
\usepackage{relsize}
\usepackage{hyperref}
\usepackage{tikz}
\usepackage{cite}
\usetikzlibrary{calc}
\usepackage{feynmp}
\hypersetup{colorlinks=true}
\hypersetup{linkcolor=blue}
\hypersetup{citecolor=blue}
\hypersetup{urlcolor=blue}
\numberwithin{equation}{subsection}

\def\T{\hat T}
\def\C{\hat C}
\def\Q{\hat Q}
\def\e{{\epsilon}}
\def\ve{{\varepsilon}}

\def\g{{\gamma}}
\def\o{{\omega}}
\def\a{{\alpha}}
\def\b{{\beta}}

\def\w{{\omega}}
\def\O{\Omega}
\def\L{\Lambda}
\def\bz{{\bar z}}

\def\bw{{\bar w}}

\def\da{{\dot \alpha}}

\def\cst{${\cal C S}^2$}
\def\CA{{\mathcal A}}

\def\CC{{\mathcal C}}

\def\CF{{\mathcal F}}
\def\CG{{\mathcal G}}
\def\CH{{\mathcal H}}

\def\ci{{\mathcal I}}

\def\CK{{\mathcal K}}
\def\CL{{\mathcal L}}

\def\CN{{\mathcal N}}
\def\CO{{\mathcal O}}
\def\CP{{\mathcal P}}

\def\CS{{\mathcal S}}

\def\CL{{\mathcal L}}

\def\SF{{\mathscr F}}

\newcommand{\pd}[2]{\frac{\partial #1}{\partial #2}}

\def\p{\partial}
\newcommand{\avg}[1]{\langle\, #1\, \rangle}
\newcommand{\bra}[1]{\langle \, #1 \,|}
\newcommand{\ket}[1]{|\, #1 \,\rangle}

\newcommand{\tr}[1]{\text{tr} \left[ #1 \right]}

\def\mcc{{\mathbb C}}

\def\mhh{{\mathbb H}}

\def\mrr{{\mathbb R}}

\def\ip{$\ci^+$}
\def\im{$\ci^-$}
\newcommand{\bd}[1]{\begin{fmffile}{#1}\begin{fmfgraph*}}
\newcommand{\ed}{\end{fmfgraph*}\end{fmffile}}

\def\half{{1\over 2}}

\newcommand{\dt}{{\mathrm d}}

\def\0{{(0)}}
\def\1{{(1)}}
\def\2{{(2)}}
\def\3{{(3)}}
\def\4{{(4)}}
\def\pmm{{(\pm)}}
\def\+{{(+)}}
\def\-{{(-)}}
\def\ch{{\cal H}^+}
\def\G{\Gamma}
\newcommand{\bea}{\begin{eqnarray}}
\newcommand{\eea}{\end{eqnarray}}
\newcommand{\be}{\begin{equation}}
\newcommand{\ee}{\end{equation}}
\newcommand{\ba}{\begin{align}}
\newcommand{\ea}{\end{align}}
\def\be{\begin{equation}}
\def\ee{\end{equation}}
\def\beq{\be\begin{array}{c}}
\def\eeq{\end{array}\ee} 
\newcommand{\cs}{\mathcal {S}} 
\newcommand{\co}{\mathcal {O}} 
\def\outst{\langle \text{out} |}
\def\inst {|\text{in}\rangle }
\newcommand{\tdh}{\tilde h }

\begin{document}

\begin{titlepage}
\unitlength = 1mm
\ \\
\vskip 3cm
\begin{center}

{\LARGE{\textsc{Lectures on the Infrared Structure}}}
\vskip.5cm
{\LARGE{\textsc{of Gravity and Gauge Theory}}}

\vspace{0.8cm}
Andrew Strominger

\vspace{1cm}

{\it  Center for the Fundamental Laws of Nature, Harvard University,\\
Cambridge, MA 02138, USA}

\vspace{0.8cm}\begin{abstract}
This is a redacted transcript of  a  course given by the author at Harvard in spring semester 2016. It contains a pedagogical overview of recent developments connecting the subjects of soft theorems, the memory effect  and asymptotic symmetries in four-dimensional QED, nonabelian gauge theory and gravity with applications to black holes. The lectures may be viewed online at  https://goo.gl/3DJdOr. Please send typos or corrections to strominger@physics.harvard.edu.
%
%
%
\end{abstract}

\vspace{1.0cm}

\end{center}

\end{titlepage}

\pagestyle{empty}
\pagestyle{plain}

\pagenumbering{arabic}

\tableofcontents
\thispagestyle{empty}
\newpage
\setcounter{page}{1}

\addtocontents{toc}{\protect\thispagestyle{empty}}
 \section{Introduction }
 \subsection{The Infrared Triangle} \label{The Infrared Triangle}
 
\begin{figure}[h] 
\begin{center}
\includegraphics[width=6.2 in]{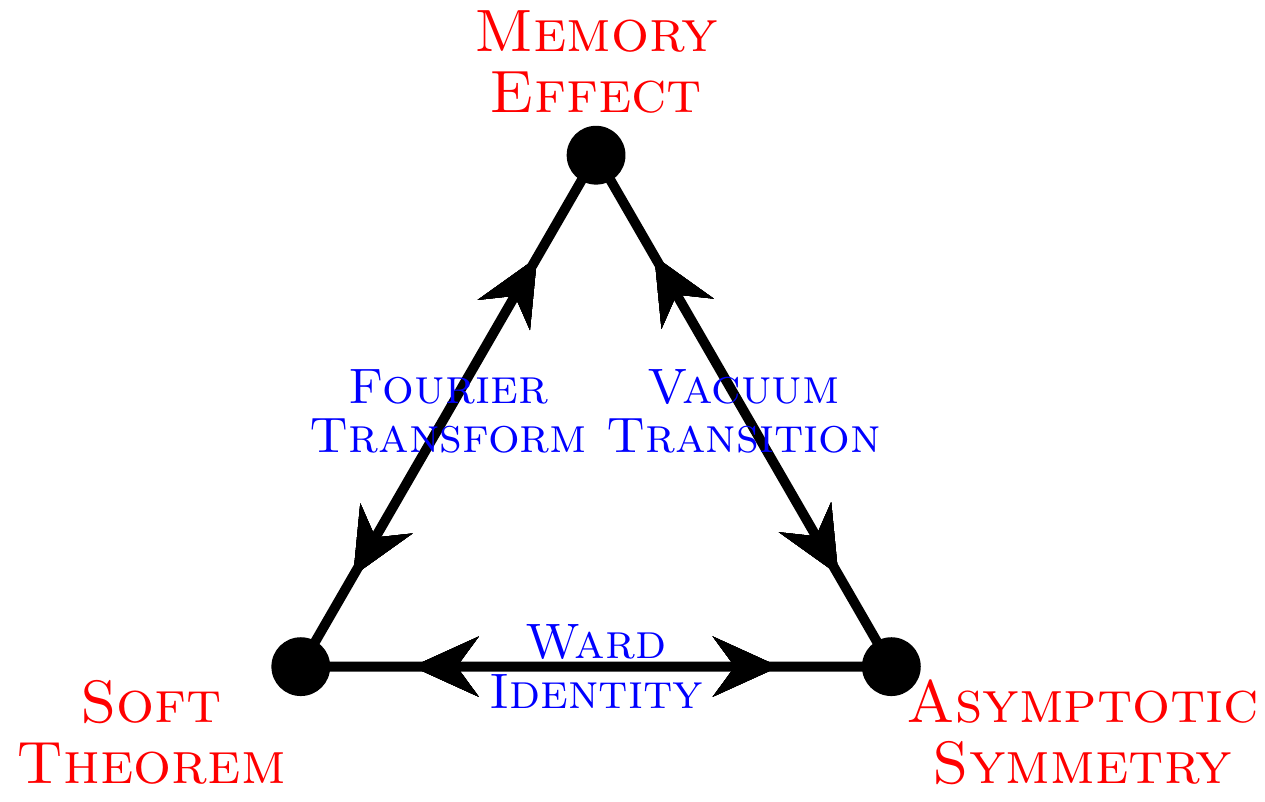}
\end{center}
\caption{ \small \bf The infrared triangle.} \label{IRtriangle}
\end{figure}

 These lectures concern a triangular equivalence relation that governs the infrared (IR) dynamics of all physical theories with massless particles. Each of the three corners of the triangle, illustrated in figure~\ref{IRtriangle}, represents an old and central subject in physics on which hundreds or even thousands (in the case of soft theorems) of papers have been written. Over the past few years we have learned that these three seemingly unrelated subjects are actually the same subject, arrived at from very different starting points and expressed in very different notations. 

The first corner is the topic of soft theorems. These originated in quantum electrodynamics (QED) in 1937 with the work of Bloch and Nordsieck \cite{Bloch:1937pw}, were significantly developed in 1958 by Low and others\cite{Low:1954kd,GellMann:1954kc,Low:1958sn, Kazes1959,Yennie:1961ad }, and were generalized to gravity in 1965 by Weinberg \cite{Weinberg:1965nx}. Soft theorems characterize universal properties of Feynman diagrams and scattering amplitudes when a massless external particle becomes \textit{soft} (i.e., its energy is taken to zero). 
These theorems tell us that a surprisingly large --- in fact, infinite --- number of soft particles are produced in any physical process, but in a highly controlled manner that is central to the consistency of quantum field theory. 

The second corner is the subject of asymptotic symmetries. 
This is the study of the nontrivial exact symmetries or conserved charges of any system with an asymptotic region or boundary. One of the earliest examples appears in the pioneering work of Bondi, van der Burg, Metzner, and Sachs (BMS) \cite{Bondi:1962px,Sachs:1962wk}, who sought to recover the Poincar\'{e} group of special relativity as the symmetry group of asymptotically flat spacetimes in general relativity (GR). Instead, in a spectacular failure of their original program, they discovered the infinite-dimensional BMS group 
whose deep implications are still being unravelled today.  
In contrast, analogous asymptotic symmetries in QED and nonabelian gauge theory were discovered only recently \cite{Strominger:2013lka,He:2014cra,He:2015zea,Kapec:2015ena,Campiglia:2015qka} and are a subject of ongoing research \cite{Barnich:2013axa, Barnich:2013sxa, Grumiller:2013swa, Mohd:2014oja, Cardona:2015woa,Adamo:2015fwa,Seraj:2016jxi,Gabai:2016kuf}. 

The third corner of the triangle is the memory effect, investigated in the context of gravitational physics in 1974 by Zel'dovich and Polnarev \cite{Zeldovich_Polnarev_1974}, and significantly developed by Christodoulou and others 
\cite{Braginsky:1986ia,1987Natur.327..123B,Christodoulou:1991cr,Wiseman:1991ss,Blanchet:1992br,Thorne:1992sdb,Favata:2010zu,Tolish:2014bka,Tolish:2014oda,Winicour:2014ska}. This is a subtle DC effect, in which the passage of gravitational waves produces a permanent shift in the relative positions of a pair of inertial detectors.  Detection of the memory effect has been proposed at LIGO \cite{Lasky:2016knh} or via a pulsar timing array \cite{vanHaasteren:2009fy,Wang:2014zls}. It is an exciting experimental prospect for the coming decades. 
Again, the gauge theory analog appeared only recently \cite{Bieri:2013hqa, Susskind:2015hpa,Pasterski:2015zua,Pate:2017vwa}. This memory corner of the triangle provides an important physical realization of the more abstractly formulated results of the other two corners, giving direct observational consequences of the infinite number of  symmetries and conservation laws.

Figure~\ref{IRtriangle} also depicts the mathematical equivalence relations connecting the three corners. Perhaps the simplest is the connection between the soft theorem and the memory effect \cite{Strominger:2014pwa}.
The former is a statement about momentum space poles in scattering amplitudes, while the latter concerns a DC shift in asymptotic data between late and early times. These are the same thing, because the Fourier transform of a pole in frequency space is a step function in time. The step function in turn can be understood as a domain wall connecting two inequivalent vacua that are related by an asymptotic symmetry \cite{Strominger:2014pwa}. Hence the memory effect both physically manifests and directly measures the action of the asymptotic symmetries. The triangle in figure~\ref{IRtriangle} is closed by noting that every symmetry has a Ward identity that equates scattering amplitudes of symmetry-related states.\footnote{Alternately, these identities can be derived as the vanishing of all matrix elements of the associated charge conservation laws.} These Ward identities turn out to be nothing but the soft theorems --- which relate amplitudes with and without soft particles --- in disguise \cite{Strominger:2013jfa,Strominger:2013lka,He:2014cra,He:2014laa}. 

It is a testimony to the unity of physics that these three seemingly disparate avenues of investigation led to the same mathematical structures. We will see in section \ref{sec:gravity} that Weinberg's 1965 soft graviton theorem reproduced the 1962 results of BMS, but in the wildly  different language and notation of Feynman diagrams as opposed to asymptotic structures at null infinity. We will further see in section \ref{sec:memory} that the 1987 Braginsky-Thorne formula for the gravitational memory effect is a Fourier transform of the Weinberg soft graviton theorem. Of course, Weinberg was scattering elementary particles, while Braginsky and Thorne were scattering black holes, but this distinction is irrelevant in the deep infrared! 

The bigger picture emerging from the triangle is that deep IR physics is extremely rich, perhaps richer than previously appreciated. Every time we breathe, an infinite number of soft photons and gravitons are produced. Quantum field theory analyses tend to treat this as a technical problem to be overcome
by resorting to carefully constructed inclusive cross sections.  IR regulators are often used that explicitly break the symmetries, and it is difficult to see them emerge as the regulator is removed.\footnote{BMS$_3$ has been recovered 
as the large radius limit of gravity in AdS$_3$\cite{Bagchi:2013toa,Andrade:2015fna,Barnich:2012aw}, but efforts \cite{Andrade:2015fna} to recover BMS$_4$ in the large radius limit of AdS$_4$ have so far not succeeded.}  GR  analyses tend to treat the BMS discovery that the deep IR limit of GR is  not special relativity as a  problem  to be avoided by the ad hoc imposition of extra boundary conditions at infinity. In fact, BMS discovered  the classical version of infinite graviton production, an equivalent indication of the richness of the deep IR. Here we will see that, far from being technicalities, these IR phenomena are associated with fundamental symmetries of nature whose fascinating  consequences we are just beginning to unravel. 

\subsection{New Developments}
New developments in this area go beyond demonstrating the equivalence of previously known phenomena. The three-dimensional perspective afforded by the triangle has led to better conceptual understanding of each old corner, discovery of new corners, and new research programs searching for and exploring new triangles. We now mention a few of the many new results that have emerged in the past few years. 

Until recently, one corner of the triangle --- asymptotic symmetries ---  was essentially unknown for any example in four-dimensional Minkowski space.  The best understood case was the BMS action in gravity. BMS showed that infinite-dimensional subgroups BMS$^\pm$ of 
the diffeomorphism group  act nontrivially  at future and past null infinity $\ci^\pm$, transforming one set of Cauchy data to a physically inequivalent one. The full group $\mathrm{BMS^+\times BMS^-}$, however, is clearly not a symmetry of gravitational scattering. For example, it includes elements that boost the past but not the future. It was widely held (in part due to misconceptions about asymptotic falloffs cleared up only recently by Christodoulou and Klainerman \cite{Christodoulou:1993uv}) that there  is no nontrivial infinite-dimensional symmetry of gravitational scattering; see \cite{Hawking:2016sgy} for further discussion. However, the equivalence to the soft graviton theorem was used in \cite{Strominger:2013jfa} to show that, in a generic physical setting, a certain `antipodal' subgroup of $\mathrm{BMS^+\times BMS^-}$ is an exact asymptotic symmetry of gravitational scattering. This discovery is detailed in section \ref{sec:supertranslations}.

The triangular perspective further led to the conclusion that the vacuum in GR (as well as in gauge theory) is not unique, as it transforms nontrivially under the antipodal symmetries.\footnote{This result was partially anticipated in prescient work of Ashtekar \cite{Ashtekar:1987tt}; see also Balachandran and Vaidya \cite{Balachandran:2013wsa}.}  In particular, distinct vacua are found to have  different angular momenta \cite{Barnich:2016lyg}. Ignoring the vacuum degeneracy leads to the false conclusion that angular momentum is not conserved (see figure \ref{amvac})! This degeneracy resolves the so-called problem of angular momentum in GR. The electromagnetic analog is discussed in section \ref{ssb}, and the gravity version in section \ref{sec:gravity}.

Another new development from the triangle involves a conjectured extension of the BMS group involving \textit{superrotations}, a Virasoro-like symmetry that acts on the sphere at $\ci$ \cite{deBoer:2003vf,Banks:2003vp,Barnich:2009se, Barnich:2011ct}. Because of certain singularities in the generators, standard asymptotic symmetry analyses could not establish whether these are bona fide symmetries. However, the result  was established    
\cite{Cachazo:2014fwa,Kapec:2014opa} using Feynman diagrams and proving the equivalent soft theorem, as discussed in section \ref{sec:superrotation}.

In abelian and nonabelian gauge theories, various soft theorems have long been known.  New symmetries in gauge theory, precise analogs of the antipodal subgroup of BMS  in gravity,  have been discovered using  the triangular equivalence, as discussed in sections \ref{qed} and \ref{nonabelian}.

Given one corner of a triangle, others can be systematically determined. Insights from gravity can be applied to gauge theory and vice versa. One or more triangles potentially exists for every type of massless particle. Let us now discuss a few of these triangular variations. 
  
\subsection{Echoing Triangles}
\begin{figure}[h] 
\begin{center}
\includegraphics{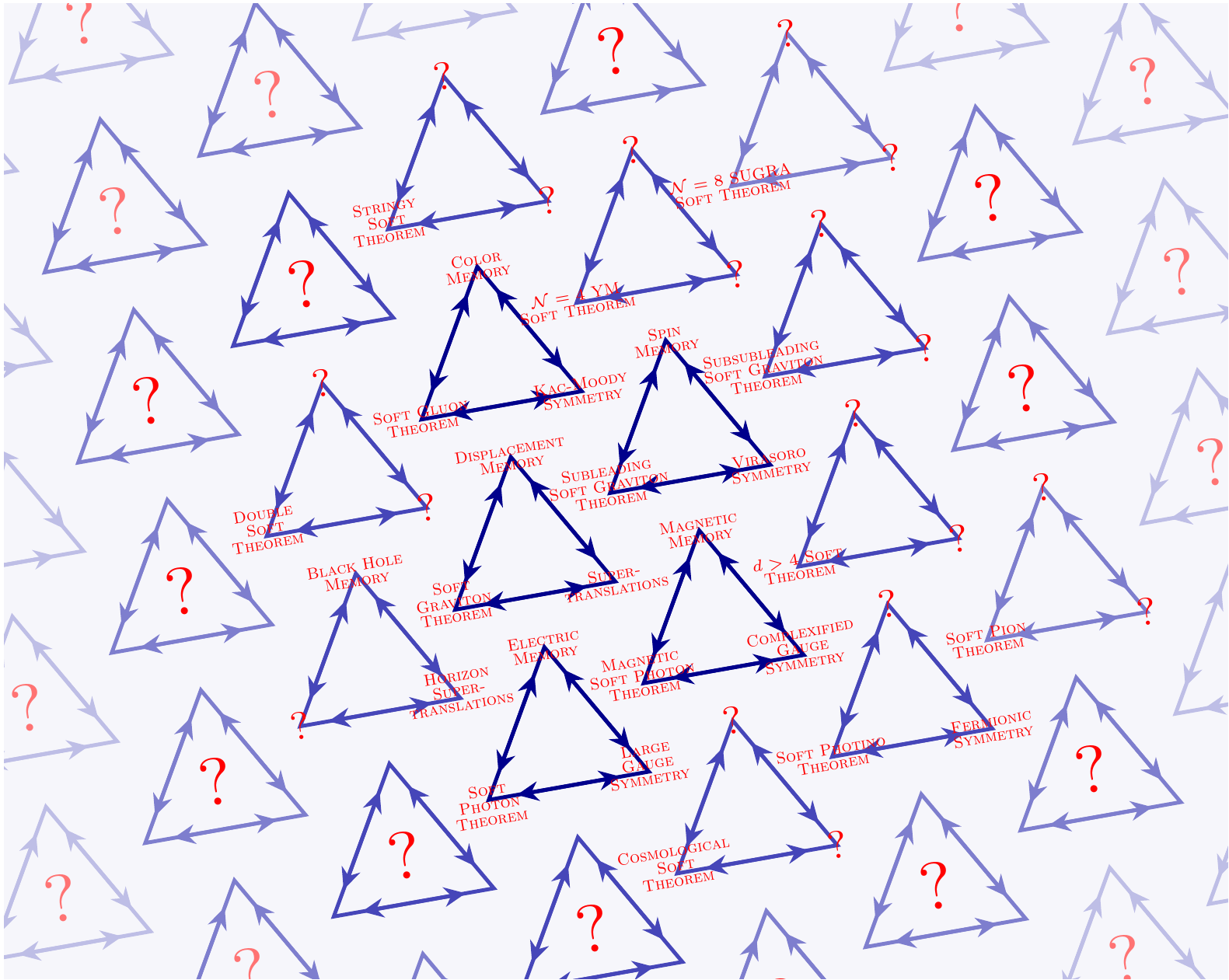}
\end{center}
\caption{ \small \bf The infrared triangle echoes throughout disparate areas of physics.} \label{echoes}
\end{figure}

The IR triangle in figure~\ref{IRtriangle} has many copies, which echo throughout much of physics, as illustrated in figure \ref{echoes}. 
Here are some examples. 
\subsubsection{\it QED, Yang-Mills Theory, Gravity, Massless Particles, ...}
The simplest cases of the triangle involve massless particles associated with local gauge symmetries, such as the photon, which will be our first example. However, whenever there is a massless particle, one can consider the soft limit of the scattering amplitude. A full triangle, including symmetry and memory, is often found to emerge, even when no underlying local gauge symmetry exists. A simple example of this discussed here  is the soft photino theorem \cite{Dumitrescu:2015fej}, which lies at the corner of a triangle including an infinite number of asymptotic fermionic symmetries. Potentially, a triangle exists for every type of massless particle, many of which have yet to be studied from this perspective. \subsubsection{\it Leading, Subleading, Sub-Subleading, ...}
   
Soft theorems are statements about the behavior of an amplitude at zero or low energy. In an expansion in powers of the soft energy, universal behavior may appear in the leading as well as the subleading or even sub-subleading orders. Indeed, Low \cite{Low:1954kd,Low:1958sn,Burnett:1967km} already discussed a subleading soft theorem in QED. Recent research derived the symmetry based on Low's subleading theorem \cite{Lysov:2014csa,Campiglia:2016hvg,Conde:2016csj}. The  associated memory effect remains to be analyzed. In gravity, a new subleading soft theorem was proven at tree-level \cite{Jackiw:1968zza,Gross:1968in,White:2011yy,Cachazo:2014fwa} and was shown to imply  \cite{Kapec:2014opa,Campiglia:2014yka,Kapec:2016jld,Cheung:2016iub} superrotation symmetry \cite{deBoer:2003vf,Banks:2003vp,Barnich:2009se, Barnich:2011ct}, a recently conjectured asymptotic symmetry. The third corner of the triangle, the spin memory effect, was derived in \cite{Pasterski:2015tva}. A proposal for measuring it using the Einstein Telescope appears in Nichols \cite{Nichols:2017rqr }.  A gravitational sub-subleading soft theorem has also been proven \cite{Cachazo:2014fwa,Du:2014eca,Zlotnikov:2014sva,Kalousios:2014uva, DiVecchia:2016amo, Campiglia:2016efb,DiVecchia:2016szw,Luna:2016idw}, and the associated symmetry has been discussed in \cite{Campiglia:2016jdj}.  A complete classification of all possible soft theorems at various orders of the soft expansion has not appeared, although some discussion of it exists \cite{Broedel:2014fsa,Elvang:2016qvq}.  New soft theorems at various orders have recently been discovered in nonabelian gauge theories \cite{Casali:2014xpa,Vera:2014tda, Adamo:2015fwa,Luna:2016idw},  supersymmetric theories \cite{Chen:2014xoa, Liu:2014vva, Bork:2015fla}, nonlinear sigma models \cite{Du:2015esa}, string theories \cite{Schwab:2014sla,Schwab:2014fia,DiVecchia:2015oba, Bianchi:2015yta, Avery:2015gxa, Bianchi:2015lnw,DiVecchia:2016szw,Sen:2017xjn}, and in other contexts \cite{Larkoski:2014hta,Schwab:2014xua,Luo:2014wea,Cachazo:2016njl, Rodina:2016jyz}.

\subsubsection{\it Double Soft Theorems, ...}

Interesting universal structures are known for limits when two or more external particles are taken to be soft. For example, for soft moduli in $\mathcal{N}=8$ supergravity, the structure constants of the hidden $E_{7(7)}$ symmetry algebra appear \cite{ArkaniHamed:2008gz}. In recent years, double-soft limits have been further examined in gravity \cite{Klose:2015xoa,Saha:2017yqi}, gauge theories \cite{Chakrabarti:2017ltl,Volovich:2015yoa,Klose:2015xoa, Cachazo:2015ksa,McLoughlin:2016uwa}, supersymmetric theories \cite{Chen:2014xoa, Chen:2014cuc, Volovich:2015yoa,Klose:2015xoa}, nonlinear sigma models \cite{Cachazo:2015ksa, Du:2015esa, Low:2015ogb}, and string theory \cite{Volovich:2015yoa,DiVecchia:2015bfa,Sen:2017xjn}. This behavior is presumably at the corner of yet another unexplored triangle and is perhaps related to structure constants or central charges of an asymptotic symmetry group. 

\subsubsection{\it Classical, Quantum, ...}
   
There is a purely classical triangle involving tree-level soft theorems and classical symmetries. This regime is where these lectures begin. In the simplest cases, the soft theorem is uncorrected, and the quantum triangle is similar to the classical one. In general, however, an interesting story involving anomalies may appear \cite{Bianchi:2014gla, Bern:2014oka,He:2014bga,Cachazo:2014dia,Bern:2014vva,Broedel:2014bza,Lipstein:2015rxa,Brandhuber:2015vhm, He:2017fsb}. 
   
\subsubsection{\it Minkowski, $D<4$, $D>4$, Cosmology, ...}
   
In these lectures, we consider only the physical case of four-dimensional Minkowski space. However, the triangle also appears in both higher dimensions \cite{Kapec:2014zla,Kapec:2015vwa,Schwab:2014xua} and in cosmology, where there are also soft theorems \cite{Kehagias:2016zry,Mirbabayi:2016xvc,Hamada:2017gdg} and asymptotic symmetries \cite{Ferreira:2016hee}.  The sub-subleading soft graviton theorem has been explored in arbitrary dimensions \cite{Zlotnikov:2014sva,Kalousios:2014uva}. 
The case of  three-dimensional gravity, where there is generally better control,  is especially interesting\cite{Bagchi:2013toa,Barnich:2014kra, Barnich:2015mui,Hartong:2015usd,Barnich:2015sca,Donnay:2015abr,Carlip:2016lnw,Afshar:2016uax,Banerjee:2016nio,Donnay:2016iyk,Donnay:2016zka,Prohazka:2017equ,Batlle:2017llu}. However, since  three-dimensional gravitons do not propagate, this case has no soft theorem. This feature gives the problem a rather different flavor, and I do not review it in these lectures. 
\subsubsection{\it Supersymmetry, $\mathcal N=1,2, ...$}
   
For many reasons it is also of interest to consider supersymmetric theories. The local $\mathcal N=1$ case has been worked out \cite{Avery:2015iix, Lysov:2015jrs}, as has the global one in \cite{Dumitrescu:2015fej}. New features should appear with soft scalars and $R$-symmetries for $\mathcal N=4$ Yang-Mills and $\mathcal N=8$ supergravity. Soft theorems in various supersymmetric theories have been worked out \cite{Bianchi:2014gla,Liu:2014vva, Brandhuber:2015vhm} and sometimes  give rise to recursion relations \cite{Luo:2015tat}. At present, even the asymptotic symmetry group for these interesting higher-$\mathcal N$ cases is unknown.
   
  \vskip.2in
 Every one of the combined possibilities mentioned here can potentially result in a new triangle. So far, each story has its own unique and surprising features, which need to be worked out case by case. Most have yet to be explored. This subject is still in its infancy, which makes it a lot of fun to talk about! 
  
\subsection{Motivation and Applications}
   
Why is this interesting? What are the motivations and applications? While the development of this subject heavily borrows some insights from modern string theory, it directly pertains to the real world. It has interwoven motivations and applications, which we now discuss. 

\subsubsection{\it Connecting Disparate Subjects} 
   
Given a concrete mathematical relation between two different subjects, results in one can be translated into what are often new results in the other. New methods of calculation, new physical phenomena and insights, and new avenues of investigation are often suggested. The current context is no exception. Soft theorems, memory effects, and asymptotic symmetries are three different ways of characterizing the behavior of the universe around us at very long distance scales. This equivalence has led to the realization that physics in the deep infrared is much richer, more subtle, and certainly much less understood than we previously believed. Notably, looking at IR structure from all three corners of the triangle has led to the realization that the vacuum in all gauge and gravitational theories is best described as infinitely degenerate.  Prior to the past several years, there was no case in which all three corners were understood. Now that we know about these triangles, once we find one corner of one triangle, we can work out the other two. As a result, numerous surprising discoveries have been made, some of which are covered in these lectures. 
   
\subsubsection{\it Flat Space Holography }
   
 A central  motivation for these IR investigations  is to understand the holographic structure of quantum gravity in four-dimensional asymptotically flat  spacetimes,  which is a good approximation to  the real world. This is how I came into the subject. A very beautiful story has unfolded over the past twenty years concerning the holographic structure of quantum gravity in anti-de Sitter space. The story begins \cite{Maldacena:1997re} with the identification of the symmetries in anti-de Sitter space with those of its proposed holographic dual. Following this successful example, the very first question we should ask in attempting a holographic formulation of flat space quantum gravity  is: ``What are the symmetries?'' Up until three years ago, the answer to this question was  unknown.  At the very least we now know \cite{Strominger:2013jfa} that the symmetry group is infinite dimensional and includes a certain subgroup of, but not all of, the BMS group on past and future null infinity. In addition, good evidence 
 \cite{Kapec:2014opa,Kapec:2016jld,He:2017fsb} indicates a conjectured Virasoro (or superrotation) symmetry \cite{deBoer:2003vf,Banks:2003vp,Barnich:2009se, Barnich:2011ct} that acts on the sphere at null infinity, referred to as the celestial sphere \cst. The Lorentz group $SL(2,\mcc )$ acts as the global conformal group on this sphere, and the Virasoro action enhances the global to the local conformal group. This suggests that the holographic dual of four-dimensional flat-space quantum gravity might be realized as an exotic two-dimensional conformal field theory (CFT) on \cst~ \cite{He:2015zea,Pasterski:2017kqt,Pasterski:2017ylz,Pasterski:2016qvg,Bagchi:2016bcd,Cardona:2017keg}.
 
\subsubsection{\it The  Gauge Theory $\cs$-Matrix}

 The study of soft particles and infrared divergences in QED, Yang-Mills theory, and collider physics is a large and central enterprise in quantum field theory. Notable recent progress includes the development of soft collinear effective theory (SCET) \cite{Bauer:2000yr}, in which infinite-dimensional symmetries likely related to those discussed here make an appearance \cite{Larkoski:2014bxa}. SCET has important practical applications for understanding jet structure and interpreting LHC and other collider events. From a more formal point of view, it is disturbing that generically no IR finite $\cs$-matrix\footnote{In massive abelian gauge theories, the Faddeev-Kulish construction \cite{Kulish:1970ut} yields some IR finite amplitudes. An analogous construction for nonabelian gauge theory was given in \cite{DelDuca:1989jt,Giavarini:1987ts} and references therein. It was recently revisited for gravity \cite{Ware:2013zja} and for QED \cite{Gabai:2016kuf}. However, Faddeev-Kulish states are not complete: for example, they exclude an incoming positron-electron pair with no incoming radiation. See section \ref{IRDiv} for further discussion.} exists in gauge theories, even if suitable inclusive cross sections are finite \cite{Weinberg:1995mt}. While this does not pose an obstacle for collider predictions, it is a big elephant in the room for mathematical quantum field theory. Surely the central object of study in these theories must somehow be mathematically well defined! Our hope is that the discovery of new symmetries in the deep IR will enable us to formulate the soft part of the $\cs$-matrix in a more satisfactory way. Moreover, the symmetries might be exploited to develop more efficient computational methods, perhaps via a connection with the SCET program.

\subsubsection{\it Miracles in $\mathcal N = 4$ Yang-Mills}
    Scattering amplitudes in $\mathcal N = 4$ Yang-Mills have been a very active subject for more than 10 years e.g.,~\cite{Witten:2003nn,Britto:2005fq, Brandhuber:2007yx, Drummond:2009fd, ArkaniHamed:2008gz, CaronHuot:2011ky,ArkaniHamed:2010kv,ArkaniHamed:2012nw,Arkani-Hamed:2013jha}. Extensive computations have revealed  miraculous cancellations that reduce very complicated expressions to  simple ones. When many cancellations occur it is natural to suspect that they are required by an underlying symmetry. It is possible that the new IR symmetries provide an explanation of those miracles or, even better, a derivation of some of the incredibly beautiful  formulas for scattering amplitudes in $\mathcal N = 4$ Yang-Mills theory.
    
  \subsubsection{\it Black Holes}
   
 Although I did not start this IR project with black holes in mind, as usual, all roads lead to black holes \cite{Strominger:2014pwa,Hawking:2016msc,Hawking:2016sgy}. The IR structure has important implications for the information paradox\cite{Hawking:1976ra}. This paradox is intertwined with the deep IR because an infinite number of soft gravitons and soft photons are produced in the process of 
black hole formation and evaporation. These soft particles carry information with a very low energy cost. They must be carefully tracked to follow the flow of information. Tracking is hard to do without a definition of the $\cs$-matrix! Moreover, their production is highly constrained by an infinite number of exact quantum conservation laws that correlate them with energetic hard particles and also with the quantum state of the black hole itself \cite{Strominger:2017aeh}. Thus black holes must carry an infinite number of conserved charges, described  as ``soft hair''  in a recent collaboration with Hawking and Perry \cite{Hawking:2016msc,Hawking:2016sgy}. The information paradox cannot be clearly stated \cite{Kapec:2016aqd}, let alone solved, without accounting for soft particles. The implications of soft hair recently have been discussed \cite{Flanagan:2015pxa,Blau:2015nee,Penna:2015gza,Donnay:2015abr, Averin:2016ybl, Compere:2016hzt, Kapec:2016aqd,Eling:2016xlx, Donnay:2016ejv,Compere:2016gwf,Compere:2016jwb, Mao:2016pwq, Mirbabayi:2016axw, Cai:2016idg,Setare:2016msj, Avery:2016zce,Shi:2016jtn, Tamburini:2017dig,CarneirodaCunha:2017sur}, for example.
\vskip.2in

These lectures both review the old material on soft theorems, asymptotic symmetries and memory as well as the new connections among them. Unsolved problems are pointed out along the way.  Some exciting developments in the past year whose proper perspective has not yet been clarified are omitted. 
These lectures are not meant to be an exhaustive review; see other sources in the References \cite{Balachandran:2013wsa,Larkoski:2014hta,Duval:2014uva,Gomis:2015ata, Andrade:2015fna, Campiglia:2015kxa, Campiglia:2015lxa, Dvali:2015rea, Banks:2014iha,Banks:2015pfi, Balachandran:2016ohv, Barnich:2016lyg,  Bousso:2016wwu,   Haco:2017ekf,Ananth:2017xpj} for interesting related developments.

\section{QED}\label{qed}	

Since this is a book  about the interconnections among many different subjects, we might start at many different places --- gauge theory or gravity, soft theorems, symmetries or memories --- and weave our way through the connections.  Here I choose the simplest  possible starting point:  classical, nineteenth-century electromagnetism.  Beginning from this point, we  will trace out these interconnections for any theory with an abelian gauge symmetry --- including, of course, QED.  

We begin in section 2.1 by recalling the Li\'enard-Wiechert solutions for a collection of charges and noting that these solutions are discontinuous at the boundary of Minkowski  space. Characterizing this discontinuity requires a careful description of the boundary, which is provided in section 2.2. In section 2.3, we show that the Li\'enard-Wiechert field obeys a boundary matching condition equating its values at the far past of \ip\ ($\ci^+_-$) with those at antipodal points of the far future of $\ci^-$ ($\ci^-_+$). Section 2.4 describes the electromagnetic Cauchy data at $\ci^\pm$. Using only the matching condition,  an infinity of conservation laws is derived in section 2.5. Section 2.6 presents the covariant  canonical formulations of the physical phase spaces at $\ci^\pm$. In section 2.7, I show for massless fields,  that the conserved charges generate (via the Dirac bracket) nontrivial large gauge transformations associated with antipodally matched angle-dependent gauge parameters that do not die off at $\ci^\pm$. A Ward identity of this symmetry   is derived in section 2.8 and shown to be equivalent to the soft photon theorem. For completeness, the standard diagrammatic derivation of the soft theorems for both photons and gravitons is reviewed in section 2.9. In section 2.10, I adapt the BMS-type analysis of asymptotic symmetries employed from studies of gravity to electromagnetism and show that it reproduces the antipodally matched large gauge symmetries. In section 2.11, it is noted that the large gauge symmetry is spontaneously broken, resulting in an  infinite vacuum degeneracy with soft photons as the Goldstone bosons.  The generalization to massive fields, and hence QED, is in section 2.12. This extension employs  a hyperbolic slicing of Minkowski space and heavily borrows results from investigations of the anti-de Sitter/conformal field theory correspondence (AdS/CFT).  In 2.13, a magnetically corrected  soft photon theorem is presented and conjectured to be nonperturbatively exact.  I derive a second infinity of conserved charges and symmetries arising from large magnetic gauge transformations.  Section 2.14 shows that for ${\mathcal N}=1$ supersymmetry, the soft photino theorem leads to an infinity of fermionic conservation laws and symmetries. These are surprising, as they are unrelated to any previously discussed local symmetry. In section 2.15, IR divergences are discussed and shown to disappear for some and possibly all fully charge-conserving amplitudes. 
\subsection{Li\'enard-Wiechert Solution}
 Maxwell's theory of electromagnetism is described by the  action 
 \begin{equation}
\begin{split}\label{qedaction}
S = - \frac{1}{4 e^2} \int d^4 x \sqrt{-g} F_{\mu\nu} F^{\mu\nu}  + S_M  ~,
\end{split}
\end{equation}
where $\mu,\nu= 0, 1,2,3$ and $S_M$ denotes a general matter action.  In form notation, the field strength $F$ is related to the gauge field  $A$ by
\be
F = \mathrm{d}A~.
\ee
The equation of motion is
\be \label{QEDeom}
\mathrm{d} \ast F = e^2 \ast j  \quad \implies \quad \nabla^\mu F_{\mu\nu} = e^2 j_\nu~,
\ee
where  $*$ is the Hodge dual, and the charge current is 
\be
j^\nu = -\frac{\delta S_M}{ \delta A_\nu}~.
\ee
The electric charge inside a  two-sphere ($S^2$) at infinity is defined by
\be\label{e_charge}
\quad  Q_E = \frac{1}{e^2} \int_{S^2} \ast F = \int_{\Sigma} \ast j \in \mathbb {Z}~.
\ee
The second equality is obtained using integration by parts and the constraint component of the equation of motion.  It says that  if we take $\Sigma$ to be any slice with the $S^2$  boundary, the electric charge is just the normal component of the charge current integrated over that slice. For convenience, we adopt here a convention in which $Q_E$ is an integer in the quantum theory. For example, the electron has charge $Q_E = 1$ (not $e$). The magnetic charge ${1 \over 2\pi}\int F$ is also integral, and it has an associated conservation law \cite{Strominger:2015bla}. We return to the interesting story of magnetic charges in section 2.13, but for now, assume they are absent. 

The theory (\ref{qedaction}) is invariant under the infinitesimal gauge transformations
\begin{equation} 
\begin{split}
\delta_\ve A = \mathrm{d} \ve \;, \qquad \delta_\ve \Phi_k = i \ve Q_k  \Phi_k \; ,
\end{split}
\end{equation} 
where $\ve \sim \ve + 2\pi$, and $\Phi_k$ is a matter field with charge $Q_k$. Since  $Q_k \in \mathbb Z$,  the finite transformation 
\be
 \Phi_k  \rightarrow  e^{i Q_k \ve} \Phi_k ~
\ee
is invariant under $ \ve \to \ve + 2\pi$. 

 Now let us solve the equation of motion. Consider a source $j$ corresponding to $n$ particles, each moving with constant four-velocity $U_k^\mu = \gamma_k \left( 1, \vec{\beta}_k \right)$, where  $U_k^2 = -1$ and 
$ \gamma_k^2 = \frac{1}{1- \beta_k^2}$. Then we have
\begin{equation}
j_\mu(x) =\sum_{k=1}^n Q_k\int d\tau  ~ U_{k\mu}\delta^{4}(x^\nu-U_k^\nu\tau)~.
\end{equation} 
Each particle is a point source with a worldline parametrized by $\tau$, so that $x_k^\mu (\tau) = U_k^\mu\tau$.	
	
Equation \eqref{QEDeom} determines the electromagnetic field due to this source.  Li\'enard and Wiechert solved this problem in 1898.  The solution for the radial component of the electric field is 
\be\label{Frt}
F_{rt}( \vec x, t) = \frac{e^2}{4 \pi }\sum_{k=1}^n  \frac{Q_k\gamma_k \left(r -t \hat{x}  \cdot \vec{\beta}_k\right)}{   \big| \gamma_k^2 \left( t- r \hat{x}  \cdot \vec{\beta}_k \right)^2 - t^2 + r^2 \big|^{3/2}}~, 
\ee
where
\be
r^2 = \vec x \cdot \vec x~, \quad  \vec x = r \hat x~.
\ee 
This formula is valid for either the advanced or the retarded potential. It has the peculiar property that it is not single valued and has a discontinuity at $r =\infty$.  Although this fact was presumably well known to many since the nineteenth century,  this basic point surprised me when I became aware of it only recently. It is crucial for all that follows, so we will go through how this happens in pedantic detail.  To fully explain it, we must first digress and review the structure of infinity in Minkowski space. 

\subsection{Minkowski Space Penrose Diagram}
	
\begin{figure}[h] 
\begin{center}
\includegraphics[width=3.8 in]{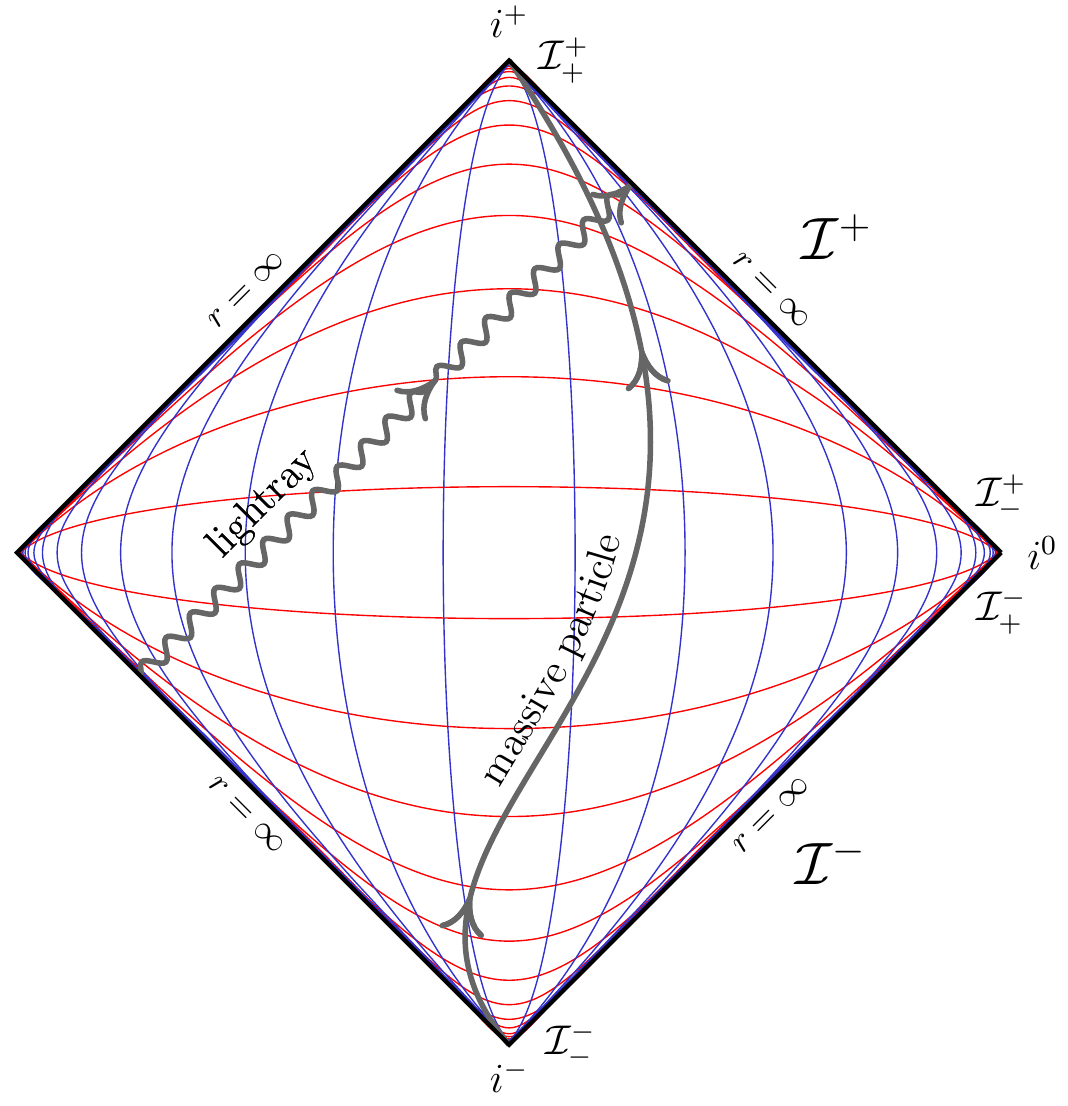}
\end{center}
\caption{ \small \bf Penrose diagram of Minkowski space. Red lines represent surfaces of constant $t$,  while blue lines represent surfaces of constant $r$.  The thick gray line is the  worldline of a massive particle moving at constant velocity, and the thick wavy gray line is a light ray.  Every two-sphere of constant $(r>0,t)$ is represented by two points, one on the left and one on the right, which are exchanged by the antipodal map.  Past and future null infinities are labelled by $\ci^\pm$, and their  four $S^2$ boundary components by $\ci^\pm_\pm$.  The points $i^\pm$ are past and future timelike infinity, while the point $i^0$ is spatial infinity. } \label{penrose}
\end{figure}

\begin{figure}[h] 
\begin{center}
\includegraphics[height=3.8in]{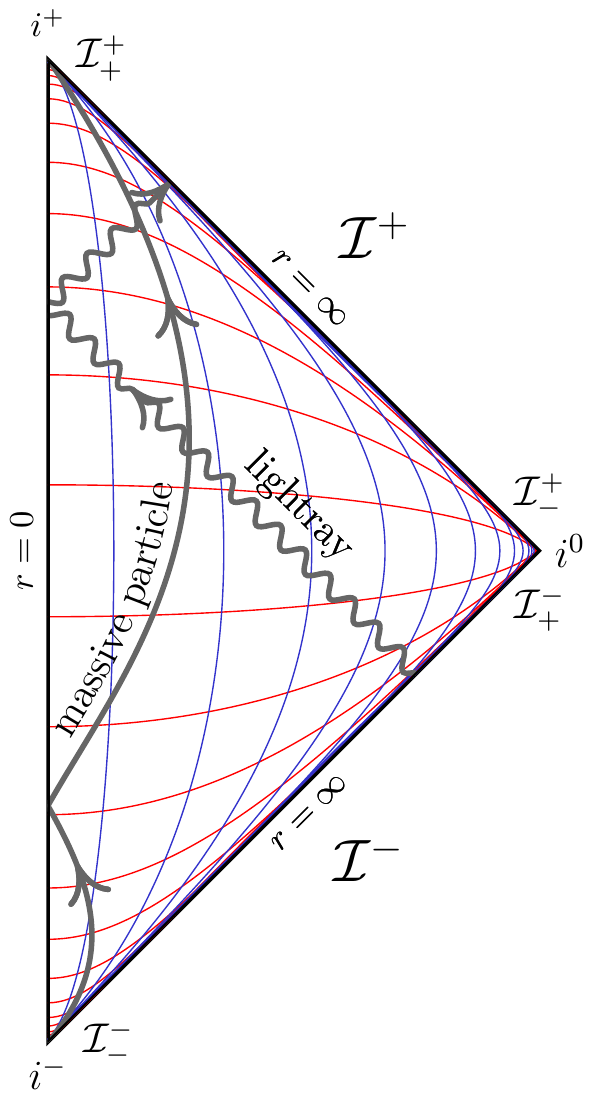}
\end{center}
\caption{ \small \bf Alternative Penrose diagram of Minkowski space in which every point except for $r=0$ is an $S^2$.} \label{penroseusual}
\end{figure}

To talk about infinity,  it is useful to introduce Penrose diagrams.  The  Penrose diagram for four-dimensional Minkowski space is shown in figure~\ref{penrose}.  In this figure all of Minkowski space is pulled into a finite region by a conformal transformation that diverges at the boundaries. Distances are not faithfully represented, but the causal structure is 
unaffected by the conformal transformation and is clearly indicated by the rule that light rays always propagate at 45 degrees.  
Every left-right pair of points at the same $(r,t)$ in figure~\ref{penrose} corresponds to an $S^2$ (except  $r=0$, which is a point), and the pair is interchanged by the antipodal map on $S^2$. 
Sometimes the picture is drawn differently, as in figure~\ref{penroseusual}. 
   Every point on this diagram represents an $S^2$ except for $r =0$, which is the origin. If you have a sequence of two-spheres that are getting smaller and smaller and you add a point in the middle,  you get $\mathbb{R}^3$. For our purposes (which include a study of the antipodal map),  figure~\ref{penrose} will be more useful.  	
	
We want to study the electric field of $n$ charges moving at constant velocity. The charges have worldlines that look like the thick gray line in figure~\ref{penrose}.  They cannot escape to $r = \infty$, because light rays move at 45 degrees on Penrose diagrams; every light ray will eventually catch up to a particle moving with constant velocity (less than $c$) and cross it to larger radius. As a result, massive charges always end up at the uppermost point, ``called future timelike infinity'' and denoted  $i^+$ in figure \ref{penrose}. Similarly, the worldlines of (massive) charges start at past timelike infinity ($i^-$).  Likewise, we call the place where light rays go to ``future null infinity'' ($\ci^+$) and the place where light rays come from ``past null infinity'' ($\ci^-$). We will be especially concerned about behavior near spatial infinity, denoted $i^0$.  This behavior turns out to be very subtle.  As we will see, almost anything, even Maxwell electromagnetism with one point particle, is singular at this point. $\ci^+$ is the product of $S^2$ with a null line.  It will be important to distinguish between the past of $\ci^+$, which we call $\ci^+_-$, and the future of $\ci^-$, which we call $\ci^-_+$.  Both these points are near, but distinct from, the point $i^0$. Figures \ref{penrose} and \ref{penroseusual} also illustrate  $\ci^-_-$, which is an $S^2$ near but distinct from the point $i^-$.  Likewise, near the point $i^+$, we have the $S^2$  denoted $\ci^+_+$. 

Now let us discuss the problem of scattering in Minkowski space, a central topic of this book. Whether in electrodynamics or gravity, classical or quantum, one starts by specifying initial data at $\ci^-$.  If there are stable massive particles, initial data at past timelike infinity $i^-$ are also needed, but we suppress that for now and return to it in section \ref{sec:massive}.  In the far past, fields and particles  disperse, and so the theory is weakly interacting.  Incoming particles or wavepackets evolve towards each other, interact in some very complicated way, and ultimately come out on $\ci^+$. The classical version of  the scattering problem is to find a classical map from the phase space defined on $\ci^-$ to the phase space defined on $\ci^+$.  The quantum version of the scattering problem is to find the $\cs$-matrix that evolves incoming states defined on $\ci^-$ to outgoing states defined on $\ci^+$.  
	
Clearly in this setting, what goes on near $i^0$ is important, because we have to know how to match on final data on $\ci^+$ given initial data on $\ci^-$. 
Specifically, we need to understand what the matching conditions are that relate the fields at $\ci^-_+$ to the fields at $\ci_-^+$. One might think that the fields should simply be taken to be equal, but this turns out to be completely wrong.  In fact, such a matching condition is not even Lorentz invariant!  We will soon see it is also not obeyed by the Li\'enard-Wiechert solution.  
	
To describe $\ci^-$, it is useful to  introduce advanced time: 
\be
v = t+r~.
\ee
To describe $\ci^+$, we introduce retarded time: 
\be \label{u}
u = t-r~.
\ee
$\ci^-$ is a three-dimensional surface, so four coordinates are not needed.  We could try to use the usual $(t, r ,\hat x)$, $\hat x$ being a unit vector labeling a point on the sphere, but this choice is awkward, because $t$ and $r$ are both infinite on $\ci$.  However, if we follow null rays backward (in time), $t+r$ is finite, and if we follow them forward, $t-r$ is finite.  So $\ci^+$ is naturally parametrized by $(u, \hat x)$, and $\ci^-$ is naturally parametrized by $(v, \hat x)$.

\subsection{Antipodal Matching Condition}

A peculiar property of the Li\'enard-Wiechert solution is that if we start at a point in the bulk,  take the limit first  to $\ci^+$ and then  to $i^0$, we get a different answer than if we take the limit  to $\ci^-$ and then to  $i^0$. 
In fact, the Li\'enard-Wiechert solution takes different values at fixed angles (or $\hat x$) on $\ci^+_-$ and $\ci^-_+$, 
but obeys an {\it antipodal} matching condition.  

To see this, first rewrite (\ref{Frt}) in retarded coordinates $u=t- r $:   
\be 
F_{rt} = F_{ru}=\frac{e^2}{4 \pi } \sum_{k=1}^n  
\frac{Q_k\gamma_k \left(r -(u+r) \hat{x}  \cdot \vec{\beta}_k\right)}{   \big| \gamma_k^2 \left( u+r- r \hat{x}  \cdot \vec{\beta}_k \right)^2 - (u+r)^2 + r^2 \big|^{3/2}}~.
\ee 
To reach $\ci^+$, hold $u$ fixed and take the limit $r\rightarrow \infty$:  
\be	 \label{Frtscri+}
F_{rt}  \big |_{\ci^+}=  \frac{e^2}{4 \pi r^2 }\sum_{k=1}^n \frac{Q_k}{  \gamma_k^2  (1- \hat{x}  \cdot \vec{\beta}_k)^2}~.
\ee
To reach $\ci^+_-$, we must further take $u \rightarrow -\infty$, but since the expression is $u$-independent, (\ref{Frtscri+}) is the final answer. 
Note, if we take  the velocity $\vec{\beta}_k = 0$, we recover the usual Coulomb field for a static charged particle.  	
	
The leading $\frac{1}{r^2}$ component of the electric field due to moving charges is not a constant: it depends on the angle of the sphere at infinity.  In standard electrodynamics texts,  such as Jackson \cite{Jackson:1998nia}, one often studies the  multipole expansion of static configurations.  The $\frac{1}{r^2}$ component is then constant and proportional to the total electric charge, while the  $\frac{1}{r^3}$ term comes from the static electric dipole moment with angular momentum $\ell = 1$.  In contrast, for  the case of a single charge moving at constant velocity, no electric dipole is in the picture, but there is a dipole moment in the $\frac{1}{r^2}$ term, in the sense that the $\ell = 1$ mode of the distribution over the sphere is nonzero.   This $\ell = 1$ mode dipole moment of the $\frac{1}{r^2}$ term  is not to be confused with what is usually called the the electric dipole moment in classical electrodynamics!  
		
To find the electric field at  $\mathcal I^-$, we work in advanced coordinates  $v = t+r$:
\be 
	 F_{rt} = F_{rv}=\frac{e^2}{4 \pi }  \sum_{k=1}^n \frac{Q_k\gamma_k \left(r -(v-r) \hat{x}  \cdot \vec{\beta}_k\right)}{   \big| \gamma_k^2 \left( v-r- r \hat{x}  \cdot \vec{\beta}_k \right)^2 - (v-r)^2 + r^2 \big|^{3/2}}~.
\ee
To reach $\ci^-$, we hold $v$ fixed and take the limit $r\rightarrow \infty$:  
\be \label{Frtscri-}	
F_{rt} \big |_{\ci^-}=  \frac{e^2}{4 \pi r^2  }\sum_{k=1}^n  \frac{Q_k}{  \gamma_k^2(1+\hat{x}  \cdot \vec{\beta}_k)^2}~.
\ee
Once again, since the expression is $v$-independent, this is also the value of $F_{rt}$ at $\ci^-_+$. Comparing equation~\eqref{Frtscri+} and equation~\eqref{Frtscri-}, we find that the values of the fields do not match.  Even for the case of one moving charge in Minkowski space, the radial component of the electric field is not continuous at spatial infinity.  Its value depends on how one approaches spatial infinity, and it is discontinuous there.  The field takes a definite value if one specifies a way of  approaching $i^0$.  Going down from $\ci^+$, or up from $\ci^-$,  one gets a definite answer, but they are not equal as long as $\vec{\beta}_k \neq0$.  
		
Since the solution is Lorentz covariant and can be obtained by boosting a static charge, the discontinuity is dictated by Lorentz invariance.  
Indeed, Lorentz transformations themselves are not smooth at  spatial infinity, because  the vector fields that generate them are singular at $i^0$.  A boost toward the north pole in the future is a boost away from the north pole in the past. The point $i^0$ is the dividing point of these two different behaviors. 

Although the fields $F_{rt} \big |_{\ci^-_+}$ and  $ F_{rt} \big |_{\ci^+_-}$ are not equal, they have an important antipodal relation:  taking $\hat x $ to $-\hat x$ takes $F_{rt} \big |_{\ci^+_-}$ to $F_{rt} \big |_{\ci^-_+}$.  Comparison of 
equation~\eqref{Frtscri+} and equation~\eqref{Frtscri-} reveals the matching condition,
\be \label{matchcond1}
 \lim _{r \rightarrow \infty} r^2 F_{ru}( \hat x) \big|_{\ci^+_-} = \lim _{r \rightarrow \infty} r^2 F_{rv}(- \hat x) \big |_{\ci^-_+}  ~,
\ee
where we have used the relation $F_{ru} = F_{rv} = F_{rt}$. That is, the leading term in the radial electric field for a collection of $n$ particles at any point on $\ci^+_-$ will be equal to the value of the field at the antipodal point on $\ci^-_+$.
		
So far  we have been considering a relatively simple nonstatic situation, in which nothing is interacting. However, it is not hard to check (see exercise 1) that equation (\ref{matchcond1}) holds in basically all reasonable circumstances. If we add electromagnetic waves on top of the $n$-particle solution (\ref{Frt}), the field strengths at $\ci^+$ and $\ci^-$  would no longer be constant.  Electromagnetic waves are radiative as opposed to Coulombic modes of the electromagnetic field. If we  want them to have finite energy,  they must die off in the infinite future and the infinite past.  For finite energy configurations, adding radiative modes will give  more complicated expressions (for example the field strength on $\ci^+$ will depend on $u$) but will not affect this basic behavior of the Coulombic, deep IR part of the electric field near spatial infinity, which is what we have been studying. It is also the case that letting the charges interact or scatter off one another preserves (\ref{matchcond1}). If one uses the retarded Li\'enard-Wiechert potential, one obtains equation (\ref{Frt}) in a neighborhood of $i^0$ with $\vec{\beta}_k$ the outgoing velocities, while the advanced Li\'enard-Wiechert gives (\ref{Frt}) in a neighborhood of $i^0$ with $\vec{\beta}_k$ the incoming velocities.  These solutions are different, but both obey antipodal matching. In section 2.9, we will also see that in quantum theory, Feynman diagrammatics implicitly assume equation (\ref{matchcond1}) in the standard weak field expansion. That might have been guessed from the fact that the matching condition is Lorentz and CPT invariant.

\begin{figure}[h] 
\begin{center}
\includegraphics{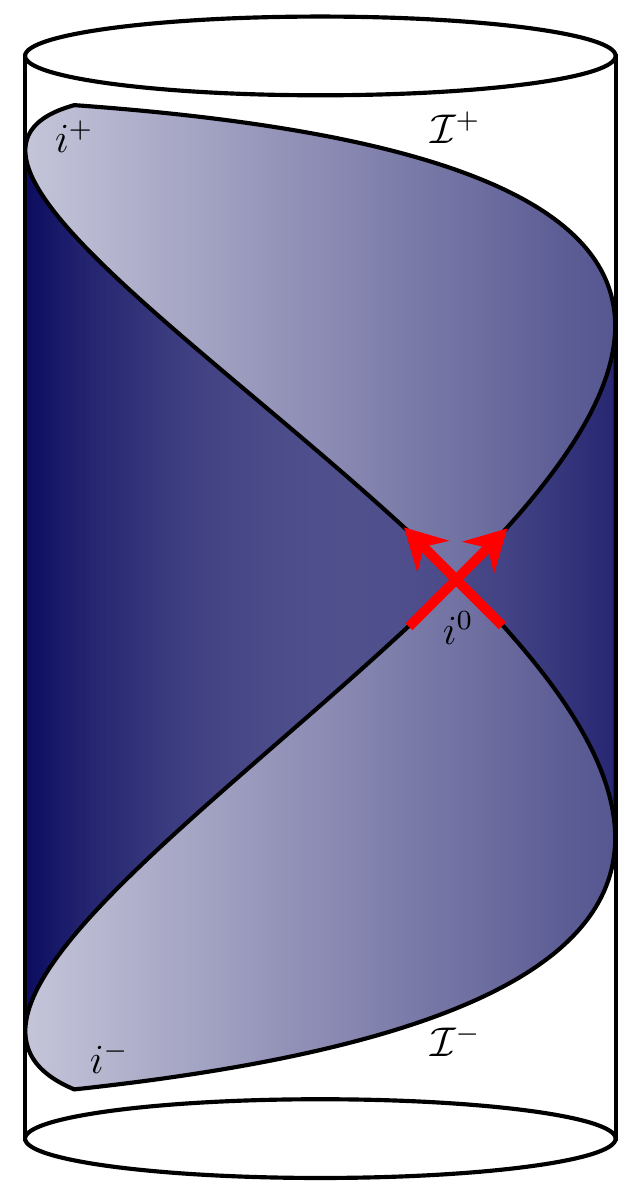}
\end{center}
\caption{ \small \bf The blue diamond represents Minkowski space conformally compactified onto the $S^3\times \mathbb{R}$ Einstein static universe. The red arrows indicate
how the generators of null infinity pass through spatial infinity.} 
\label{conformalcompactification}
\end{figure}

Another way of thinking about the matching condition, using the conformal invariance of  electromagnetism, is to map or ``conformally compactify'' Minkowski space to the cylinder using a conformal rescaling of the metric (figure~\ref{conformalcompactification}). 	 
This mapping is especially simple in electromagnetism, which is invariant under such rescalings, but it is also very useful in gravity, where it formed the basis of Penrose's analysis of null infinity \cite{Penrose:1962ij,Penrose:1964ge}.  The cross sections of the cylinder are $S^3$, and Minkowski space is represented by the blue diamond.   
Null infinity $\ci$ is  the lightcone of spatial infinity $i^0$.  The future lightcone wraps around the cylinder and meets itself at $i^+$, while the past lightcone meets itself at $i^-$. The Minkowski diamond is wrapped on the cylinder and touches itself at $i^0$.   The null generators of  $\ci$ start at past null infinity $\ci^-$ and pass through spatial infinity $i^0$  to future null infinity \ip.  The matching condition (\ref{matchcond1}) then states that the fields are continuous along the generators of $\ci$, including when they cross $i^0$, even though it is generically a singular point.  It is singular, because, for example, even just one charge in the interior needs an image charge at $i^0$, which will cause the electric field to diverge there.   One certainly cannot  demand that the fields be smooth at spatial infinity. In the presence of  multiple moving charges, the singularity can become very complicated and in general requires arbitrarily many parameters to describe. However, we are still able to consistently require, without violating Lorentz invariance (which is a conformal symmetry of the compactified geometry),  that the fields are continuous along the null generators passing through $i^0$ as prescribed by equation (\ref{matchcond1}). 
\subsection{Asymptotic Expansions }
\begin{figure}[h] 
\begin{center}
\includegraphics{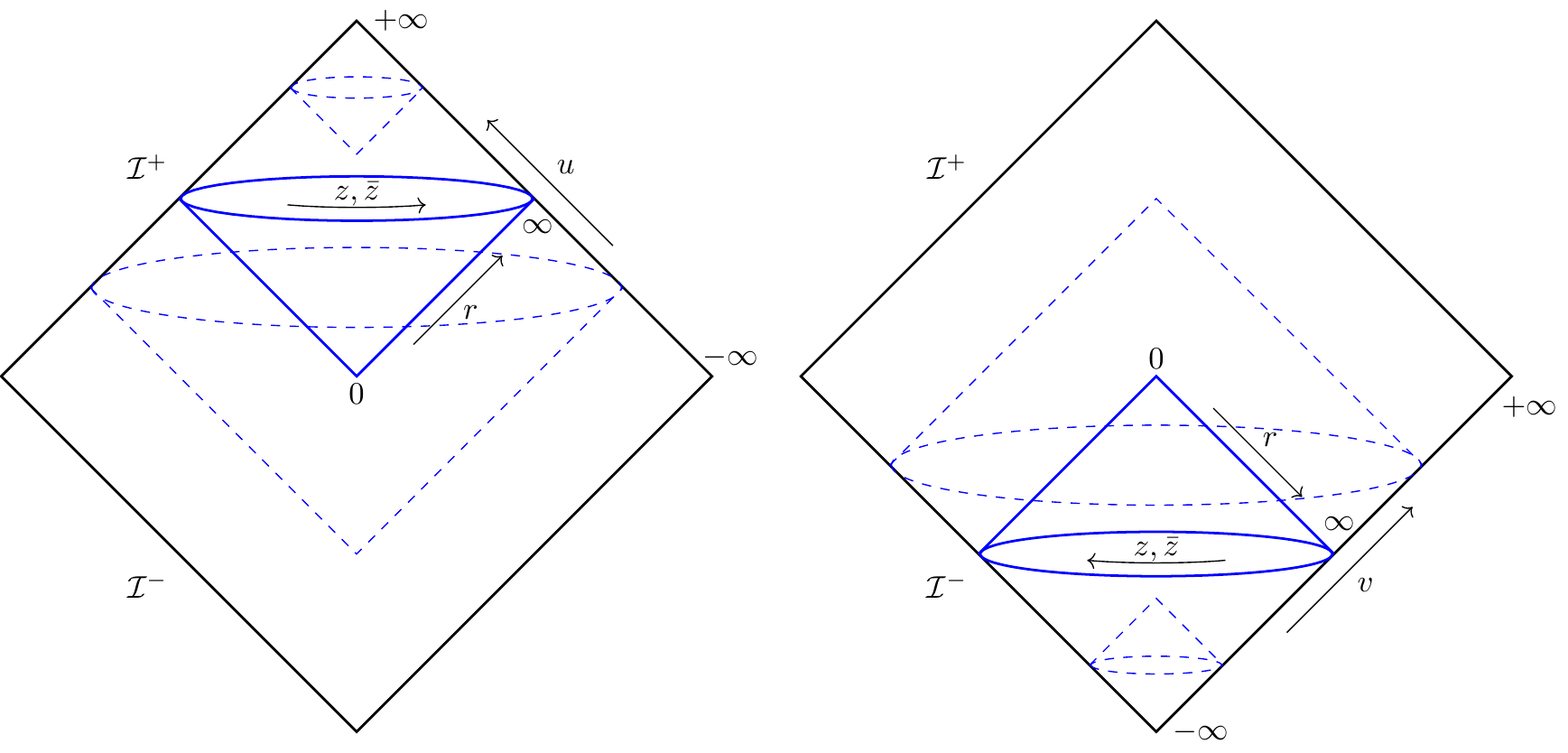}
\end{center}
\caption{ \small \bf In the left diagram, \ip\ is parametrized by retarded time $u$ and spherical coordinates $(z,\bz)$ in retarded Bondi coordinates, while in the right diagram, $\ci^-$ is parametrized by advanced time $v$ and spherical coordinates $(z,\bz)$ in advanced Bondi coordinates. The advanced and retarded $(z,\bz)$ coordinates are chosen so that they are related by an antipodal map on spheres of constant $(u,r)$. } \label{bondi}
\end{figure}

  Before going any further, it is convenient to introduce specific advanced and retarded coordinates, illustrated in figure \ref{bondi}, and to be more precise about our large-$r$ expansions around $\ci^\pm$.

In retarded coordinates $(r,u,z,\bz)$, the Minkowski line element  is
\be \label{retmetric}
ds^2=-du^2 -2du dr + 2r^2\gamma_{z\bz}dz d\bz~.
\ee These coordinates will be used in the neighborhood of $\ci^+$. Here, $u=t-r$ is the retarded time coordinate introduced in equation (\ref{u}),  $r$ is the radial coordinate, and $z$ is a complex coordinate on the unit sphere with metric\be
 \quad \gamma_{z \bz} = \frac{2}{(1+z \bz)^2}~.
\ee   If we keep $(u,z , \bz)$ fixed and take the limit $r \rightarrow \infty$, we move  out along a null line to \ip.  One can see that this is a null line, because along this path, $du = dz = d \bz =0$, which implies $ds^2 = 0$. The standard metric on Minkowski space,   
\be
ds^2=-dt^2+ (d\vec x)^2~,
\ee
is related to the metric in equation~\eqref{retmetric} by the coordinate transformations
 \be\label{ret}
(\vec x)^2=r^2~, \quad t=u+r~, \quad  x^1+i  x^2={2rz\over 1+z\bz}~, \quad { x}^3 =r \frac{1-z\bz}{1+z\bz} ~.  
\ee 
The inverse transformation is 
\be \label{cartret}r^2=(\vec x)^2~, \quad  u=t-r~, \quad z={x^1+ix^2 \over x^3+r}.\ee
Here, $z$ runs over the entire complex plane; $z = 0$ is the north pole,  $z = \infty$ is the south pole, $z \bz =1$ is the equator, and $z\to -{1/ \bz}$ is the antipodal map. This is a convenient coordinate system to work in near $\ci^+$ because, as we will see, everything falls off near $\ci^+$, so we can   expand fields in powers of $\frac{1}{r}$.  However, we cannot easily use these coordinates near $\ci^-$, because $u=-\infty$ there.  
To work in a neighborhood of $\ci^-$, we introduce advanced coordinates.  The advanced line element is
\be
ds^2=-dv^2+2dvdr +2r^2\gamma_{z\bz}dz d\bz~.
\ee
This metric can be obtained from the standard Cartesian metric on Minkowski space by means of the coordinate transformations
 \be\label{adv}
(\vec x)^2=r^2~, \quad t=v-r~, \quad  x^1+i  x^2=-{2rz\over 1+z\bz}~, \quad { x}^3 =-r \frac{1-z\bz}{1+z\bz} ~.  
\ee 
Crucial minus signs introduced into the last two terms of (\ref{adv}) 
imply that the $z$ in the advanced coordinates denotes the antipodal point on the sphere to the $z$ in the retarded coordinates (the sign reverses under  $z\to -1/\bz$).  If we take a light ray  which crosses Minkowski space, then the value of $z$  at which it starts in advanced coordinates will be the same as the value of $z$ at which it ends in retarded coordinates. Moreover, $z$ is constant along the null generators of \im \  as they pass through $i^0$ to \ip.   This perhaps odd-seeming  choice of notation  simplifies subsequent formulas. 

Now we wish to expand around $\ci^+$.  Given any field---say, the $z$-component of the vector potential---we can write an expansion for it as a sum 
\be
\begin{split}
A_z(u,r,z,\bz) =\sum_{n=0}^{\infty} \frac{A_z^{(n)}(u,z, \bz)}{r^n} ~,
\end{split}
\ee
where the coefficients depend only on the coordinates $(u, z, \bz)$ parametrizing $\ci^+$. In exercise 2, it is shown that $A_z^{(0)}$ is the local Cauchy data.  
The superscript $(n)$ will be used to  denote the order in the expansion  (the power of $1/r$) about $r = \infty$.
In general, we are going to be expanding many fields  about both \ip \ and \im . Using our new notation, we can finally rewrite the matching condition (\ref{matchcond1}) as   
\be \label{mcone}
F_{ru}^{(2)} (z, \bz) \Big|_{\ci^+_-} = F_{rv}^{(2)} (z, \bz) \Big|_{\ci^-_+}~, 
\ee
where  $F_{ru}^{(2)}$ is the $\frac{1}{r^2}$ term in the expansion of the $ru$ component of the field strength around $\ci^+$. Evaluating it at $\ci^+_-$ is equivalent to taking $u = -\infty$:
\be
F_{ru}^{(2)} (z, \bz) \Big|_{\ci^+_-}  = F_{ru}^{(2)} ( -\infty, z, \bz)  ~.
\ee
The simplicity of expression (\ref{mcone}) motivated our definition of the $z$ values on $\ci^+$ as antipodally related to those on $\ci^-$.  		
\subsection{An Infinity of Conserved Charges}

In this section we will show that the matching condition (\ref{mcone}) immediately implies, without further ado, the existence of an 
infinite number of conserved charges in all electromagnetic theories in Minkowski space. 

Consider any function $\ve$ on Minkowski space obeying the boundary condition
\be \label{parammatch}
\ve(z,\bz)|_{\ci^+_-}=  \ve(z,\bz)|_{\ci^-_+}~.
\ee
Note that  $\ve(z,\bz)$ is not smooth near spatial infinity, but instead is antipodally identified.   Now we define future and past charges:
\be \label{chargeconserve}
Q_\ve^+ = \frac{1}{e^2} \int_{\ci^+_-} \ve \ast F,\quad Q^-_\ve = \frac{1}{e^2} \int_{\ci^-_+}  \ve \ast F~.
\ee
It then  immediately follows from (\ref{mcone}) that any theory involving electromagnetism has an infinite number of conservation laws, one for every function $\ve$:
\be \label{ccv}Q_\ve^+ = Q^-_\ve.\ee
  For example, we could take  $\ve|_{\ci^-_+}=Y_{\ell m}$ to be a spherical harmonic, in which case we have one conservation law for every value of the angular momentum $(\ell, m)$.  One way of stating this conservation law is that all the  incoming multipole moments  of the electromagnetic field are equal to  the antipodal outgoing multipole moments. All of these moments are nonzero in a generic time-dependent situation. In this basis, the antipodal map acts as $(-)^\ell $. The $\ell=0$ mode is of course total charge conservation, which equates the net incoming to the net outgoing charge. Here we discover  an infinite family of higher harmonics  of this familiar global conservation law. 
  
    The conservation laws  look even more interesting and nontrivial when we use the Gauss law to write the surface integral expression for the electric field as a volume integral expression of  the incoming and outgoing fields.  At this point, we make the  simplifying assumption that all stable charged particles are massless, deferring the more intricate discussion of the (more realistic) massive case to section \ref{sec:massive}. We then  find
\begin{equation} \label{Q+Gauss}
\begin{split}
Q^+_\ve = \frac{1}{e^2} \int_{\ci^+} \mathrm{d} \ve \wedge \ast F  + \int_{\ci^+} \ve  \ast j +  \cancelto{0}{\frac{1}{e^2} \int_{\ci^+_+} \ve \ast F} ~,
\end{split}
\end{equation}
where we recall $j$ is the charged matter current.
This last term arises because $\ci^+$ has two boundaries: we began in (\ref{chargeconserve}) with an integral over $\ci^+_-$ and used integration by parts to get a volume integral 
over $\ci^+$ plus a surface integral over $\ci^+_+$.  However, we can set this last surface integral to zero under our assumption that there are no massive charged particles, which implies that the electric field will vanish at $\ci^+_+$.
Likewise, for $\ci^-$,
\begin{equation} \label{Q-Gauss}
\begin{split}
Q^-_\ve = \frac{1}{e^2} \int_{\ci^-} \mathrm{d} \ve \wedge \ast F  + \int_{\ci^-} \ve  \ast j +  \cancelto{0}{\frac{1}{e^2} \int_{\ci^-_-} \ve \ast F} ~.
\end{split}
\end{equation}
		
Charge conservation, equation~\eqref{ccv}, is true for any $\ve$ obeying equation~\eqref{parammatch}.  However, we restrict to the special case in which   
\be
\p_u\ve=\p_v\ve=0~.
\ee
Our formulas are all true for the more general case, but consideration of those cases turns out not to yield anything new.  A particular case of equation~\eqref{parammatch} is $\ve =  \text{constant}$, in which case the first term on the RHS  of the $Q^\pm_\ve$ equation vanishes. 
Equation~\eqref{ccv}  reduces to, for $m$ ($n$) incoming (outgoing) particles
with charges $Q^{\text{in}}_k~(Q^{\text{out}}_k)$
\be \label{usu}
\sum_{ k = 1 }^m Q^{\text{in}}_k = \sum_{k = 1}^{n} Q^{\text{out}}_k~.  
\ee
This is just the  familiar statement that  the sum of all the incoming charges must be equal to the sum of all the outgoing charges. 
		
Now suppose $\ve$ is not constant.  Then we get an infinite number of conserved charges that are on exactly the same footing as the usual conservation law (\ref{usu}) but with both terms in (\ref{Q-Gauss}) and (\ref{Q+Gauss}) contributing. The second term is still a sum over electric charges but is now weighted by an arbitrary function of the angle.  The first is an additional mysterious term linear in the electromagnetic field $F$. We will call this a \textit{soft photon term}, because, as we show later, when promoted to an operator it creates and annihilates zero-energy photons; for now it is just a classical integral of $F$.  Unlike the constant $\ve$ case, whose associated charge involves only incoming matter fields, these more general conservation laws mix the charged matter and the electromagnetic field.  We have an exact statement that relates various integrals, weighted by $\ve$, of the incoming initial data to the outgoing final data. The conservation law says that, for any $\ve$, the number obtained from the incoming integral exactly equals the number obtained from the outgoing integral. 
		
Thus far, we have written the charges in form notation in \eqref{Q+Gauss} and \eqref{Q-Gauss}. For many computations, explicit coordinate representations are useful. In retarded coordinates \eqref{retmetric},
\begin{equation}
\begin{split}
Q^+_\ve= \frac{1}{e^2} \int_{\ci^+_-} \ve \ast F = \frac{1}{e^2} \int_{\ci^+_-} d^2 z \gamma_{z\bz} \ve F_{ru}^{(2)}  ~.  
\end{split}
\end{equation}
Note there is an $r^2$ in the Hodge dual, which cancels the $1/r^2$ that comes with the $F_{ru}^{(2)}$ term, resulting in an integral which is finite for $r\to \infty$. This is equal, by virtue of the boundary condition \eqref{parammatch}, to
\be
	Q^-_\ve=\frac{1}{e^2} \int_{\ci^-_+} d^2 z \gamma_{z\bz} \ve F_{rv}^{(2)}~.
\ee
To integrate these formulas by parts,  we need to use the constraint equations on the null surfaces $\ci^\pm$. The constraint equation near $\ci^+$ has an expansion in powers of $1/r$, with  leading term 
\begin{equation}
\begin{split}\label{abelianconstrainteq}
\partial_u F_{ru}^{(2)} + D^z F^{(0)}_{uz} +  D^\bz F^{(0)}_{u\bz}   + e^2   j_u^{(2)} = 0~.
\end{split}
\end{equation}
 $D_z$ is the covariant derivative with respect to the unit round $S^2$ metric $\gamma_{z \bz}$, and we also define $D^z=\gamma^{z \bz}D_\bz$. Choosing $\p_u \ve |_{\ci^+} = 0$, integrating the boundary expression for $Q^+_\ve$ by parts, and using the constraint equation gives two terms,
\be \label{qex}
Q^+_\ve  = \underbrace{ - \frac{1}{e^2} \int_{\ci^+} du d^2 z \left( \p_z \ve F_{u\bz}^{(0)} + \p_\bz \ve F_{uz}^{(0)} \right) }_{Q_S^+}+
\underbrace{ \int_{\ci^+} du d^2 z \ve \gamma_{z\bz} j^{(2)}_u  }_{Q_H^+}~, 
\ee
where we have defined a ``soft'' charge $Q_S^+$, which is linear in the electromagnetic field,  and a ``hard'' charge $Q_H^+$, which is typically linear in the charge current but quadratic or higher in the charged matter fields.  We have already seen the equation in form notation in (\ref{Q+Gauss}).  
The term ``soft'' refers to objects with zero energy, while ``hard'' denotes energetic excitations. $Q_H^+$ is referred to as the hard charge, because $j$ is the matter current for energy-carrying matter fields. If $\ve$ is constant, then $Q_S^+$ vanishes, and $Q_\ve^+$ collapses to the total charge flux through future null infinity. If $\ve$ is not constant, then the hard term is still there, but it has an unfamiliar form, because the charges that pierce null infinity are weighted by an arbitrary function that depends on the angle at which they exit the spacetime. However, we still have an overall  conserved quantity, provided  we compensate by including the soft term. The soft charge involves a term of the form
\be\label{fre}
\int_{-\infty}^\infty du~  F_{u z}^{(0)}\equiv N_z~, ~
\ee
convoluted with $\p_\bz\ve$ and integrated over the sphere.
Equation (\ref{fre}) is the $\omega\to 0$ limit of 
\be
\label{photonop}
\int_{-\infty}^\infty du~ F_{u z}^{(0)} e^{i \omega u}~,
\ee
which is  a Fourier component with nonzero energy and frequency $\omega$ of the electromagnetic field. Moreover, it is a component of the electromagnetic field that is transverse to \ip\ , as the $z$ index is  transverse to a light ray passing through \ip .  Promoted to a quantum operator, \eqref{photonop}  creates and annihilates outgoing photons of energy $\omega$. However, the expression we are interested in appearing in (\ref{qex}) has no $\omega$; rather it is an $\omega \rightarrow 0$ limit.  This term creates and annihilates soft particles with zero  energy, which is why we call it the soft term. Equation (\ref{qex}) indicates that these soft photons have polarization $\p_\bz\ve$. 

The soft photon mode $N_z$ is simply related to  gauge transformations at $\ci^+_\pm$. To see this, consider the curl
\be \label{noBfield}
\begin{split}
\p_\bz N_z - \p_z N_\bz &= \int_{-\infty}^\infty du ~   \left[\p_\bz F_{uz}^{(0)}- \p_z F_{u\bz}^{(0)}\right] ~\\
&=- \int_{-\infty}^\infty du~ \p_u F^{(0)}_{z \bz}
= -F^{(0)}_{z \bz} \Big|_{\ci^+_-}^{\ci^+_+}  ~,
\end{split}
\ee
where we have used the Bianchi identity in the second line.  At this point we assume that there are no asymptotic states with magnetic charges (i.e.,~no magnetic monopoles), and no long-range magnetic fields, so that $F_{z \bz}\big|_{\ci^+_\pm} = 0$.  Then the curl~\eqref{noBfield} vanishes, and we can define a real scalar $N$ by
\be\label{ndef}
N_z   \equiv e^2 \p_z N~.
\ee
Imposing the gauge condition $A_u^{(0)} = 0$, it follows that 
\begin{equation}
\begin{split}
	e^2 \p_z N = \int_{-\infty}^\infty du F_{uz}^{(0)}  = A_z^{(0)} \big|_{\ci^+_+} - A_z^{(0)} \big|_{\ci^+_-}  ~. 
\end{split}
\end{equation} 
The soft photon mode \eqref{fre} is the difference between the $z$-component of the gauge field at the future of $\ci^+$ and the past of $\ci^+$. If we want to have finite energy, the gauge field better be pure gauge at both the beginning and the end of \ip , and the relative shift better also be pure gauge. Indeed, we have just seen that the shift is a gauge transformation with parameter  $e^2 N$.  The charge has a simple expression in terms of the soft photon mode:
\be \label{qexz}
Q^+_\ve  =2\int  d^2 z N\p_z \p_\bz\ve + \int_{\ci^+} du d^2 z \ve \gamma_{z\bz} j^{(2)}_u  ~. 
\ee

\subsection{Canonical Electrodynamics at $\ci$}
 \label{sec:Symplectic}
 We have found an infinity of conserved charges in abelian gauge theories. One expects an associated infinity of symmetries.  In a canonical Hamiltonian formalism, these symmetries are easily  identified as the Dirac bracket action of the charges on the phase space. (To go the other way, from symmetries to charges, one uses Noether's theorem.) In this  subsection we develop this formalism, and in the next we determine the symmetry. 
	
In a Hamiltonian formulation, one has a phase space $\Gamma$ with coordinates $x^I = \{ q^i , p_j \}$, where $i,j = 1, \dotsc, N/2$, and $N$ is the (even) dimension of the phase space. For example, the theory of a particle in three dimensions has a six-dimensional phase space with $i,j = 1, 2, 3$. In field theory,  the index $I$ becomes continuous, and the phase space is infinite-dimensional.  The symplectic two-form $\Omega$,  \be\begin{split}
\Omega = \frac{1}{2} \Omega_{IJ} dx^I \wedge dx^J   ~,
\end{split}
\ee
is needed to define the Hamiltonian dynamics. 
$ \Omega_{IJ} $ is antisymmetric and must be invertible.  Once the symplectic form is known, quantum commutators are constructed as 
\be
\label{formalcommutator}
\left[ A , B \right] =  i \Omega^{IJ} \p_I A \p_J B ~.
\ee
These are related to classical Poisson brackets by a factor of $i$. Much  in these lectures is relevant at the purely classical level, but we will always use quantum commutators  to avoid  writing two sets of formulas that differ only by a factor of $i$. 	\subsubsection{\it Symplectic Form}

The classical phase space in electromagnetism can be defined as the allowed initial data on any Cauchy surface $\Sigma$.  For a Cauchy surface $\Sigma$ in Minkowski space, the symplectic form for free electrodynamics can be written as (see exercise 2 and \cite{Crnkovic:1986ex,Ashtekar:1987tt,Lee:1990nz,Wald:1999wa})
\begin{equation}\label{smp}
\begin{split}
\Omega_\Sigma = - \frac{1}{e^2} \int_\Sigma \delta ( \ast F ) \wedge \delta A ~ . 
\end{split}
\end{equation} Here, $\delta A$ is an on-shell variation of the gauge field $A$ and also a 1-form on the infinite-dimensional phase space $\Gamma$ on which $\wedge$ denotes the wedge product.
$\delta ( \ast F ) \wedge \delta A $ is a closed 3-form in spacetime, and its integral $\Omega_\Sigma$ is a 2-form on $\Gamma$. Equation \eqref{smp} gives the same value for any pair of 3-surfaces $\Sigma$ with the same boundary. In particular, $\Sigma$ can be chosen either to be a spacelike slice or pushed out to a null surface at $\ci^\pm$.   For $\Omega_\Sigma$ to be invertible, it must have no zero modes. This requires that $\delta A$ be restricted by gauge conditions, constraints, and boundary conditions. Specifying all of these can be quite subtle (and may entail boundary corrections to \eqref{smp}).    Writing out the indices,   
\be
\Omega_\Sigma =   \frac{1}{e^2} \int_{\Sigma} d\Sigma^\mu \delta F_{\mu\nu} \wedge \delta A^\nu  ~,
\ee
where $d\Sigma^\mu$ is the induced measure times the unit normal vector to $\Sigma$. 

Now we take $\Sigma = \ci^+$.  This choice is very nice, because at $\ci^+$ everything spreads out, becomes very weakly interacting, and essentially reduces to free field theory.  The symplectic form becomes, very simply,  
\begin{equation}
\begin{split}
\Omega_{\ci^+} &=  \frac{1}{e^2} \int du d^2 z \left( \delta F^{(0)}_{uz} \wedge \delta A^{(0)}_\bz + \delta F^{(0)}_{u\bz} \wedge \delta A^{(0)}_z \right)~.  
\end{split}
\end{equation}
This expression typically remains exact even in interacting theories (as long as the IR theory is free), because interactions are weak near \ip. 
	
Now we come to a tricky point, which was incorrectly treated in some of the literature until recently.  We need to be very careful about what is happening at the boundaries  of $\ci^+$ and $\ci^-$. We begin by separating out the constant ($u$-independent) part of  $A^{(0)}_z$  \cite{He:2014cra,He:2015zea},
\begin{equation}\label{defAhat}
\begin{split}
A^{(0)}_z (u, z, \bz) = {\hat A}_z(u, z ,\bz) +   \p_z \phi (z, \bz)  ~,
\end{split}
\end{equation} 
where
\begin{equation}
\begin{split}
\p_z \phi &\equiv \frac{1}{2}  \left[ A_z^{(0)} \big|_{\ci^+_+} + A_z^{(0)}  \big|_{\ci^+_-}  \right]  ~. 
\end{split}
\end{equation}
 The constant piece $\p_z \phi $  is pure gauge, because we are demanding that the magnetic field vanish at the boundaries $\ci^+_\pm$.\footnote{The symplectic form in the presence of magnetic charges and fields has so far not been  constructed.} This boundary condition is crucial in the following: without it, the charges below would fail by a factor of 2 to generate the proper symmetries. Now we can substitute \eqref{defAhat} into the expression for the symplectic form and find 
\begin{equation}
\begin{split}
\Omega_{\ci^+} =  \frac{2}{e^2} \int du d^2 z ~\p_u \delta {\hat A}_z \wedge \delta {\hat A}_\bz -2  \int d^2 z ~\p_z \delta \phi  \wedge \p_\bz \delta N    ~, 
\end{split}
\end{equation}
where $N$ was defined in \eqref{ndef}.
We see that the radiative components of $A^{(0)}_z$ are paired with each other in the symplectic form,\footnote{Since $\hat A_z$ has no constant term, the sum of its boundary values vanishes, and $N$ therefore does not appear in the first term.} and the soft photon mode is paired with the field $\phi$, which is the sum of the boundary values.  

Note that the very existence of a Hamiltonian formulation requires that we keep the mode $\phi$, which is absent in a standard Fourier decomposition. Without it, there is no symplectic partner for the soft photon zero-mode $N$. 
	
	
\subsubsection{\it Commutators}

The next thing to discuss is the commutators.  As we see from equation~\eqref{formalcommutator}, to derive the commutators one must invert the symplectic form.  
Since it is a sum of terms, one of which only involves $\hat A$ and the other which only involves the boundary fields $N$ and $\phi$, we can invert them separately.  Consider the first one.  It implies the commutator
\be
\left[\p_u {\hat A}_z \left( u , z , \bz \right) , {\hat A}_\bw \left( u' , w , \bw \right) \right] = - \frac{ie^2}{2}  \delta  \left( u - u' \right) \delta^2 \left( z - w \right) ~.
\ee
Integrating with respect to $u$, we find 
\be\label{dto}
\left[ {\hat A}_z \left( u , z , \bz \right) , {\hat A}_\bw \left( u' , w , \bw \right) \right] = - \frac{ie^2}{4} \Theta \left( u - u' \right) \delta^2 \left( z - w \right) ~, 
\ee
where
\begin{equation}
\begin{split}
\Theta(u) = \frac{1}{\pi i } \int \frac{d\omega}{\omega} e^{i \omega u}  ~, 
\end{split}
\end{equation}
with $\Theta(u<0) = -1$ and $\Theta(u>0) = 1$, and a potential integration constant in \eqref{dto} is fixed by antisymmetry. 
Equation \eqref{dto} is a standard lightcone commutator. We can also compute the commutator between the boundary fields $\phi$ and $N$, which does not depend on $u$: 
\be\label{dtp1}
\left[ \phi (z,\bz) , N ( w , \bw  ) \right]  = - \frac{i}{4\pi  } \log | z - w |^2 + f(z,\bz) +g(w,\bw)~.
\ee
At various points, one must be careful about the integration functions $ f$ and $g$, but we are going to ignore them for the present as   the computations below require only 
\be\label{dtp2}
	\left[ \p_z \phi(z,\bz) , \p_\bw N ( w , \bw  ) \right]  = - \frac{i}{4\pi  }\p_z \p_\bw \log | z - w |^2  =   \frac{i}{2 } \delta^2 (z-w)~,
\ee
which follows from
\be
\p_\bz \frac{1}{z-w} = 2 \pi \delta^2 (z-w)~.
\ee
\subsection{Large Gauge Symmetry}\label{lgs}Now we can easily compute the commutator action  of $Q^+_\ve$. 
$Q^+_\ve$ has a term involving the matter field and a linear term involving $\p_z N$ and $\p_\bz N$ (i.e.,~the soft photon terms).  The soft terms do not commute with $A^{(0)}_z$.   We find 
\begin{equation}
\begin{split}
 \left[  Q_\ve^+ , A_z^{(0)} (u,z,\bz) \right]  = i \p_z \ve(z,\bz)  ~, 
\end{split}
\end{equation}
while a similar calculation on 	
$\ci^-$ yields\begin{equation}
\begin{split}
 \left[  Q_\ve^- , A_z^{(0)} (v,z,\bz) \right]  = i \p_z \ve(z,\bz)  ~. 
\end{split}
\end{equation}
Some other useful commutators are	
\begin{equation}\label{dsat}
 \left[  Q_\ve^+ , N  (z,\bz) \right]  =0 ~,~~
 \left[  Q_\ve^+ , \hat A_z (u,z,\bz) \right]  =0 ~,~~
 \left[  Q_\ve^+ ,  \phi  (z,\bz)\right]  = i \ve(z,\bz)   ~.
\end{equation}

This leads to a remarkable conclusion in the free Maxwell theory: {\it the infinite number of symmetries  generated by  the conserved charges $Q^+_\ve$ in a canonical formalism are  just gauge transformations with parameter $\ve$!}
Of course, we have long known that for  constant $\ve$,  the total charge $Q^+_{\ve=1}$ generates constant gauge transformations on all the fields.  Here, we see that even when $\ve$ is an arbitrary  function, $Q^+_\ve$ transforms $A_z$ by a  non-trivial, ``large'' gauge transformation that does not die off at infinity.  Rather, the gauge parameter goes to an angle-dependent, but $u$-independent, function at $\ci^+$.  At  $\ci^-$,  $Q^-_\ve$ generates a transformation for which the gauge parameter  approaches  the antipodally transformed function. These are pretty weird gauge transformations!  The fact that the condition $A_z=0$ is not invariant under  these symmetries implies that they are spontaneously broken and the vacuum is infinitely degenerate, but more on that story in Section 2.11.

Now suppose our theory includes charged matter.  The electromagnetic field commutators  above are unchanged as long as the coupling approaches a constant in the infrared. We showed that the commutator of the soft charge with the gauge field itself is a large gauge transformation.  With matter added, we must check that $Q^+_\ve$ properly generates the gauge transformations on the matter fields. This is the role of   the  hard term.  The charged matter current $j_u$ is the conserved Noether current associated with the global U(1) matter symmetry. By the Noether construction, $j_u$  canonically generates this symmetry on \ip:\footnote{Here as elsewhere in this section, we make  the  assumption that the charged fields are massless, so that the charge flux all goes through \ip.  It remains true in the massive case that $Q^+_\ve$ generates gauge transformations, but there are very interesting new twists; see section \ref{sec:massive}.}
\begin{equation}
\left[  j^{(2)}_u (u',w,\bw) , \Phi_k (u,z,\bz)  \right] = - Q_k  \Phi_k  (u,z,\bz)\g^{z\bz}\delta^2(z-w)\delta(u-u')  ~,
\end{equation} 
where $\Phi_k$ is a matter field of charge $Q_k$. This commutator  implies
\begin{equation}
\begin{split}
\left[ \int_{\ci^+}  \ve \ast j  , \Phi_k (u,z,\bz)  \right] = - Q_k  \ve(z,\bz) \Phi_k  (u,z,\bz)  = i \delta_\ve \Phi_k (u,z,\bz)~
\end{split}
\end{equation} 
and \be
\big[ Q^+_\ve , \Phi_k (u,z,\bz)  \big]   = i \delta_\ve \Phi_k (u,z,\bz)~.
\ee
Hence, we conclude that the sum of the hard and soft term properly generates  local angle-dependent gauge transformations on $\ci^+$.  

Historically, the story ran the other way around. First it was discovered \cite{He:2014cra}, through an asymptotic analysis, that QED had an infinite number of nontrivially acting gauge symmetries generated by the antipodally matched functions $\ve$. Then the generators were constructed via the Noether method and identified as conserved charges. However, the asymptotic symmetry analysis --- which we will see some of in Section 2.10 --- requires assumptions about asymptotic falloffs near $\ci$.  Such assumptions are unnecessary in the treatment here starting from the conserved charges. As we have seen, their existence follows directly from nineteenth-century electromagnetism.

To summarize, so far we have found that there are an infinite number of conserved charges in electromagnetism.  We have further showed that these charges canonically generate an infinite number of large gauge symmetries that act nontrivially on the physical phase space.  
	
\subsection{Ward Identity}
In this subsection we derive Ward identities, which relate quantum scattering amplitudes.  These identities express the dynamical consequences of the fact that the conserved charges commute with the Hamiltonian, or equivalently, with the $\cs$-matrix,  since $\cs\sim \exp (i H T)$ for  $T \rightarrow \infty$. 
 \subsubsection{\it Symmetries of the $\cs$-matrix}
  Quantum scattering amplitudes can be written in the form 
\be
\outst \cs \inst~.
\ee
Charge conservation is then \begin{equation}
\begin{split}\label{qedwardid}
\outst\left( Q^+_\ve  \cs - \cs Q_\ve^-  \right)  \inst= 0  ~. 
\end{split}
\end{equation}By the matching condition, $Q^+_\ve$ is equal to $Q^-_\ve$, but we use $Q^+_\ve$ when acting on out-states and $Q^-_\ve$ when acting on in-states, because they are expressed in variables  appropriate for acting on the out- and  in-states, respectively. 
Exponentiating the charge to generate a finite symmetry,  \eqref{qedwardid} becomes the statement:  given an in-state $X$ that evolves to an out-state $Y$,  a large-gauge-transformed in-state $X$ evolves to a large-gauge-transformed  out-state $Y$. 	

The action of $Q_\ve^- $ on the in-state is 
\begin{equation}
\begin{split}\label{Qinaction}
Q_\ve^- \inst= - 2 \int d^2 z  \p_\bz\ve \p_z N^-(z,\bz) \inst+ \sum_{k=1}^m Q^{\text{in}}_k \ve(z^{\text{in}}_k,\bz^{\text{in}}_k ) \inst~, 
\end{split}
\end{equation}
where $N^-(z,\bz) $ denotes the incoming soft photon field on $\ci^-$.  We have assumed that the in-state can be described by $m$ hard particles that are coming in at points on the asymptotic sphere denoted by $z^{\text{in}}_k$.  The first term is the action of the soft charge, and the second term is the action of the hard charge.  Similarly, the action of $Q_\ve^+ $ on the out-state takes the form 
\begin{equation}
\begin{split}\label{Qoutaction}
\outst Q_\ve^+ = 2 \int d^2 z  \p_z \p_\bz \ve   \outst   N (z,\bz) + \sum_{k=1}^n Q^{\text{out}}_k \ve(z^{\text{out}}_k,\bz^{\text{out}}_k )    \outst  ~. 
\end{split}
\end{equation}
Finally, we can write the Ward identity as
\begin{equation}
\begin{split}\label{qedsoftth}
& 2 \int d^2 z \p_z\p_\bz \ve  \outst \left( N(z,\bz)   \cs - \cs N^-(z,\bz) \right)  \inst\\
&\qquad \qquad \qquad\qquad\qquad = \left[ \sum_{k=1}^m Q^{\text{in}}_k \ve(z^{\text{in}}_k,\bz^{\text{in}}_k ) 
- \sum_{k=1}^n Q^{\text{out}}_k \ve(z^{\text{out}}_k,\bz^{\text{out}}_k )  \right] \outst\cs \inst ~. 
\end{split}
\end{equation}
	
Equation \eqref{qedsoftth} is an infinite number of Ward identities, one for every function $\ve$ on the sphere.  They relate any $\cs$-matrix element between any pair of incoming and outgoing states multiplied by the soft factor in square brackets,  to the same $\cs$-matrix element with the insertion of certain soft photon modes. This is a very general relation. It would be incredibly surprising if,  after 90 years of studying QED, we discovered an infinite number of new relationships between scattering processes.  We have not done that. Rather, we have  rediscovered some relations that have been known for a long time but were derived using very different methods, namely, Feynman diagrams and mode expansions, and go under the name of ``soft theorems.''  What we will do next is  show that these identities, which we derived from charge  conservation, are precisely the same thing as the well-known soft theorems in abelian gauge theories. 

Before proceeding with this demonstration, as a little teaser for section \ref{nonabelian}, it is interesting to consider the  special case in which $\ve$ is taken to be 
\be
\ve(w,\bw) = \frac{1}{z-w} ~.
\ee
Using the fact that
\be
\p_\bz \frac{1}{z-w} = 2\pi \delta^2 ( z - w ) ~
\ee
to easily perform the integral on the LHS of equation~\eqref{qedsoftth},   we can rewrite the Ward identity as
 \begin{equation}
\begin{split}\label{curr}
4\pi  \outst \left( \p_z N   \cs - \cs \p_zN^- \right)\inst
 =   \left[ \sum_{k=1}^m \frac{ Q^\text{in}_k }{ z - z_k^{\text{in}} } -\sum_{k=1}^n \frac{ Q^\text{out}_k }{ z - z_k^{\text{out} } }   \right] \outst {\cal S} \inst~.
\end{split}
\end{equation}
 In two-dimensional CFT with a current algebra, when one inserts a $U(1)$ Kac-Moody current into a correlation function of operators of charge  $Q_k$, one gets exactly this kind of Ward identity.  More generally, it was recently shown \cite{Nande:2017dba} that the entire soft factor of the $\mathcal{S}$-matrix is computed by a current algebra with level determined by the cusp anomalous dimension. We will see in section \ref{nonabelian}, in the more general nonabelian setting,  that this is in fact not a coincidence and that the large gauge symmetry is equivalent to a $U(1)$ Kac-Moody symmetry on the sphere at null infinity. 

 \subsubsection{\it Mode Expansions}
To show that the Ward identity \eqref{qedsoftth} is in fact a soft theorem, we have to transcribe notation. Until this section,  the presentation has been  rather  different than what is found in most quantum field theory textbooks.  We  have characterized particles by the points at which they came in at null infinity, used advanced and retarded coordinates, and derived conservation laws from antipodal matching conditions.
	

In traditional quantum field theory,  one works in a basis of plane waves. To compare notations we must rewrite \eqref{qedsoftth} in terms of a plane wave basis, using a conventional mode expansion for $A_z$.  It is not manifestly obvious that the commutation relations \eqref{dto} and \eqref{dtp1}  are equivalent to the conventional ones. One way to see that they must nevertheless be the same is to note that  both follow from the slice-independent covariant symplectic form \eqref{smp}. We evaluated it by pushing the slice up to $\ci^+$, as opposed to the conventional use of the $t =0$ slice and a plane wave basis. Such bases employ the standard Cartesian coordinates for Minkowski space  
 \begin{equation}
\begin{split}
ds^2 = - dt^2 + d \vec{x} \cdot d \vec{x} ~.
\end{split}
\end{equation} 
Near \ip, $A_\nu$ has the on-shell outgoing plane wave  mode expansion 
\begin{equation}
\begin{split}\label{modexp}
A_\nu (x) = e \sum_{\alpha=\pm} \int \frac{d^3q}{(2\pi)^3} \frac{1}{2\omega} \left[ \ve_\nu^{*\alpha} (\vec{q}) a_\alpha^{\text{out}} (\vec{q}) e^{i q \cdot x }  
+ \ve_\nu^{\alpha} (\vec{q}) a_\alpha^{\text{out}} (\vec{q})^\dagger e^{- i q \cdot x }  \right]~,
\end{split}
\end{equation}
where $q^2=0$, the two polarization vectors satisfy a normalization condition $\ve^\nu_\alpha \ve^*_\beta{}_\nu  = \delta_{\alpha \beta }$, and 
\begin{equation}
\begin{split}
\left[ a_\alpha^{\text{out}} (\vec{q}\hspace{2pt}) , a_\beta^{\text{out}} (\vec{q}\hspace{2pt}')^\dagger \right] = \delta_{\alpha\beta} (2\pi)^3   (2 \omega_q) \delta^3 \left( \vec{q} - \vec{q}\hspace{2pt}' \right) ~ .
\end{split}
\end{equation}
This is the standard textbook formula for commutators of modes of the free  electromagnetic field.  

The next step is to rewrite asymptotic quantities on \ip\ in terms of the familiar creation and annihilation operators appearing in \eqref{modexp}. Near \ip\ it is convenient to use retarded coordinates, in which  the metric takes the form 
\be
ds^2 = -du^2 - 2 dudr + 2 r^2 \gamma_{z \bz }dz d \bz~. 
\ee
The transformation to retarded from Cartesian coordinates was given in \eqref{cartret}. 
A null vector $q^\mu$, satisfying $q^\mu q_\mu = 0$,  is labelled by a point on the sphere, up to its overall magnitude.  Hence, there is a natural map from null vectors $q^\mu$ to points $(z,\bz)$ on the sphere toward which the null vector is directed. 
We can write this as 
\begin{equation}\label{nullmomentum}
\begin{split}
q^\mu = \frac{\omega}{1+z\bz} \left( 1+ z\bz, z+\bz,-i(z-\bz),1-z\bz\right) = (\o, q^1, q^2, q^3)~.
\end{split}
\end{equation}
	As an example of this, let us suppose that $z$ is taken to be the north pole at $z = 0$. Then we find $q^\mu = \omega (1, 0, 0, 1)$ (i.e.,~a null vector pointing along  the $x^3$-axis).  We may further choose the polarization vectors orthogonal to $q^\mu$ as 
\begin{equation}
\begin{split}
\ve^{+\mu}(\vec{q}\hspace{2pt}) &= \frac{1}{\sqrt{2} } \left( \bz , 1 , - i , - \bz \right) ~, ~~ 	\ve^{-\mu}(\vec{q}\hspace{2pt}) = \frac{1}{\sqrt{2} } \left(z , 1 , i , -z \right)~, \quad q_\mu \ve^{\pm \mu}(\vec{q}\hspace{2pt}) = 0~,\quad \ve^\mu_\alpha \ve^*_\beta{}_\mu  = \delta_{\alpha \beta } ~.
	\end{split}
\end{equation}
Now  let us  consider the \ip\  field $A_z^{(0)} (u,z,\bz)$.  By definition, 
\begin{equation}\label{sx1}
\begin{split}
A_z^{(0)} (u,z,\bz) = \lim_{r \to \infty} A_z(u,r,z,\bz) ~. 
\end{split}
\end{equation}
Using the above formulas, we can take all the $q^\mu$s in the mode expansion and rewrite them in terms of points on the asymptotic sphere.   Since $A_z(u,r,z,\bz)$ has an expansion in terms of creation and annihilation operators, $A_z^{(0)} (u,z,\bz)$ must also have such an expansion. In particular, we expect  $A_z^{(0)} (u,z,\bz)$ to create and annihilate photons that land at the point $(z, \bz)$, since it is an operator that is localized at that point on the asymptotic sphere. Moreover, rotating about the point $(z,\bz)$, $A_z^{(0)}$  gets one phase, while  $A_\bz^{(0)}$  gets the opposite phase. Hence we  expect $A_z^{(0)}$  to  create one photon helicity and annihilate the other, while $A_\bz$ does the opposite.  These expectations are realized when one evaluates \eqref{sx1} in a large-$r$ saddle-point approximation (see exercise 3):
\begin{equation}
\begin{split}\label{modexp1}
A^{(0)} _z (u,z,\bz) = - \frac{i}{8\pi^2}  \frac{\sqrt{2} e }{1+ z \bz }  \int_0^\infty d\omega
 \left[  a_+^{\text{out}} ( \omega {\hat x} ) e^{-i \omega u }   - a_-^{\text{out}} ( \omega {\hat x} )^\dagger e^{i \omega u }   \right]  ~,~~~~~\hat x  = \hat x (z, \bz)~.
\end{split}
\end{equation}
The ``out'' creation and annihilation operators involve the three-momentum $\o\hat x$; $\hat x$ is a unit vector that points to $(z,\bz)$ on the sphere, with the relationship between $\hat x_1$, $\hat x_2$, and $\hat x_3$ and the point $(z,\bz)$ as in equation~ \eqref{nullmomentum}. So $A^{(0)} _z (u,z,\bz)$ annihilates a  positive helicity photon that is headed to the  point $\hat x (z, \bz)$ and creates a negative helicity photon that is headed to the same point. It can create and annihilate photons of all different frequencies. 
Equation \eqref{modexp1} is the basic relationship between the out fields in the $1/r$ expansion at $\ci^+$  and the standard  outgoing creation and annihilation operators in the plane wave expansion usually employed in quantum field theory.

The Ward identity involves $\p_z N$, so we need to determine its mode expansion.  To be precise  about the zero-momentum limit, define
\be
	\p_z N = \frac{1}{2e^2} \lim_{\omega \to 0^+ } \int_{-\infty}^\infty  du \left( e^{i\omega u} + e^{- i \omega u } \right) F_{uz}^{(0)}~.
\ee
This definition ensures that  $\p_z \p_\bz N$ is Hermitian. Had our definition only involved either $e^{i\omega u}$ or $e^{-i\omega u}$, this would not be the case.  
Using \eqref{modexp1}, we find 
\be \label{Nzmode}
\p_z N   = -  \frac{1}{8\pi e}  \frac{\sqrt{2}  }{1+ z \bz } \lim_{\omega \to 0^+ } \left[  \omega a_+^{\text{out}} ( \omega {\hat x} )  + \omega a_-^{\text{out}} ( \omega {\hat x} )^\dagger \right]~.
 \ee
One might wonder why we care about this operator, since we are taking the limit $\omega \rightarrow 0$ while  multiplying by $\omega$.  One might think it is just identically zero.  However, if it were zero, the Ward identity would not make sense, because it would imply that the RHS of equation~\eqref{curr} would vanish. The  reconciliation is that the $\cs$-matrix elements with insertions of these operators have  poles that cancel the explicit factor of $\omega$ appearing in equation~\eqref{Nzmode}.
	 
There is a similar formula for $\p_z N^-$:
\begin{equation}
\begin{split}\label{mdef}
\p_z N^-  =  - \frac{1}{8\pi e}  \frac{\sqrt{2} }{1+ z \bz } \lim_{\omega \to 0^+ } \left[  \omega a_+^{\text{in}} ( \omega {\hat x} )  + \omega a_-^{\text{in}} ( \omega {\hat x} )^\dagger \right] ~ .
\end{split}
\end{equation}
The Ward identity in the form \eqref{curr} can be reexpressed as
 \begin{equation}
\begin{split}\label{weinbergsoftth}
\lim_{\omega \to 0 } \Big[\omega \outst \left(\vphantom{a^{\dagger}}a_+^{\text{out}} ( \omega {\hat x} )    {\cal S} \right.&\left.- {\cal S}  a_-^{\text{in}} ( \omega {\hat x} )^\dagger\right) \inst\Big] \\&
 =   {\sqrt{2}} e  (1+z\bz) \left[  \sum_{k=1}^n \frac{ Q^\text{out}_k }{ z - z_k^{\text{out} } }  - \sum_{k=1}^m \frac{ Q^\text{in}_k }{ z - z_k^{\text{in}} } \right] \outst {\cal S} \inst~.
\end{split}
\end{equation}
 This is beginning to look very much like the standard soft photon theorem.  There is a soft pole in the matrix element on the LHS, which we have rendered finite via explicit multiplication by $\omega$.  There are also collinear poles on the RHS for $z\to z_k$.

 \subsubsection{\it Soft Theorems} 
 
 Now in principle to show that \eqref{weinbergsoftth} is exactly equivalent to the standard soft theorem, I could take the RHS of the equation~and rewrite the $z$s and $z_k$s in terms of the momenta of the particles.  However, it is easier and equivalent  to proceed in reverse. That is, we begin with the textbook formula for the soft photon theorem and show that it is equivalent to \eqref{weinbergsoftth}  by going from momentum space to points on a sphere. For the reader's benefit, the next section reviews  the standard textbook derivation of the soft photon theorem. The soft theorem states 
\begin{equation}
\begin{split}\label{spthstandform}
\lim_{\omega \to 0 } \left[  \omega\outst a_+^{\text{out}} ( \vec{q}\hspace{2pt})     \cs  \inst  \right] &=
 e\lim_{\omega \to 0}\left[  \sum_{k=1}^m \frac{ \omega Q_k^{\text{out}} p^{\text{out}}_k  \cdot \ve^+  }{ p^{\text{out}}_k  \cdot q   } 
 - \sum_{k=1}^n \frac{ \omega Q_k^{\text{in}}   p^{\text{in}}_k  \cdot \ve^+  }{ p^{\text{in}}_k  \cdot q }  \right]   \outst \cs \inst ~\\
&=  - \lim_{\omega \to 0 } \left[  \omega\outst \cs  a_{-}^{\text{in}\dagger} ( \vec{q}\hspace{2pt}) \inst\right]~,
\end{split}
\end{equation}
where $q^\mu = ( \omega, \vec q \hspace{2pt})$ is the momentum of the soft photon, and  we are taking the in-state and out-state in plane wave bases: 
\be 
\inst=|p_1^{\text{in}},\dotsc,p_n^{\text{in}}\rangle~,~~~~~~\outst = \langle p_1^{\text{out}},\dotsc,p_m^{\text{out}}|~.
\ee
The equality of the matrix elements involving in and out soft photons is a consequence of $\cal CPT$ invariance. 
Equation \eqref{spthstandform} is not quite the standard form, as we have multiplied both sides of the equation by a factor $\omega$ in order to have a finite limit.  In the limit where $\omega \rightarrow 0$ (the soft limit), the ratios appearing on the RHS are finite and depend only on the direction of $\vec q$, but not its magnitude. When one writes the  formula without this factor of $\omega$,  there is a pole as $\omega \to 0$, which is called the ``soft pole'' or ``Weinberg pole.''  
 	
To make the comparison between  equation~\eqref{weinbergsoftth} and equation~\eqref{spthstandform}, we note that in addition to the $\omega \to 0$ pole, there is a collinear pole, since $q^\mu$ and $p^\mu_k$ are both  null vectors. When $p_k^\mu$ is proportional to $q^\mu$, $p_k \cdot q = 0$, and $\frac{1}{p_k \cdot q}$ diverges.  In \eqref{weinbergsoftth}, collinearity implies  $z_k=z$, which again gives  a pole.  Hence the pole structure of equation~\eqref{weinbergsoftth} and equation~\eqref{spthstandform} agrees.

Now we want to  turn equation~\eqref{spthstandform} into equation~\eqref{weinbergsoftth} by writing the momentum of each hard particle in terms of the energy $E_k$ and a point on the sphere $(z_k, \bz_k)$. For example, 
\begin{equation}\label{wc}
\begin{split}
(p^{\text{in}}_k)^\mu &=  E^{\text{in}}_k \left(1 ,  \frac{ z^{\text{in}}_k + {\bar z}^{\text{in}}_k}{ 1 + z^{\text{in}}_k {\bar z}^{\text{in}}_k },
 \frac{- i \left(  z^{\text{in}}_k - {\bar z}^{\text{in}}_k \right) }{ 1 + z^{\text{in}}_k {\bar z}^{\text{in}}_k }, \frac{1- z^{\text{in}}_k {\bar z}^{\text{in}}_k}{1+ z^{\text{in}}_k {\bar z}^{\text{in}}_k} \right) ~.
\end{split}
\end{equation}
Similar formulas apply to the soft photon momentum $q^\mu$ (see \eqref{nullmomentum}) and outgoing hard particle momenta $p^{\text{out}}_k$.
In other words, knowing that the massless charged particle is going to land at the point $(z_k, \bz_k)$ determines what its momentum must be. This enables us 
 to  replace all momenta in equation \eqref{spthstandform} with their expressions in terms of energies and points on a sphere. Summing the contribution from an outgoing positive helicity soft photon and an incoming negative helicity soft photon then exactly reproduces equation~\eqref{weinbergsoftth}. The straightforward but somewhat intricate algebra is left to exercise 5. The standard soft photon formula reduces to the Ward identity following from invariance under large gauge symmetries, or equivalently, the matrix element of the charge conservation law.  This establishes the central result connecting the  soft photon theorem to the large gauge symmetry of electromagnetism. 
 
 The story might have been told backward. Starting with the soft photon theorem, we could reverse engineer and deduce the fact that abelian gauge theories have an infinite number of conservation laws associated to antipodally identified large gauge transformations. They are mathematically equivalent statements. 


 \subsection{Feynman Diagrammatics}\label{diagrams}
 
In this section, we  review the standard field theory derivation of the leading photon and graviton soft theorems in the form given by Weinberg \cite{Weinberg:1965nx,Weinberg:1995mt}.    
	
\subsubsection{\it Soft Photons }
The soft photon theorem states that any $\cs$-matrix element with an additional soft ($q^\mu\to0$) photon is equal to the original matrix element multiplied by the soft factor plus corrections of order $q^0$:
\begin{equation}\label{mft}
\begin{split} 
\outst a_+^{\text{out}} ( \vec{q}\hspace{2pt})     \cs  \inst  \  = 
   e\left [\sum_{k=1}^m \frac{  Q_k^{\text{out}} p^{\text{out}}_k  \cdot \ve^+  }{ p^{\text{out}}_k  \cdot q   } 
 - \sum_{k=1}^n \frac{  Q_k^{\text{in}}   p^{\text{in}}_k  \cdot \ve^+  }{ p^{\text{in}}_k  \cdot q }  \right]   \outst \cs \inst    + \co (q^0) ~.
\end{split}
\end{equation}
The leading order term in the soft expansion is a pole. 

\begin{figure}[h] 
\begin{center}
\includegraphics[width=6.2in]{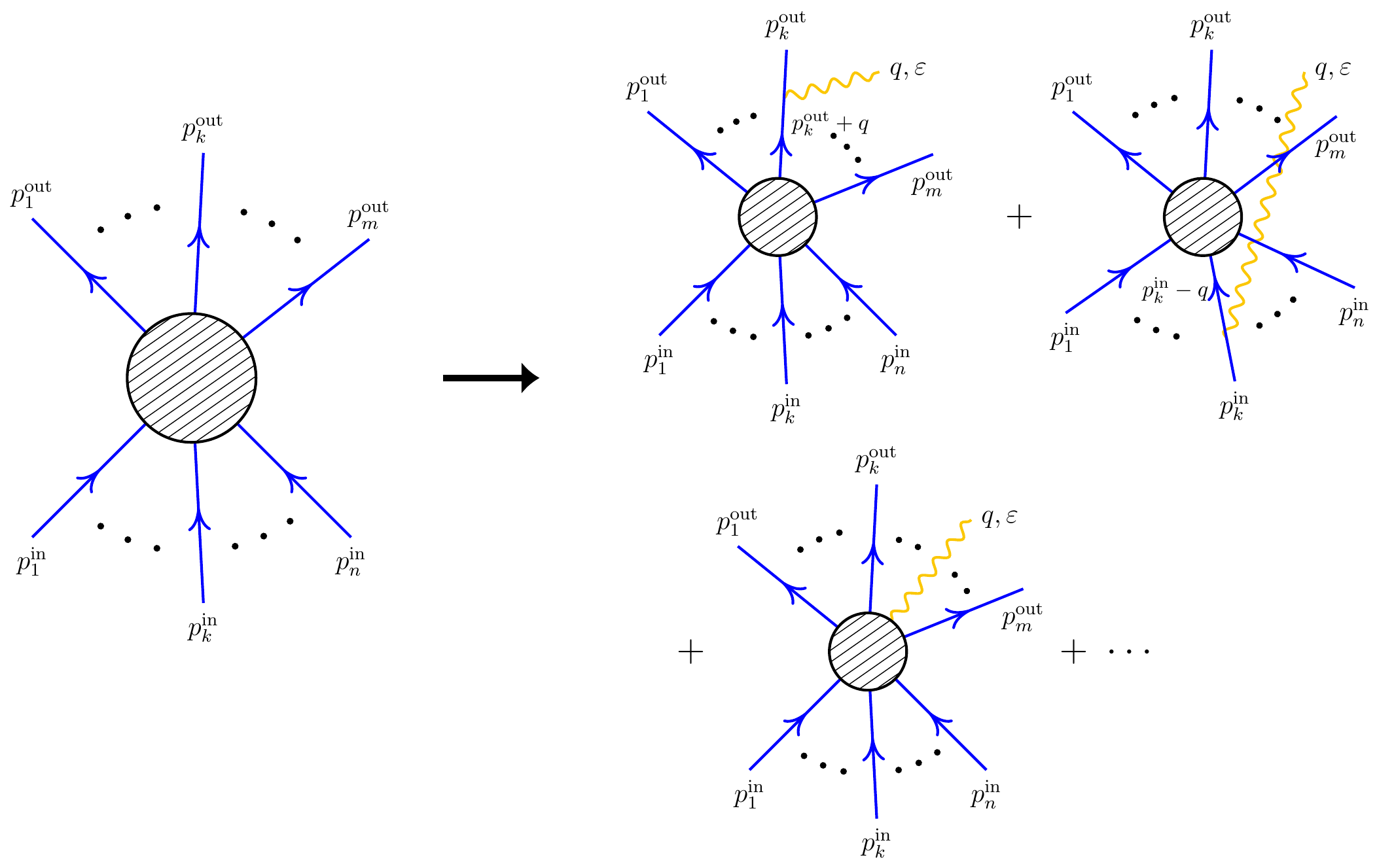}
\end{center}
\caption{ \small \bf On the left is a Feynman diagram representing $n\to m$ scattering.  On the right the effect of adding  an outgoing  soft photon (or graviton) with momentum $q$ and polarization $\ve$ is illustrated.   In the upper diagrams  the soft particle attaches to an external propagator, while in the lower one it  attaches to an internal propagator.}
\label{feynmandiagram}
\end{figure}

To derive this formula, let us take any scattering process with $n$ incoming and  $m$ outgoing particles and then consider adding to it one outgoing photon, denoted by a wavy line in figure \ref{feynmandiagram}, with momentum $q$. (The derivation for an incoming photon is similar.) 	 
 In the soft limit, we can write the amplitude as a sum of two types of terms, ones in which the soft photon attaches to an external line and others in which the soft photon attaches to an internal line.  The soft photon can attach to any one of the $n+m$ external lines, so we must include a sum over all such terms.  
The full amplitude has a Laurent expansion in $q$ with  an infinite number of terms whose detailed form depends on what theory we are talking about.  For the pole we need not specify what theory we are studying except that it has a photon. That is one of the beauties of this formula. 
	 
The LSZ rule for computing scattering amplitudes starts out by computing the time-ordered Green's functions using the Feynman $i \epsilon$ prescription and then amputating the external legs. The  Feynman diagrams have factors for vertices and propagators.  What happens when we attach the extra photon to an external leg is, since external legs are amputated, we need  only add  a vertex and propagator for the particle to whose external leg the photon is added. The difference between the diagram with and without the attached external soft photon is just the vertex and propagator.
 
Now I have to say a little bit about the interaction vertex.  Let us take the interaction  to be
\be
\mathcal{L}_{\rm int}=-A^\mu j_\mu~.
\ee
For a scalar field of charge $Q$, the charge current is 
\be
j_\mu= i Q( \phi \p_\mu\phi^*- \phi^*\p_\mu \phi)~.
\ee For a plane wave, this is just
\be
j_\mu \sim  2 Q p_\mu~,
\ee
where we have used the normalization for single-particle states
\be
\langle p | p'\rangle = 2 \omega_p (2 \pi)^3 \delta^3 (p-p')~.
\ee
This is the electromagnetic current associated to a scalar field of  charge $Q$,  meaning  that under gauge transformations, it acquires a phase $e^{i Q\ve}$.  
Moreover, we  take the polarization of the photon to obey $\ve^\mu q_\mu = 0$, where $q^\mu$ is the momentum of the photon.   
The propagator for a scalar particle of mass $m$ gives a factor 
\be\label{weinbergpole}
\frac{-i}{(p+q)^2 +m^2} = \frac{-i}{p^2 + 2 p \cdot q + q^2 + m^2} =  \frac{-i}{2 p \cdot q }~,
\ee
where we have used the fact that in a scattering amplitude, all the external lines must be on-shell, so $q^2 = 0$ and $p^2 = -m^2$.
The vertex factor is, up to $\mathcal{O}(q)$ corrections, 
\be\label{qedvertexfactor}
i e  \ve^\mu 2 Q p_\mu~,
\ee
where $\ve^\mu$ comes from $A^\mu$ and $2 Q p_\mu$ comes from $j_\mu$.  
The total contribution is
\be
i  e\ve^\mu   (2 Q  p_\mu)  \frac{ - i}{  (p+q)^2+m^2 } \to  \frac{ e Q \ve\cdot p }{  q\cdot p}~.
\ee
There is one such term for every outgoing particle, while for the incoming particles there is an additional minus sign. Altogether these give 
\be \label{totalsoftfactor}
\sum_{k=1}^m \frac{eQ_k^{\text{out}} p^{\text{out}}_k  \cdot \ve  }{ p^{\text{out}}_k  \cdot q   } 
 - \sum_{k=1}^n \frac{eQ_k^{\text{in}}   p^{\text{in}}_k  \cdot \ve   }{ p^{\text{in}}_k  \cdot q } ~.
\ee
At the end of the day, if we just consider the diagrams in which the photon attaches to an external leg, we simply multiply the $\cs$-matrix element by the factor in equation~\eqref{totalsoftfactor}, sometimes called the ``soft factor'', which you may recognize from equation~\eqref{spthstandform}. 

We  have not yet considered the terms coming from the photon attaching to an internal leg. The key point is  that the internal propagators are never on-shell (i.e.,~they never have $p^2 = -m^2$). In the propagator, one then never has the cancellation between $p^2$ and $m^2$, so if we take $q^\mu \to 0$, the difference between $p^2$ and $m^2$ will dominate, and we will not get a pole.  These types of diagrams are most certainly nonzero, but they do not contribute to the pole, so we can forget about them in the soft limit. This is an extremely simple derivation. Up to some signs, one finds the same thing for a soft incoming photon.  

Now we note  an important feature of this formula. The condition $\ve^\mu q_\mu = 0$ defines the polarization vector only up to shifts of $\ve^\mu$ by $q^\mu$, because $q^2 = 0$.  The physical amplitude with the soft photon should be invariant if we shift $\ve^\mu$ by any multiple of $q^\mu$.  Now it is interesting to see what happens to the soft factor  \eqref{totalsoftfactor}.  If we shift $\ve^\mu$ by $q^\mu$, it shifts by 
\be
\sum_{k=1}^m eQ_k^{\text{out} }- \sum_{k=1}^n eQ_k^{\text{in} }=0~.
\ee
In other words, global charge conservation guarantees that this soft factor is gauge invariant.  This observation was in fact the basis of Low's derivation of the soft formula in 1958 \cite{Low:1958sn}.

We have only worked out the soft theorem for the case of a scalar.  For a fermion or some other kind of charged particle, it is a little more complicated, but it works out to the same expression  \cite{Weinberg:1995mt}. One way of seeing that this must be so is that \eqref{totalsoftfactor} is the only formula with the right dimensions that is invariant under shifts $\ve^\mu\to \ve^\mu+q^\mu$.

\subsubsection{\it Soft Gravitons}

So far  we have largely deferred any discussion of gravity. However, having  just derived the soft formulas for quantum electrodynamics, the generalization to gravity is short and simple, so we do it now.  We do not need to redraw diagrams, because we can just imagine that the wavy lines are  gravitons of momentum $q^\mu$ rather than photons. We do need to specify the interaction
\begin{equation}
\begin{split}
\mathcal{L}_{\rm int} &=  \sqrt{8 \pi G} h^{\mu\nu} T_{\mu\nu} ~, 
\end{split}
\end{equation}
which couples  the graviton field $h^{\mu\nu}$ in the expansion  $g^{\mu\nu}=\eta^{\mu\nu}-\sqrt{32 \pi G} h^{\mu\nu}$ to matter.  This field is normalized so that there are no factors of $G$ in the kinetic term.  
The graviton has polarization tensor $\ve_{\mu\nu}$ satisfying
\be
\ve_{\mu\nu} q^\mu = 0~, \quad \ve^{\mu\nu} \eta_{\mu\nu} = 0~.
\ee
For a scalar field,
\begin{equation}
\begin{split}
T_{\mu\nu} =    \p_\mu \phi \p_\nu \phi - \frac{1}{2}   \eta_{\mu\nu}   \p^\rho \phi \p_\rho \phi  ~.
\end{split}
\end{equation}
Now, we apply the same argument, which is that we multiply the $\cs$-matrix element by a factor of the propagator and the vertex.  The propagator is the same, 
\be
\frac{-i }{2 p \cdot q}~.
\ee
Whether one has  added an external photon or graviton, these poles, which become the holomorphic poles on the sphere at null infinity, are ubiquitous in  soft limits. For a scalar field the interaction is
\be
T_{\mu\nu}  \sim 2 p_\mu p_\nu~,
\ee
where the term in $T_{\mu\nu}$ proportional to $\eta_{\mu\nu}$ does not contribute due to $\ve_{\mu\nu}$ being traceless.  
In total, the product of the vertex and propagator gives a factor
\begin{equation}
\begin{split}
i\sqrt{32 \pi G}\ve^{\mu\nu}   p_\mu p_\nu  \frac{- i}{  ( p  + q  )^2 +m^2 }   \to \sqrt{8 \pi G}\frac{ \ve^{\mu\nu} p_\mu p_\nu }{ p \cdot q } ~.
\end{split}
\end{equation}
This is just the result for one external particle.  
More generally, if we have $m$ outgoing and $n$ incoming particles, we have a soft factor 
\be
 \sqrt{8 \pi G}\sum_{k = 1}^m   \frac{ \ve^{\mu\nu} p^{\text{out}}_{k\mu} p^{\text{out}}_{k\nu}}{ p^{\text{out}}_k \cdot q } 
-  \sqrt{8 \pi G}\sum_{k = 1}^n  \frac{ \ve^{\mu\nu} p^{\text{in}}_{k\mu} p^{\text{in}}_{k\nu} }{ p^{\text{in}}_k \cdot q } ~.
\ee
We get a very similar-looking formula to the gauge theory case with  $p_k$ replacing $Q_k$. Just as in the gauge theory case, we cannot get poles by coupling the soft graviton to internal lines. 

The soft factor must  be invariant under the shift $\ve^{\mu\nu} \to \ve^{\mu\nu} +  \Lambda^\mu (q) q^\nu$, the analog of gauge invariance in the gravity case.  Performing such a shift, we can pull the $\Lambda^\mu$ out from under the sum and  find that the soft factor shifts by
\be
\Lambda^\mu \left[ \sum_{k = 1}^m  p^{\text{out}}_{k\mu} -  \sum_{k = 1}^n  p^{\text{in}}_{k\mu} \right] = 0~,
\ee
due to  global energy-momentum conservation. We saw above that global charge conservation was required for the consistency and gauge invariance of the soft photon theorem.  Here we see that global energy-momentum conservation is required for the consistency and gauge invariance of the soft graviton theorem. So far we have not used global angular momentum conservation, which we will see in section \ref{sec:superrotation} is responsible for a new soft theorem and set of asymptotic symmetries called ``superrotations''.
\subsection{Asymptotic Symmetries}
Our discussion so far began with   the derivation of  conserved charges.  These were then shown, via a canonical formalism, to generate asymptotic symmetries.  We might have instead begun (as was the case historically) with a direct analysis of the asymptotic symmetries. Such analyses are more of an art than a science, in part because of ambiguities  in the choice of asymptotic falloffs and gauge conditions, which are often only a posteriori justified.  However, despite the lack of rigor, this approach has been extraordinarily fruitful and the source of many physical insights. Once one has an idea of how it should go, more rigorous methods can then be developed.  The fact that it is an art and not a science is not a derogatory statement.  In fact, maybe art is better than science! 
	
An asymptotic symmetry group (ASG) is defined as
 \be\label{asgdef}
 {\rm ASG}= {\text{allowed gauge symmetries} \over \text{trivial gauge symmetries}}~.
\ee
 One studies the theory 
on some spacetime and imposes boundary conditions that describe how the field components behave near the boundary. The boundary conditions should be weak enough so that all physically reasonable solutions are allowed, but strong enough so that relevant  charges are finite and well defined.  The $allowed~ gauge~ symmetries$ are any ones that respect the boundary conditions.  The $trivial~gauge ~symmetries$ are the ones that act trivially on the physical data of the theory.  

It can be quite subtle to decide what  kind of behavior is  allowed at infinity or which data is physical. Indeed, this problem remains unresolved in full generality for asymptotically flat spacetimes in GR.   In many cases,  part of the definition of the theory is specifying the  boundary conditions. Moreover, we shall encounter examples in sections \ref{mnetic} and \ref{stry} for which the ASG is not even  a subset of the local symmetry group, indicating that  the definition \eqref{asgdef} is too narrow. Nevertheless,  despite the vagaries of this procedure, it has proven extraordinarily useful.

Oddly, asymptotic symmetry analyses were first carried out in the context of gravity rather than gauge theory.  An early application was the seminal work  of Bondi, van der Burg, Metzner and Sachs \cite{Sachs:1962wk,Bondi:1962px} (BMS), who wanted to find the subgroup of diffeomorphisms of asymptotically flat spacetimes that act nontrivially on the asymptotic data. It seemed intuitively obvious that if one has an isolated system that is flat near  infinity, it  should be acted on non-trivially by the  Poincar\'e group.   For an asymptotically flat black hole, a boost is a diffeomorphism that should certainly be allowed but must be  nontrivial, because it changes the energy.  Similarly, translations move the black hole to a different place. BMS expected the asymptotic symmetry analysis would reproduce the isometries of flat spacetime itself, namely, the Poincar\'e group.  To everyone's great surprise and consternation, what they got instead was an infinite-dimensional group, now called the BMS group. This contains the finite-dimensional Poincar\'e group as a subgroup but has an additional infinity of generators known as ``supertranslations.''  This result implies that GR does \textit{not} reduce to special relativity for  weak fields and long distances, as na\"{i}vely expected. As we will see below, it is now realized that the BMS group is not the whole story: it is both too big and too small, but much more on this in section 5. 

	Another famous asymptotic symmetry analysis appeared in the work of Brown and Henneaux in the late 1980s on $\mathrm{AdS}_3$ \cite{Brown:1986nw}.  They  found a result  that no one was expecting:  the asymptotic symmetry algebra is two copies of the Virasoro algebra with a computable central charge --- a harbinger of AdS/CFT. 	This beautifully demonstrated the power of the method. It is not just a way of confirming what we already know.  
	
	What would we expect for the asymptotic symmetry group of electrodynamics? This question has been asked only in the past few years \cite{Strominger:2013lka,He:2014cra}.  In Minkowski  space electromagnetism, we are interested in boundary conditions at $\ci^+$ or $\ci^-$. At these null boundaries, one does not have the same freedom in choosing the boundary conditions as one does, for example, for the timelike boundary of a box. Rather one must derive the asymptotic behavior from the field equations. If we consider a sphere at large $r$, its surface area grows like $r^2$, so for the energy flux at any moment to be finite, $T_{uu} \sim \co (1/r^2)$.  Note that we are not talking about the integrated energy flux, whose finiteness requires that fields fall off in a certain way at late and early times on $\ci^\pm$.  In electromagnetism,  
\be
T_{uu} \sim F_{uz}F_{u \bz} \frac{\gamma^{z \bz}}{r^2} + \dotsc ~,
\ee
where the factor of $\frac{1}{r^2}$ comes from inverting the metric on the sphere of radius $r$.  This suggests that $F_{uz}\sim \co(1)$.  However, $F_{ru}$ is the long-range electric field, so it should go as $\co(1/r^2)$ to have finite charge configurations.  Similarly, $F_{rz} \sim \co (1/r^2)$.  To summarize: 
\be \label{ffo}
F_{uz} \sim \co(1),~~F_{ur},F_{zr}  \sim \co\left({1 \over r^2}\right)~.
\ee
This suggests we choose the boundary falloff conditions for the gauge fields to be\be\label{afo}
A_z \sim \co(1)~,\quad A_r\sim \co \left({1 \over r^2}\right)~, \quad  A_u \sim \co\left({1 \over r}\right)~.
\ee
If we try to impose stronger boundary conditions for $A_\mu$, we would teleologically disallow some physically reasonable  initial data configurations.	

Now we ask: what kind of gauge transformations do these falloffs allow?   Generic gauge transformations take the form 
\be
\delta A_z = \p_z \ve ~, \quad \delta A_u = \p_u \ve~,\quad \delta A_r = \p_r \ve~.  
\ee
These boundary conditions are consistent with 
\be\label{gt}
\ve  = \ve (z, \bz) + \co \left(\frac{1}{r}\right)~.
\ee
Note, the $\co(1)$ piece of $\ve$ cannot depend on $u$ without violating the boundary conditions \eqref{afo}. The transformations \eqref{gt} are the exact electrodynamic analog of the BMS transformations found much earlier for gravity. The additional necessity of an antipodal matching condition for transformations on $\ci^+$ and $\ci^-$ may  be deduced from the need for a well-defined Lorentz invariant scattering problem. 

There are several ways to see that the transformations \eqref{gt} are nontrivial as well as allowed. One is to show that their Ward identities have nontrivial physical content and are equivalent to the soft photon theorem. Another is to note that they impart distinct phases to charged particles at different positions on \ip. This leads to an Aharonov-Bohm type interference pattern, which could be observed for electrons stationed near \ip. Details of how this ``electromagnetic memory effect'' can be observed are discussed in the literature \cite{Bieri:2011zb, Susskind:2015hpa, Pasterski:2015zua}. We will discuss the similar gravitational analog in  detail in section \ref{sec:memory}.

Equation \eqref{gt} together with antipodal matching  are of course exactly the large gauge symmetries discussed in section \ref{lgs}. This very simple asymptotic analysis quickly leads to the result that we previously arrived at  in a lengthier  manner starting from conserved charges.  In fact, this is how the symmetries were first discovered \cite{Strominger:2013lka}.    However, if one thinks about it a little bit harder, one might ask, for example, why we did not allow $A_u$s that are $\co(r^0)$ but  pure gauge, since they would maintain the falloff behavior of the field strengths. Indeed, interesting new developments \cite{Campiglia:2016hvg,Conde:2016rom,Conde:2016csj} suggest that pure gauge modes for which \eqref{ffo} is preserved but \eqref{afo} is not provide an effective way to understand the subleading soft theorem \cite{Low:1954kd,Low:1958sn} and associated symmetries \cite{Lysov:2014csa}. 
So while some basics of asymptotic symmetry analyses are firmly understood, much remains to be learned.
\subsection{Spontaneous Symmetry Breaking, Vacuum Degeneracy, and Goldstone Bosons}\label{ssb}

As we shall now elucidate, the  preceding results imply that abelian gauge theories have an infinite vacuum degeneracy and undergo spontaneous symmetry breaking.  However, because of the unusual angular dependence of the symmetry parameters, the nature of the symmetry breaking and relation of the degenerate vacua are different than what we are used to.\footnote{Large gauge symmetries are unlike any previously discussed symmetry both in their asymptotic angle dependence and in the fact that the action is described  at null, not spatial, infinity. Phrases like ``spontaneous symmetry breaking,'' ``Goldstone boson,'' ``superselection sector,'' and even ``conservation law'' are used with slightly different meanings in different physical contexts.  
In importing those words to the present context, I have necessarily adapted and refined their meanings. I have done so in the way I thought most natural, but other adaptations might be possible.}  
Let us begin by recalling  the standard case of spontaneous breaking of a global symmetry.

A classic example of spontaneous symmetry breaking is a scalar field theory with the  ``Mexican hat'' potential
\begin{equation}
\begin{split}
V(\Phi) = - m^2  \Phi^* \Phi + {\lambda}\big( \Phi^* \Phi \big)^2 ~, 
\end{split}
\end{equation}
illustrated in figure \ref{hat}. 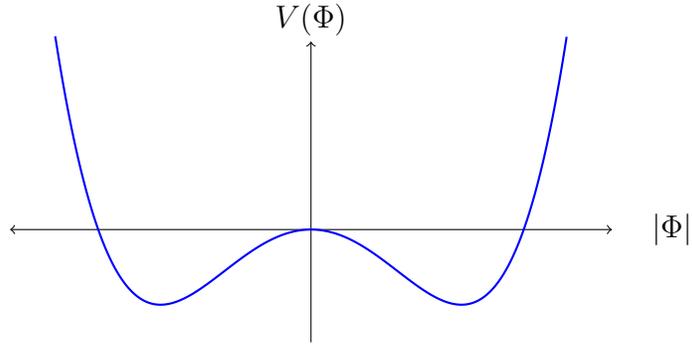
\begin{figure}[ht!]
\begin{center}
\begin{tikzpicture}
\draw [<->] (-4,0) -- (4,0);
\draw [->] (0,-1.5) -- (0,2.5);
\draw[domain=-3.4:3.4,variable=\r,blue,thick,samples=200,smooth] plot (\r,-\r*\r/2+\r*\r*\r*\r/16);
\node at (0,2.8) {$V(\Phi)$};
\node at (4.8,0) {$|\Phi|$};
\end{tikzpicture}
\end{center}
\caption{\small \bf The Mexican hat potential.}
\label{hat}
\end{figure}
This theory has a global symmetry, under which 
\begin{equation}
\begin{split}\label{globalsymmetry}
\Phi \to e^{i\theta} \Phi ~, \qquad \Phi^* \to e^{-i\theta} \Phi^*~. 
\end{split}
\end{equation}
The minimum of the potential occurs when 
\begin{equation}
\begin{split}
\Phi_{\text{vac}}(x) =   \frac{m}{\sqrt{2\lambda}}  e^{i\rho} ~ ,
\end{split}
\end{equation}
for arbitrary constant $\rho$. If we want to study excitations of this theory, we expand around $\Phi = \Phi_{\text{vac}}$, not $\Phi = 0$.  In the quantum theory, a particular choice $\ket{\rho = \rho_0}$ of vacuum state is made. Clearly, this state is not invariant under the global symmetry transformations \eqref{globalsymmetry}, that is, if $Q_\theta$ generates \eqref{globalsymmetry}, then it acts on the vacuum state as
\begin{equation}
\begin{split}\label{Qthetaaction}
e^{i Q_\theta} \ket{\rho_0} \sim \ket{ \rho_0 + \theta} \neq 0~. 
\end{split}
\end{equation}
Hence we have a symmetry of the Lagrangian but not a symmetry of the vacuum state. In this case, we say that the symmetry \eqref{globalsymmetry} is spontaneously broken. 

Goldstone's theorem  
states that whenever a continuous global symmetry is spontaneously broken, there exists a massless excitation about the spontaneously broken vacuum. Decomposing $\Phi(x)=|\Phi(x) |e^{i\rho(x)}$,  $\rho$ transforms as $\rho(x) \to \rho(x) + \theta$. Hence the Lagrangian can depend on $\rho $ only via the derivative $\p_\mu \rho$; there cannot be any mass term for  $\rho$, and it is a massless field.  $\rho$ --- identified as the field that  transforms inhomogeneously under the broken symmetry --- is referred to as the Goldstone boson. 

Let us now relate this elementary example to our discussion in the previous sections. We have a charge $Q^+_\ve$ that generates a symmetry of the Lagrangian of any abelian gauge theory. However, this charge does not annihilate the vacuum. Instead, it creates an extra soft photon mode $\phi$, which, according to \eqref{dsat}, transforms inhomogeneously under the broken symmetry. Hence the soft photons are the Goldstone bosons of spontaneously broken large gauge symmetry. There is an infinite  vacuum degeneracy, since we can add any number of soft photons to any vacuum state and obtain another vacuum state with the same zero energy. Classically, the infinite-dimensional space of vacua can be labeled by flat abelian connections $\p_z \ve(z,\bz)$ on the sphere.

There is a crucial difference between the usual Mexican hat story of spontaneous global symmetry breaking and the spontaneous breaking of the large gauge symmetries. In the usual story, the different vacua form superselection sectors (i.e., no physical finite energy operator exists that can move us from one vacuum to another). In other words, the charge $Q_\theta$ that would in principle move us between various vacua is not a normalizable operator in the Hilbert space. To see this, consider a system in a vacuum $\rho_0$. Now, suppose we have a bubble with a different value of $\rho$,  say $\rho_1$, in the interior. Then along the surface  of this bubble, $\Phi$ must have nonzero gradient.  Hence the bubble carries energy proportional to its area. To change the vacuum from $\rho_0$ to $\rho_1$ everywhere, we need this bubble to grow in size and cover the entire volume of the system. In infinite volume systems, this process will take an infinite amount of energy and therefore cannot be achieved. Thus, there are no finite-energy physical processes that change the vacuum state in systems of infinite size, and the $Q_\theta$ appearing in \eqref{Qthetaaction} is not a finite energy operator on the Hilbert space.  This argument relied on the area of the bubble growing with its volume. This is only true in more than one space dimension. In two spacetime dimensions the bubble wall is a  point, and the area does not grow. Finite energy processes may then give rise to a change in the vacuum. 
In spacetime dimensions $d>2$, scattering processes break up into superselection sectors, but in $d=2$ they do not. 

Such superselection sectors clearly do not arise for the large gauge symmetry. The vacuum state is changed by soft photon creation, which occurs in nearly all scattering processes.\footnote{In the gravitational version of this story, the change in vacuum is classically measured by the gravitational memory effect, which one hopes to  measure at LIGO \cite{Lasky:2016knh} or elsewhere.}  The $\mathcal{S}$-matrix elements do not factorize into superselection sectors. 
One way to understand this is that, even though we are working in $d=4$, the symmetry action in some regards mimics the $d=2$ case. 
  The parameter that generates the spontaneously broken symmetry is an arbitrary  local function on the sphere --- one parameter for every point --- as opposed to the  constant parameter in \eqref{globalsymmetry}. 
Hence vacuum-to-vacuum transitions need not occur simultaneously everywhere at once on the sphere at infinity. Restricting to just one point,  finite energy transitions are possible.  

\begin{figure}[ht!]
\begin{center}
\includegraphics[width=3.0in]{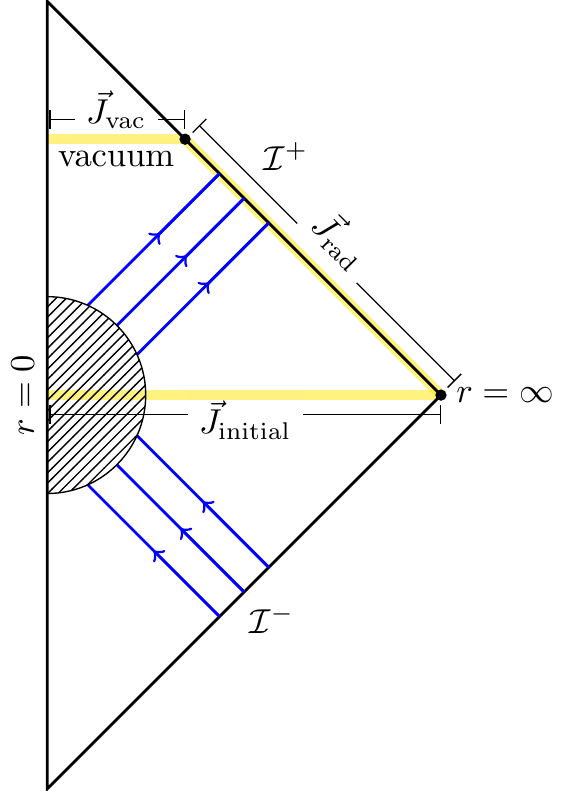}
\end{center}
\caption{\bf \small The existence of degenerate vacua in general relativity can be deduced from angular momentum conservation.  According to the Einstein equation, an initial configuration with angular momentum $\vec J_{initial}$ can decay to flat space while emitting total angular momentum flux $\vec J_{rad}\neq \vec J_{initial}$. Angular momentum is nevertheless conserved, because the final vacuum carries angular momentum. }
\label{amvac}
\end{figure}

At the classical level, the degenerate electromagnetic vacua connected by large gauge transformations all have vanishing $F_{\mu\nu}$ and hence carry no large gauge charge. This is because the large gauge group is abelian, and hence its action commutes with the charge. As discussed in section \ref{sec:gravity}, an analogous classical vacuum degeneracy appears in gravity. However in this case, the asymptotic  symmetry group is nonabelian, and the different vacua are classically distinguished by the charges they carry (e.g., angular momentum $\vec J$). The existence of classically degenerate vacua is then more obvious, as it can then be deduced from charge conservation. Consider, for example, the situation depicted in figure \ref{amvac},
 in which an initial configuration with angular momentum $\vec J_{initial}$
radiates angular momentum $\vec J_{rad}$ and decays to a  vacuum spacetime with vanishing Riemann tensor but angular momentum $\vec J_{vac}$.\footnote{Explicit solutions of this general type are studied below in section \ref{chair}. Formulas for the infinitesimal change in the  vacuum angular momentum under a supertranslation are in  \cite{Barnich:2011mi}, and the finite case is treated in \cite{Barnich:2016lyg}. Vacua with nonzero $\vec J$ are also discussed in \cite{Chen:2014uma}.} Angular momentum conservation implies
\be \vec J_{initial}=\vec J_{rad}+\vec J_{vac}~.\ee
 Insisting on vacuum uniqueness in gravity leads to the nonsensical conclusion that angular momentum is not conserved. This is sometimes referred to as the problem of angular momentum in GR, but it is neatly resolved by the existence of degenerate vacua with nonzero $\vec J$.

\subsection{Massive QED}\label{sec:massive}

In the previous sections, we have seen that massless electrodynamics has infinitely many conserved charges, and moreover that the quantum matrix elements of the conservation laws are  the soft photon theorem for amplitudes with massless charged external states.  Of course, in the real world, charged particles like the electron have mass. In this 
subsection we generalize the previous results to massive particles. 

The basic obstacle is that our analysis so far has been based on $\ci$, but massive particles never reach $\ci$ (for discussion, see \cite{Winicour:1988aq}). For theories that only have massless particles, $\ci$ forms a Cauchy surface --- that is, specification of initial (or final) data on $\ci^-$ (or $\ci^+$) is sufficient to obtain the complete evolution of the field into the interior. This no longer works for massive particles, since they do not arrive at $\ci$. Rather, they arrive from $i^-$ and asymptote to  $i^+$.  Interestingly, resolving this problem essentially forces one to the use of hyperbolic slices of Minkowski space employed for holographic reduction by de Boer and Solodukhin \cite{deBoer:2003vf}. The massive formulas are a bit more complicated than the massless ones, but they have a beautiful structure and make suggestive connections to AdS/CFT. 

The first step is  to understand how the symmetries act near $i^\pm$. This has been discussed in two papers \cite{Kapec:2015ena,Campiglia:2015qka}. In the first paper, the authors present a gauge invariant description of the action of the asymptotic symmetries on outgoing massive particles and show that the Ward identities correctly imply the soft theorem. In the second paper, the authors worked in Lorenz gauge. Despite making a specific gauge choice, the discussion of the second paper is  more illuminating. We now review it. 

Most of the requisite  machinery has already been set up in the massless analysis. The main step left is to understand how the large gauge symmetries act on massive states. Given this knowledge, we know the action of the hard part of the charge on asymptotic  massive states.  With this extra term in hand, we can proceed as before by transforming into momentum space to reproduce the soft theorem. 

For this purpose, a  prescription is needed to extend the asymptotic gauge parameter $\ve(z,\bz)$ into the bulk interior of Minkowski space and $i^\pm$. While this bulk extension can be done in many different ways, due to gauge invariance, any extension should give rise to the same final result. It turns out to be very  convenient to extend the gauge parameter from the boundary into the bulk using  Lorenz gauge
\begin{equation}
\begin{split}
\nabla^\mu A_\mu = 0 ~,
\end{split}
\end{equation}
which implies that the gauge parameter must satisfy
\begin{equation}
\begin{split}\label{gaugecond1111}
\Box \ve = 0 ~. 
\end{split}
\end{equation}
We need  to solve this equation with the boundary condition that it asymptote  to a specific function $\ve(z,\bz)$ on $\ci^+$.  We might try to solve it by Fourier transforming. However, this gives scattering solutions that fall off like ${1\over r}$ near \ip\ and therefore cannot satisfy our boundary condition.  Fortunately, \eqref{gaugecond1111} also has solutions that have no Fourier transform and do not die off at \ip\ (see exercise 6). The solution that approaches $\ve(z,\bz)$ can be written as
\begin{equation}
\begin{split}\label{vedefgreen}
\ve(x) = \int d^2 {\hat q} G \big( x , {\hat q} \big) \ve({\hat q}) \; ,
\end{split}
\end{equation}
where we have parametrized the point on the asymptotic sphere by a unit vector ${\hat q}$ that points toward $(z,\bz)$, and $G$ is a Green's function that must satisfy
\begin{equation}
\begin{split}
\Box G(x,{\hat q}) = 0 ~, \qquad \lim_{r\to\infty\atop u~\text{fixed}} G(x,{\hat q}) = \delta^2 ( {\hat x}  - {\hat q}  ) ~. 
\end{split}
\end{equation} This has the solution
\begin{equation}
\begin{split}\label{gfd}
G \big( x,{\hat q} \big)  = - \frac{\sqrt{\g({\hat q})}}{4\pi}  \frac{  x^\mu x_\mu }{ ( q \cdot x )^2 } ~, 
\end{split}
\end{equation} where $q^\mu = \big( 1 , {\hat q} \big)$.
The first property is easy to show:
\begin{equation}
\begin{split}
\Box G \big( x , {\hat q} \big)  &= 
- \frac{\sqrt{\g({\hat q})}}{4\pi}  \p_\nu \left[  \frac{  2 x^\nu   }{ ( q \cdot x )^2 }   -  \frac{  2 x^\mu x_\mu }{ ( q \cdot x )^3 }   q^\nu \right] \\
&=  - \frac{\sqrt{\g({\hat q})}}{4\pi}   \left[  \frac{  8   }{ ( q \cdot x )^2 }   -   \frac{  4     }{ ( q \cdot x )^2 }    -  \frac{  4   }{ ( q \cdot x )^2 }     +  \frac{  6 x^\mu x_\mu }{ ( q \cdot x )^4 }   q^2  \right]  = 0  ~. 
\end{split}
\end{equation}
The second property is exhibited by writing $G$ in retarded coordinates: \begin{equation}
\begin{split}
G \big( x,{\hat q} \big)  =  \frac{\sqrt{\g({\hat q})}}{4\pi}  \frac{  u ( u + 2 r ) }{ \big[   u + r \left( 1 -  {\hat q} \cdot {\hat x}  \right) \big]^2   }  
\end{split}
\end{equation}
which at large $r$ vanishes if ${\hat q} \neq {\hat x}$ and diverges if ${\hat q} = {\hat x}$. The overall normalization of the delta function can then be determined by integrating over ${\hat q}$. 

Another important property of $G$ can be seen by considering the limit toward $\ci^-$,  $r\to\infty$, keeping $v=u+2r$ constant. In this limit, $G$ localizes to the point ${\hat q} = - {\hat x}$. Thus, we find that in the $\ci^-$ limit, the large gauge parameter is antipodally related to the one obtained in the $\ci^+$ limit. In other words, $G$ ``knows'' about the antipodal map, which is a very convenient feature of Lorenz gauge. In hindsight, this may not be altogether surprising, since the gauge chosen here is Lorentz invariant, and the antipodal map is required by Lorentz invariance. 
\subsubsection{\it Hyberbolic Slices}
\begin{figure}[ht!]
\begin{center}
\includegraphics[width=3.2 in]{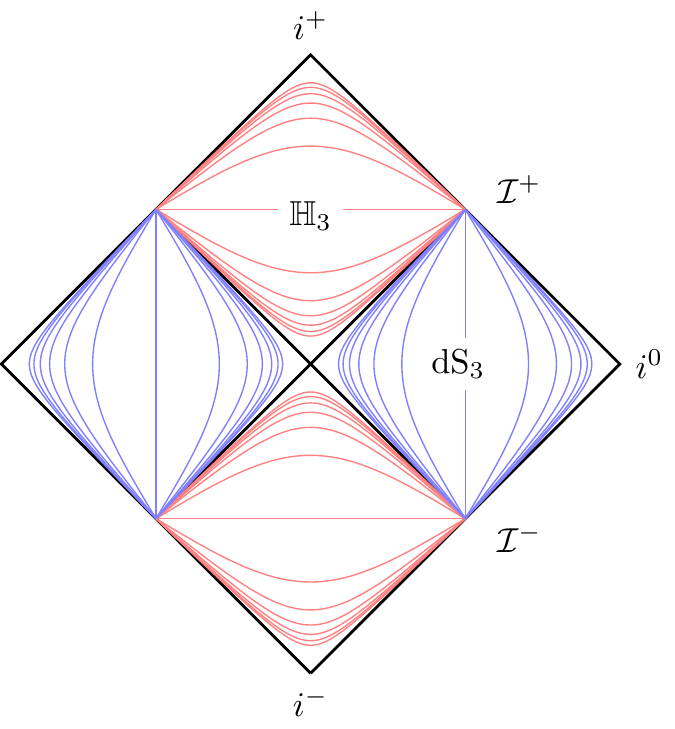}
\end{center}
\caption{\small \bf Hyperbolic slicing of Minkowski space. The slices correspond to constant $\tau^2=t^2-r^2$ surfaces. The red lines correspond to $\mhh_3$ slices and have $\tau^2 > 0$, whereas the blue lines correspond to the dS$_3$ slices with $\tau^2 < 0$. }\label{hyperbolicslicing}
\end{figure}

Now we are in a position to find the limit of this bulk gauge parameter onto $i^+$. For this purpose, we introduce a new set of coordinates. The retarded $(u,r,z,\bz)$ or advanced $(v,r,z,\bz)$ coordinates that we have employed so far are natural when one is interested in null infinity. However, they are not good near $i^\pm$. To discuss these boundaries, consider a hyperbolic slicing of Minkowski space, with slices labeled by
the coordinate 
\begin{equation}
\begin{split}
\tau^2 = - x^\mu x_\mu = t^2 - r^2~. 
\end{split}
\end{equation}
Hypersurfaces corresponding to constant $\tau$ are hyperbolic spaces. For $\tau^2 > 0$, the hypersurfaces are $\mhh_3$ (or Euclidean $\text{AdS}_3$), whereas for $\tau^2 < 0$ the hypersurfaces are three-dimensional de Sitter space $\text{dS}_3$, as illustrated in figure \ref{hyperbolicslicing}. 
Here we are interested in resolving the structure of $i^\pm$, for which we focus on the $\mhh_3$ slices. The $\text{dS}_3$ slices were used to resolve the structure of spatial infinity $i^0$  by Ashtekar and Romano \cite{Ashtekar:1991vb} in a related context. 

On the $\mhh_3$ slices, we introduce the coordinate $\rho$:
\begin{equation}
\begin{split}
\rho = \frac{r}{ \sqrt{ t^2 - r^2 } } ~. 
\end{split}
\end{equation}
The Minkowski metric then takes the form
\begin{equation}
\begin{split}
ds^2 = - d\tau^2 + \tau^2 \left[ \frac{d\rho^2}{1+\rho^2} + \rho^2 d\Omega_2^2 \right] ~. 
\end{split}\label{hyperboliccoord}
\end{equation}
The metric on the $\tau =$ constant hypersurfaces is that of $\mhh_3$, whose isometry group is  $SL(2,\mcc)$. These are  Lorentz transformations, which  in Minkowski space  map each of these slices into themselves. Similarly, $SL(2,\mcc)$ maps  the dS$_3$ slices  into themselves. 

The Green's function \eqref{gfd} in these coordinates takes the form
\begin{equation}\label{wet}
\begin{split}
 G(\tau,\rho,\hat x;\hat q)={\sqrt{\g(\hat q)} \over 4\pi  \left[ \sqrt{1+\rho^2}-\rho \hat q\cdot \hat x \right] ^2} ~. 
\end{split}
\end{equation}
Note that $\tau$ has completely dropped out of this equation (i.e., $\p_\tau G = 0$). This implies that the gauge parameter \eqref{vedefgreen} is independent of $\tau$. Viewed as a  three-dimensional Green's function, \eqref{wet} has been studied extensively  in the context of AdS$_3$ and is known as the bulk-to-boundary propagator for a massless scalar, which relates quantities on the boundary to those in the bulk.  We are beginning to see  the holographic structure of AdS$_3$  echoing in Minkowski space. Of course, this Green's function seems quite special since it does not depend on $\tau$ at all, but in other applications to flat space holography \cite{Pasterski:2016qvg,Bagchi:2016bcd}, $\mhh_3$ Green's functions (or bulk-to-boundary propagators) for fields of any conformal weight can arise and are accompanied by $\tau$-dependence in the full Minkowski space solutions.

\begin{figure}
\begin{center}
\includegraphics[width=3.2 in]{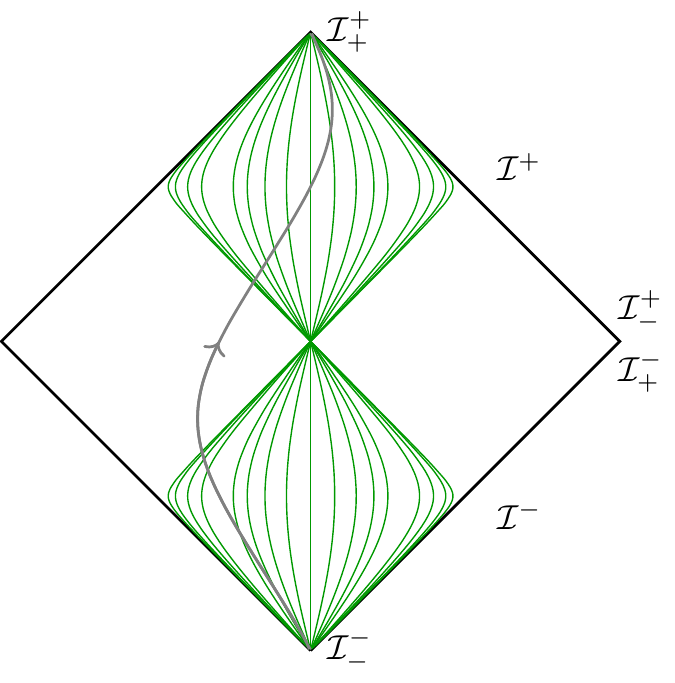}
\end{center}
\caption{\small \bf Hypersurfaces of constant $\rho$ are shown in green. The grey line is the worldline of a massive particle moving at constant velocity. The worldline of a particle asymptotes to a surface of constant $\rho$ as $\tau \to \infty$.}
\label{hyperbolicpenrose}
\end{figure}

Now consider a massive particle moving with constant momentum and following the trajectory $\vec{r} = \frac{1}{E} \vec{p}\, t + \vec{r}_0$ for some fixed $\vec{r}_0$, as illustrated in figure \ref{hyperbolicpenrose}. 
$(\tau,\rho,{\hat x})$ are given by
\begin{equation}
\begin{split}
\rho &= \frac{r}{\sqrt{t^2-r^2} } = \frac{ | \frac{1}{E} \vec{p} \, t + \vec{r}_0  | }{ \sqrt{ t^2 - \left( \frac{1}{E}\vec{p}\,t + \vec{r}_0 \right)^2  } }  ~, \\
\tau &= \sqrt{ t^2 - r^2 } =  \sqrt{ t^2 -  \left( \frac{1}{E}\vec{p}\,t + \vec{r}_0 \right)^2  }~, \\
 {\hat x} &= \frac{\vec{r}}{r} = \frac{ \vec{p}\, t + E \vec{r}_0  }{ |  \vec{p}\, t + E \vec{r}_0  | }  ~. 
\end{split}
\end{equation}
At late times, $t \to \infty$, 
\begin{equation}\label{pos}
\begin{split}
\rho   \to   \frac{|\vec{p}\,|}{m} ~, \qquad \tau   \to   \frac{m}{E} t ~, \qquad {\hat x} \to {\hat p} ~. 
\end{split}
\end{equation}
Hence $\rho$, and $\hat x$ asymptote to constants, and a massive particle approaches a fixed location on the unit hyperboloid at $i^+$. Hence in this resolution, $i^+$ is the hyperbola $\mhh_3$, much as \ip\ is $\mathbb{R}\times S^2$. The symplectic form for massive particles on $\mhh_3$ is worked out in exercise 7. 

This construction readily gives the large gauge symmetry action on such a one-particle state at asymptotically late times. As $\ve$ is independent of the parameter $\tau$ labelling the hypersurfaces, we simply evaluate it at \eqref{pos}. Writing the hard part of the charge in a vicinity of $i^+$  as an integral over the asymptotic late-time $\mhh_3$, one concludes that
\begin{equation}
\begin{split}
Q^{+H}_\ve \ket{ \vec{p} }  =  Q\ve \left(   \frac{|\vec{p}\,|}{m} , {\hat p} \right) \ket{ \vec{p} }  ~, 
\end{split}
\end{equation} 
where $\ve$ is given by \eqref{vedefgreen}. To summarize, the action of the large gauge charge on a massive particle is proportional to the Lorenz gauge value of the gauge parameter at the $\mhh_3$ point to which the massive particle asymptotes. 
\subsubsection{\it Soft Theorem}
We are now set to use this result to derive the quantum Ward identity.  The derivation is essentially identical to the massless case. Equations \eqref{Qinaction} and \eqref{Qoutaction} now include additional terms on the RHS from massive states:\footnote{For notational brevity, we  introduce here a simplified notation for  $n$-point amplitudes with $m$ outgoing particles and $n-m$ incoming particles. We drop the superscripts ``in'' and ``out'' on the momenta and charges and simply take the particles labeled $k=1,\dotsc,m$ to be outgoing and particles labelled $k=m+1,\dotsc,n$ to be incoming. We further adopt a  convention in which an incoming negative helicity photon of energy $p^0$ is described as a outgoing positive helicity photon with energy $-p^0$. Similarly, an incoming matter state of helicity $h$, charge $Q$, and energy $p^0$ is described as an outgoing matter state of helicity $-h$, charge $-Q$, and energy $-p^0$. }
\begin{equation}
\begin{split}
\bra{\text{out}} Q^+_\ve &=  - 2 \int d^2 w \p_\bw \ve \bra{\text{out}}  \p_w N \\
&\qquad \qquad   + \sum_{k\in\text{massless}}  Q_k  \ve(z_k ,\bz_k ) \bra{\text{out}} + \sum_{k\in\text{massive}} Q_k  \ve \left(   \frac{|\vec{p} _k|}{m _k} , {\hat p _k} \right)    \bra{\text{out}}  ~. 
\end{split}
\end{equation}
The Ward identity becomes  
\begin{equation}
\begin{split}
&- 2 \int d^2 w \p_\bw \ve \p_w \bra{\text{out}} \big[ N(w,\bw) \CS - \CS N^-(w,\bw) \big] \ket{\text{in}} \\
& \qquad \qquad \qquad  = - \left[     \sum_{k\in\text{massless}} Q_k  \ve(z_k ,\bz_k ) + \sum_{k\in\text{massive}} Q_k   \ve \left(   \frac{|\vec{p}_k|}{m _k} , {\hat p _k} \right) \right] \bra{\text{out}} \CS \ket{ \text{in}}~.  
\end{split}
\end{equation}
To verify that this is the soft photon theorem, set, as before,   $\ve(w,\bw) = \frac{1}{z-w}$ and obtain
\begin{equation}
\begin{split}\label{massivewardid}
& \frac{\sqrt{2} }{1+z\bz}  \lim_{\omega \to 0^+ }  \left[  \omega \bra{\text{out}} a_+  \big( \omega {\hat x}(z,\bz) \big)  \CS \ket{\text{in}}  \right] \\
& \qquad\qquad\qquad\qquad =   e \left[   \sum_{k\in\text{massless}} \frac{ Q_k  }{  z - z_k  }  + \sum_{k\in\text{massive}} Q_k   \ve \left(   \frac{|\vec{p}\, _k|}{m _k} , {\hat p _k} \right)   \right] \bra{\text{out}} \CS \ket{ \text{in}} ~.  
\end{split}
\end{equation}
Using the property 
\begin{equation}
\begin{split}
G \left( \frac{ |\vec{p_k}\,|}{m} ,\hat p_k; w,\bw \right) = \frac{1}{2\pi}  \p_\bw \left[ \frac{\sqrt{2}}{1+w\bw} \frac{ p_k \cdot \ve^+}{ p_k \cdot {\hat q} } \right] ~ ,
\end{split}
\end{equation}
we find 
\begin{equation}
\begin{split}
\ve \left( \frac{ |\vec{p_k}\,|}{m} ,\hat p_k  \right) &= \int d^2 w  G \left( \frac{ |\vec{p_k}\,|}{m} ,\hat p_k; w , \bw  \right) \frac{1}{z-w}  = \frac{\sqrt{2}}{1+z\bz}  \frac{ p_k \cdot \ve^+(z,\bz)}{ p_k \cdot {\hat q}(z,\bz) } ~.
\end{split}
\end{equation}
Plugging this into \eqref{massivewardid} and multiplying both sides by $\frac{1+z\bz}{\sqrt{2}}$, we reproduce the  soft photon theorem for massive charged particles in the  standard momentum space form \eqref{mft}. It is rather satisfying to see this detailed equivalence emerge from the properties of bulk-to-boundary propagators on $\mhh_3$!

\subsection{Magnetic Charges}\label{mnetic}
In this section, we derive a new nonperturbative soft theorem and generalize the analysis to include magnetic charges \cite{Strominger:2015bla}. 

 In perturbation theory, it has been proven using Feynman diagrams  \cite{Weinberg:1995mt} that the soft photon theorem is tree-level exact (i.e.,~it does not receive any loop corrections). It is natural to ask whether there can be nonperturbative corrections. To even talk about nonperturbative contributions to the soft theorem, one first needs a theory that exists nonperturbatively.  QED  ---  the theory of photons and electrons ---  does not exist due to the Landau pole. It must  be embedded in some bigger theory --- maybe one that is asymptotically free --- that does exist nonperturbatively. To the best of my knowledge, all examples of such bigger  theories contain magnetic monopoles. 
 
This leads us to ask whether the soft photon theorem is corrected when magnetic monopoles are present in the asymptotic states. It is not hard to see  that there must indeed be such a correction.  In low-energy effective field theory, a magnetic monopole is represented by a point particle with magnetic charge. A soft photon will certainly couple to such an object, and the  pole due to the nearly 
on-shell propagator \eqref{weinbergpole} is still present. The only thing that changes is the coupling of the magnetic particle to the gauge field \eqref{qedvertexfactor}. This coupling gives rise to a different soft factor, but the pole remains. 

To derive the nonperturbative corrections to the soft theorem, we begin with the perturbative electric soft factor
\begin{equation}
\begin{split}
S^\a_0 = \sum_k \frac{e Q_k p_k \cdot \ve^\a}{q \cdot p_k} ~, 
\end{split}
\end{equation}
for a soft photon of polarization $\ve_\a$. 
Consider an amplitude involving asymptotic particles with integral magnetic charges $M_k$ defined by
\begin{equation}
\begin{split}
M_k = \frac{1}{2 \pi } \int_{S^2_k} F ~,
\end{split}
\end{equation}
where $S^2_k$ is a sphere surrounding the $k$th particle.  Since the pole in the soft factor is unchanged, we need to simply determine the vertex factor. A trick for this purpose is to  transform to  dual variables
\begin{equation}
\begin{split}\label{dualitytransform}
{\tilde F} = - \frac{2\pi}{e^2} \ast F ~, \qquad {\tilde e} = \frac{2\pi}{e} ~, \qquad {\tilde Q}_k = \frac{1}{{\tilde e}^2} \int_{S^2_k} \ast {\tilde F}  = M_k  ~, \qquad {\tilde M}_k = \frac{1}{2 \pi } \int_{S^2_k} {\tilde F} = - Q_k ~. 
\end{split}
\end{equation}
This duality interchanges the electric and magnetic charges as $({\tilde Q}_k , {\tilde M}_k) = ( M_k , - Q_k )$. Note that  we are not assuming here that the theory has any kind of duality symmetry; we are simply rewriting it in terms of the dual field strength as a shortcut to derive magnetic terms in the soft factor. 
It is useful to define a dual gauge potential that couples to magnetic charges in the same way that the usual gauge potential couples to electric charges:
\begin{equation}
\begin{split}\label{Adualdef}
{\tilde F} = \dt {\tilde A} = - \frac{2\pi}{e^2} \ast \dt A~, 
\end{split}
\end{equation}
which determines ${\tilde A}$ in terms of $A$ (up to magnetic gauge transformations). In the bulk four dimensions, this is a highly nonlocal  relationship, but as we shall see, it  becomes local on $\ci^+$. Note that \eqref{Adualdef} also relates  the polarization $\ve_\a$ to its dual ${\tilde \ve}_\a$ as defined by \begin{equation}
\begin{split}
A_\mu(q) = e \ve_\mu^\a (q)  ~, \qquad {\tilde A}_\mu(q) = {\tilde e} {\tilde \ve}_\mu^\a (q)  ~. 
\end{split}
\end{equation}
Since $\tilde A$ couples to magnetic particles in exactly the same way that $A$ couples to electric particles, we immediately conclude that magetic charges must correct the soft formula to 
\begin{equation}
\begin{split}\label{modifiedsoftfactor}
S_0^\a =  \sum_{k} \frac{ p_k  \cdot \left(  Q_k  e \ve^\a + M_k  \tilde{e}{\tilde \ve}^\a \right) }{q \cdot p_k}  ~. 
\end{split}
\end{equation}
Electric and magnetic gauge invariance implies that the amplitude should remain invariant under separate shifts of $\ve^\a$ and ${\tilde \ve}^\a$. This indeed follows from the fact that  electric and magnetic charge are each conserved. It is easy to check  that this formula is invariant under electric-magnetic duality. It is conjectured to be nonperturbatively exact \cite{Strominger:2015bla}.

It is illuminating to describe the duality transformations \eqref{dualitytransform}  in retarded coordinates near \ip.  Expanding near $\ci^+$, they simplify to 
\begin{equation}
\begin{split}\label{dualtransformexplicit}
{\tilde F}_{z\bz}^\0 = \frac{2\pi i}{e^2} \g_{z\bz} F_{ru}^\2 ~, \qquad {\tilde F}_{ru}^\2 = \frac{2\pi i}{e^2} \g^{z\bz} F_{z\bz}^\0 ~, \qquad {\tilde F}_{uz}^\0 = \frac{2\pi i}{e^2} F_{uz}^\0 ~. 
\end{split}
\end{equation}Integrating the last equation of \eqref{dualtransformexplicit} and choosing the integration constant 
to get a linear relation between $A_z$ and $\tilde A_z$, we find
\begin{equation}\label{ser}
\begin{split}
{\tilde A}^\0_z = \frac{2\pi i}{e^2} A_z^\0~. 
\end{split}
\end{equation}
This formula is quite interesting. It states that the duality transformation is simply multiplication by $i$. This is  unsurprising, since $F_{uz}$ corresponds to the radiative mode of the electromagnetic field with one  polarization. A multiplication by $i$ corresponds to a phase shift by $\frac{\pi}{2}$, which is equivalent to exchanging electric and magnetic fields. The soft factor may then finally be written in the simple form for positive helicity
\begin{equation}
\begin{split}\label{modifieftfactor}
S_0^\a =  \sum_{k} \frac{( eQ_k+\frac{2 \pi i }{e}M_k) p_k  \cdot   \ve^+ }{q \cdot p_k}  ~, 
\end{split}
\end{equation}
while the combination $ eQ_k-\frac{2 \pi i }{e}M_k$ appears for the opposite helicity. 

In the presence of magnetic charges, there is a second infinity of conserved charges. Let us define an outgoing charge by \begin{equation}
\begin{split}
{\tilde Q}^+_\ve = \frac{1}{2\pi} \int_{\ci^+_-} \ve F = \frac{i}{2\pi} \int_{\ci^+_-} d^2 z \ve F_{z\bz}^\0 .
\end{split}
\end{equation}
The Lorentz-invariant matching condition 
\be \label{sxd}F_{z\bz}^\0|_{\ci^+_-}=-F_{z\bz}^\0|_{\ci^-_+} \ee
 then implies that this outgoing magnetic charge is equal to the incoming magnetic charge on $\ci^-$: 
\begin{equation}
\begin{split}
{\tilde Q}^+_\ve={\tilde Q}^-_\ve = - \frac{i}{2\pi} \int _{\ci^-_+} d^2 z \ve F_{z\bz}^\0 ~. 
\end{split}
\end{equation}
The extra minus sign in  \eqref{sxd} is present because the $(z,\bz)$ coordinates used on $\ci^-$ are antipodally mapped to the ones on $\ci^+$, reversing the orientation of the sphere. 

To what symmetry is this new charge associated?  In the case of the electric charges, this question was answered by constructing the asymptotic  Dirac brackets and working out the action of the charge on the fields. In doing so, we assumed that $F_{z\bz}^\0$ vanished on the boundaries of $\ci^+_\pm$. This boundary condition is consistent in the absence of  magnetically charged particles but obviously cannot be imposed here. In fact, the correct boundary conditions and Dirac brackets in the presence of both electric and magnetic charges have not yet been worked out. It remains an important open problem. 

 However, without going through this long procedure, duality covariance gives an obvious guess for the final result:  ${\tilde Q}^\pm_\ve$ should generate {\it magnetic} gauge transformations on the dual gauge fields, namely
\begin{equation}
\begin{split}
{\tilde \delta}_\ve {\tilde A}^\0_z = \p_z \ve ~. 
\end{split}
\end{equation}
Using \eqref{ser}, this can be written in terms of the gauge field $A_z$ as \begin{equation}
\begin{split}
{\tilde \delta}_\ve A_z^\0 = - \frac{i e^2}{2\pi } \p_z \ve ~. 
\end{split}
\end{equation}
These may be interpreted as electric  gauge transformations but with an imaginary gauge parameter ${\tilde \ve} = - \frac{ie^2}{2\pi} \ve$. The original real $U(1)$ symmetry is now enhanced to a complex $U(1)$. This complexification enables us to simultaneously and $locally$ realize both the electric and magnetic gauge symmetry on the asymptotic fields! From this point forward, showing that the Ward identity of this complexified large gauge  symmetry is precisely the full nonperturbative electric+magnetic soft photon theorem follows through exactly as before. We will not work through the derivation here. 

It is instructive to note here that in the standard electric  presentation of abelian gauge theory, the large electric symmetries are manifest as a nontrivial subgroup of the gauge symmetry, whereas magnetic symmetries are not. In the dual magnetic description, the opposite is true. The lesson  here is that not all asymptotic symmetries can be thought of as subgroups of some bulk gauge symmetry. As truly physical symmetries with measurable dynamical implications, they have a more fundamental status. Since trivial gauge symmetries correspond to redundant degrees of freedom, one may have many different presentations of a theory in which different redundant degrees of freedom appear. However, since the asymptotic symmetries are physical, they must exist in each presentation, although not necessarily describable as a subgroup of a local gauge symmetry. Indeed, we will see in the next section that the soft photino theorem in $\CN=1$ QED implies the existence of infinitely many fermionic symmetries. These are not a subgroup of a local gauge symmetry in any known presentation of the theory.

\subsection{Supersymmetry}\label{stry}
 In supersymmetric theories, every particle has a superpartner. The partner  of the photon is called the photino. One might expect  a soft photino theorem as a superpartner  of the soft photon theorem. This is indeed the case. For every soft theorem, one expects an asymptotic symmetry. Furthermore, since the soft theorem can be thought of as one relationship between two $\mathcal{S}$-matrix amplitudes for every direction $(z,\bz)$ of the soft particle, the associated asymptotic symmetry should be infinite dimensional. The soft photino theorem is thus expected to give rise to an infinite number of fermionic symmetries of the $\cs$-matrix. 

This simple conclusion is quite surprising as, in the simplest case of $\CN=1$, the global supersymmetry is generated by four real fermionic parameters. Clearly, the infinitely many fermionic symmetries alluded to above cannot possibly be a subgroup of these finitely many supersymmetries! This is therefore another example, following the magnetic one,  of an asymptotic symmetry that is not realized as a subgroup of any manifest local symmetry of the full bulk theory. However, in the magnetic case, we were able to construct a reformulation of the theory in terms of dual variables in which the magnetic symmetry was manifest as a subgroup of the local gauge symmetry of the dual magnetic potentials. In contrast, there is no known  description of the fermionic symmetries in $\CN=1$ theories in which a nontrivial infinite-dimensional symmetry is manifest.\footnote{In superspace formalisms, the gauge parameter sometimes has a fermionic superpartner that is gauge fixed to zero. However,  the fermionic symmetries we find here are unrelated to these superpartners. It is of course possible that alternate superspace formulations exist in which our fermionic symmetries are superpartners of gauge symmetries.} Hence, the fermionic symmetries that we uncover here are completely new and different.

In this section, we present the soft photino theorem and rewrite it as the matrix element of an infinite set of fermionic conservation laws, with some details referred to  the original work \cite{Dumitrescu:2015fej}. In section \ref{sft},  we review the soft photino theorem. Section \ref{fcl} derives the equivalent fermionic conservation laws. Further directions involving supersymmetric asymptotic symmetries are discussed  in section \ref{dscc}.

\subsubsection{\it Soft Photino Theorem}\label{sft}

First we review  the soft photino theorem. The Lagrangian for $\CN=1$ supersymmetric QED with photino $\Lambda$ and generic matter  supercurrent $\CK$ (the superpartner of the charged matter current $j_\mu$) is
\begin{equation}
\begin{split}
\CL = - \frac{1}{4e^2} F_{\mu\nu} F^{\mu\nu} - \frac{i}{e^2} {\bar \Lambda} {\bar \sigma}^\mu D_\mu \Lambda + i {\bar \CK} {\bar \Lambda} - i \Lambda \CK + \CL_{\text{matter} } + \ldots , 
\end{split}
\end{equation}
where we follow the spinor conventions of Wess and Bagger \cite{Wess:1992cp}. Here, $\CL_{\text{matter}}$ is the full interacting matter Lagrangian and ``$\ldots$'' denotes any higher derivative interactions. Our discussion of the soft photino theorem will not depend on the precise nature of these terms. The only important term is the linear coupling of the supercurrent to the photino. The structure of this interaction is fixed completely by gauge invariance and supersymmetry. 

For clarity, we work with a specific example, but  the discussion is completely general and applies to any supersymmetric abelian gauge theory. Consider the case where the matter content of this theory contains a single, massless charged chiral multiplet with charge $Q$. This multiplet contains a complex scalar field $\Phi$ and its superpartner $\Psi_\a$. The fermionic current is
\begin{equation}
\begin{split}
\CK_\a = \sqrt{2} Q {\bar \Phi} \Psi_\a , \qquad {\bar \CK}_\da = \sqrt{2} Q \Phi {\bar \Psi}_\da. 
\end{split}
\end{equation}

To describe the soft theorem, consider an amplitude with $n$ particles with momenta $p_k$, $k=1,\dotsc,n$, and a photino with momentum $q$. Since all momenta are null, we can write them in terms of spinors as 
\begin{equation}
\begin{split}
p_{k\mu} \sigma^\mu_{\a\da} = \eta_{k\a} {\bar \eta}_{k\da} , \qquad q_\mu \sigma^\mu_{\a\da} = \eta_\a {\bar \eta}_\da. 
\end{split}
\end{equation}
The soft photino theorem (a derivation is in  \cite{Dumitrescu:2015fej}) is then simply 
\begin{equation}
\begin{split}\label{softphotinothm}
\lim_{\eta \to 0} \bra{\text{out}} a_+ (\eta)  \CS \ket{\text{in}} = \sqrt{2}ie \sum_k \frac{1}{\eta \cdot \eta_k } \bra{\text{out}} \SF_k \CS \ket{\text{in}} , 
\end{split}
\end{equation}
where $\SF$ is the operator whose action on states is given by
\begin{equation}
\begin{split}\label{sfdef}
\bra{ \Phi} \SF = - Q \bra{ \Psi}  , \qquad \bra{ \bar \Psi} \SF = Q \bra{\bar \Phi } , \qquad \bra{ \bar \Phi} \SF = \bra{ \Psi } \SF = 0. 
\end{split}
\end{equation}
Note that the soft theorem here takes on a qualitatively different form than the examples considered so far. In the soft photon theorem, the soft factors were all simply numbers in a plane wave basis, whereas here the soft factor is an \emph{operator} that changes external bosons into external fermions and vice versa.   Going forward, we will encounter  many more examples in which the soft factor is an operator acting on the external states. 

Note that $\SF$ is a fermionic operator (i.e., it takes bosonic states into fermionic states and vice versa). This is to be expected, as all nonvanishing amplitudes must have an even number of external fermions. If  the LHS of \eqref{softphotinothm} is nonvanishing, the $\CS$-matrix $\bra{\text{out}} \CS \ket{\text{in}}$ must itself vanish.  Note also that $\SF$ only acts on charged states, and even then it does not act on all charged states, only on $\bra{ \Phi}$ and on $\bra{ \bar \Psi}$. This action is quite distinct from that of the supercharge itself, which acts on all states; see Exercise 8. 

We will not derive the soft photino theorem in detail,  but we can understand qualitatively why the soft photino theorem must involve the operator $\SF$ as follows. The derivation of the soft theorem for the photino follows through in exactly the same way as for the photon, with the ${1 \over p_k\cdot q}$ pole coming from the untruncated, nearly on-shell  propagator. The only new element is the coupling of the photino with the chiral fields. 
For an outgoing positive helicity photino, the relevant interaction term is $\bar{\CK}_\da = \sqrt{2} Q \Phi {\bar \Psi}_\da$. The factor of $Q$  indicates that one only gets a contribution to the  soft factor from particles with charge, as seen in \eqref{softphotinothm}. Moreover, one only gets a nonzero soft factor from states created by fields ${\bar \Phi}$ and $\Psi_\a$, which are precisely the states on which the operator $\SF$ has a nonzero action.\footnote{The notation used here is that the ket state $\ket{\Phi}$ is created by the field $\Phi$. Thus, its adjoint $\bra{\Phi}$ is created by the field ${\bar \Phi}$.}

\subsubsection{\it Fermionic Conservation Laws}\label{fcl}

Following the theme of these lectures, we strongly suspect that \eqref{softphotinothm} can be recast as the quantum matrix element of a charge conservation law.  There are not too many sensible possibilities for what that charge could be. Our approach will be  to simply motivate  an Ansatz and then establish the conservation law by checking that the corresponding Ward identity is equivalent to \eqref{softphotinothm}.

Note that the generic structure of this fermionic charge, $\SF[\chi]$ ($\chi$ is an arbitrary angle-dependent  fermionic parameter), has to be similar to the structure of the large gauge charge --- namely, that there is a soft charge linear in the photino field and a hard charge containing the matter particles. Thus, we guess that the new infinity of conserved charges $\SF[\chi]$ loosely takes the form of a $\ci^+_-$ integral:
\begin{equation}
\begin{split}
\SF [ \chi] \sim \int \chi \Lambda + \int \chi \CK . 
\end{split}
\end{equation}
To refine this guess, we require that somehow supersymmetry should play a role in this story. This is because the soft photino theorem itself was obtained as a supersymmetrization of the soft photon theorem. Supersymmetry commutes with all gauge transformations. Thus, it cannot be possible to obtain $\SF[\chi]$ directly by the action of supersymmetry on $Q^+_\ve$. However, the reverse might be (and as we shall see, is) true: the action of supersymmetry on $\SF[\chi]$ could give us $Q^+_\ve$. Another motivation for suspecting such a relation is that it  implies $\SF[\chi]$ as an operator has mass dimension $-\frac{1}{2}$, consistent with the mass dimension of $\SF$ as defined in \eqref{sfdef}.

To implement this guess, note that the supersymmetry transformations of the photino field are
\begin{equation}
\begin{split}\label{photinosusytransform}
\delta^{\text{susy}}_\a \Lambda_\b = - F_{\a\b} + \ldots. 
\end{split}
\end{equation}
Here, $\delta^{\text{susy}}_\a$ is a supersymmetry transformation with a spinor index $\a$, and $F_{\a\b} = \sigma^{\mu\nu}_{\a\b} F_{\mu\nu}$. The corrections ``$\ldots$'' depend on the precise form of the Lagrangian and of the theory under consideration. In the off-shell superfield formalism, these corrections are  simply $\ve_{\a\b}D$, where $D$ is an auxiliary field. In any case, these terms will not matter for our  purposes,  since they die off rapidly  near $\ci^+$.
Equation \eqref{photinosusytransform} then suggests that  for the supersymmetry transformation of the fermionic charge,
\begin{equation}
\begin{split}
\zeta^\a \delta_\a^{\text{susy}} \SF[\chi] = Q^+_{\zeta \chi} , \qquad {\bar \zeta}^\da \delta_\da^{\text{susy}} \SF[\chi] = 0 . 
\end{split}
\end{equation}
This equation can be used  to determine the precise form of $\SF[\chi]$. One might anticipate at this point that it is just a $\ci^+_-$ integral of $\chi\Lambda$. This is almost right, except that $\Lambda$ has a term that grows in the far past of \ip  , which we must take some care to project out.

To do so  explicitly, let us describe in more detail the structure of spinors near $\ci^+$.  Define a spinor basis by
\begin{equation}
\begin{split}
\sigma_z{}^z \xi^\pmm = \pm \frac{1}{2} \xi^\pmm, \qquad \xi^\+ \xi^\- = 1 . 
\end{split}
\end{equation}
Near $\ci^+$, we can then expand the photino and fermionic current as
\begin{equation}
\begin{split}
\Lambda_\a = \frac{1}{r} \lambda_\+ \xi^\+_\a + \frac{1}{r^2} \lambda_\- \xi^\-_\a + \CO \big( r^{-3} \big) , \\
{\bar \CK}_\da = \frac{1}{r^2} {\bar k}_\- {\bar \xi}^\-_\da + \frac{1}{r^3} {\bar k}_\+ {\bar \xi}^\+_\da + \CO \big( r^{-4} \big) . 
\end{split}
\end{equation}
The constraint equation on $\ci^+$ is 
\begin{equation}
\begin{split}\label{photinoconstrainteq}
 (1+z\bar{z})D_\bz \lambda_\+ - 2 \p_u \lambda_\- = e^2 {\bar k}_\- . 
\end{split}
\end{equation}
This is the superpartner of the leading-order Maxwell constraint equation \eqref{abelianconstrainteq}, and it will play the same role as that constraint in converting boundary expressions for charges into bulk ones. The supersymmetry transformation of $\lambda_\pmm$ can be determined by taking the $\ci^+$ limits of \eqref{photinosusytransform}. It is convenient to define the following supersymmetry transformation
\begin{equation}
\begin{split}
\delta_\- = \xi_\-^\a  \delta_\a^{\text{susy}}  . 
\end{split}
\end{equation}
Then, near $\ci^+$, \eqref{photinosusytransform} simplifies to
\begin{equation}
\begin{split}\label{ipsusy}
\delta_\- \lambda_\+ = - 2 (1+z\bar{z})F^\0_{uz} , \qquad \delta_\- \lambda_\- =  F_{ur}^\2 - \g^{z\bz} F_{z\bz}^\0  + \ldots, 
\end{split}
\end{equation}
where ``$\ldots$'' are the $D$-term contributions. In general, $\lambda_\+$ goes to a finite value at large $|u|$. Equation \eqref{photinoconstrainteq} then implies that $\lambda_\-$ may diverge linearly in $u$. Thus, to define a fermionic charge as an integral over $\ci^+_-$ of $\lambda_\-$, we must project out the term linear in $u$:
\begin{equation}
\begin{split}
\SF[\chi] = \frac{1}{e^2} \int d^2 z \g_{z\bz} \chi \xi^\- \left( 1 - u \p_u \right) \lambda_\- \big|_{\ci^+_-} . 
\end{split}
\end{equation}
The supersymmetry transformation \eqref{ipsusy} then immediately implies
\begin{equation}
\begin{split}
\zeta^\a \delta_\a^{\text{susy}} \SF[\chi] = Q^+_{\zeta^\a \chi_\a } . 
\end{split}
\end{equation}
This fermionic charge satisfies all the properties discussed above. Appropriate photino matching conditions imply that it commutes with the $\mathcal{S}$-matrix.  The corresponding Ward identity states
\begin{equation}
\begin{split}
\bra{\text{out}} \left( \SF[\chi] \CS - \CS \SF^-[\chi] \right) \ket{ \text{in}} = 0 .
\end{split}
\end{equation}
Then, using formulas  from the previous sections for rewriting boundary expressions in a plane wave basis, we find that this is indeed equivalent to the soft photino theorem \eqref{softphotinothm}.

The symmetry corresponding to this charge follows with the help of the symplectic form for fermions on $\ci^+$. The action of the charge on the photino is found to be
\begin{equation}
\begin{split}
\left\{ \SF[\chi] , \lambda_\+ (u,z,\bz) \right\} = 0 , \qquad \left\{ \SF[\chi] , \bar \lambda_\- (u,z,\bz) \right\} = -\p_\bz \chi . 
\end{split}
\end{equation}
We see that  the photino gets an inhomogeneous shift,  implying a new kind of ``fermionic gauge transformation.''
This new and infinite set of symmetries exists for all supersymmetric abelian gauge theories, yet is not a subset of any previously considered local fermionic symmetry. 

\subsubsection{$\CN=4,8$}\label{dscc}

It is an important and outstanding problem to extend this construction to gauge theories
with $\CN=4$ supersymmetry or gravity theories with $\CN=8$ supersymmetry. Clearly, conceptually new features will arise from  supermultiplets of soft theorems with nonabelian $R$-symmetry groups. Multiple soft limits are known not to commute in this more general context\cite{ArkaniHamed:2008gz}. At this time, there is not even a conjecture for what the asymptotic symmetry group should be!

The fact that the action of supersymmetry on the fermionic charges/symmetries gives the 
large gauge charges/symmetries is closely related to an observation of Larkoski \cite{Larkoski:2014hta} that the leading and subleading soft theorems in QED are related by conformal symmetry. Going forward to higher supersymmetry groups, we may encounter a rich structure in which hierarchies of asymptotic symmetries related to leading, subleading, and sub-subleading soft theorems merge into a unified infinite-dimensional  symmetry group. 
This will be interesting indeed to explore. 

In Barnich et al. \cite{Barnich:2015mui}, a charge-weighted vacuum partition function that encodes the BMS-degeneracy 
of the vacuum in three-dimensional gravity was constructed. It would be of interest in supersymmetric theories to define and compute charge-weighted vacuum indices characterizing the degeneracy. 

\subsection{Infrared Divergences}\label{IRDiv}
    A salient and much-studied feature of QED in the deep infrared is the appearance of IR divergences in the integral over momenta in internal loops \cite{Weinberg:1995mt}. These divergences set all conventional Fock-basis $\cs$-matrix elements to zero. Often they are dealt with by restricting to inclusive cross sections in which physically unmeasurable photons below some IR cutoff are traced over. The trace gives a divergence, which offsets the zero and yields a finite result for the physical measurement \cite{Yennie:1961ad,Kinoshita:1962ur,Lee:1964is,Weinberg:1965nx}. While this is adequate for most experimental applications, for some purposes it is nice to have an $\cs$-matrix.  For example, discussions of unitarity or asymptotic symmetries require an $\cs$-matrix.

\begin{figure}
\begin{center}
\includegraphics[width=3.0 in]{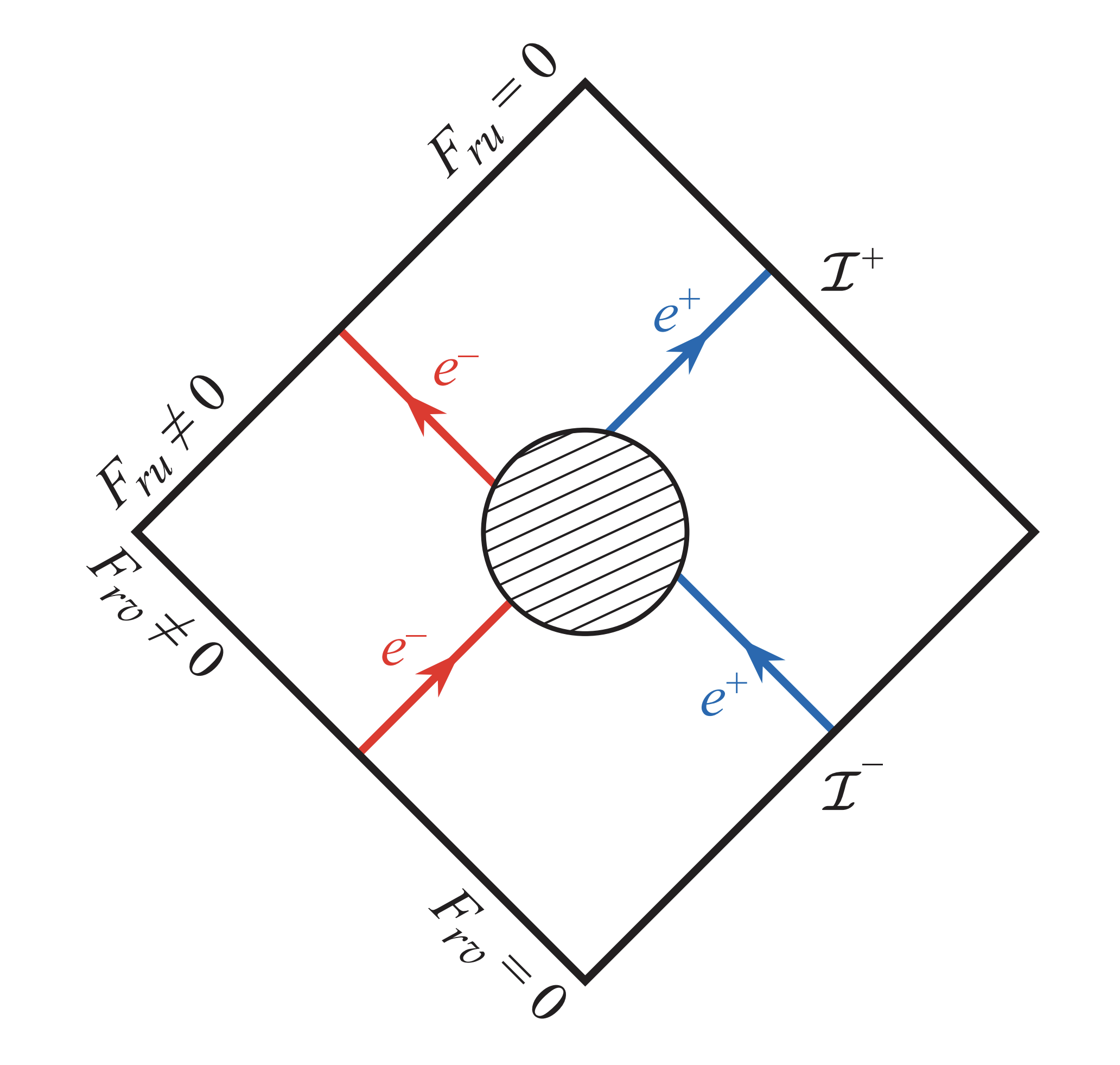}
\end{center}
\caption{\bf\small Bhabha scattering, $e^+e^-\to e^+e^-$, involves incoming and outgoing electron-positron pairs unaccompanied by radiative photons. In the massless case depicted,  the Maxwell equation implies Coulomb fields on $\ci$ to the future of the incoming pair and the past of the outgoing pair. If the scattering angle is nonzero, these Coulomb fields cannot obey the required matching conditions where \ip\ and $\ci^-$ meet.  The amplitude for Bhabha scattering therefore vanishes identically. 
} \label{Bhabha}
\end{figure}

 It is natural to ask whether the considerations of this book bear on the issue of IR divergences. We shall see that they indeed give a satisfying picture: imposing the infinity of conservation laws on amplitudes  eliminates some and possibly  all  IR divergences.  
 To see this,  consider Bhabha scattering as illustrated in figure \ref{Bhabha},
 in which an incoming electron-positron pair scatters to an outgoing electron-positron pair, with no incoming or outgoing radiation.  To restrict attention to $\ci$ (rather than $i^\pm$), we  take the charged particles to be massless.\footnote{We suppress here the effects of loop corrections in the massless case, which make the coupling run in the deep IR. The qualitatively similar and more realistic  massive case is treated in  \cite{Kapec:2017tkm}.} In this case the constraint equations on \ip\ imply (see \eqref{abelianconstrainteq})
\begin{equation}\label{cpeq}
\partial_u F^{(2)}_{ru} + \cancelto{0}{D^z F^{(0)}_{uz}} +  \cancelto{0}{D^\bz F^{(0)}_{u\bz}}   =-e^2   j^{(2)}_u,
\end{equation} 
while on $\ci^-$,
\begin{equation}\label{cmeq}
\partial_v F^{(2)}_{rv} - \cancelto{0}{D^z F^{(0)}_{vz}} - \cancelto{0}{D^\bz F^{(0)}_{v\bz}}   = e^2   j^{(2)}_v.
\end{equation}
Here  we have set 
$F^{(0)}_{uz}=F^{(0)}_{vz}=0$, as this process involves no radiative modes. 
 $j^{(2)}_v$ and $j^{(2)}_u$ are the electron and positron charge fluxes at $\ci$. For example, an outgoing positron at $(u_0,z_0,\bz_0)$ creates a flux
 \be \label{cr} j^{(2)}_{u}=\gamma^{z\bz}\delta(u-u_0)\delta^2(z-z_0).\ee
 Since the incoming (outgoing) state is taken to be the vacuum in the far past (future), we should solve  \eqref{cmeq} (\eqref{cpeq}) with the boundary condition that the Coulomb field vanishes in the far past (future). One then finds that, as long as the scattering angle is nontrivial, the fields cannot obey the matching conditions \eqref{mcone} 
\be \label{mconeb}
F_{ru}^{(2)} (z, \bz) \Big|_{\ci^+_-} = F_{rv}^{(2)} (z, \bz) \Big|_{\ci^-_+}~ 
\ee  at the boundary of $\ci$ near $i^0$.
Therefore Bhabha scattering must vanish identically. 

\begin{figure}
\begin{center}
\includegraphics[width=5.0 in]{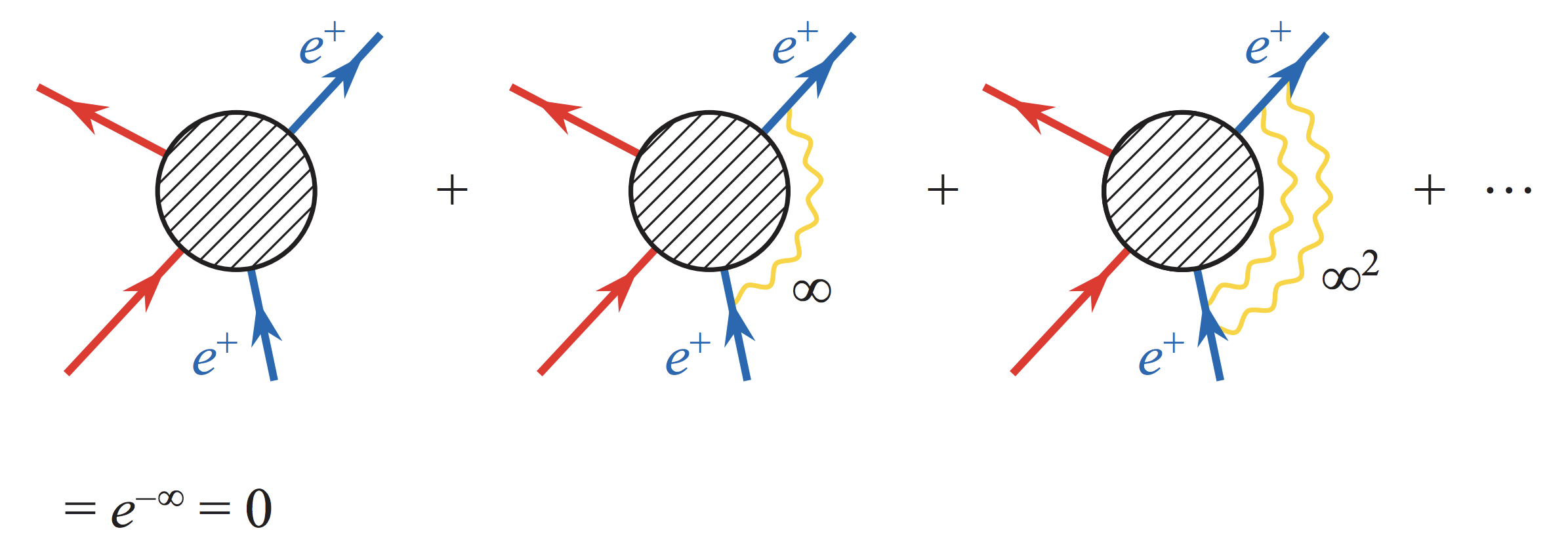}
\end{center}
\caption{\bf\small The resummation of IR divergent  processes, in which pairs of external charged legs exchange soft photons, sets the amplitude for Bhabha scattering to zero. 
} \label{resum}
\end{figure}

This may sound bizarre at first, but in fact the result is well known and is usually attributed to IR divergences. As depicted in figure \ref{resum},
 the tree level one-photon exchange is finite, but the one loop correction is infinite due to IR divergences in pairwise exchanges of soft photons between charged external legs. The source of this divergence is the $\omega\to 0$ pole in the coupling of a soft photon to a charged particle. Including arbitrary numbers of soft photon exchanges and resumming, this divergence exponentiates with a minus sign and sets the full amplitude to zero. 
Here we have provided a simple physical interpretation of that well-known result. IR divergences are not 
``real", rather they are  a clever trick employed  by the Feynman rules to set conservation-law-violating amplitudes to zero! 

Finding a new way of computing zero, however, is not enough: we would like to use our insights to construct finite amplitudes.  The problem above was the mismatched Coulomb fields produced by the 
asymptotic charge fluxes. This mismatch can be avoided by using radiative, rather than the Coulombic,  modes to satisfy the constraints:
\bea\label{cpezq}
\cancelto{0}{\partial_u F^{(2)}_{ru}} + D^z F^{(0)}_{uz} + D^\bz F^{(0)}_{u\bz}   &=&-e^2   j^{(2)}_u,
\cr
 \cancelto{0}{\partial_v F^{(2)}_{rv}} - {D^z F^{(0)}_{vz}} -{D^\bz F^{(0)}_{v\bz}}   &=& e^2   j^{(2)}_v.
\eea
For a single outgoing positron, this has the  radiative shock wave solution\footnote{This solution is singular at $z=\infty$, but in the case of Bhabha scattering with zero net global charge the singularity is cancelled.}   \be\label{rdr} A^{(0)}_z=-{ e^2 \over 4\pi (z-z_0)}\theta(u-u_0).\ee 
The field strength has support only at  at $u=u_0$, but the gauge potential is shifted by the shock wave, indicating a transition between the degenerate vacua. Its frequency space Fourier transform  has the signature ${ \o \to 0}$ pole. The corresponding 
``dressed" coherent quantum state for the outgoing positron, 
\be\label{jui} |j_0\rangle_{\rm dressed}= e^{({i\over 2 \pi}\int {d^2z \over z_0-z}A^{(0)}_\bz(u_0,z,\bz)-h.c.)}|j_0\rangle \; , \ee
is annihilated by the \ip\ constraint equation \eqref{cpezq} (with the fields promoted to operators). 
This state can be described as a positron surrounded by a cloud of soft photons. Dressing all 
the positrons and electrons in this manner, one finds for ``dressed" Bhabha scattering\footnote{When there is net global charge, this procedure sets the leading radial component of the electric field to a nonzero but angle-independent constant.} 
\be \label{mconebs}
F_{ru}^{(2)} (z, \bz) \Big|_{\ci^+_-} =0= F_{rv}^{(2)} (z, \bz) \Big|_{\ci^-_+}~  ,
\ee 
and the matching conditions \eqref{mconeb} are trivially satisfied. Of course, there are many other ways to satisfy the matching condition, but the example \eqref{jui} is particularly simple and illustrates the main point.

\begin{figure}
\begin{center}
\includegraphics[width=6.0 in]{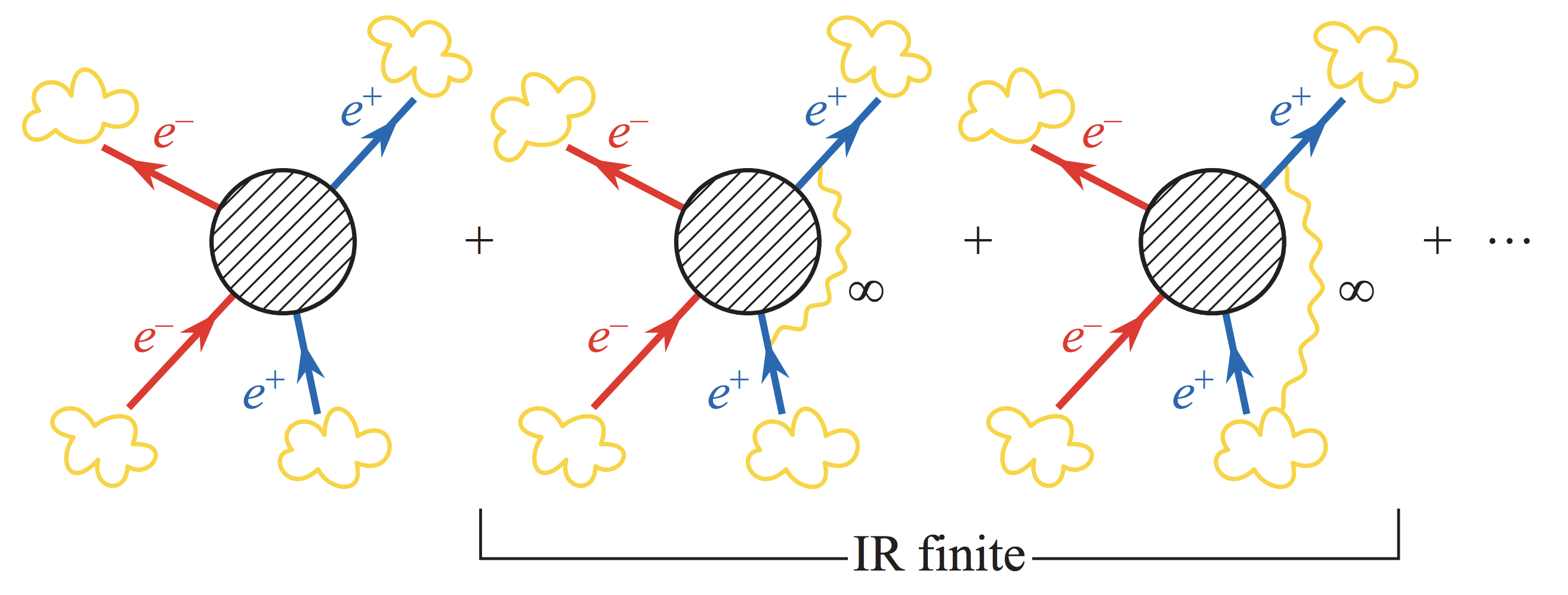}
\end{center}
\caption{\bf\small The Faddeev-Kulish mechanism. IR divergences from soft exchanges between external charges are pairwise cancelled by exchanges between external charges and soft photon clouds surrounding each particle.} \label{rfk}
\end{figure}

Since the conservation laws are satisfied, there is no reason for dressed Bhabha amplitudes to vanish and no need for IR divergences. In fact, these amplitudes are known to be completely  IR finite!\footnote{There are UV divergences in this particular example, as  the positrons, electrons, and shock waves are all infinitely localized. These divergences can be eliminated by smearing but are not our interest here.} The dressed state here is a special case of the ones shown by Kulish and Faddeev  
in 1970 \cite{Kulish:1970ut} to have  IR finite scattering amplitudes. As shown in figure \ref{rfk}, there is a pairwise cancellation of IR divergences in soft photon exchanges between pairs of external charges and external charges and soft clouds.\footnote{More generally, cloud-cloud exchanges also participate in the cancellations.}    
We can now understand the role of Faddeev-Kulish clouds as inserting radiative photons to satisfy the conservation laws.

This is progress, but Faddeev-Kulish states are highly nongeneric.  When the net charge is zero, they all have vanishing leading radial electric fields near spatial infinity as in \eqref{mconebs}. When it is not zero, the leading electric field is nonzero but is independent of the angle, because  the soft photon cloud shields  all the angle-dependent components. Such states are unphysical. When pairs of protons are thrown at one another at the LHC, they are not  followed up with a finely tuned cloud of soft photons to shield the angle-dependent part of the electric field. The $\cs$-matrix $\cs_{FK}$ restricted to Faddeev Kulish states is not unitary: rather $\cs_{FK}^\dagger \cs_{FK}={\cal P}_{FK}$  is a projection operator projecting onto such shielded states. ${\cal P}_{FK}$ projects out the states we scatter in accelerator experiments. 

\begin{figure}
\begin{center}
\includegraphics[width=6.0 in]{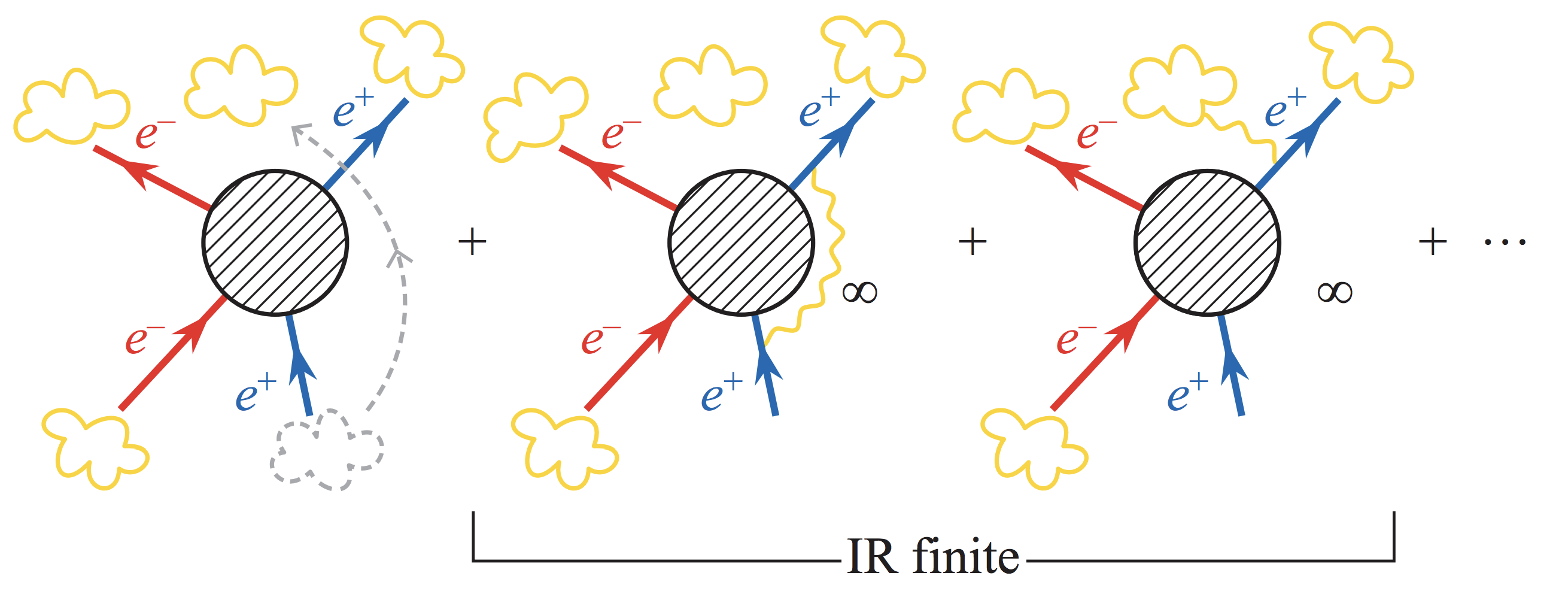}
\end{center}
\caption{\bf\small  IR divergences from soft exchanges between external charges are pairwise cancelled by exchanges between external charges and soft photon clouds associated with, but not necessarily surrounding, each particle. } \label{cjt}
\end{figure}

It is desirable to  have IR finite scattering amplitudes for all physical states. The discussion here suggests that IR divergences arise only to enforce charge conservation.
Accordingly, it was conjectured in \cite{Kapec:2017tkm} that all amplitudes allowed by charge conservation are free of IR divergences.\footnote{This conjecture was recently proven by R. Akhoury and S. Choi \cite{Choi:2017ylo}.} As depicted in figure \ref{cjt}, a 
 leading order analysis corroborates this conjecture. Let us move the  soft cloud surrounding an incoming positron to an antipodal outgoing position in a manner such that all the constraints on $\ci$ are still satisfied. This can be done by omitting the incoming positron  dressing operator and acting with its hermitian conjugate on the out state. When we move the radiative cloud from $\ci^-$ to \ip, the constraints imply nontrivial but matched Coulomb fields will be created near $i^0$. The original  IR divergence from soft photon exchange between the outgoing positron and the soft cloud around the incoming positron is traded for a divergence  from exchange between the outgoing positron and the new antipodally positioned outgoing cloud.  The soft photon theorem, which takes the same form (up to a sign) for ingoing and outgoing photons,  implies that the IR divergent parts of these diagrams are equal, and hence the pairwise cancellation of these leading divergences is still in effect  with the repositioned cloud \cite{Kapec:2017tkm}.

It is possible to antipodally reposition all the incoming clouds to outgoing ones  while maintaining the leading Faddeev-Kulish cancellation of IR divergences.  Physical  scattering processes at accelerators are described  by such incoming charges with no incoming soft clouds. The resulting outgoing soft clouds are then simply identified as the soft radiation inevitably produced in the scattering process.

 %
%
 %
 %
 %
 %

\section{The $\CS$-matrix as a Celestial Correlator}	
\label{smcc}
So far, the framework for our discussion has been the conventional 
one of  an $\CS$-matrix that maps an incoming to an outgoing Hilbert space. Multiparticle states in the in- and out-Hilbert spaces are described as  asymptotic, noninteracting  energy-momentum eigenstates. In this section, we present an alternate description, which for some purposes is both computationally and conceptually more convenient.

\begin{figure}[h] 
\begin{center}
\includegraphics[width=6.0 in]{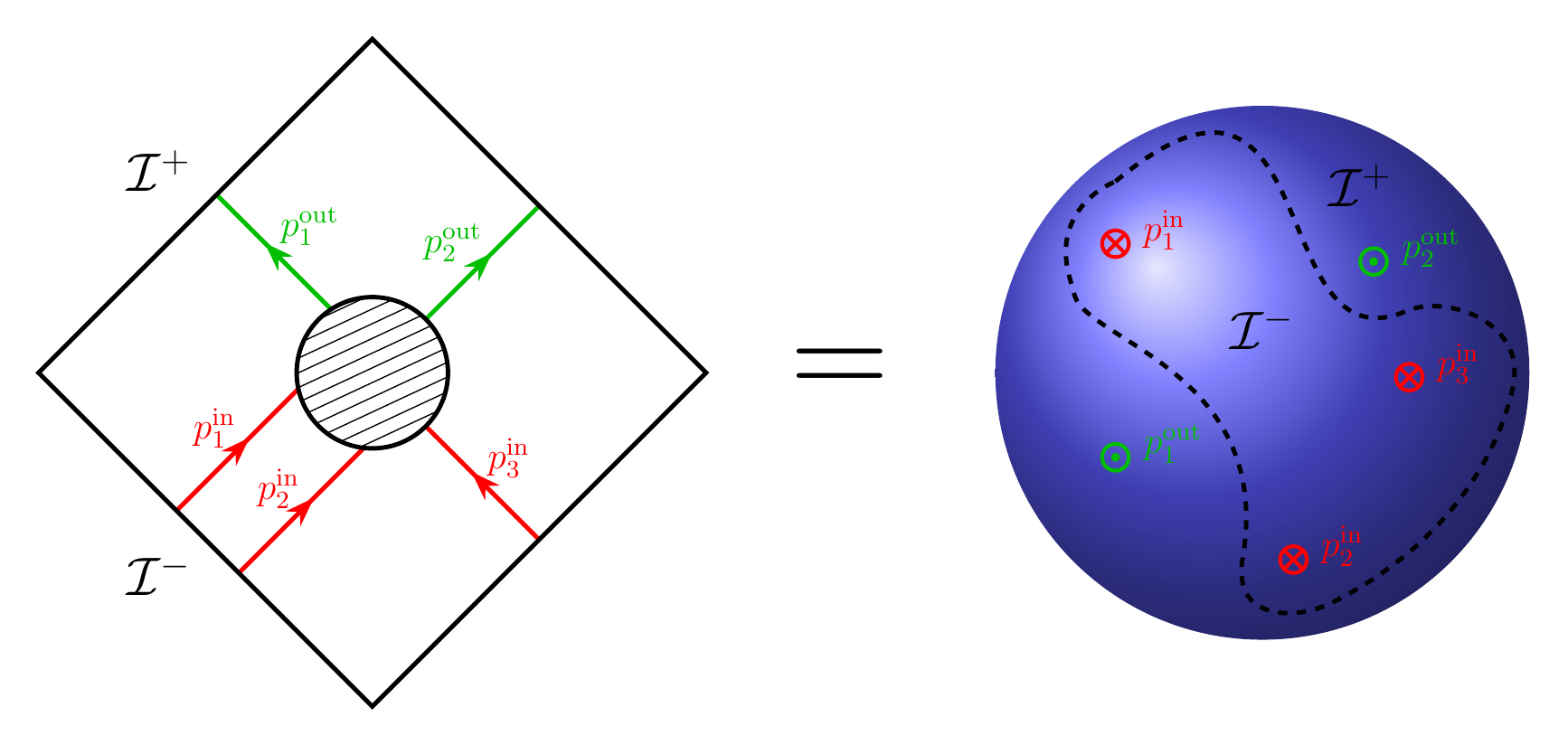}
\end{center}
\caption{\bf\small The four-dimensional Minkowski $\cs$-matrix for any theory can be rewritten as a two-dimensional correlator on the celestial sphere \cst, which is parametrized by the asymptotic angle $(z,\bz)$ on $\ci$. Incoming or outgoing massless particles are represented by operators at the location where they enter $\ci^-$ or exit \ip, labeled by quantum numbers such as the energy or boost charge. The angles on $\ci^+$ and $\ci^-$ are antipodally identified so that a free massless particle enters and exits at the same point on \cst.  The dashed line on \cst\ separates the regions associated with incoming and outgoing particles. The four-dimensional $SL(2, \mathbb{C})$ Lorentz invariance acts as the global two-dimensional conformal group on \cst.} \label{cstwo}
\end{figure}

The alternate description of scattering \cite{He:2015zea} is as  a type of correlation function on a sphere, depicted in figure \ref{cstwo}. 
 In the massless case, ``in'' and ``out'' massless particles are labelled by operators denoted
\be O_k(z,\bz).\ee
Here,  \be z={x^1+ix^2 \over r+x^3}~ \ee denotes the point on the sphere at past or future null infinity where the particle of type $k$ enters or exits the spacetime. This sphere is referred to as the celestial sphere and is denoted \cst. $O_k$ may depend on other quantum numbers, such as the angular momentum, boost weight, energy, or electromagnetic charge, but we often suppress these labels.  
The asymptotic angle  $(z,\bz)$ and either energy or boost weight replace the conventional three-momentum $\vec p$ characterizing a massless particle.

Massive particles in energy-momentum eigenstates, in contrast, do not enter or exit spacetime at a definite point on \cst . Rather, as discussed in section \ref{sec:massive},  they enter or exit $i^\pm$ at a definite point on the hyperboloid $\mhh_3$ in \eqref{pos}. They correspond to operators that are smeared on \cst\ with the weighting given by \eqref{wet}. Alternately, boost-eigenstate wave functions for massive particles associated to a point on \cst\ can be constructed using the bulk-to-boundary propagator on $\mhh_3$ \cite{Pasterski:2016qvg}.

It is then natural to express the $n$-particle scattering amplitudes in the form of  a celestial
correlator on  \cst:
\begin{equation}\label{sew}
\begin{split}
\bra{\text{out}} \CS \ket{\text{in}} \to \avg{ \CO_1(z_1,\bz_1) \cdots \CO_n (  z_n , \bz_n ) } .
\end{split}
\end{equation}
Note that no assumptions are being made here; we are simply rewriting the $\CS$-matrix in a different notation. 

We are interested in Lorentz invariant theories.   The four-dimensional Lorentz symmetry is $SL(2, \mathbb{C})$, which  acts  on \cst\ as \be
z \to \frac{az + b}{cz +d}, \quad ad-bc = 1.
\ee
This formula is very familiar to anyone who has studied two-dimensional CFT, because $SL(2, \mathbb{C})$ is the global conformal group of the two-sphere. 
 In quantum field theory, scattering amplitudes must transform covariantly under the Lorentz group, so these correlation functions must transform covariantly under the global conformal group. Hence they must look a lot like the familiar correlation functions in a two-dimensional CFT on the sphere. This connection  becomes manifest when the external particle wave functions are taken to be $SL(2, \mathbb{C})$ primaries 
 labeled by their conformal dimensions and location on \cst\ rather than the traditional plane waves \cite{Cheung:2016iub,Pasterski:2016qvg,Cardona:2017keg}. 
 
This connection becomes even more interesting later on when we get to quantum gravity. We will see that --- up to currently unresolved issues involving IR finite quantum anomalies --- the global conformal group gets enhanced to the full local conformal group $z \to w(z)$ that appears in two-dimensional CFT.  There is even an operator constructed from soft gravitons \cite{Kapec:2016jld,Cheung:2016iub,He:2017fsb} whose Ward identities are those of a two-dimensional stress tensor. This likely places powerful constraints on the structure of scattering amplitudes that are currently under investigation \cite{Pasterski:2016qvg,Cardona:2017keg}. More ambitiously, one may attempt to find a two-dimensional CFT whose correlators on \cst\ reproduce the $\cs$-matrix of 
a four-dimensional quantum theory of gravity.  This would provide  a microscopic realization  of the holographic principle in four-dimensional flat space quantum gravity, or flat space holography.

At this point in the lectures, we are still considering QFT and have not started to discuss gravity in any depth. Nevertheless, if the QFT can be coupled to gravity, it suggests that the celestial correlators \eqref{sew} are those of a  two-dimensional CFT, possibly in unusual representations.\footnote{The work of Pasterski et al. and Cardona and Huang \cite{Pasterski:2017kqt,Pasterski:2017ylz,Pasterski:2016qvg,Cardona:2017keg} suggests that the principal series plays an important role.} Indeed, as a simple example we will see in the next section that tree-level celestial correlators involving soft gluons are those of  a nonabelian two-dimensional Kac-Moody algebra. A brief discussion of the abelian case  already appeared around equation \eqref{curr}.

\section{Nonabelian Gauge Theory}\label{nonabelian}

Now we turn to another type of soft theorem, namely, the soft gluon theorem.  This theorem applies to theories with a nonabelian gauge group $\CG$ which is neither confined nor higgsed. Otherwise, there are no soft gluons! Unlike the (leading) soft photon theorem, the soft gluon theorem has corrections at the loop level. These are  related to the running of the gauge coupling in the infrared, which affects the coupling constant appearing  in front of the hard part of the charges. However, this is not the whole story of quantum corrections. Even in  $\CN=4$ Yang-Mills, where the coupling does not run, there are still 
corrections to the soft gluon theorem that are one-loop exact \cite{Bern:2014oka, He:2014bga,Cachazo:2014dia,Bern:2014vva}. I have a little bit to say about this below, but the consequences of these anomalies for the asymptotic symmetries remain to be analyzed. 
A related issue in gravity was recently addressed in He et al.\cite{He:2017fsb}.

Soft theorems are essential for 
controlling IR divergences and getting IR finite inclusive cross sections. The absence of an uncorrected quantum soft theorem in the nonabelian case is intertwined with the fact that there is no known unitary $\CS$-matrix for quantum nonabelian gauge theories,
although a large class of IR finite dressed states were constructed in \cite{DelDuca:1989jt,Giavarini:1987ts} and earlier references therein. For a general
state, the best one can do is to define finite inclusive cross sections with an IR
cutoff. 
This is perfectly acceptable for most or all experimental applications, which in practice have an IR cutoff from limits on detector sensitivity. However, it is a  problem, for example, for the theoretical studies of $\CN=4$ Yang-Mills scattering amplitudes, which do not actually exist.  
It  is hard to analyze  unitarity or symmetries of the $\CS$-matrix without an $\CS$-matrix!  The lack of an $\CS$-matrix for these theories is an  elephant in the room. We hope that understanding  the IR symmetries will enable the construction of a finite unitary $\cs$-matrix or a suitable replacement thereof.

In this section, which largely reviews He et al. \cite{He:2015zea}, we sidestep these issues simply by restricting the discussion to tree level.  
The presentation will proceed a bit differently than that for the abelian case. Rather than discuss a canonical formalism with an in- and out-Hilbert space, it is more efficient to illustrate the physics in the language of celestial correlators discussed in section \ref{smcc}. In this language, the scattering of a soft gluon becomes the  insertion of a current into a correlation function on \cst. In section \ref{kmoody}, it is shown that this current obeys the Ward identities of a $\CG$-current algebra,  which provides  a  useful alternate representation of the asymptotic symmetry group. In section \ref{ccharge}, we  construct the infinity of nonabelian conserved charges.

\subsection{ $\CG$-Kac-Moody Algebra}\label{kmoody}
\label{kma}
Following the discussion around expression \eqref{sew}, we adopt  the language of celestial correlators, in which asymptotic particles are represented by operators $\CO_k$ on \cst :
\begin{equation}
\begin{split}
\bra{\text{out}} \CS \ket{\text{in}} =  \avg{ \CO_1(E_1,z_1,\bz_1) \cdots \CO_n ( E_n , z_n , \bz_n ) } .
\end{split}
\end{equation}
The operators $\CO_k$ are taken to be in the $k$th representation
of the  nonabelian gauge group $\CG$, whose generators  satisfy
\begin{equation}
\begin{split}
\big[ T^a_k , T^b_k \big] = i f^{abc} T^c_k ,
\end{split}
\end{equation}
where the $f^{abc}$ are the real, completely antisymmetric structure constants normalized so that 
\begin{equation}
\begin{split}
f^{acd} f^{bcd} = \delta^{ab} = \text{tr} \big[ T^a T^b \big] , 
\end{split}
\end{equation}
with $T^a$ the generators in the adjoint representation, and the trace is over the suppressed color index. The field strength is constructed from the gauge field $\CA_\mu=\CA_{\mu}^aT^a$ as \begin{equation}
\begin{split}
\CF_{\mu\nu} = \p_\mu \CA_\nu - \p_\nu \CA_\mu - i \big[ \CA_\mu , \CA_\nu \big] = \CF^a_{\mu\nu} T^a 
\end{split}
\end{equation}
and obeys the equations of motion
\begin{equation}
\begin{split}
\nabla^\nu \CF_{\nu\mu} - i \big[ \CA^\nu , \CF_{\nu\mu} \big] = g_{YM}^2 j^M_\mu . 
\end{split}
\end{equation}
Here $g_{YM}$ is the gauge coupling, and $j^M$ is the matter color current.
The nonabelian gauge transformations act on the gauge and matter fields as
\begin{equation}
\begin{split}
\delta_\ve \CA_\mu = \p_\mu \ve - i \big[ \CA_\mu , \ve \big] ~, \qquad \delta_\ve \phi_k = i \ve^a T^a_k \phi_k~, \qquad \delta_\ve j_\mu^M = - i \big[ j_\mu^M , \ve \big] . 
\end{split}
\end{equation}
Near $\ci^+$, the gauge field has a large-$r$ expansion as in the abelian case:
\begin{equation}
\begin{split}\label{Afalloff}
\CA_z(u,r,z,\bz) &= A_z(u,z,\bz) + \CO \left( \frac{1}{r} \right) , \\
\CA_r(u,r,z,\bz) &= \frac{1}{r^2} A_r(u,z,\bz) + \CO \left( \frac{1}{r^3 } \right) , \\
\CA_u(u,r,z,\bz) &= \frac{1}{r} A_u(u,z,\bz) + \CO \left( \frac{1}{r^2} \right) . \\
\end{split}
\end{equation}
The field strength then has the expansion
\begin{equation}
\begin{split}
\CF_{ur} = \frac{1}{r^2} F_{ur} + \CO \left( \frac{1}{r^3} \right) , \qquad \CF_{uz} =  F_{uz} + \CO \left( \frac{1}{r} \right) , \qquad \CF_{z\bz} = F_{z\bz} + \CO \left( \frac{1}{r} \right) ,
\end{split}
\end{equation}
where the leading components are given by
\begin{equation}
\begin{split}
F_{ur} = \p_u A_r + A_u , \qquad F_{uz} = \p_u A_z , \qquad F_{z\bz} = \p_z A_\bz - \p_\bz A_z - i \big[ A_z , A_\bz \big] . 
\end{split}
\end{equation}
The asymptotics  \eqref{Afalloff} allow large gauge transformations infinitesimally generated by
\begin{equation}
\begin{split}
\delta_\ve A_z(u,z,\bz) = D_z \ve(z,\bz) . 
\end{split}
\end{equation}
A finite transformation yields an arbitrary flat $\CG$-connection over \cst. These flat connections label the inequivalent vacua. 

In classical Yang-Mills theory, the scattering problem is to find  a map from initial data on $\ci^-$ to final data on $\ci^+$. These data include the connection $A_z$, 
as the gauge field explicitly appears in the equations of motion and therefore must be specified  as part of the initial Cauchy data on $\ci^-$. However, the maximal Cauchy development of incoming data on $\ci^-$ via the field equations cannot by itself fully determine the outgoing data on \ip, because given any one solution, another can always be obtained by acting with a large gauge transformation on \ip.  As in the abelian case, when there are no long-range magnetic fields,\footnote{Although it has  not been carefully worked out, a  generalization that drops this restriction presumably exists, as it is not required in the derivation of the Kac-Moody symmetry which follows.} the ambiguity is eliminated, and the scattering problem becomes well defined if we impose the antipodal boundary condition 
\begin{equation}\label{roi}
\begin{split}
A_z \big|_{\ci^+_-} =  A_z \big|_{\ci^-_+} ,
\end{split}
\end{equation}
along with $F_{ru}|_{\ci^+_-}= F_{rv}|_{\ci^-_+}$. This boundary condition is preserved by large gauge transformations provided the parameter on $\ci^+$ is related to the one on $\ci^-$ by
\begin{equation}
\begin{split}
\ve  (z,\bz)\big|_{\ci^+_-} = \ve (z,\bz) \big|_{\ci^-_+} . 
\end{split}
\end{equation}
Hence given any one solution of the scattering problem, infinitely more may be generated by acting on both the in- and the out-data with such a transformation. 

Note that the boundary condition \eqref{roi} is required simply for the existence of a well-defined classical scattering problem.
The maximal Cauchy development of data on $\ci^-$ gives the data on $\ci^+$ only up to large gauge transformations. Without an additional condition such as \eqref{roi}, we cannot even determine whether the final state is a color singlet. As in the abelian case, the infinity of conservation laws is an immediate consequence of the need for a matching condition, although in this section, we adopt a different approach based on celestial correlators.

The tree-level nonabelian soft theorem can be written \begin{equation}
\begin{split}\label{softgluonthm}
\avg{ \CO_1(p_1) \cdots \CO_n(p_n) \CO^a(q,\ve)}_{U=1} = g_{YM} \sum_{k=1}^n \frac{p_k \cdot \ve}{ p_k \cdot q } \avg{ \CO_1(p_1) \cdots T^a_k \CO_k(p_k) \cdots \CO_n(p_n) }_{U=1}  + \CO(q^0) . 
\end{split}
\end{equation}
Here, $\CO^a(q,\ve)$ is the soft gluon operator of momentum $q$ and polarization $\ve$ with color index $a$ displayed. 
In textbook QFT, a state with a red gluon at the south pole and an anti-red gluon at the north pole is usually said to have zero total color charge. This implicitly assumes that the color connection $A_z=U^{-1}\p_zU$ with $U\in G$ on the $S^2$ is trivial ($U=1$ or constant) so that the gluon does not change color under parallel transport from the north to the south pole.  This assumption is made explicit  in \eqref{softgluonthm} by the $U=1$ subscript. 

In fact, given the boundary condition \eqref{Afalloff}, it is not possible in all contexts to choose the color frame  $U=1$ everywhere. Consider a generic classical solution with an initially trivial flat connection $A_z=0$ with $U=1$ and a pulse of radiation passing through \ip. The final connection will be flat if the system decays to zero energy,  but the equations of motion imply that the final flat connection will not be trivial: $A_z=U^{-1}\p_zU\neq 0$.\footnote{The boundary condition \eqref{Afalloff} prevents us from gauging this away with a time dependent gauge transformation.}  A pair of quarks near \ip\ initially in a color singlet  will generically not be a singlet at late times, due to a change in the color frame.  This is  the ``color memory" effect.  
An explicit formula for the color memory in terms of  a Green's function convoluted with the color flux at \ip, analogous to the one derived for gravity in  \cite{Strominger:2014pwa}, is given in \cite{Pate:2017vwa}.  As in QED, the nonabelian gauge theory vacua are infinitely degenerate and labeled classically by flat connections.  

We define the soft gluon operator at \ip\ just as in the abelian case:
\begin{equation}
\begin{split}
N_z = \int_{-\infty}^\infty du F_{uz} = A_z \big|_{\ci^+_+} - A_z \big|_{\ci^+_-} . 
\end{split}
\end{equation}
Similarly, at $\ci^-$, 
\begin{equation}
\begin{split}
N^-_z = \int_{-\infty}^\infty dv F^-_{vz} = A^-_z \big|_{\ci^-_+} - A^-_z \big|_{\ci^-_-} . 
\end{split}
\end{equation}
Finally, we define the soft gluon current
\begin{equation}
\begin{split}
J_z = - \frac{4\pi }{ g_{YM}^2 }  \left( N_z - N_z^- \right) . 
\end{split}
\end{equation}
Trading  the operator momenta $p_k$ in \eqref{softgluonthm} for $z_k$ and $q$ for $z$,  the soft theorem takes the suggestive form
\begin{equation}
\begin{split}\label{softgluonthm1}
	\avg{ J^a_z \CO_1 (z_1,\bz_1)\cdots \CO_n (z_n,\bz_n)}_{U=1} = \sum_{k=1}^n \frac{1}{z - z_k} \avg{  \CO_1(z_1,\bz_1)  \cdots T^a_k \CO_k(z_k,\bz_k) \cdots \CO_n(z_n,\bz_n)  }_{U=1} .
\end{split}
\end{equation}
This is a very familiar formula in two-dimensional CFT. It is the Ward identity of a holomorphic Kac-Moody symmetry.\footnote{ There is also an  antiholomorphic current $J_\bz$, whose insertions are related to negative helicity gluons. One may consider the double soft limits involving either  $J_z J_w$ or $J_z J_\bw$.  It turns out that the former is unambiguous, whereas the latter depends on the order of limits \cite{He:2015zea}: see exercise 9.  This was not the case with photons and begs for interpretation.}

Consider now the weighted integral of the current around a  contour $\CC$ on \cst\ 
\begin{equation}
\begin{split}
J_\CC [\ve] = \oint_\CC \frac{dz}{2\pi i} \tr{ \ve J_z } , 
\end{split}
\end{equation}
where $\ve(z)$ is any holomorphic function in the interior of $\CC$. 
Insertions of these operators obey
\begin{equation}
\begin{split}\label{Jcepinsert}
	\avg{ J_\CC [\ve]  \CO_1 \cdots \CO_n }_{U=1} =\sum_{k\in\CC}  \avg{  \CO_1  \cdots \ve^a(z_k)T^a_k\CO_k \cdots \CO_n  }_{U=1} . 
\end{split}
\end{equation}
That is, they generate gauge transformations $\ve$ inside $\CC$.

So far, all the amplitudes have been constructed with the trivial connection $U=1$ on \cst. 
Large gauge transformations change this connection, so we would like to understand the more general case. The Ward identity \eqref{Jcepinsert} relates $U=1$ amplitudes to the more general case. Consider, for instance, an infinitesimal change in flat connection
\begin{equation}
\begin{split}
\delta U(z,\bz) =  i \ve(z,\bz) + \dotsc. 
\end{split}
\end{equation}
Suppose we choose $\ve$ so that $\ve = 0$ outside $\CC$ and is nonzero and holomorphic in $\CC$. Then, the change in the correlator under this shift is simply given by a large gauge transformation of the operators themselves: 
\begin{equation}
\begin{split}\label{gaugetransform1}
\delta_\ve \avg{ \CO_1 \cdots \CO_n }_{U=1} &\equiv  \avg{ \CO_1 \cdots \CO_n }_{U=1+i\ve} -\avg{ \CO_1 \cdots \CO_n }_{U=1}\\ &= i \sum_{k\in\CC}\avg{ \CO_1 \cdots \ve^a(z_k)T^a_k \CO_k \cdots \CO_n }_{U=1}  \\
&= i  \avg{ J_\CC [\ve]  \CO_1 \cdots \CO_n }_{U=1}.
\end{split}
\end{equation}
We can characterize this relation by
\begin{equation}
\begin{split}
J_\CC [\ve]  = \int_{R_\CC} d^2 z \g_{z\bz} \ve^a \frac{ \delta }{ \delta U^a }, 
\end{split}
\end{equation}
where $R_\CC$ is the region inside the contour $\CC$. 
In other words, insertions of $J_\CC[\ve]$ generate locally holomorphic large gauge transformations on the boundary that are characterized by a general  flat connection $A_z=U^{-1}\p_zU$ .

\subsection{Conserved Charges}\label{ccharge}
In section \ref{kma}, I presented the  soft gluon theorem as  a Kac-Moody Ward identity. It is also possible to formulate the entire problem as an infinity of charge conservation laws.  The nonabelian analogue of \eqref{chargeconserve} is\footnote{As in section \ref{mnetic}, the magnetic generalization is  obtained by 
replacing $\ast F$ with $F$ in \eqref{der}. It would be interesting to understand the action of 
$S$-duality on the electric and magnetic charges.} 
 
\begin{equation}\label{der}
\begin{split}
Q_\ve^+ = \frac{1}{g_{YM}^2} \int_{\ci^+_-}  \tr{ \ve \ast F }. 
\end{split}
\end{equation} As before, by integrating by parts and using the nonabelian constraint equations, this can be written as an integral over all of $\ci^+$.  The final expression contains a soft term $Q^{+S}_\ve$ and a hard term $Q^{+H}_\ve$. Similar constructions apply on $\ci^-$. The soft term is, of course, related to the current defined here. The precise relationship turns out to be
\begin{equation}
\begin{split}
J_\CC [\ve]   = Q^{-S}_\ve - Q^{+S}_\ve.
\end{split}
\end{equation}
The left-hand side depends only on the value of $\ve$ on the curve $\CC$. This has a meromorphic extension to \cst. On the right-hand side, we take $\ve$ to be given by this extension inside $\CC$ and zero outside $\CC$. Actually, this procedure gives two possible expressions,
depending on which side of $\CC$ we regard as the interior. It is illuminating to check that the equivalence of these two expressions is a consequence of the soft theorem.  

I close with a comment on loop corrections. Rather than using the soft theorem, the nonabelian conservation law $Q_\ve^+ =Q_\ve^-$ can be derived directly  from the nonabelian matching conditions \eqref{roi}, as we did for the abelian case. As discussed, such matching conditions are classically required to determine the future color frame from the past one, and without them the scattering problem makes no sense. This suggests that the conservation laws somehow survive the one-loop quantum corrections \cite{Bern:2014oka, He:2014bga,Cachazo:2014dia,Bern:2014vva} to the soft theorem, perhaps in a modified or shifted form along the lines discussed for gravity in \cite{He:2017fsb}.

\section{Gravity}\label{sec:gravity}Finally  we turn to gravity, where the implications of the IR structure become even more far reaching. Oddly, many of the ideas herein were first understood in the gravitational rather than the seemingly simpler gauge-theoretic context. These lectures have presented the material in ahistorical order.

Section \ref{sec:asymptoticallyflatspacetimes} reviews the basics of asymptotically flat spacetimes and Bondi coordinates. Section \ref{sec:supertranslations} begins with a review of the BMS analysis of asymptotic data and the discovery of supertranslations of \ip\ and $\ci^-$. Then I present the antipodally matched subgroup of past and future supertranslations, which is a symmetry of gravitational scattering. Following the abelian gauge theory case, the conserved charges are derived from the matching condition, shown to generate the symmetry and to imply the leading soft graviton theorem. In section \ref{sec:superrotation}, conserved charges are derived from a matching condition for subleading metric components and their relation to superrotation symmetry (an extension of BMS symmetry), and a new subleading soft theorem are discussed.

\subsection{Asymptotically Flat Spacetimes}
\label{sec:asymptoticallyflatspacetimes}

In the previous sections,  flat Minkowski space in retarded coordinates near $\ci^+$ was described by the metric 
\begin{equation}
\begin{split}
ds^2 = - du^2 - 2 du dr + 2r^2 \g_{z\bz} dz d\bz. 
\end{split}
\end{equation}
We would now like to study gravitational theories in which the metric is asymptotic to, but not exactly equal to, the flat metric.  We will work in Bondi coordinates $(u,r,z,\bz)$, and we abbreviate $\Theta^A=(z,\bz)$. In this gauge, the most general four-dimensional metric takes the form
\begin{equation}
\begin{split}\label{bondimetric}
ds^2 = - U du^2 - 2 e^{2\beta} du dr + g_{AB} \left( d\Theta^A + \frac{1}{2} U^A du  \right)\left( d\Theta^B + \frac{1}{2} U^B du  \right), 
\end{split}
\end{equation}
where
\begin{equation}
\begin{split}\label{detcond}
\p_r \det \left(  \frac{g_{AB}}{r^2}  \right) = 0. 
\end{split}
\end{equation} Equation
\eqref{detcond} implies that $r$ is the luminosity distance.
Note that we have completely fixed here the local diffeomorphism invariance by the conditions $g_{rr} = g_{rA} = 0$ together with \eqref{detcond}. We adopt these coordinates largely because they are used in most of the literature on asymptotically flat spacetimes.  However, important insights have arisen in other gauges, such as the harmonic gauge \cite{ Campiglia:2015kxa, Campiglia:2015lxa}, which may ultimately prove more useful. 

So far, we have not imposed any sort of asymptotic flatness condition. Any geometry can be described locally by the metric \eqref{bondimetric}. Imposing  asymptotic flatness at large $r$ with fixed $(u,z,\bz)$ leads to falloff conditions on the metric components. There is no a priori preferred method of determining what these falloffs must be, and the literature discusses various options. They are typically chosen to be weak enough to allow for all interesting solutions, including  gravity waves, but strong enough to rule out unphysical solutions, such as those  with infinite energy. For the natural  choice made by Bondi, van der Burg, Metzner, and Sachs (BMS) \cite{Sachs:1962wk,Bondi:1962px}, the large-$r$ structure of the metric is constrained to be (see exercise 10)
\begin{equation}\label{bondimetricangularmomentum}
\begin{split}
ds^2 &= - du^2 - 2 du dr + 2 r^2 \g_{z\bz} dz d\bz \\
&+ \frac{2m_B}{r}du^2 + r C_{zz} dz^2 + r C_{\bz\bz}d\bz^2 +D^zC_{zz}d u d z+D^\bz C_{\bz\bz} d u d\bz \\
&  + \frac{1}{r}\left(\frac{4}{3}(N_z+u\p_z m_B)-\frac{1}{4}\p_z(C_{zz}C^{zz})\right)d u d z+c.c.+\ldots, 
\end{split}
\end{equation}
where  $D_z$ is the covariant derivative with respect to  $\g_{z\bz}$;  $C_{zz}$, $m_B$ and $N_z$ depend on $(u,z,\bz)$ but not on $r$. The first three terms in \eqref{bondimetricangularmomentum} are simply the flat Minkowski metric, and the remaining terms are the leading corrections. The ellipsis ``$\ldots$'' involves further subleading terms at large $r$, whose precise structure we will not need. Near $\ci^+$, spacetime is  flat to leading order, and we can relate the retarded Bondi coordinates to the standard Minkowski coordinates  $u = t - r + \ldots$.
Equation \eqref{bondimetricangularmomentum} corresponds to  the large $r$ falloffs:
\begin{equation}
\begin{split}\label{bdyfalloff}
g_{uu} &= - 1 + \CO \left( \frac{1}{r} \right), \qquad g_{ur} = - 1 + \CO \left( \frac{1}{r^2} \right)  , \qquad g_{uz} = \CO(1), \\
g_{zz} &= \CO(r) , \qquad g_{z\bz} = r^2 \g_{z\bz} + \CO(1) ~, \qquad g_{rr} = g_{rz} = 0. 
\end{split}
\end{equation}
So far, no mention of Einstein's equations has been made. \eqref{bondimetricangularmomentum} is simply a geometric constraint that defines the class of spacetimes that we are interested in studying.

The quantity $m_B$ is known as the Bondi mass aspect. For Kerr spacetimes, $m_B=GM$ is 
constant and proportional to the mass. However, in a generic spacetime, it will depend on both the retarded time $u$ and on the angle $(z,\bz) $.  Its integral over the sphere is the total Bondi mass. $N_z$ is known as the angular momentum aspect, because its integrals over the sphere, contracted with a rotational vector field, give the total angular momentum. Our definition here of this term is shifted relative to some of the literature to simplify some formulas below. 
$C_{zz}$ describes gravitational waves.
It is ${1 \over r}$-suppressed relative to the leading metric component and is transverse to \ip.
The ``Bondi news tensor" is defined by 
\begin{equation}\label{nws}
\begin{split}
N_{zz} = \p_u C_{zz}. 
\end{split}
\end{equation}
It is the gravitational analogue of the field strength $F_{uz} = \p_u A_z$, and its square is proportional to the energy flux across \ip. 

Eventually we will need to supplement  \eqref{bdyfalloff} with additional boundary conditions near the boundaries  $\ci^+_\pm$ of \ip. If $N_{zz}$ does not fall off fast enough for $u\to\pm\infty$, the solution will have infinite energy. We discuss this in Section \ref{bms}.

\subsection{Supertranslations}
\label{sec:supertranslations}
We now turn to the  asymptotic symmetries of gravitational theories in asymptotically flat spacetimes. As argued by BMS \cite{Sachs:1962wk,Bondi:1962px}, these are generated by diffeomorphisms that preserve both the Bondi gauge \eqref{bondimetric} as well as the boundary falloffs \eqref{bdyfalloff}.  Since researchers were looking for symmetries that act in the asymptotic region where spacetime is almost flat, it was expected that one would reproduce the isometries of flat spacetime itself, namely, the Poincar\'e group. Had this been the case,  GR would reduce to special relativity at large-distances and weak fields.  Surprisingly, as we now review, what they got instead was an infinite-dimensional group, now called the BMS group. This contains as a subgroup the finite-dimensional Poincar\'e group. However, the four global translations are elevated to a whole function's worth of ``supertranslations" that act independently on each point of the asymptotic sphere. Moreover, as we shall see, GR does not reduce to special relativity at large distances and weak fields. Instead, a large space of degenerate vacua remains. 

\subsubsection{\it BMS Analysis}\label{bms} We will not reproduce here the entire calculation of BMS \cite{Sachs:1962wk,Bondi:1962px}, which is a bit lengthy. We instead make a simplifying assumption that eliminates six Lorentz generators. Namely, we restrict consideration to  diffeomorphisms that have the large-$r$ falloffs:
\begin{equation}\label{fof}
\begin{split}
\xi^u , \xi^r \sim \CO(1) ~, \qquad \xi^z , \xi^\bz \sim \CO \left( \frac{1}{r} \right). 
\end{split}
\end{equation}
This condition is equivalent to the statement that the vector field is $\CO(1)$ at large $r$ in an orthonormal frame, thereby eliminating boosts and rotations that grow with $r$ at infinity.  We return to a discussion of these in section \ref{sec:superrotation}. 

The Lie derivative of the metric components at large $r$ are then
\begin{equation}
\begin{split}
\CL_\zeta g_{ur} &= - \p_u \zeta^u + \CO \left( \frac{1}{r} \right), \\
\CL_\zeta g_{zr} &= r^2 \g_{z\bz} \p_r \zeta^\bz - \p_z \zeta^u + \CO \left( \frac{1}{r} \right),  \\
\CL_\zeta g_{z\bz} &= r\g_{z\bz} \left[ 2 \zeta^r + r D_z \zeta^z + r D_\bz \zeta^\bz \right] + \CO (1), \\
\CL_{\zeta} g_{uu} &= - 2 \p_u \zeta^u - 2 \p_u \zeta^r + \CO \left( \frac{1}{r} \right). 
\end{split}
\end{equation}
Then, requiring that the Bondi gauge conditions \eqref{bondimetric} and falloffs \eqref{bdyfalloff} are both  preserved implies that at large $r$,
\begin{equation}
\begin{split}\label{stransdiff}
\zeta = f \p_u - \frac{1}{r} \left( D^z f \p_z + D^\bz f \p_\bz \right)   + D^z D_z f \p_r + \ldots, 
\end{split}
\end{equation}
where $f(z,\bz)$ is any function of $(z,\bz)$. It turns out that \eqref{stransdiff} also preserves all the remaining conditions. We note here that the last term in \eqref{stransdiff} is derived from the condition that $g_{ur} = - 1 + \CO \left( \frac{1}{r^2} \right)$. In other formulations of asymptotically flat spacetimes (for instance, the one considered by Newman and Unti \cite{Newman:1962cia}, where $g_{ur} = -1$), this term takes on a different form. The first two terms are universal and are directly measured by the gravitational memory effect, which we discuss in Section \ref{sec:memory}.
\begin{figure}[h] 
\begin{center}
\includegraphics[width=3.2 in]{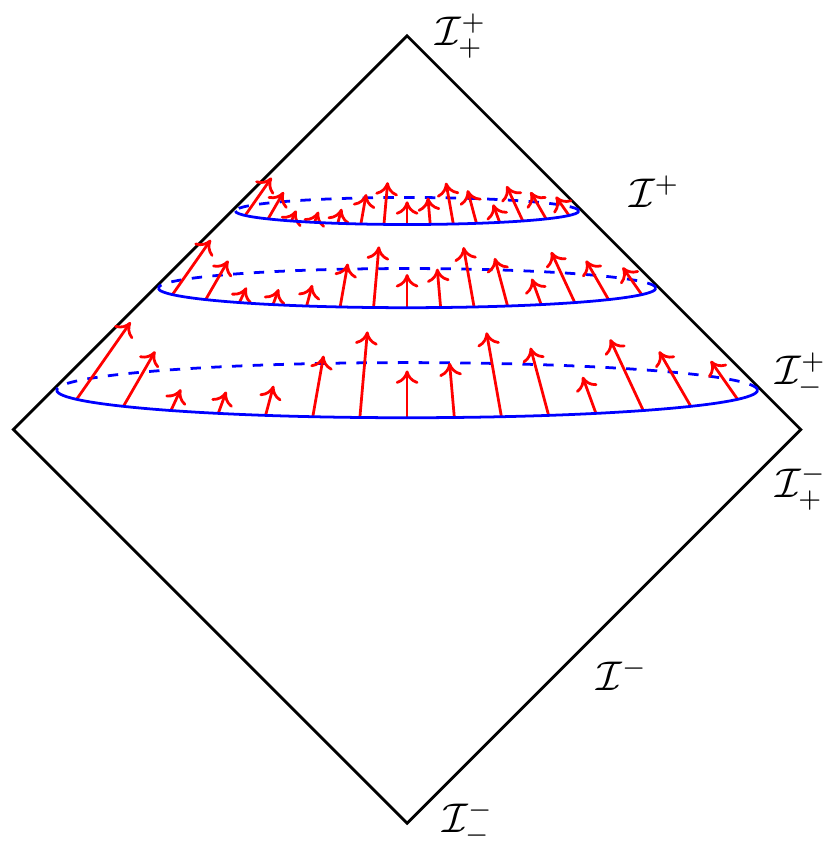}
\end{center}
\caption{\bf \small Under a supertranslation, retarded time $u$ is shifted independently at every angle on $\ci$.} \label{supertranslation}
\end{figure} 
 The transformations generated by \eqref{stransdiff} are called {\it supertranslations}, depicted in figure \ref{supertranslation}. They are  generalizations of the four translations in Minkowski space.  For instance, for $f(z,\bz)=$ constant, \eqref{stransdiff} generates $u$-translations. What is less obvious, though, is that if $f(z,\bz)$ is taken to be the $\ell=1$ harmonic on the sphere, we retrieve the three spatial translations.  The most general $f$ effectively allows separate translations along every null generator of \ip.

Supertranslations transform one geometry into a new, physically inequivalent geometry, despite the fact that they are diffeomorphisms. To see this, consider a solution where an outgoing  pulse of gravitational or electromagnetic waves crosses the south pole of $\ci^+$, and another pulse crosses the north pole of  \ip, both at retarded time  $u=100$. 
Now supertranslate this solution with a function $f(z,\bz)$ that has the property that $f$(south pole)=100 and $f$(north pole)=0. The new solution now has one outgoing pulse at the north pole only at  $u=100$ and one at the south pole only at $u=200$. The outgoing data are measurably changed by the supertranslation.

This simple example shows that the effect of a supertranslation on a solution can be discerned even at the classical level. While the structure  here follows the abelian gauge theory case quite closely,  the large abelian gauge transformations only modified phases of states and therefore are somewhat quantum in nature. For this reason, it can be  less confusing to discuss asymptotic symmetries in the case of gravity.  Although supertranslations were not understood to be a symmetry of gravitational scattering until recently \cite{Strominger:2013jfa,He:2014laa}, the fact that they act nontrivially on the phase spaces at $\ci^\pm$  was understood  fifty years ago by BMS \cite{Bondi:1962px,Sachs:1962wk}. It is odd that this symmetry of gravity was at least partially understood fifty years ago, whereas the precisely analogous symmetries in electromagnetism were only understood in the past few years \cite{Strominger:2013lka,He:2014cra,Kapec:2015ena,Campiglia:2015qka}. This is perhaps in part due to the quantum nature of the electromagnetic symmetries.

The  action of supertranslations on the $\ci^+$ data  $m_B$, $C_{zz}$, and $N_{zz}$  can be determined by computing the Lie derivative of the appropriate component of the metric and then extracting the appropriate coefficient in the large-$r$ expansion. This process gives (see exercise 11)
\begin{equation}\label{dse}
\begin{split}
\CL_f N_{zz} &= f \p_u N_{zz}, \\
\CL_f m_B &= f \p_u m_B + \frac{1}{4} \left[ N^{zz} D_z^2 f + 2 D_z N^{zz}  D_z f + c.c. \right], \\
\CL_f C_{zz} &= f \p_u C_{zz} - 2 D_z^2 f. \\
\end{split}
\end{equation}
The last equation above is especially interesting. Suppose we supertranslate flat Minkowski spacetime described by $m_B = N_{zz} = C_{zz} = 0$. Equation \eqref{dse} implies that the supertranslated spacetime will still have zero Bondi mass and Bondi news and vanishing Riemann tensor. This is consistent with the fact that a diffeomorphism cannot change the physical mass squared or create gravitational waves. However, the supertranslated spacetime does have a nonzero $C_{zz}$. One may check that the vanishing of the curvature in fact requires 
\be\label{sx2} C_{zz}=-2D_z^2C,\ee
 for some function $C(z,\bz)$. Under a supertranslation, 
 \be \CL_fC=f.\ee 
 Hence $C$ is the Goldstone boson of spontaneously  broken supertranslation invariance. It parametrizes  the classically inequivalent  gravitational vacua.  Since the $\ell=0,1$ modes of $C$ are annihilated by $D_z^2$, the four rigid spacetime translations are not broken. 

If one drops the overly restrictive falloffs \eqref{fof} on $\zeta$, one obtains the larger BMS$^+$ group which is a semidirect product of supertranslations with Lorentz transformations on \ip. The four spacetime translations are an ideal of
BMS$^+$ and so may be canonically identified in any BMS$^+$ frame. In general, there is no preferred
Poincar\'{e} subgroup of BMS$^+$. Different subgroups are transformed into one another via conjugation by supertranslations. In particular, this implies that there is no BMS$^+$ invariant definition of angular momentum. This is sometimes referred to as ``the problem of 
angular momentum in general relativity,"  although I view it as a feature, not a bug. However, given any particular classical choice of vacuum, there is a unique Poincar\'{e} subgroup of BMS$^+$ under which it is invariant.\footnote{Thus, classically, there is always an unbroken Poincar\'{e} subgroup of BMS$^+$ in flat space. In the quantum theory, a generic superposition of  vacua will have only the four global translations as an unbroken symmetry. }

So far, we have not utilized any equations of motion. Assume that the geometry  is governed by the Einstein equations
\begin{equation}
\begin{split}
R_{\mu\nu} - \frac{1}{2} g_{\mu\nu} R = 8 \pi G T^M_{\mu\nu} . 
\end{split}
\end{equation}
Since we are here interested in the structure of null infinity, we assume that $T_{\mu\nu}^M$ is a matter stress tensor corresponding to massless modes.\footnote{The massive case is treated by Campiglia and Laddha  \cite{Campiglia:2015kxa, Campiglia:2015lxa}.}
Plugging in the explicit form of the metric \eqref{bondimetricangularmomentum} and expanding in large $r$, we find that the leading $uu$ component of Einstein's equations is
\begin{equation}
\begin{split}\label{mbconstraint}
\p_u m_B = \frac{1}{4} \left[ D_z^2 N^{zz} + D_\bz^2 N^{\bz\bz} \right] - T_{uu} ,  
\end{split}
\end{equation}
where
\begin{equation}
\begin{split}
\label{Tuu}
T_{uu} = \frac{1}{4} N_{zz} N^{zz} + 4\pi G \lim_{r\to\infty} \left[ r^2 T^M_{uu} \right]  . 
\end{split}
\end{equation}
Equations \eqref{mbconstraint} and \eqref{Tuu} constrain the leading data at \ip. 
There is an additional constraint involving $N_z$ from the $uz$ component of the Einstein equation. However, \eqref{mbconstraint} suffices for our current discussion on supertranslations. 

The traceless Bondi news $N_{zz}(u,z,\bz)$ comprises two unconstrained real functions on \ip\ as expected for the two helicities of the massless graviton. We assume that near the past and future boundaries of \ip, $\ci^+_-$ and $\ci^+_+$, the news falls off faster than $1\over |u|$. 
These (and stronger) asymptotic boundary conditions were proven by Christodoulou and Klainerman \cite{Christodoulou:1993uv} to hold in a finite neighborhood of flat space: here we shall consider spacetimes with this asymptotic behavior but do not require them to be near flat space in the deep interior.\footnote{If the news is only required to fall off slowly enough to make the total energy finite, $C_{zz}$ is not  well defined at the boundaries $\ci^+_\pm$. This is one of the reasons it was thought impossible to match the BMS frames of $\ci^\pm$. However, the results of \cite{Christodoulou:1993uv} demonstrate that the region of phase space in which the boundary values of $C_{zz}$ are well defined and the matching (given below) is possible is a physically reasonable one.} The news then trivially determines $C_{zz}$ up to an integration 
function by integrating  (\ref{nws}).  Moreover, the vanishing of the Weyl tensor at $\ci^+_-$ requires $C_{zz}|_{\ci^+_-} =-2D_z^2C|_{\ci^+_-}$
\cite{Strominger:2013jfa}. Therefore we may take the integration function to be $C|_{\ci^+_-}$.
Given the news tensor and these initial data at $\ci^+_-$, the constraints may be integrated to give the mass  $m_B$ and $N_z$ everywhere on \ip. Hence the Cauchy data include 
\be \label{des1} \{N_{zz}(u,z,\bz), C(z,\bz)|_{\ci^+_-}, m_B(z,\bz)|_{\ci^+_-}\}.\ee
At higher orders in $1\over r$, more data are needed, including 
$N_z|_{\ci^+_-}$. This will be important when we consider superrotations below, but is suppressed for now. 

There is an analogous story at $\ci^-$. The metric  in advanced Bondi coordinates $(v,r,z,\bz)$ has the expansion
\begin{equation}
\begin{split}
ds^2 &= - dv^2 + 2 dv dr + 2 r^2 \g_{z\bz} dz d\bz  + \frac{2m_B}{r}dv^2 + r C_{zz} dz^2 + r C_{\bz\bz} d\bz^2 + \dotsc ,
\end{split}
\end{equation}
where $m_B$ and $C_{zz}$ depend on $(v,z,\bz)$ rather than on $(u,z,\bz)$. The angular $(z,\bz)$ coordinates on $\ci^-$ are as usual antipodal to the ones on $\ci^+$, and $v = t + r + \dotsc$. Supertranslations act on $\ci^-$ as
\begin{equation}
\begin{split}
\CL_f N_{zz} = f \p_v N_{zz} , \qquad \CL_f C_{zz} = f \p_v C_{zz} + 2 D_z^2 f,
\end{split}
\end{equation}
which may be enlarged with boosts and rotations to give the action of BMS$^-$ on $\ci^-$. 
The constraint equation takes the form
\begin{equation}
\begin{split}
\p_v m_B = \frac{1}{4}  \left( D^2_z N^{ zz}+  D^2_\bz N^{\bz\bz}\right) + T_{vv} , \qquad T_{vv} = \frac{1}{4} N_{zz} N^{zz} + 4\pi G \lim_{r\to\infty} \left[ r^2 T_{vv}^M \right] . 
\end{split}
\end{equation}
Defining $C_{zz}|_{\ci^-_+}=2D_z^2C|_{\ci^-_+}$, the analog of the Cauchy data \eqref{des1} is 
\be \label{des2} \{N_{zz}(v,z,\bz), C(z,\bz)|_{\ci^-_+}, m_B(z,\bz)|_{\ci^-_+}\} .\ee

\subsubsection{\it The scattering problem}\label{sec:Matching}
The scattering problem in classical general relativity is, roughly speaking, to find the map from Cauchy data on $\ci^-$ to that on $\ci^+$.\footnote{Of course, if a black hole is formed, we need Cauchy data on \ip$\cup \CH$, where $\CH$ is the future horizon, but for now, we assume that black holes are absent.} Such a map is not even formally determined from the maximal Cauchy development of the $\ci^-$ data (\ref{des1}) with the Einstein equation. This determines the data on \ip\  at most up to a supertranslation, or more generally, a BMS$^+$ frame. A prescription is needed to  attach \ip, choose a future BMS$^+$ frame, and determine the initial values for integrating $m_B$ (and $N_z$) along \ip\ using the constraints. {{\it Without such a prescription, the scattering problem in GR is not defined.}}  In  \cite{Strominger:2013jfa} 
it was proposed that the BMS$^+$ frame should be determined by the Lorentz- and CPT-invariant matching conditions 
\be  \label{cmm}C(z,\bz)|_{\ci^+_-} =C(z,\bz)|_{\ci^-_+}   ,~~~m_B(z,\bz)|_{\ci^+_-}=m_B(z,\bz)|_{\ci^-_+} .\ee
The matching condition for $N_z$ is discussed below. Conditions \eqref{cmm} break the combined $\mathrm{BMS^+\times BMS^-}$ action on \ip\ and  $\ci^-$ down to the diagonal subgroup that preserves these conditions, namely,
\be\label{diq}  f(z,\bz)|_{\ci^+_-}=f(z,\bz)|_{\ci^-_+}~ .\ee This condition fixes the BMS$^+$ frame in terms of the BMS$^-$ frame. The diagonal subgroup generated by \eqref{diq} remains as a symmetry of gravitational scattering. Given one solution of the scattering problem, this group generates infinitely many more. 

 With our conventions, (\ref{cmm}) {\it antipodally} equates past and future fields near spatial infinity.  We expect that it is the only Lorentz and CPT invariant choice, and it is implicit in most or all GR computations in asymptotically flat spacetimes. The matching condition (\ref{cmm}) was proven \cite{He:2014laa} to be implicit to all orders in standard weak field perturbation theory by demonstrating its equivalence to Weinberg's soft graviton theorem \cite{Weinberg:1965nx}. 
This perturbative analysis motivates the proposal  that (\ref{cmm}) is part of the definition of the scattering problem whenever the fields are sufficiently weak near spatial infinity, even if the interior contains a black hole.

\subsubsection{\it Conserved Charges}
\label{sec:WeinbergSoft}
The mere existence of infinitely many matching conditions, one for every point on the celestial sphere, implies  an infinite number of conserved charges. 
Following closely the gauge theory discussion, the supertranslation charges are \begin{equation}
\label{SupertranslationCharges}
\begin{split}
Q^+_f &= \frac{1}{4\pi G} \int_{\ci^+_-} d^2 z \g_{z\bz} f m_B , \\
Q^-_f &= \frac{1}{4\pi G} \int_{\ci^-_+} d^2 z \g_{z\bz} f m_B .  
\end{split}
\end{equation} 
The matching conditions \eqref{cmm} immediately  imply the conservation law
\begin{equation}
\begin{split}\label{Qfeq}
Q^+_f = Q^-_f~. 
\end{split}
\end{equation}
Integrating by parts, using the constraint equation, and assuming $m_B$ decays to zero  in the far future, we can write
\begin{equation}
\begin{split}
Q^+_f &=\frac{1}{4\pi G} \int_{\ci^+} du d^2 z \g_{z\bz} f \left[ T_{uu} - \frac{1}{4}  \left(D^2_z N^{zz}+  D^2_\bz N^{ \bz\bz} \right) \right]  , \\
Q^-_f &=\frac{1}{4\pi G} \int_{\ci^-} dv d^2 z  \g_{z\bz}f \left[  T_{vv} + \frac{1}{4}  \left( D^2_z N^{ zz}+  D^2_\bz N^{ \bz\bz}\right) \right]  . \\
\end{split}
\end{equation}

To understand the conservation law better,  suppose we take $f(z,\bz) = \delta^2(z-w)$. Then \eqref{Qfeq} implies that the integrated (over $u$) energy flux at a point $w$ on $\ci^+$ is equal to the integrated energy flux at the antipodal point $w$ on $\ci^-$:\begin{equation}\label{angl}
 \int_{\ci^+} du \g_{z\bz}  \left[ T_{uu} - \frac{1}{4}  \left(D^2_z N^{zz}+  D^2_\bz N^{ \bz\bz} \right) \right] = \int_{\ci^-} dv  \g_{z\bz} \left[ T_{vv} + \frac{1}{4}  \left( D^2_z N^{ zz}+  D^2_\bz N^{ \bz\bz}\right) \right]  . 
\end{equation} Here, the local energy at a point includes not only the usual term involving the stress tensor, but also a crucial extra term that is linear in the gravitational field and is also a total $u$ derivative (i.e., it is a soft graviton contribution to the local energy). Thus, the conservation law \eqref{Qfeq} is the statement that {\it energy is conserved at every angle}.

 In the quantum theory, conserved charges commute with the $\mathcal{S}$-matrix:
\begin{equation}
\begin{split}
Q^+_f \CS - \CS Q^-_f = 0 .
\end{split}
\end{equation}
We can now sandwich this between in- and out-states. The resulting Ward identity is of course equivalent to Weinberg's soft graviton theorem, \begin{align}\label{stw}
	\bra{\mathrm{out}}a_\pm\mathcal{S}\ket{\mathrm{in}}= \sqrt{8 \pi G}\sum_k\frac{\ve^{\pm \mu\nu}p_{k\mu }{p_{k\nu}}}{q\cdot p_k}\bra{\mathrm{out}}\mathcal{S}\ket{\mathrm{in}},
\end{align}
where $a_\pm$  annihilates a helicity $\pm$ graviton. 
We already derived this formula using Feynman diagrams in section \ref{diagrams}. 
 The proof \cite{He:2014laa} that \eqref{stw} and \eqref{angl} are equivalent is almost exactly the same as in the  gauge theory case and is not repeated here.

One expects that, in a Hamiltonian formulation, supertranslation symmetries corresponding to $f$ are generated via Dirac brackets or commutators with $Q^+_f$: \be\label{rm} [ Q^+_f,\ldots]=i\delta_f,~~~~~[Q^+_f,H]=0,~~~~H=Q^+_{f=1}.\ee  
This guess turns out to be correct, but, as in the gauge theory case, verification is quite subtle \cite{He:2014laa}.
It requires computing the terms in the symplectic form involving the Goldstone mode $C$ and the soft graviton mode $N$. On the physical phase space one must impose the vanishing of the Weyl tensor as a constraint at the boundaries $\ci^+_\pm$ of \ip. This changes various commutators by important factors of 2, yielding exactly \eqref{rm}.   These factors of 2 are, as in the gauge case,  related to the fact that soft gravitons and Goldstone modes are both described by a single real scalar field, despite the existence of two polarizations of the former \cite{He:2014laa}.

The existence of these conserved charges is in principle experimentally verifiable.  Indeed, proposed  tests of the gravitational memory effect \cite{Lasky:2016knh,vanHaasteren:2009fy,Wang:2014zls}, although not initially recognized as such, are tests of supertranslation charge conservation \cite{Strominger:2014pwa}. 

\newcommand{\ab}[1]{\left|#1\right|}
\newcommand{\av}[1]{\left\langle#1\right\rangle}
\newcommand{\br}[1]{\left[#1\right]}
\newcommand{\cu}[1]{\left\{#1\right\}}
\newcommand{\pa}[1]{\left(#1\right)}
\renewcommand{\bra}[1]{\left\langle#1\right|}
\renewcommand{\ket}[1]{\left|#1\right\rangle}
\renewcommand{\ed}{\,\mathrm{d}}
\renewcommand{\pd}{\partial}

\renewcommand{\e}{\epsilon}
\newcommand{\wb}{{\bar{w}}}
\renewcommand{\H}{\mathcal{H}}
\newcommand{\scri}{\mathcal{I}}
\renewcommand{\O}{\mathcal{O}}
\renewcommand{\L}{\mathcal{L}}

\subsection{Superrotations}\label{sec:superrotation}

Our analysis so far has centered around $C_{zz}$ and the Bondi mass $m_B$, which are the first nontrivial corrections to the flat metric near $\ci$. Proceeding in the large-$r$ expansion, at the next order we encounter the angular momentum aspect $N_z$, which bears the same relation to the total angular momentum that the $m_B$ bears to the total mass. In this subsection, we see that the matching condition for $N_z$ leads to conserved superrotation charges, just as the matching condition for $m_B$ led to conserved supertranslation charges.

\subsubsection{\it Conserved Charges}

The angular momentum aspect $N_z$ is subject to the constraint equation $G_{uz}=8\pi GT_{uz}^M$. The leading  ${uz}$ component of the Einstein equations is (compare to \eqref{mbconstraint} for the analogous constraint on $m_B$)
\begin{align}
	\label{AngAspectConstraint}
	\pd_uN_z=\frac{1}{4}\pd_z\pa{D_z^2C^{zz}-D_\bz^2C^{\bz\bz}}-u\pd_u\pd_zm_B-T_{uz} .
\end{align}
Here we have introduced the (rescaled) momentum density in the $z$-direction of the gravitational field (compare to \eqref{Tuu} for $T_{uu}$):
\begin{align}\label{tuz}
	T_{uz}\equiv 8\pi G\lim_{r\to\infty}\br{r^2 T_{uz}^M}-\frac{1}{4}\pd_z\pa{C_{zz}N^{zz}}-\frac{1}{2}C_{zz} D_zN^{zz}.
\end{align}
$N_z$ is constrained in relation to  a momentum density $T_{uz}$, in contrast to $m_B$, which was constrained in relation to  the energy density $T_{uu}$.
Once $m_B$ and $C_{zz}$ are fully specified, then $\pd_u N_z$ is also specified by \eqref{AngAspectConstraint}. This fixes $N_z$ up to an integration function. 
We fix this function by the matching condition \cite{Kapec:2014opa} \be  \label{nm}N_z(z,\bz)|_{\ci^+_-}=N_z(z,\bz)|_{\ci^-_+}, \ee 
which is just like the matching condition for $m_B$.  A new subleading soft graviton theorem was proven  \cite{Cachazo:2014fwa,Geyer:2014lca,White:2014qia}, to all orders using Feynman tree diagrams,\footnote{Quantum mechanically, (\ref{nm}) is possibly deformed by an anomaly at one-loop \cite{Bern:2014oka,Cachazo:2014dia,He:2014bga,Bern:2014vva,Broedel:2014fsa,Broedel:2014bza}. Since $some$ matching relation of the form (\ref{nm}) must exist for gravitational scattering to be defined, this reasoning suggests that these one-loop corrections deform rather than eliminate the conserved charges. This is an important open problem on which recent progress has been made \cite{Kapec:2016jld,He:2017fsb,Kapec:2017tkm}. } and then shown to imply (\ref{nm}) \cite{Kapec:2014opa,Campiglia:2015yka}.  Motivated by this perturbative analysis, we propose  that (\ref{cmm}) and (\ref{nm}) are part of the definition of the scattering problem whenever the fields are sufficiently weak near spatial infinity, even if the interior contains a black hole.

 Condition \eqref{nm} implies a second infinity of conserved charges, which can be constructed from an arbitrary vector field $Y^z$ on the sphere.  Using (\ref{nm}), we find 
 \be \label{asd} Q^+_Y={1 \over 8 \pi G}\int_{\ci^+_- }d^2z(Y_{\bar{z}}N_z+Y_z N_\bz)={1 \over 8  \pi G}\int_{\ci^-_+}d^2z(Y_{\bar{z}}N_z+Y_z N_\bz)=Q^-_Y.\ee
Equation \eqref{asd} expresses conservation of superrotation charge. The special cases for which $Y^z$ is one of the 6 global conformal Killing vectors on $S^2$ are conservation of ADM angular momentum and boost charge, sometimes referred to as the BORT (Beig-O'Murchada-Regge-Teitelboim) \cite{Regge:1974zd,Beig:1987zz} center-of-mass. Choosing the vector field to be a delta function, 
 these new conservation laws equate net ``in'' and ``out'' angular momentum flux for every angle. 
A recent proposal to test superrotation charge conservation via the gravitational spin memory effect \cite{Pasterski:2015tva} was made by Nichols \cite{Nichols:2017rqr}.

\subsubsection{\it Symmetries}

Given an infinite number of conserved charges that generalizes angular momentum and boost charges, one naturally expects an infinite number of symmetries generalizing Lorentz transformations. Indeed, conjectures that such a superrotation symmetry exists have been made from several different perspectives, as reviewed in this subsection.  We will see that  subtleties arise in part because a finite symmetry action changes the asymptotics, the consequences of which remain to be analyzed. 

Aspects of superrotation symmetry were implicit in early work of Penrose \cite{Penrose:1972xrn}, but were first conjectured in their modern form in the prescient work of de Boer and Solodukhin \cite{deBoer:2003vf}, who were trying to understand holography in Minkowski space. 
These authors considered the hyperbolic slicing of Minkowski space, which is discussed in section \ref{sec:massive}, and argued that the parameter $\tau$ that labels the slices should be viewed as the coordinate of an internal space. This describes  four-dimensional Minkowski space as a ``noncompactification'' to the three-dimensional hyperbolas $\mathbb{H}_3$.  Because the internal space is noncompact, the effective field theory on $\mathbb{H}_3$ contains a continuum of representations of the Lorentz isometry group.  De Boer and Solodukhin \cite{deBoer:2003vf} conjecture that standard holographic arguments nevertheless apply. In particular, the asymptotic analysis of Brown and Henneaux for AdS$_3$\cite{Brown:1986nw} implies that Lorentz symmetry should be enhanced to the Virasoro symmetry, or equivalently, to superrotations. 

A second line of reasoning that leads to superrotations arose while  revisiting the original BMS analysis. It was suggested by Banks \cite{Banks:2003vp} in a footnote, and independently in greater detail by Barnich and Troessaert \cite{Barnich:2009se,Barnich:2011ct,Barnich:2010eb,Barnich:2011mi}, that some of the falloff assumptions in BMS are overly restrictive. We now review this illuminating perspective.

 In the course of the derivation of the supertranslations in section \ref{sec:supertranslations},  an important restriction was imposed: we required that the components of the vector field $\zeta$ be bounded in an orthonormal frame as the latter approaches $\scri^+$. However, in general this is clearly too strong an assumption, because it rules out both boosts and rotations. One should (and BMS did \cite{Bondi:1962px,Sachs:1962wk}) do an analysis without that assumption. We now summarize the salient points of this analysis and show along the way how the supertranslations naturally generalize to superrotations.

Lorentz Killing vectors are of the general  form
\begin{align}
	\zeta_Y=\pa{1+\frac{u}{2r}}Y^z\pd_z-\frac{u}{2r}D^\bz D_zY^z\pd_\bz-\frac{1}{2}(u+r)D_zY^z\pd_r+\frac{u}{2}D_zY^z\pd_u+c.c.,
\end{align}
where $(Y^z,Y^\bz)$ is a two-dimensional vector field  on \cst. At null infinity $\zeta_Y$ simplifies to
\begin{align}
	\zeta_Y|_{\scri^+}=Y^z\pd_z+\frac{u}{2}D_zY^z\pd_u+c.c..
\end{align}
The claim is (see exercise 12) that for  flat  Minkowski space,  described by the first three terms of  \eqref{bondimetricangularmomentum},  if we take
\begin{align}
	Y^z= 1,z,z^2,i,iz,iz^2,
\end{align}
then the  six real vector fields $\zeta_Y$ generate  the Lorentz transformations. 

Instead of restricting ourselves to $Y^z\sim1,z,z^2$, let us instead compute, for a general $Y^z$, the Lie derivative with respect to $\zeta_Y$ (denoted by $\L_Y$) of various components of the metric. We find that 
\begin{align}
	\L_Yg_{ur}&=\O\!\pa{\frac{1}{r^2}},\\
	\L_Yg_{zr}&=\O\!\pa{\frac{1}{r}},\\
	\L_Yg_{z\bz}&=\O(r),\\
	\L_Yg_{uu}&=\O\!\pa{\frac{1}{r}},\\
	\L_Yg_{\bz\bz}&=2r^2\gamma_{z\bz}\pd_\bz Y^z+\O(r).
	\label{eq:Holomorphic}
\end{align}
Let us focus now on the last equation, which is the most important one. The falloff conditions \eqref{bdyfalloff} are obeyed if the first $\O (r^2)$ term vanishes. This requires that $Y^z$ be a holomorphic vector field. Locally this is solved by  $Y^z=z^n$ for any integer $n$. However, only the restricted choice $Y^z\sim1,z,z^2$ leads to globally defined vector fields, the six global conformal Killing vector fields of the sphere. BMS discarded all the rest on the grounds that equation \eqref{eq:Holomorphic} is violated at isolated points in the more general meromorphic case. For example, if $Y^z=\frac{1}{z-w}$, then
\begin{align}
	\pd_\bz Y^z=2\pi\delta^2(z-w)\neq0,
\end{align}
and the falloff condition is violated at $z=w$.\footnote{This violation can be physically interpreted as due to cosmic strings \cite{Strominger:2016wns}.}
The Lie bracket algebra of $Y^z=z^n$ for any $n$ is the centerless Virasoro algebra.

Interestingly, exactly the same equation \eqref{eq:Holomorphic} was encountered some 20 years later by Belavin, Polyakov, and Zamolodchikov \cite{Belavin:1984vu} in analyzing the symmetries of two-dimensional CFTs. However Belavin et al. chose to allow $Y^z$ to be any meromorphic function on the sphere, allowing for analytic singularities. This was a very good idea. While every practitioner of two-dimensional CFT will agree that this is the right thing to do, different explanations are given to support it.  One is that the symmetries can be analyzed in a local patch, and singularities outside the patch do not affect the local identities obtained with the symmetry. Another is that the Ward identities of these symmetries provide quick derivations of relations among correlators that can be independently derived using other, more tedious methods.  This justification readily adapts to four-dimensional quantum gravity where, as shown in the next subsection,  the Ward identities are equivalent to the new subleading soft theorem   \cite{Cachazo:2014fwa,Geyer:2014lca,White:2014qia}, as well as to the superrotation conservation laws of the previous subsection.
 
 In any case, since allowing meromorphic vector fields was such a good idea in two-dimensional CFT, perhaps we should go beyond the six global vector fields that were retained by BMS in their original analysis and instead consider the whole infinite-dimensional local conformal group generated by $Y^z=z^n$ for all $n$. This is a tantalizing conjecture, because we have learned an incredible amount about two-dimensional CFTs simply by unraveling the powerful implications of the local  conformal = Virasoro group.  One may hope for similarly powerful insights into four-dimensional quantum gravity. Perhaps we have even gotten our foot in the door of the long-sought holographic description of quantum gravity in Minkowski space, in which bulk quantum gravity in Minkowski space is dual to an exotic CFT on \cst. The conjectured  superrotation symmetry will be verified at linearized and tree level in section \ref{sst}. 

\subsubsection{\it Canonical Formalism}
It is natural to ask whether superrotation charges generate superrotation symmetries in a canonical formalism.  We will see that the answer to this question is yes, but only at linearized order. 
As discussed at the end of this subsection, singularities prevent the exponentiation of the infinitesimal transformations. 

As with supertranslations, we begin by investigating how the boundary data specifying the geometry change under superrotations; see Barnich and Troessaert \cite{Barnich:2011mi} for further details. The Lie derivative with respect to $Y$ (denoted $\delta_Y$) of the $C_{zz}$ component of the metric is given by
\begin{align}
	\delta_{Y}C_{zz}&=\frac{u}{2}D\cdot Y N_{zz}+Y\cdot D C_{zz}-\frac{1}{2}D\cdot Y C_{zz} + 2 D_zY^z C_{zz} - u D_{z}^3 Y^{z}.
\end{align}
Taking the $u$ derivative yields
\begin{align}
	\label{eq:SuperrotatedBondiNews}
	\delta_{Y}N_{zz}&= \frac{u}{2}D\cdot Y  \pd_u N_{zz}+Y\cdot D N_{zz} +2D_z  Y^z N_{zz}-  D_{z}^3 Y^{z}.
\end{align}
Note that if we sit at $u=0$, then the last three terms in \eqref{eq:SuperrotatedBondiNews} are exactly the infinitesimal transformation law of a stress tensor in a two-dimensional CFT (i.e., the linearization of the Schwarzian derivative). 

Now consider the conserved superrotation charge\begin{align}
	Q_Y^+=\frac{1}{8\pi G}\int_{\scri_-^+} d^2z [Y_\bz N_z + Y_zN_{\bz}]~.
\end{align}
Integrating by parts and using the constraints \eqref{AngAspectConstraint} gives 
\begin{align}\label{dfr}
	Q_Y^+&=Q_H^++Q_S^+,\nonumber\\	Q_S^+&=-\frac{1}{16\pi G}\int_{\scri^+} d u d^2z [D_z^3Y^z u{N^z}_\bz+D_\bz^3Y^\bz u{N^\bz}_z] ,\nonumber \\
	Q_H^+&=\frac{1}{8\pi G}\int_{\scri^+} d u d^2z\pa{Y_\bz T_{uz}+Y_z T_{u\bz}+u\pd_z Y_\bz T_{uu}+u\pd_\bz Y_z T_{uu}},
\end{align}
where the graviton stress-energy component $T_{uz}$ was defined in \eqref{tuz}. The soft charges are linear in the metric flutuation $C_{zz}$, while the hard charge is quadratic.
To compute commutators, we need \cite{Ashtekar:1978zz,Ashtekar:1981bq,Ashtekar:1981sf,Ashtekar:1987tt}
\begin{align}
	\label{eq:NewsComm}
	\br{N_{\bz\bz}(u,z,\bz),C_{ww}(u',w,\wb)}=16\pi Gi\gamma_{z\bz}\delta^2(z-w)\delta(u-u').
\end{align}
It follows that 
\begin{align}
	\label{eq:SoftTrans}
	\br{Q_S^+,C_{zz}}&=-iuD_z^3Y^z,\\
	   \br{Q_H^+,C_{zz}}&=\frac{iu}{2}D\cdot Y N_{zz}+iY\cdot D C_{zz}-\frac{i}{2}D\cdot Y C_{zz} + 2 iD_zY^z C_{zz}.
\end{align}
Putting this together, we conclude that 
\be \br{Q_Y^+,\ldots}=i\delta_Y.\ee
The conserved charge generates the symmetry as expected. 

However, something is not quite right.  Commutators with $Q^+_Y$ shift the news by a function that goes to a 
constant at $\ci^+_\pm$, while $C_{zz}$ diverges linearly. Such behavior violates the boundary conditions we started off with and hence maps points in the initially defined phase space to points outside that phase space. To make good sense of this behavior, we need to define a larger phase space. This is related to the observation \cite{Strominger:2016wns}  that singularities produced by superrotations correspond to cosmic strings, which destroy asymptotic flatness. Hence, to fully understand superrotations, we will at a minimum need to consider a larger phase space that allows for such defects. This is a topic for future investigations.

\subsubsection{\it Subleading Soft Theorem}\label{sst}

Given the  infinities of conserved superrotation charges and symmetries, one expects an associated soft theorem. There is of course 
Weinberg's soft graviton theorem, but we have already seen that this  is equivalent to supertranslations. Invoking the IR triangle (Section \ref{The Infrared Triangle}) leads to the conclusion that there must be a second soft theorem in gravity. Indeed, motivated by this observation, a new subleading soft theorem was proven diagrammatically \cite{Cachazo:2014fwa,Jackiw:1968zza,
Gross:1968in,White:2011yy}  and shown \cite{Kapec:2014opa,Campiglia:2014yka} to be equivalent to the superrotation charge conservation equation \eqref{asd}. This logic can be turned around and the subleading soft theorem used to prove \eqref{asd}.  

At the level of the quantum  $\cs$-matrix, superrotation charge conservation states
\begin{align}\label{dtp2}
	\bra{\mathrm{out}}\pa{Q_Y^+\mathcal{S}-\mathcal{S}Q_Y^-}\ket{\mathrm{in}}=0.
\end{align}
From the  decomposition  \eqref{dfr} it is clear that this identity equates a soft graviton insertion to a hard term. Equation \eqref{dtp2}  is most directly expressed in terms of the $(z_k,\bz_k)$ coordinates at which asymptotic particles pierce \cst. After quite a lengthy calculation \cite{Kapec:2014opa} using relations such as \eqref{wc} and setting $Y^z={1 \over z- (q^1+iq^2 /q^0+q^3)}$, one finds this equation can  be reexpressed in momentum space as \be \begin{split}
	\lim_{\omega\to0}\pa{1+\omega\pd_\omega}\bra{p_{n+1},p_{n+2},\ldots}a_-(q)\mathcal{S}\ket{p_1,p_2,\ldots}=\\
	\sqrt{8\pi G} S^{(1)-}\bra{p_{n+1},p_{n+2},\ldots}\mathcal{S}\ket{p_1,p_2,\ldots},
\end{split}\ee
where 
$a_-(q)$ is the annihilation operator for a negative helicity graviton of four-momentum $q=\omega(1,\hat{q})$, and the subleading soft factor is 
\begin{align}
    \label{Subleadingfactor}
	S^{(1)-}=-i\sum_k\frac{p_{k\mu}\varepsilon^{-\mu\nu}q^\lambda J_{k\lambda \nu}}{p_k\cdot q},\quad J_{k\mu\nu}\equiv L_{k\mu\nu}+S_{k\mu\nu},
\end{align}
with $L_{k\mu\nu}$ the orbital angular momentum and $S_{k\mu\nu}$ the helicity of the internal spin of the $k$th particle. In reference \cite{Cachazo:2014fwa} it was directly  shown, through some diagrammatics far more tedious than Weinberg's, that this subleading expression is valid at tree level. This confirms the classical  existence of an infinite number of  conserved superrotation charges.

Comments about this soft relation are now in order. Recall from our discussion of supertranslations that the addition of a soft graviton to a scattering amplitude produces a Weinberg pole proportional to $1/\omega$. But here, the prefactor $\pa{1+\omega\pd_\omega}$ projects out this pole, eliminates the leading divergence and leaves the  finite, subleading $\O(1)$ piece (note the factors of $q$ in both the numerator and denominator of \eqref{Subleadingfactor}).  We see there is a universal form not just for the leading but also for the subleading part of the amplitude in the soft limit.  Although we have not discussed it, a similar (and much older) story exists for QED: there is also a subleading soft theorem  discovered in 1958 by Low \cite{Low:1958sn}. Following the triangle, it  is also related to a new symmetry in gauge theory, as discussed in \cite{Lysov:2014csa,Campiglia:2016hvg}.

 The subleading soft factor  $S^{(1)-}$ can be obtained from  the leading soft factor by replacing $p_\nu$ with $q^\mu J_{k\mu\nu}$, in other words, by replacing translations with rotations about $q$. This makes sense, because the first term in the hard piece $Q_H^+$ is $Y^z T_{uz}$, which generates a rotation around the axis through which the particle is emerging, whereas the hard part of the supertranslation-generating charge involves $f T_{uu}$. This relation can be made more precise by looking at details of geodesics with nonzero angular momentum, for which it can be shown that the replacement $f T_{uu}\to Y^z T_{uz}$ is akin to the replacement $p_\nu\to q^\mu J_{k\mu\nu}$, with $q$ the direction around which the rotation is performed. Indeed, this is how the soft  formula  was first guessed before it was properly derived.

While I do not reproduce the diagrammatic proof of the subleading soft graviton theorem here, I present a simple but important consistency check. Recall that the Weinberg soft graviton theorem involves the leading soft factor
\begin{align}\label{wsf}
	\sum_k\frac{\varepsilon^{\mu\nu}p_{k\mu}p_{k\nu}}{p_k\cdot q},
\end{align}
which for consistency had better vanish for pure gauge gravitons, that is, when
\begin{align}
    \varepsilon^{\mu\nu}=\Lambda^\mu q^\nu.
\end{align}
Indeed, with this choice, the soft factor \eqref{wsf} becomes
\begin{align}
	\Lambda^\mu\sum_k\frac{p_{k\mu}p_k \cdot q }{p_k\cdot q}=\Lambda^\mu\sum_kp_{k\mu}=0,
\end{align}
where the last equality follows from energy-momentum conservation.

Consistency demands that a similar property hold for our new expression for $S^{(1)}(\e_\Lambda)$. Since it is not automatically symmetric, let us now insert
\begin{align}
	\varepsilon_\Lambda^{\mu\nu}=q^\mu\Lambda^\nu+q^\nu\Lambda^\mu~
\end{align}
and hence find that
\begin{align}
    \label{SoftFactorCheck}
	iS^{(1)}(\varepsilon_\Lambda)=q^\mu\Lambda^\nu\sum_k J_{k\mu\nu}+\sum_k\frac{p_k\cdot\Lambda q^\mu q^\nu J_{k\mu\nu}}{p_k\cdot q}=0,
\end{align}
as required. In the last step, the second term vanishes just because $J_{k\mu\nu}$ is antisymmetric while  $q^\mu q^\nu$ is symmetric. Meanwhile, the first term vanishes by angular momentum conservation. We again see that angular momentum is to superrotations as  energy-momentum is to supertranslations. 

A powerful reformulation of superrotational invariance has recently been achieved in the celestial language of section \ref{smcc}. References \cite{Kapec:2016jld,Cheung:2016iub} construct a four-dimensional soft operator whose amplitudes are precisely those of a two-dimensional CFT stress tensor.

\section{The Memory Effect}\label{sec:memory}
\begin{figure}\begin{center}
\includegraphics[width=3.4 in]{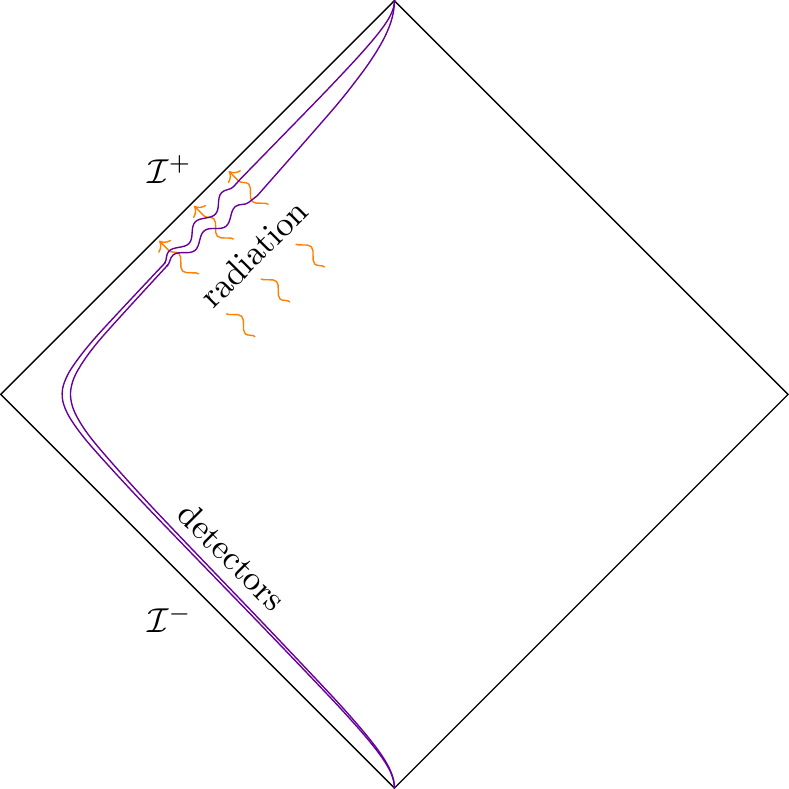}
\end{center}
\caption{\bf\small The memory effect. The passage of gravitational radiation past a pair of inertial detectors stationed near \ip\ causes a temporary oscillation in their relative positions followed by a permanent displacement which ``remembers'' certain moments of the energy flux.  } \label{memdet}
\end{figure}

In this section, we describe the third corner of the IR triangle, the memory effect. 
In the case of gravity, the memory effect characterizes  a pair of inertial detectors stationed near \ip\  in a region with no Bondi news at both late and early times. At intermediate times, gravity waves may pass through, causing oscillating distortions in their relative separations, which we denote $(s^z, s^\bz)$ and depict in figure \ref{memdet}.  
 The equation of geodesic deviation implies (see exercise 13)
\be r^2\g_{z\bz}\p_u^2s^\bz=-{R_{uzuz}} s^z~,\ee
where 
\be R_{uzuz}=-{r \over 2}\p_u^2C_{zz}~.\ee
Integrating this equation reveals a DC effect. Namely, the initial and final separations differ by (in retarded coordinates)
\be \label{dis} \Delta s^\bz={\gamma^{z\bz} \over 2r} \Delta C_{zz}s^z~.\ee
This is the gravitational memory effect \cite{Zeldovich_Polnarev_1974,Braginsky:1986ia,1987Natur.327..123B,Christodoulou:1991cr,Wiseman:1991ss,Blanchet:1992br,Thorne:1992sdb,Favata:2010zu,Tolish:2014bka,Tolish:2014oda,Winicour:2014ska}. The difference $\Delta C_{zz}$ between initial and final transverse metric components need not vanish, as flatness does not require $C_{zz}=0$. Proposals to measure the gravitational memory effect with a variety of methods are given in Lasky et al. and van Haasteren and Levin \cite{Lasky:2016knh,vanHaasteren:2009fy}. 

The effect is harder to see than gravity waves themselves but has a decent chance of being  measured in the coming decades. The connection of the memory  effect with basic symmetries of gravity makes this an exciting prospect.

A quick way \cite{Strominger:2014pwa} to see the equivalence of  gravitational memory to the soft graviton theorem is to compare formulas in the original literature.  An excerpt with the central formula for the outgoing metric fluctuation sourced by elementary particle collisions from  Weinberg  \cite{Weinberg:1965nx} is
\vskip.1in
\includegraphics{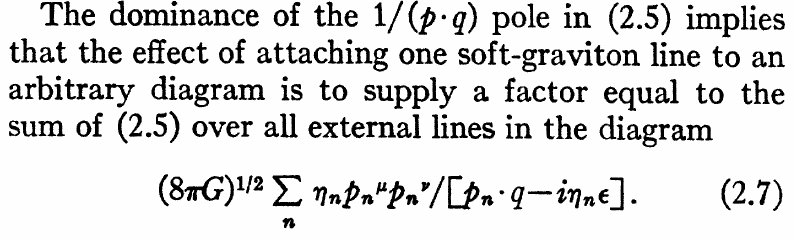}

\noindent It is interesting to compare this to the central equation for the DC shift in the  outgoing metric sourced by black hole or neutron star collisons in Braginsky and Thorne \cite{1987Natur.327..123B}:
\vskip.1in
\includegraphics{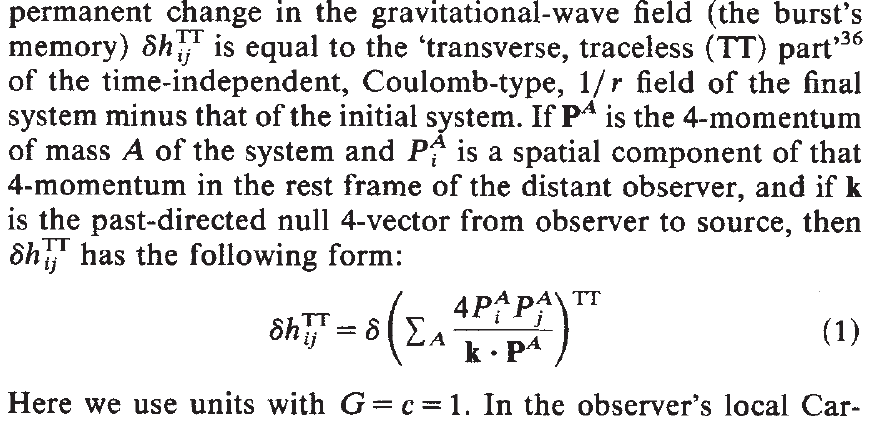}

\noindent A glance reveals that the two equations  are indeed strikingly similar!
In fact, to see  that they are identical, not just similar,  requires a conversion of notation and a Fourier transform, as follows \cite{Strominger:2014pwa}:

\begin{itemize}
\item Replace the four-momenta $P^A_i$ of colliding stars or black holes with the  four-momenta $p_n^\mu$ of elementary particles. 
\item Account for the different conventions for Newton's constant $G$ and normalization.
\item Substitute the graviton momentum $q$ with its energy $\omega$ times the unit null vector $k$ via $q=\omega k$.
\item Act with a  Fourier transform $\int dt e^{i\omega t} $ on the  Weinberg momentum-space formula to obtain the Braginsky-Thorne formula for the difference between the initial and final transverse metric components.
\end{itemize}

The fact that identical formulas were independently derived, using wildly different methods, for both the collisions of black holes and the collisions of elementary particles is a testimony to the universality of the IR phenomena discussed in these lectures.   At very long distances, astrophysical black holes and elementary particles are both effectively pointlike! 

Universal phenomena are often related to symmetries.  Since the memory effect is a Fourier transform of the soft theorem, and the soft theorem is the Ward identity of supertranslation symmetry, there should be a direct connection between memory and supertranslations that does not involve soft theorems.  To derive this connection, note that 
if  there is no energy flux or retarded time dependence of the asymptotic data at late and early
times, $C_{zz}$ must vanish up to a supertranslation:
\be C^{\text{early}}_{zz}=-2D_z^2C^{\text{early}},~~~~C^{\text{late}}_{zz}=-2D_z^2C^{\text{late}}~,\ee
where $\p_uC^{\text{early}}=\p_uC^{\text{late}}=0~$. The early and late geometries are related by a supertranslation
\be \Delta C=C^{\text{late}}-C^{\text{early}}~.\ee
Hence we can think of a pulse of radiation passing through \ip\ as a domain wall separating 
diffeomorphic but  BMS-inequivalent vacua. Using the constraint equation \eqref{mbconstraint}, we find 
\be D_z^2\Delta C^{zz}=2\Delta m_B+2\int du T_{uu}~,\ee
where the $u$-integral extends from the early to the late regions. Solving this differential equation, it follows that 
\be \Delta C(z,\bz)=-\int d^2w \g_{w\bw}G( z,\bz;w,\bar{w})
\left(\int du T_{uu}(w,\bw)+\Delta m_B \right)~,\ee
where the Green's function is 
\be \label{mq} G(z,\bz;z',\bz')={1\over \pi}\sin^2{\Delta \Theta \over 2}\log \sin^2{\Delta \Theta \over 2} ~,\ee
with $\Delta \Theta$ the angle on \cst\ between $(z,\bz)$ and $(z',\bz')$. 
 This is an explicit  formula for the supertranslation induced by waves (gravity or otherwise) passing through \ip. 
One of its unusual characteristics is that it is highly nonlocal on \cst. Indeed, if a gravity wave passes through the north pole, the effect vanishes both there and at the south pole but is large near  the equator. 

\begin{figure}[h] 
\begin{center}
\includegraphics[width=3.5 in]{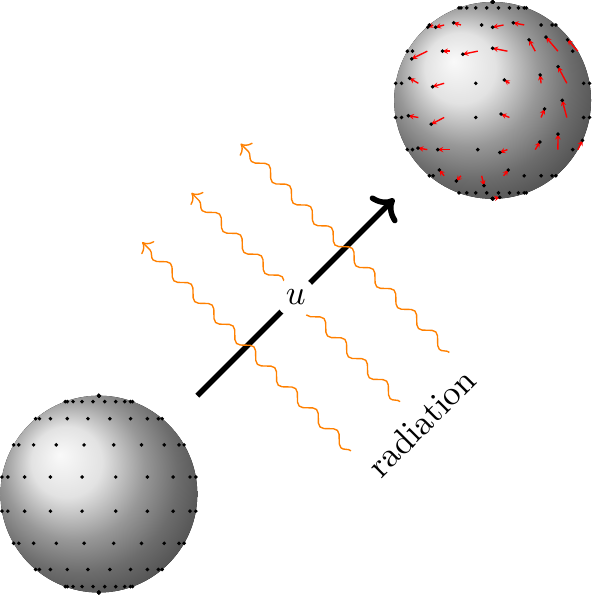}
\end{center}
\caption{\bf\small An array of evenly spaced inertial detectors (black dots) on the sphere near \ip\ will be permanently displaced (red arrows) by the passage of gravitational radiation. The displacements are a measurement of the BMS diffeomorphism, which relates the vacua before and after the passage of the radiation.   } \label{membms}
\end{figure}

The relative displacements \eqref{dis} of inertial detectors have a simple explanation in this framework. Imagine tiling \cst\ with  initially evenly spaced inertial detectors. Passage of a gravity wave will adjust the relative $(z,\bz)$ positions by amounts of order $1 \over r$, defining a diffeomorphism on \cst, as depicted in figure \ref{membms}.  
This was shown \cite{Strominger:2014pwa} to be exactly the  angular diffeomorphism appearing in the subleading term of the supertranslation vector field \eqref{stransdiff}. It would be exciting indeed to observe these supertranslations in the sky!

The gravitational memory effect imparts a physical meaning to the soft graviton theorem. Soft gravitons may seem  a bit unphysical, because it takes longer and longer to measure them as the energy goes to zero. Surprisingly, despite this, the memory effect can be measured in a finite time, because the Fourier transform of the Weinberg pole is a step function in retarded time. 

This completes the third leg of the triangle containing Weinberg's soft graviton theorem and asymptotic BMS symmetries. Memory is connected to the soft theorem by a Fourier transform and measures transitions between BMS-inequivalent vacua. There are other kinds of memory effects associated with other triangles. Taking the Fourier transform of the subleading soft theorem led to a new gravitational memory effect called spin memory \cite{Pasterski:2015tva}, which gives relative time delays to counterorbiting objects and might be measurable \cite{
Nichols:2017rqr}.
 In (unconfined and  unhiggsed) nonabelian gauge theory, the color memory effect rotates the relative colors of nearby quarks \cite{Pate:2017vwa}.  If a pulse of gluons passes a pair of initially singlet quarks, it will generically no longer be in a singlet. In abelian gauge theories, such as QED, the electromagnetic memory effect gives relative phases to adjacent charged particles, which can be measured by quantum interference or other experiments, as  recently discussed in the literature \cite{Bieri:2013hqa, Bieri:2011zb, Susskind:2015hpa, Pasterski:2015zua}. 

\section{Black Holes}

So far, black holes have barely been mentioned. It turns out  that the IR triangle is  highly relevant to the behavior of black holes and to the famous information paradox \cite{Hawking:1976ra}. While  many of the implications of the IR symmetries for black hole physics are topics of current investigation and remain to be fully understood, others follow straightforwardly from  the groundwork  laid in these lectures. Some of the latter already well understood material is  described in this  section, with occasional pointers to open problems.   

We begin in section \ref{info} with a lightning review of the information paradox.  In section \ref{shai}, it is shown that the infinity of conservation laws requires  that black holes must carry corresponding ``soft hair'', invalidating Hawking's  argument \cite{Hawking:1976ra} that information is destroyed. In section \ref{chair}, the properties of soft hair are perturbatively described at the classical level. Section \ref{hcha} constructs the horizon contribution to the linearized supertranslation charge and shows that this charge generates horizon supertranslations. Quantum aspects of soft hair and its relation to soft gravitons on the horizon are discussed in section \ref{qhai}.  The electromagnetic version involving soft photons is in section \ref{ehai}. Future outlook and the status of the information paradox are discussed in section \ref{disc}. 

\subsection{The  Information Paradox}\label{info}

\begin{figure}[ht!]
\begin{center}
    \includegraphics[width=.6\textwidth]{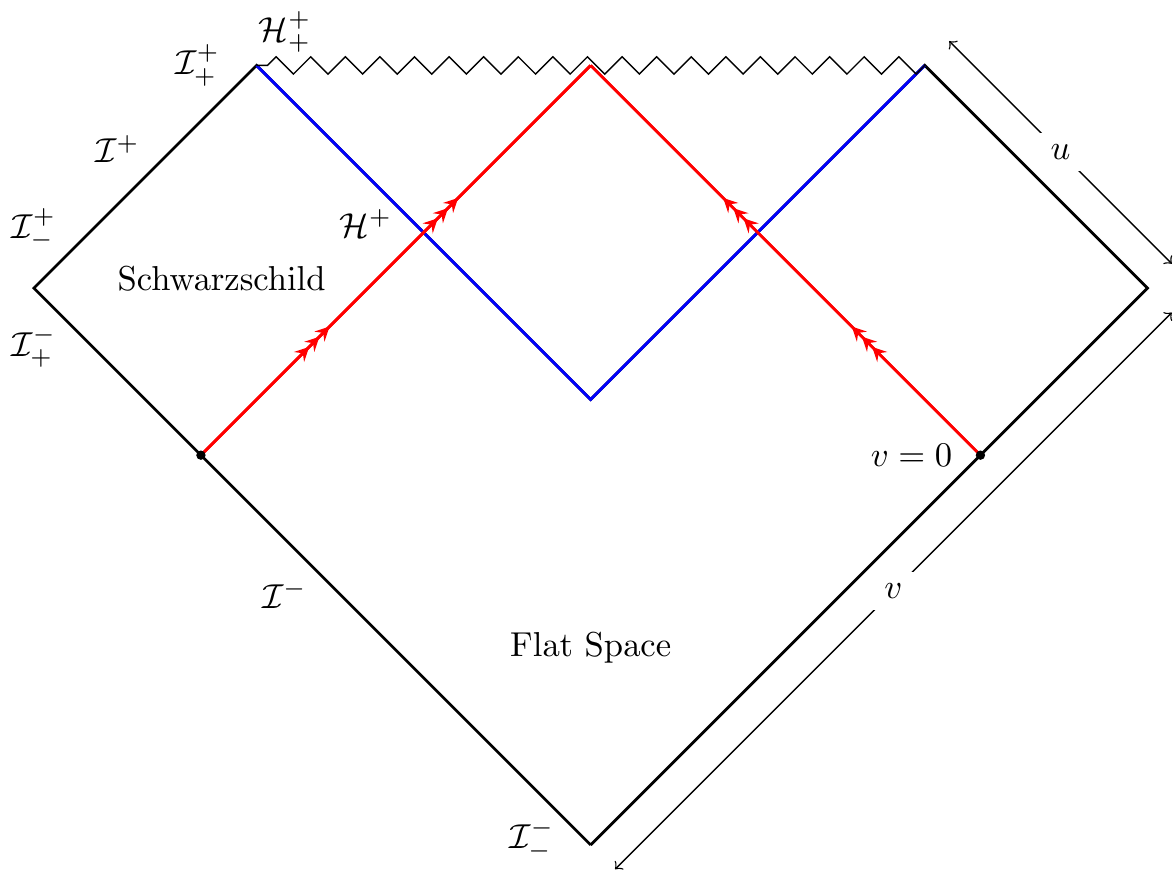}
\end{center}
\caption{ \small \bf Vaidya geometry. The red lines denote a null spherical shockwave imploding at advanced time $v=0$. When the shockwave crosses its  Schwarzschild radius, the (blue) event horizon forms.  The causal future of the event horizon is the black hole and ends at the jagged singularity.   }
\label{bhf}
\end{figure}

There are many ways to characterize the information paradox, but perhaps the most striking is in the formation and total evaporation of a black hole in an asymptotically flat space. We begin with  the process of black hole formation, which is classically described  by the Penrose diagram in figure \ref{bhf}.
One could consider a more generic collapsing geometry,  but it is useful to be specific. In this picture, the far past is empty flat space. Then at advanced time $v=0$, a null shockwave with total energy $M$ is sent in. This infinitely thin collapsing shell of matter eventually forms a Schwarzschild black hole of mass $M$. The resulting spacetime geometry can be described by the Vaidya metric, which in advanced coordinates takes the form
\begin{align}
	ds^2=-\pa{1-\frac{2M G\theta(v)}{r}} d v^2+2 d v d r+2r^2\gamma_{z\bz}d zd\bz~,
\end{align}
where $\theta$ denotes the step function
\begin{align}
	\theta(v)=\begin{cases}
		0&\text{if }v<0~,\\
		1&\text{if }v\ge0~.
	\end{cases}
\end{align}
This metric evidently describes flat space before the passing of the shockwave and a Schwarzschild black hole after. The two regions are glued along the shockwave of infalling matter, whose collapse results in the creation of an event horizon, that is, a region from which light rays cannot escape to infinity (depicted in blue). The spacelike singularity is hidden behind the horizon. 

More generally, the collapsing matter may have arbitrary multipole moments and may not be spherically symmetric. However, according to the usual interpretation (which we question below) of the no-hair theorem \cite{Carter:1971zc}, regardless of how it is formed, after a rather short time scale (roughly speaking, the light-crossing time of the black hole), it will settle down to a  stationary state that is characterized by only ten conserved Poincar\'e charges:  the energy-momentum $(E, \vec P)$, the total angular momentum $\vec J$ and the boost charge or BORT center of mass $\vec K$.  The standard statement is that  a black hole in Einstein gravity has ten hairs, or equivalently, the phase space is ten-dimensional.\footnote{Of these, only two ---  the magnitude of the intrinsic spin $J$ and mass $M=\sqrt{E^2-\vec P^2}$ --- are invariant under Poincar\'{e} transformations.} Of course, in the presence of gauge fields, there may be more, such as the total electric charge, but we ignore these for the moment.

\begin{figure}[ht!]
\begin{center}
    \includegraphics[width=.6\textwidth]{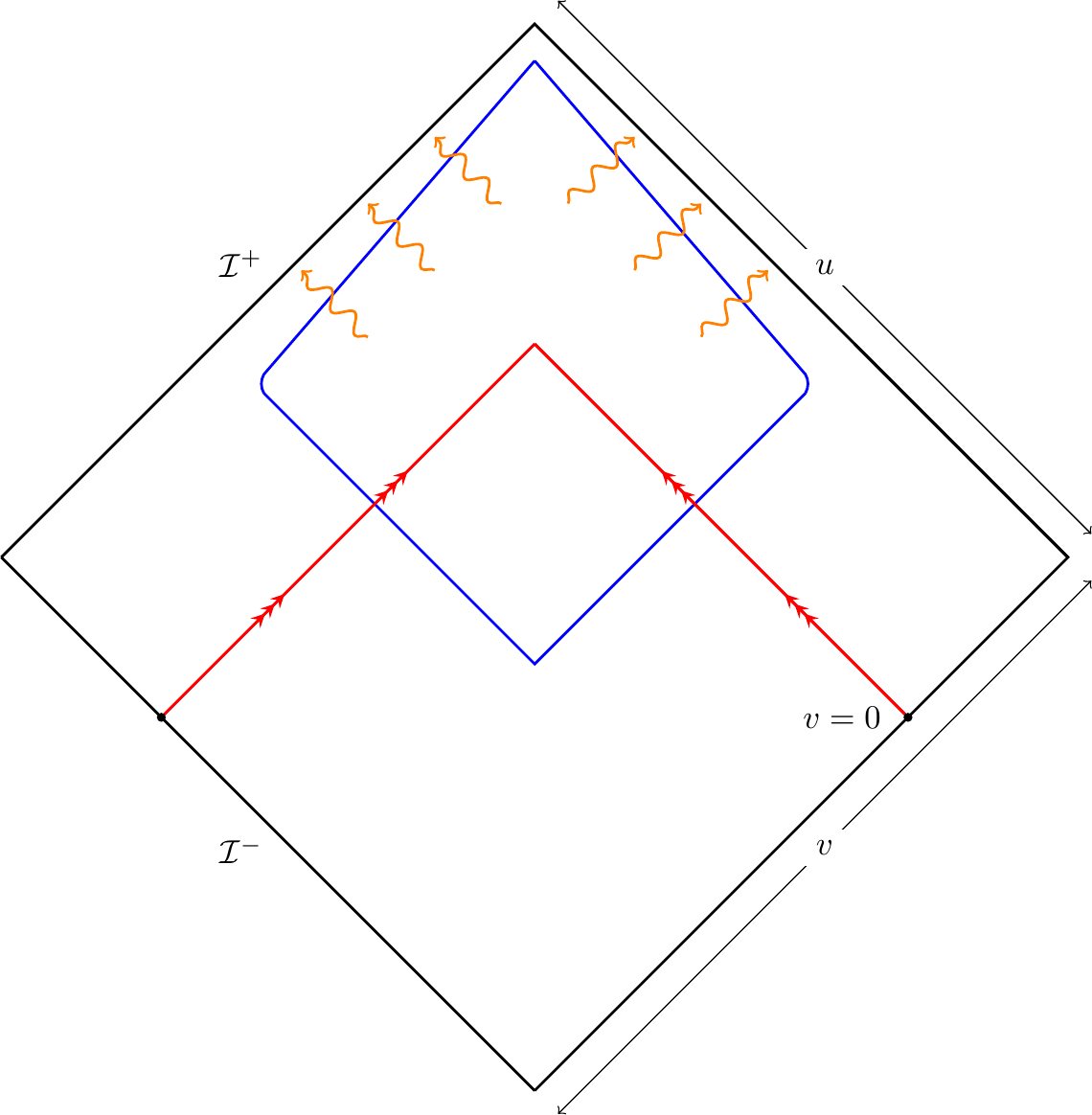}
\end{center}
\caption{ \small \bf Evaporating black hole.  The red collapsing shell makes a black hole, which at early times, looks just like the classical black hole of figure \ref{bhf}. However, at late times the black hole evaporates. This process causes the horizon, depicted in blue,  to become timelike, and the black hole eventually disappears.  In the far future the system reverts to a vacuum state.    }
\label{bhq}
\end{figure}

Quantum mechanically, the situation differs radically.  If we wait a long time, the black hole will eventually evaporate \cite{Hawking:1975iha}. This leads to  the  qualitatively different picture in figure \ref{bhq}. 
As before, we first throw in a shockwave to create the black hole. Then, after a very long time, the black hole begins to evaporate. Since classical black holes do not evaporate, this must take a period of time that goes to infinity as $\hbar\to0$. As long as $\hbar>0$, however, the black hole eventually starts to Hawking evaporate, becoming smaller and smaller until (assuming no remnants) it ultimately disappears altogether, as illustrated at the top of the diagram.

Hawking gave a semiclassical derivation of the spectrum of radiated particles in his famous 1975 paper \cite{Hawking:1975iha}. He found  a blackbody spectrum with a temperature given by the Hawking temperature,
\begin{align}
	\label{HawkingTemperature}
	T_H=\frac{\hbar }{8\pi GM}~.
\end{align}
Moreover, he argued that the outgoing Hawking radiation does not carry any information about what formed the black hole. The radiation results from a pair production process outside the horizon, in which a  negative energy particle goes into the black hole and is correlated with a positive energy particle going out to infinity. This process is insensitive to all details concerning  the interior of the black hole. The no-hair theorem is invoked in this argument to conclude that  the exterior contains no information about the formation process. We could have made the black hole of red, green, or blue  matter, but since the final geometry is the same in all cases, so  must be the outgoing Hawking radiation.
Finally, the black hole is presumed to eventually decay to the vacuum, traditionally assumed to be unique (also questioned below)  and incapable of storing information.  Hence, Hawking argued \cite{Hawking:1976ra}, all information falling into the black hole has no place to go and is destroyed in the formation/evaporation process.

Both in Hawking's argument and in figure \ref{bhq},  a  divide is placed  between the periods of black hole formation and evaporation.  It is assumed that quantum effects are negligible during formation and that during evaporation the black hole can be approximated as stationary. These assumptions are justified if the mass $M$ is large enough in Planck units. Ignoring constants, the Hawking temperature \eqref{HawkingTemperature} varies like 
\begin{align}
	T\sim\frac{\hbar}{GM}~.
\end{align}
Thus, $T\to0$ as $\hbar\to0$ or as the black hole gets bigger. The Stefan-Boltzmann law then implies that the power radiated by the black hole varies  as
\begin{align}
	P\sim\frac{AT^4}{\hbar^3}\sim\frac{\hbar}{G^2M^2}~,
\end{align}
where $A$ is the area of the horizon.  It follows that the rate of mass loss also obeys  
\begin{align}
	\frac{dM}{dt} \sim\frac{\hbar}{G^2M^2}~.
\end{align}
The typical time for the black hole to lose a small fraction, say $1\%$, of its total mass is \begin{align}
	t_\mathrm{typ}\sim.01\frac {G^2M^3}{\hbar}~.
\end{align}
So we have to wait a long time of order $G^2M^3/\hbar$ to see the geometry change appreciably due to Hawking emission. Any classical transient effects that occur during the matter shell infall, such as those associated with emission of radiation to $\scri^+$, will all clear out and settle down in timescales that cannot involve $\hbar$. Hence we have a clean division between the classical black hole formation phase and the onset of Hawking evaporation. The argument for information loss \cite{Hawking:1976ra} applies only  when $M$ is large and such a clean division can be made.

We now revisit this argument in light of the new insights into the IR structure of gravity. 

\subsection{Soft Hair}\label{shai}

In the standard analysis,  it is \textit{not} claimed that the Hawking radiation is {\it completely} uncorrelated with the quantum state of the black hole. For example, even in the semiclassical regime, it has always been presumed that the total mass is   exactly conserved.  If the black hole emits a quantum of Hawking radiation of energy $\Delta E$, then it must necessarily decrease its mass by an amount exactly equal to $\Delta E$. This holds at the level of quantum states and has never been  thought to be a probabilistic statement. Instead there should exist an exact correlation for all conserved quantum numbers (such as the energy) of the black hole and the conserved quantum numbers of the outgoing radiation.  This correlation is not manifest in the leading-order Hawking calculation. It must be enforced by hand when computing the backreaction of the outgoing Hawking quanta on the geometry. The result will be exact correlations between the quantum states of the outgoing Hawking radiation and of the black hole. Not much attention has been paid to these correlations, even if they do modify the thermal spectrum and imply that some information is carried out by the radiation. The reason is they are too few in number --- one for each of the ten black hole hairs --- to rescue the macroscopic quantities of lost information.  

This story is, however, significantly modified \cite{Strominger:2014pwa,Hawking:2016msc,Hawking:2016sgy} by the infinite number of conserved charges discussed in these lectures.  First, it is a common misconception, originating in a misinterpretation of the no-hair theorem,  that black holes (in the absence of gauge fields) have only ten hairs, corresponding to the ten conserved Poincar\'{e} charges. The no-hair theorem states that all stationary solutions are diffeomorphic to Kerr spacetimes.  While this statement is mathematically correct, these diffeomorphisms, if they are within BMS,  need not act trivially on the black hole spacetime.  Generically they map one black hole spacetime to a second, physically inequivalent one.  A simple example is a  boost  that changes both the energy and momentum and so clearly changes the quantum state. However, we have learned that there is no canonically preferred ``boost'' element of the asymptotic BMS symmetry group. For the same reason that two black hole spacetimes that differ by a boost are physically inequivalent, two black hole spacetimes which differ by any BMS transformation (except Killing time translations) are also physically inequivalent. This simple observation implies that there are not  merely ten, but an infinite\footnote{In the quantum theory, it may not be sensible to discuss hairs localized to a region smaller than the Planck length, which would render the number of hairs finite\cite{Hawking:2016msc}.} number of ``soft hairs''  \cite{Hawking:2016msc} on classical black holes.  We will see below that superrotation charges serve as a diagnostic distinguishing the different black hole ``hairdos'' and that  
 a supertranslated Schwarzschild black hole generically carries nonzero angular momentum. 

\begin{figure}[ht!]
\begin{center}
    \includegraphics[width=.6\textwidth]{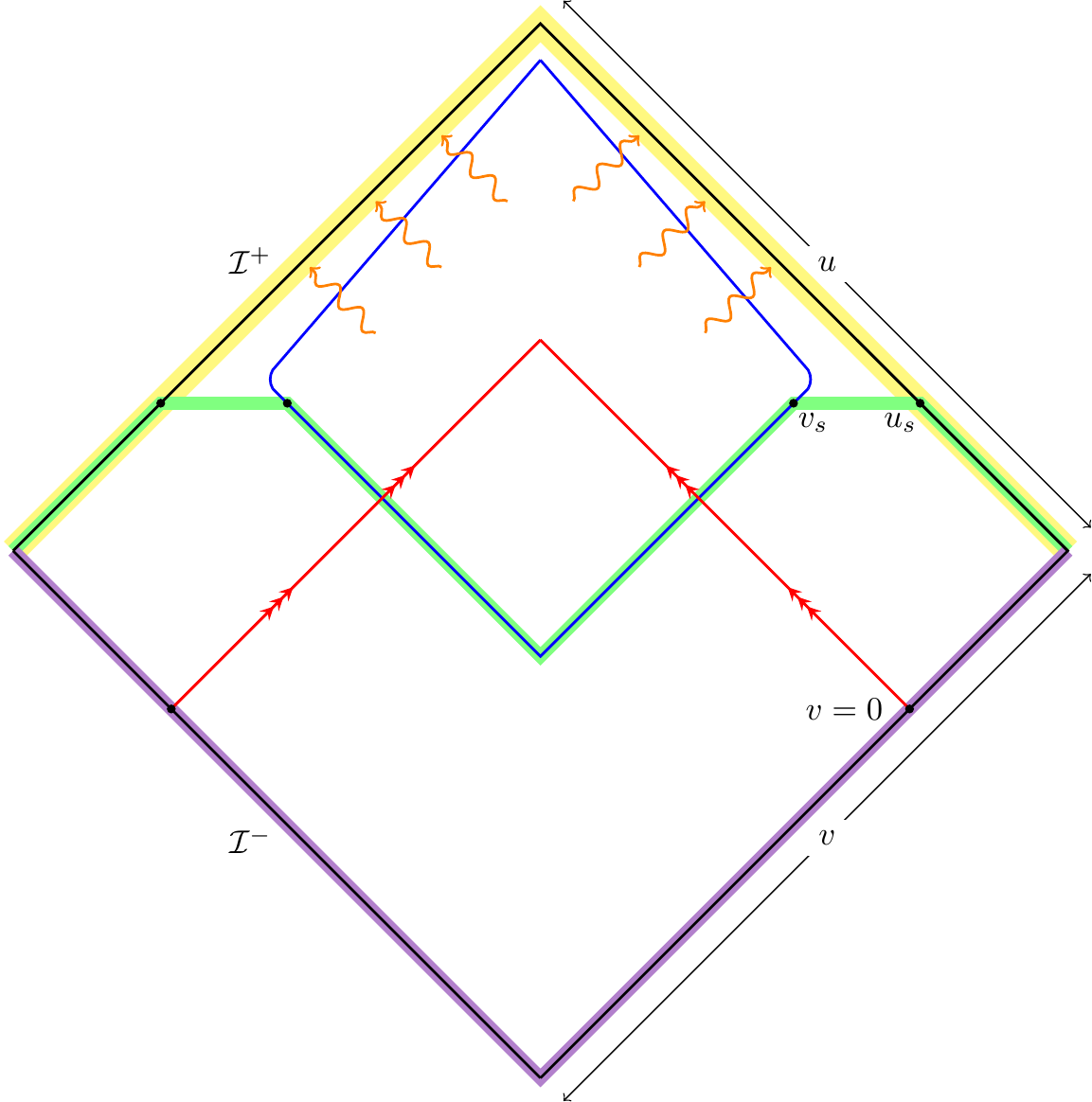}
\end{center}
\caption{ \small \bf An infinite number of conserved charges can be defined as surface integrals at $\ci^+_-$ or $\ci^-_+$. Integrating by parts and using the constraints, they can be written as an outgoing integral over \ip\ (yellow surface), an incoming integral over $\ci^-$ (purple surface), or an intermediate time integral over the green surface that includes a black hole contribution. The equality of these three expressions is possible only if the black hole itself can carry an infinite number of charges, or soft hair.    }
\label{bhs}
\end{figure}

The presence of a lush head of soft hair on black holes is required for  supertranslation and superrotation charge conservation.  Since the conservation laws follow from the long-distance behavior of fields near spatial infinity, they should not be affected by the presence of black holes --- classical or quantum. The conserved charges can be expressed as bulk integrals over any Cauchy surface. In the presence of  a classical black hole, \ip\ is no longer a Cauchy surface. One must include a contribution from the future event horizon $\CH^+$, which is the black hole contribution to the conserved charge. In  the quantum theory, black holes evaporate, but one may still choose to consider a Cauchy surface --- such as the green one in figure \ref{bhs} --- which crosses the spacetime before the black hole has disappeared. 
The integrand on such a surface will contain a contribution both from the black hole and from \ip. 

A generic massless Hawking quantum from an evaporating black hole will  carry nonzero amounts of all the supertranslation, superrotation, and  (if it is charged) large-gauge charges across \ip.  Exact conservation of these charges requires that the black hole decrease its supertranslation, superrotation, and large-gauge charges by an exactly compensating amount.  This enforces an  infinite number of correlations between the quantum state of the black hole and the quantum state of the outgoing Hawking radiation.  A mere ten constraints from Poincar\'{e} charges forcing correlations  cannot have a significant effect on how macroscopic quantities of information flow in and out of the black hole; an infinite number of constraints gives a very different story. It potentially significantly modifies the information flow, or equivalently, the entanglement, between the portions of the bulk quantum state on \ip\ and on the horizon. Motivated by this observation, in the next few subsections we begin a detailed perturbative analysis of soft hair. 
We note that it has been suggested that soft hair is an alternate description of the edge modes  discussed in \cite{Donnelly:2014fua,Donnelly:2015hxa,Harlow:2015lma,Maldacena:2016upp,Harlow:2016vwg}.

\subsubsection{\it Classical Hair}\label{chair}

The first and simplest thing to understand about soft hair, and the subject of this section,  is   classical, linearized supertranslation hair on a Schwarzschild black hole.

Our first problem is to extend the action of supertranslations from a neighborhood of $\ci$ to the entire spacetime. 
The extension of an asymptotic gauge symmetry into the interior is gauge dependent. 
In a general time-dependent situation, there is unlikely to be a  canonical choice of gauge. 
Quantum fluctuations further diminish the utility of specific choices.  For quantum gravity in asymptotically flat spacetimes, we expect the only fully well-defined observables are supported at the boundary at infinity.
It is nevertheless sometimes possible, armed with a gauge choice,  to define interior quantities, such as local gravitational 
energy densities, at first nontrivial order in perturbation theory around the Schwarzschild solution. 
This can be useful in developing a picture and intuition for the behavior of the spacetime away from its boundary. For example, one may show at leading order in perturbation theory that, at both the classical and quantum levels, the total energy comprised of linearized perturbations plus the mass of the black hole itself is conserved. Moreover, this perturbative conservation law is the linearization of an exact, nonperturbative conservation law, which can only be exactly phrased in terms of asymptotic quantities.  It is in this spirit that we study the linearized action of supertranslations in the Schwarzschild solution. 

\vskip.2in
\noindent{\it \bf \textit{Supertranslated Schwarzschild Solution.}}
In advanced Bondi coordinates, the Schwarzschild metric is  \begin{equation}
  \label{eq:2}
  ds^2=-V dv^2+2 dvdr+r^2
\g_{AB}d\Theta^Ad\Theta^B\,~,~~~~V\equiv1-{2m_B \over r}~,
\end{equation}
where $m_B=GM$. It is not hard to show that, for the Schwarzschild geometry, the supertranslation vector field,  
\be\label{ssy} \zeta_f=f\p_v+{1 \over r}D^A f \p_A -\half D^2f \p_r ~
,~~~ f=f(z,\bz)~,\ee 
preserves Bondi gauge for all $r$, not just to leading order. Taking the Lie derivative, we find 
 \bea\label{sf} ds^2+\CL_f ds^2&=-\left(V -\frac{\displaystyle m_B D^2f}{\displaystyle r^2}\right)dv^2+2 dvdr- dvd\Theta^AD_A(2Vf+D^2f) \cr\cr &~~~~~~+
(r^2\g_{AB}+2rD_AD_Bf-r\g_{AB}D^2f)d\Theta^Ad\Theta^B ~.\eea
The event horizon is at $r=2m_B+\half D^2f$.  This geometry describes a black hole with linearized supertranslation hair. Horizon supertranslations have been studied in other gauges 
\cite{Koga:2001vq,Hotta:2002mq,Ashtekar:2004gp}.
\vskip.2in
\noindent{\it \bf \textit{Superrotation Charges.}}
Supertranslating a black hole does not add supertranslation charges to the black hole, just as an ordinary translation of a black hole does not add momentum.  This result follows from the fact that the supertranslation group is abelian and may also be seen at linear order directly, because $m_B$ is unshifted in \eqref{sf}. However, as supertranslations and superrotations do not commute, a supertranslated black hole can and does carry superrotation charges, already at the classical level. In (\ref{asd}), we derived the conserved superrotation charges 
 \be Q^-_Y =\frac{1}{8\pi G } \int_{\ci^-_+} d^2\Theta\sqrt{\g}Y^A N_A~,\ee where $Y^A$ is any smooth vector field on the sphere. 
We are interested in the differential  superrotation charges carried by an infinitesimally supertranslated Schwarzschild black hole of the type considered in the previous paragraph. As seen from (\ref{sf}), under a supertranslation $\delta_fg_{\mu\nu}=\CL_f g_{\mu\nu}$ of the Schwarzschild geometry,
 \be \delta _{f} N_A=-3 m_B \p_A f~ .\ee
 It follows immediately  that \cite{Flanagan:2015pxa,Hawking:2016sgy}
\bea \label{dsk} \Q^-_Y(g,h=\delta _fg)=-{3 \over 8 \pi}\int_{\ci^-_+} d^2\Theta\sqrt{\g}  Y^A M \p_Af~. \eea
The hat here on $\Q^-_Y$ and on other quantities below are to remind the reader that we are doing a linearized analysis. 
Equation \eqref{dsk} is nonzero for a generic vector field $Y^A$ and supertranslation $f$. An infinite number of superrotation charges can be independently added to the black hole by different choices of $f$.  Hence superrotation charges classically distinguish differently supertranslated black holes. 
Classical black holes sport an infinite head of ``supertranslation hair.''

\vskip.2in\noindent{\it \bf \textit{Hair implants.}}
In the previous paragraphs we described a supertranslated eternal Schwarzschild black hole. To be certain such objects really exist, in this section 
we describe how one physically implants hair, or rearranges the hairdo,  by throwing something into a Schwarzschild black hole. Similar conclusions were reached using the membrane paradigm \cite{Penna:2015gza,Eling:2016xlx}.

At advanced time $v_0$ in the Schwarzschild geometry, we send in a  linearized shockwave with energy-momentum density   
\be \T_{vv}={\mu+\T(\Theta)\over 4\pi  r^2}\delta(v-v_0) \ee near $\ci^-$. Here $\mu$ is the total mass of the shockwave, and we take $\T$ to have only $\ell>2$ components for simplicity. We wish to solve for the linearized metric in such a way that the solution is diffeomorphic to the Schwarzschild solution both before and after the shockwave. Stress-energy conservation, $\nabla_\mu\T^{\mu\nu}=0 $,
then mandates subleading-in-${1 \over r}$ corrections 
to the stress tensor for shockwaves that are not spherically symmetric. 
Defining 
\be \C(\Theta)\equiv G \int d^2\Theta^\prime G(\Theta,\Theta')\T(\Theta'),\ee
where $G$ is the Green's function \eqref{mq} arising in the memory effect, 
the solution is \begin{equation}\label{src}
\begin{split}
\T_{vv} &= \frac{1}{4\pi r^2} \left[   \mu + \frac{1}{4G} D^2 \left( D^2 + 2 \right) \C - \frac{3 M }{ 2r } D^2 \C  \right] \delta \left( v - v_0 \right)  ~, \\ \\
\T_{vA} &= - \frac{3 M }{ 8\pi r^2 } D_A \C\delta \left( v - v_0 \right) ~. 
\end{split}
\end{equation}
Solving the linearized Einstein equation with this shockwave source then gives  (see exercise 15)  \bea\label{mp}  h_{vv}&=& \theta(v-v_0) \left(\frac{2G \mu }{r}  - {m_BD^2\C \over r^2}\right)~, \cr\cr h_{AB}&=&-2r\theta(v-v_0)\left(D_AD_B\C -\frac{1}{2}\gamma_{AB} D^2\C\right)~,\cr\cr 
h_{vA}&=& \theta(v-v_0)\p_A \left( 1-\frac{\displaystyle 2m_B}{\displaystyle r}+\half D^2\right)\C~.\eea Comparing with the formula (\ref{sf}) for a supertranslation of the Schwarzschild metric, we  find that (\ref{mp}) can be written 
\be h_{\mu\nu}=\theta(v-v_0)\left(\CL_{f=-\C}g_{\mu\nu}+{2 G\mu \over r}\delta_\mu^v\delta_\nu^v\right)~.\ee
Hence the shockwave is a domain wall interpolating between two 
BMS-inequivalent Schwarzschild vacua, whose mass parameters differ by $\mu$. 

The shockwave induces a shift in the transverse components of the metric perturbation on the horizon.  Integrating  over a null generator of the horizon, we have
\be \label{svx} \Delta h_{AB}(r=2m_B,\Theta)=\int^\infty_{-\infty} dv \p_vh_{AB}(r=2m_B,v,\Theta)=-4m_B\left(D_AD_B\hat C (\Theta)-\frac{1}{2}\gamma_{AB}D^2\hat C(\Theta)\right)~.\ee
Hence the shockwave induces a supertranslation on the Schwarzschild horizon, which is registered  by a nonzero value for the 
zero mode of the transverse, traceless fluctuations of the horizon metric. 
In the physical phase space, two black hole spacetimes that differ by any element of BMS correspond to different points. For the case of boosts, the two spacetimes have different energies. For supertranslations, they are energetically degenerate but carry different angular momenta and other superrotation charges.  An important difference between boosts and supertranslations is that, as discussed in section \ref{sec:supertranslations},  the latter act nontrivially on all the zero-energy vacua  as well, imparting angular momentum and superrotation charges at quadratic order \cite{Barnich:2016lyg}. Hence the phase space of asymptotically flat geometries with nonzero energy and four Killing vectors is not a simple product of vacuum and  black hole phase spaces.\footnote{We consider here supertranslations that act in unison on the horizon and null infinity. As discussed at the end of this subsection, one may try  to consider separate actions, but we do not do so here.   }

The formula (\ref{dsk}) of superrotation charges requires only the asymptotic behavior of the black hole and hence pertains to any geometry with nonzero mass.  We would like to understand how a classical black hole compares in this regard to a configuration with the same mass and many internal states, such as a star, or an elementary particle with few or no internal states.  Suppose we send the supertranslating shockwave into a star or a collection of stars instead of into a black hole. The wave will excite and rearrange the interior structure of the star and, in the case of multiple stars, their relative motions. Generically gravitational radiation will carry some, but not all, of the superrotation charge back out to infinity, while some will be retained by the star(s). It is unsurprising that a star or a collection of stars (which has many internal degrees of freedom and possible interior states) can carry many superrotation charges. There is no no-hair theorem for a star! Now consider instead replacing the black hole by a massive, stable, ``bald'' elementary particle with no internal degrees of freedom. Such an object cannot carry arbitrary superrotation charges: the pre- and post-superrotation charges are generically the same (except for the $\l=1$ component). To leading order, the supertranslating shockwave will simply be reflected through the origin and scatter back up to future null infinity. The elementary particle has no mechanism to absorb  all the superrotation charges.  The outgoing wave will cancel the superrotation charges induced by the ingoing wave and, in the far future, the superrotation charges will revert to their initial incoming values.

So we see that, in its ability to absorb superrotation charges, a black hole acts more like a complex ``hairy'' star with many internal degrees of freedom than
a massive ``bald'' elementary particle. The observer at infinity  can confirm this by sending in shockwaves and watching what comes out. This is a classical signal that black holes are not bald. 

However, there are also intriguing differences, yet to be fully understood,  between the supertranslation hair on a black hole as described above and  the internal states of a star.  Bondi gauge supertranslations change the metric of the spacetime everywhere, including both at the event horizon and at infinity, while internal states of a star can be changed with possibly\footnote{I add the caveat ``possibly'' because  in full quantum gravity, the holographic principle may ultimately  imply that the internal states of the star are recorded at the boundary at infinity. However, ignoring this, it is  at least  true semiclassically.}  no effect at infinity. It is interesting  to ask whether horizon supertranslations can be untethered
from supertranslations of null infinity. There has been discussion of this point  \cite{Donnay:2015abr,Penna:2015gza,Hawking:2016sgy,Donnay:2016ejv,Mao:2016pwq},  but as it remains unsettled, I do not 
go into detail here. A related point is that
 we do not know how to describe  supertranslation hair (one proposal is  edge  modes    \cite{Donnelly:2014fua,Donnelly:2015hxa,Harlow:2015lma}) on a classical stationary black hole in AdS, whose asymptotic boundary  does not support supertranslation symmetry. In contrast, a star in AdS clearly retains many internal degrees of freedom, which, however,  AdS/CFT  tells us  are  recorded on the boundary. Evidently this issue goes to the heart of the holographic structure of spacetime. We will not settle it here. The next subsection sheds some light on it by constructing the  intrinsic  horizon contribution to the supertranslation charge  without reference  to infinity. 

\subsubsection{\it Horizon Charges}\label{hcha}
In the absence of eternal black holes or massive fields, we saw in section \ref{sec:WeinbergSoft} that the linearized supertranslation charges $\Q^+_f$ can be written as volume integrals of local operators over $\ci^+$. However, for the Schwarzschild geometry this is clearly impossible, as \ip\ is not a Cauchy surface. Instead, for massless fields, 
$\ci^+\cup\ch$ is a Cauchy surface. Hence one expects  a relation of the form \be \label{sro} \Q^+_f=\Q_f^{\ci^+}+\Q_f^{\ch}.\ee
The second term, $\Q_f^{\ch}$, is the horizon supertranslation charge. 
In general (for example, with Hawking radiation), supertranslation charge can be exchanged between the horizon and \ip,  and one does not expect them to be separately conserved.

 The literature \cite{Ashtekar:1981bq,Crnkovic:1986ex,Zuckerman:1989cx,Lee:1990nz,Wald:1999wa,Barnich:2001jy,Avery:2015rga} on the covariant canonical formalism, symplectic forms, and conserved charges in GR contains general expressions for the charges associated with diffeomorphisms of a surface $\Sigma$ as an integral over its boundary $\p \Sigma$. In our case,  $\ch$ has past and future boundaries, $\ch_\pm$, which  is the key reason  it is possible to have a nontrivial horizon charge.\footnote{In the quantum theory, the black hole evaporates, and the horizon (however it is defined) may not have a boundary.  This  suggests there cannot be a quantum exact notion of horizon charges independently of those at infinity.} The precise form of the horizon contribution $\Q_f^{\ch}$ will depend on the coordinate choice used to extend the supertranslations in from infinity  to the horizon. Adopting  the Bondi gauge, one finds the boundary expression  \cite{Hawking:2016sgy}
\bea  \Q_f^{\ch }&=&{M\over 8\pi }\int d^2\Theta\sqrt{\g} f\left[ D^A\p_r h_{Av}+2h_{vv}+D^2h_{vr}\right]^{\ch_+}_{\ch_-}~ .\eea
Integrating by parts and using the constraints gives the horizon integral
\bea  \label{saz} \Q_f^{\ch}&=&{1\over 8\pi G }\int_{\ch}d^2\Theta\sqrt{\g} dvf\left( \frac{1}{4m_B}\right.D^AD^B\p_vh_{BA} \cr&&+ \half D^2h_{vv}  +m_BD^2\p_r h_{vv}+{1 \over 2}D^2h_{vr}-\left.\frac{1}{4m_B}D^Ah_{Av} \right)~. \eea

As a check, we would like to see that \eqref{saz} generates supertranslations of the horizon components of the metric via the inverted symplectic form. However, this is still not quite possible,  because the Bondi gauge allows residual trivial diffeomorphisms that vanish at the boundaries $\ch_\pm$ of the horizon and are zero modes of the symplectic form. We must  first eliminate these modes and reduce to the  physical horizon phase space $\G_{\ch}$, which is parametrized by $h_{AB}$.  After some effort \cite{Hawking:2016sgy},  
 $\Q_f^{\ch}$ reduces to  \be\label{hch}  \Q_f^{\ch}={1\over 32 \pi G m_B }\int_{\ch} dv d^2\Theta\sqrt{\g} D^AD^Bf\p_vh _{AB}~
.\ee This charge is nothing but  a soft graviton on $\ch$, just as the linearized charge  $\Q_f^{\ci^+}$ is a soft graviton on \ip. The symplectic form can now be inverted, and the commutator is  \be \left[\Q^{\ch}_f,h_{AB}\right]=i2m_B(2D_AD_Bf-\gamma_{AB}D^2f)~. \ee
This relation is the desired linearized supertranslation action on the horizon metric fluctuations.

This  construction makes sense only in leading-order perturbation theory: in the general case the classical horizon is defined only nonlocally, and  in the quantum case it evaporates altogether. In the presence of interactions it is unlikely that a clean separation can be made between the two terms on the right-hand side of (\ref{sro}).  Nevertheless, it provides a starting point for an analysis of horizon supertranslations. 

We note that the construction requires the charges $\Q_f^{\ch}$ (i.e., the soft graviton modes) to be incorporated as symplectic partners of the $v$-independent  part of $h_{AB}$ in the physical phase space $\G_{\ch}$. Moreover, the analysis is fully intrinsic to the horizon and does not require any properties of asymptotic infinity. Because the classical horizon has boundaries, it is possible for pure diffeomorphisms that do not vanish at these boundaries to be nonzero eigenvectors of the symplectic form. This is the basic reason that supertranslations act nontrivially on the black hole horizon. 
\subsubsection{\it Quantum Hair}\label{qhai}
 Let $|M\rangle$ denote a  quantum state corresponding to a Schwarzschild black hole. Then an infinitesimally supertranslated black hole corresponds to the quantum state 
 \be |M_f\rangle=|M\rangle+iQ_f^+|M\rangle~.\ee
In  section \ref{chair} it was noted that, because the 
supertranslation group is abelian, a classically supertranslated black hole does not carry classical supertranslation hair. The quantum version of this statement is that to linear order in $f$
\be \langle M_f|Q^+_{f'}| M_f\rangle=0 \ee
for any non-constant $f'$. However, it does  not imply that  the supertranslation charge $Q^+_f$ annihilates $|M\rangle$. Instead, $Q^+_f$ adds a soft graviton to $|M\rangle$, leaving its energy unchanged but adding angular momentum and more general  superrotation charges. According to  the decomposition 
\eqref{sro}, the soft part of the charge has two terms, one that acts on $\ch$ and the other on \ip. Hence the soft graviton may be either a soft mode on $\ch$ or a soft mode on \ip. 

Similar statements apply to the quantum vacuum. The supertranslated vacuum carries no  classical supertranslation charge, as $m_B=0$. However, the soft part 
$Q_f^{+S}$ acts nontrivially on the vacuum, turning it into an orthogonal vacuum with an extra soft graviton and different superrotation charges. 

\subsubsection{\it Electric Hair }\label{ehai}
In this section, we repeat the discussion in the context of abelian gauge theories, replacing supertranslations with large gauge transformations and soft gravitons with soft photons. Technically, the discussion is much simpler, both because there are fewer indices and because there is less ambiguity in the extension of boundary quantities into the bulk. However, conceptually it is 
more subtle,  because the action of large gauge transformations on charged particles is only a phase, which is most naturally detected by quantum interference experiments. The quantum nature of the large gauge symmetry action contrasts with that of supertranslations, which has a 
clear classical interpretation as advancing or delaying the arrival times of massless particles at \ip. Moreover, in the gauge theory case, there is no analog of superrotation charge that provides  a classical measurement  of the supertranslation hairdo on the black hole. Because the large gauge symmetry is abelian, classical large gauge charges of large gauge transformed black holes must vanish. It is for these conceptual reasons that we discussed the gravity case before the gauge theory one. However, the gauge theory case is also illuminating, as it displays many of the basic concepts in simpler forms that are uncluttered by proliferating indices. 

To begin, let us briefly recall the argument for conservation of  the total electric charge. The total charge may be written as a surface integral at infinity,
\begin{align}
\label{TotalElectricCharge}
	\Q^+_{\ve=1}= \frac{1}{e^2}\int_{\scri^+_-}\ast F~,
\end{align}
where again the hat on $\Q$ serves to remind the reader that in this section, we linearize around a fixed background. 
Equivalently, using integration by parts, we may write it as an integral of the charged matter current on a bulk slice $\Sigma$ that ends at $\ci^+_-$: \begin{align}
\label{BulkElectricChargeSimple}
	\Q^+_{\ve=1}=\int_\Sigma\ast j_M~.
\end{align}
The result must be the same regardless of which slice $\Sigma$ through the bulk we choose, as illustrated in the context of  black hole formation and evaporation depicted in figure \ref{bhs}. 
For instance, we could compute $\Q^+_{\ve=1}$ by integrating over the purple slice hugging $\ci^-$ at the bottom of the diagram before we create the black hole. Alternately, we could use the yellow slice at the top of the diagram after the black hole has evaporated or  the green slice that runs  into the horizon of the black hole. In the latter case, we can either continue the slice along the horizon as depicted, or we could end the slice at the horizon with an extra boundary term that tells us about the charge carried by the black hole itself. Either way, the black hole typically contributes to the  electric charge. This contribution is required to ensure that the total $\Q^+_{\ve=1}$ is conserved throughout the entire bulk evolution.

This argument may be applied to the  infinite number of such conserved electric charges that generalize the total charge. That is, given any function $\ve$ on \cst, there is a conserved charge \eqref{chargeconserve}, \begin{align}
\label{BulkElectricChargeGeneral}
	\Q^+_\ve= \frac{1}{e^2}\int_{\scri^+_-}\ve\ast F=
\frac{1}{e^2}\int_\Sigma\pa{ \mathrm{d}\ve\wedge\ast F+e^2\ve\ast j_M}~,
\end{align}
where $\Sigma$ again is an arbitrary three-dimensional slice of the bulk ending on $\ci^+_-$, and $\ve$ on the slice is any function with the prescribed boundary value at $\ci^+_-$. Clearly, as we vary this slice and let it end at different surfaces on the black hole horizon, there will be different amounts of large gauge charge that pass through the horizon portion of the slice. This amount will vary in a complicated way as we change the time at which the slice meets the horizon. Therefore, to ensure charge conservation, it must be the case that black holes are capable of storing $\Q_\ve^+$  charge, just as they store global charge. That is, the black hole must carry an infinite amount of soft electric hair.

We wish to analyze the linearized charges in perturbation theory about a fixed black hole background. 
For the case of supertranslations, we linearized around the eternal Schwarzschild black hole. It should also be possible to do a linearized analysis around the more physical geometry  of a black hole formed by gravitational collapse. In practice, for supertranslations, the equations become  unwieldy in a time-dependent background.  However, for the large gauge transformations, the algebra simplifies and such an analysis   is both instructive and possible, as we shall now see.

\begin{figure}[ht!]
\begin{center}
    \includegraphics[width=.6\textwidth]{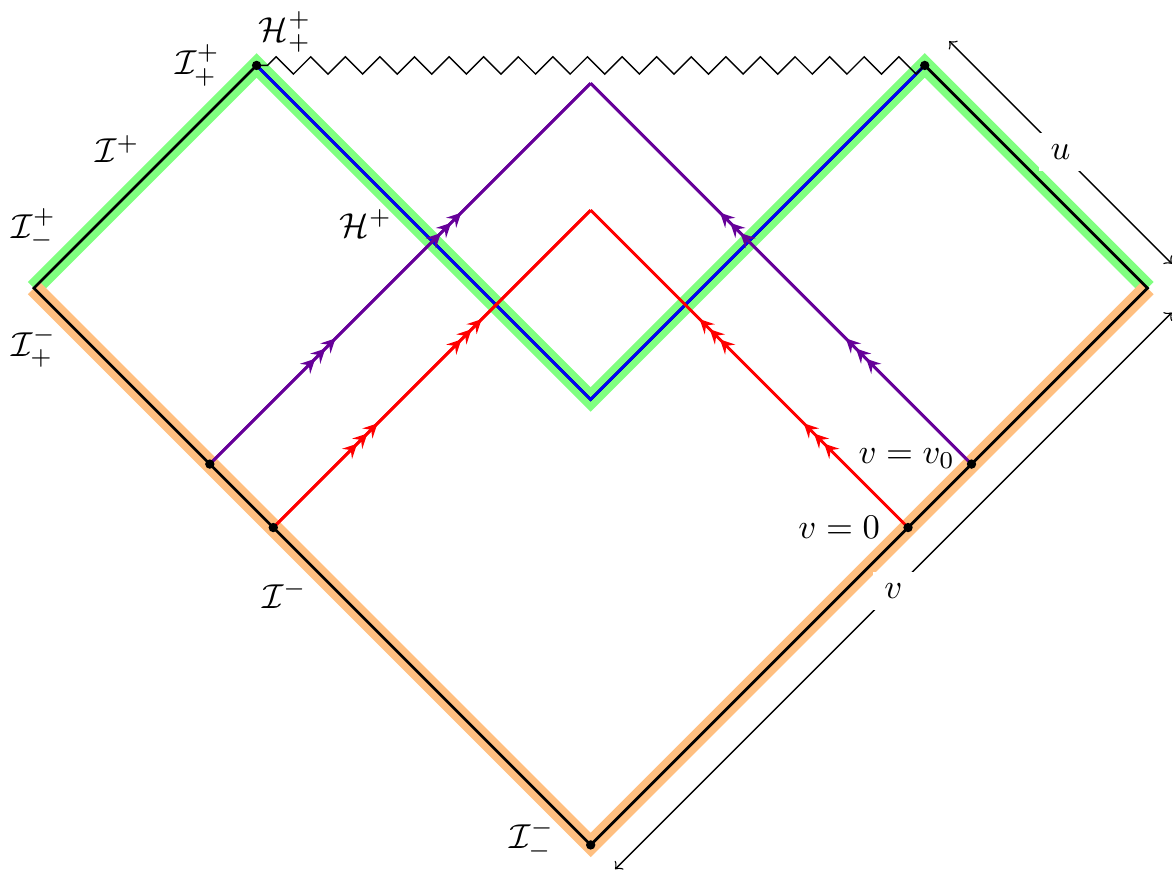}
\end{center}
\caption{ \small \bf A black hole formed by the spherically symmetric red  shockwave at $v=0$ is classically supertranslated at a later time $v=v_0$ by the asymmetric purple shockwave. The solution is diffeomorphic to the Schwarzschild geometry or vacuum away from the shockwaves but generically carries arbitrary superrotation charges in the region $v>v_0$.  The charges computed on the orange and green surfaces must agree, but the latter adds contributions from the horizon and null infinity.}
\label{bhcc}
\end{figure}

Consider a Vaidya black hole formed by a neutral null incoming shockwave at $v=0$. Then at some later time $v_0$, we send in a linearized null shockwave with an asymmetric null charge current $j$, so that the higher large gauge charges are excited. This situation is depicted in figure \ref{bhcc}. 
 To be explicit, let us take
\begin{align}
	j_v=\frac{Y_{\ell m}(z,\bz)}{r^2}\delta(v-v_0)~.
\end{align}
It takes only one line to check that this current is conserved. We are going to examine the effects of this shell of charge in linearized perturbation theory, neglecting its backreaction on the geometry. Note that since it has an angular profile $Y_{\ell m}$, the net charge is zero. If the net charge were nonzero, we would need to discuss a transition from a Schwarzschild to a Reissner-N\"ordstrom black hole instead of just a Schwarzschild to a Schwarzschild one. 

Now we solve the sourced Maxwell equation. The leading  constraint equation at $\ci^-$  \eqref{abelianconstrainteq} says that
\begin{align}
	\pd_vF_{rv}^{(2)}+\gamma^{z\bz}\pa{\pd_zF_{\bz v}^{(0)}+\pd_{\bz}F_{zv}^{(0)}}= e^2Y_{\ell m}(z,\bz)\delta(v-v_0)~.
\end{align}
There are many different ways of solving this constraint equation, as it is one equation for the four components of the vector potential. The one that most simply illustrates the physics is obtained by stipulating that the radial electric field vanish on $\ci^-$,
\begin{align}\label{Stip}
    F_{vr}^{(2)}=0~, 
\end{align}
and the initial connection is $A_z|_{\ci^-_-}=0$. 
These initial data on $\ci^-$ then determine the full solution. 
Explicitly in $A_v=0$ gauge, we have
\begin{align}\label{dsg}
	A_z=\pd_z\br{\theta(v-v_0)\frac{e^2}{\ell(\ell+1)}Y_{\ell m}(z,\bz)}~,
\end{align}
as in equation (\ref{rdr}). It is easy to check that this potential is indeed a solution by using $\p_v\theta(v-v_0)=\delta(v-v_0)$ and 
\begin{align}
	D^2Y_{\ell m}(z,\bz)=-\ell(\ell+1)Y_{\ell m}(z,\bz)~.
\end{align}
If we ignore the factor $\theta(v-v_0)$, $A_z$ is manifestly just a pure gauge transformation, so $F$ is identically zero both before and after the shockwave. We sent in a complicated angle-dependent charge distribution but dressed it with photons in just such a way that no electric field was produced. After the shockwave passes, the entire spacetime  (including both the black hole and null infinity) is acted on by a gauge transformation $A\to A+\mathrm{d}\ve$, with gauge parameter
\begin{align}\label{CDS}
	\ve=\frac{e^2}{\ell(\ell+1)}Y_{\ell m}(z,\bz)~.
\end{align}
Hence, the effect of this shockwave  is to induce a large  gauge transformation. It is a domain wall connecting a black hole spacetime to a large gauge transformation thereof. 

The large gauge transformation \eqref{CDS} depends only on the data of the charge current and not the further choice of electric field data \eqref{Stip}. In Hawking et al. \cite{Hawking:2016msc}, for instance, it was not assumed that $F_{vr}^{(2)}=0$, but the same conclusion was reached. To see why, integrate the constraint  along the horizon:
\begin{align}\label{sxo}
	\int_{\ch}dv\br{\pd_vF_{rv}^{(2)}+\gamma^{z\bz}\pa{\pd_zF_{\bz v}^{(0)}+\pd_{\bz}F_{zv}^{(0)}}}= e^2Y_{\ell m}(z,\bz)~.
\end{align}
If the radial electric field eventually decays near $\ch_+$,  $\int_{\ch} dv\pd_vF_{vr}^{(2)}=0$. In the more general case,  the gauge field is  a Heun function and does not have a simple analytic expression. We nevertheless learn that the zero-modes of the transverse components of the field strength are always excited  when charge is asymmetrically thrown in.

Equation \eqref{sxo} can be rewritten as
\be\label{swe} \int_{\ch}dvd^2z Y_{\ell' m'}\pa{\pd_zF_{\bz v}^{(0)}+\pd_{\bz}F_{zv}^{(0)}}=e^2\delta_{\ell,\ell'}\delta_{m,m'}~.\ee
This expresses the fact that the classical soft photon mode is excited by the asymmetric charge shell or, for that matter, by any charged object that is thrown into the black hole. In the quantum theory, the quantum 
state of the soft photons on the horizon will be a coherent state in which the expectation value of the transverse electric field obeys 
\eqref{swe}.  We note that, in contrast, horizon soft photons cannot be excited by throwing in photons from $\ci^-$, because the relevant  greybody factors vanish at zero momentum. Nevertheless, it follows from \eqref{swe} that soft photons on the horizon can be excited by throwing in charges and are needed for a perturbative description of black hole dynamics. 

The incoming charge $\Q_\ve^-$ can be written as an integral over the orange slice hugging $\scri^-$ in figure \ref{bhcc}. This slice is punctured by $i^-$, past timelike infinity. In general, one must be careful about contributions from the $i^-$ puncture, but for the present case of massless fields they vanish. This conclusion follows from the mathematical fact that $\scri^-$ by itself constitutes a complete Cauchy surface for massless fields. Physically, it means massless fields decay before reaching timelike infinity. 

Likewise, the integral for the charge $\Q_\ve^+$ can be extended up along $\scri^+$, giving an integral $\Q_\ve^{\scri^+}$ along \ip. Clearly, this term is no longer on its own  equal to $\Q_\ve^-$, because $\scri^+$ no longer constitutes a Cauchy surface. To apply Gauss's law, we need a complete Cauchy slice. We may obtain one by following the green slice hugging  the event horizon $\ch$. This additional portion of the surface gives the black hole contribution to the  charge.  Again if we do not include massive particles,  $\ch$ and null infinity $\scri^+$ together form a Cauchy surface for massless fields, so there should be no contributions from $i^+$.

The conservation law thus takes the form
\begin{align}
	\Q_\ve^-=\Q_\ve^{\scri^+}+\Q_\ve^{\ch}~.
\end{align}
This equation says that the large gauge charge computed along the green slice has two pieces: one from null infinity (the outgoing charge $\Q_\ve^{\scri^+}$) and one from the black hole (the horizon charge $\Q_\ve^{\ch}$). Moreover, the sum of these two charges equals the single term obtained by computing $\Q_\ve^-$ on the single incoming orange slice.

We have already seen in section \ref{sec:Symplectic} that in a canonical formalism, $\Q_\ve^{\scri^+}$ generates large gauge transformations on the data at \ip. 
It is not hard to see that $\Q_\ve^{\ch}$ similarly generates large gauge transformations on the data at $\ch$. Integrating by parts, we find
\begin{align}
	\Q_\ve^{\ch}=\frac{1}{e^2}\int_{\ch_+}\ve \ast F=\frac{1}{e^2}\int_{\ch}\pa{ \mathrm{d}\ve\wedge\ast F+e^2\ve\ast j_M}~.
\end{align}
In exercise 14, the interested reader is  invited to show, following the same steps as in section ~\ref{sec:Symplectic}, that in   advanced gauge $A_v=0$,\footnote{ $A_v=0$ gauge is natural for analyzing how excitations from $\scri^-$, such as our shockwave, affect $\ch$, because advanced Bondi  coordinates cover both regions. However, they do not cover  \ip. For this reason, it would be  interesting to analyze the problem in harmonic gauge, as considered by Campiglia and Laddha \cite{Campiglia:2015yka}.}
\begin{align}
	\br{\Q_\ve^{\ch},A_z}_{\ch}=i\pd_z\ve|_{\ch}~,
\end{align}
which  are large gauge transformations on the horizon. This equation also states that the soft part of the horizon charge --- the soft photons --- are symplectically paired with the Goldstone bosons of spontaneously broken large gauge transformations on the horizon. 

In general, a necessary condition for a gauge mode to be physical is that it have a nonvanishing symplectic form. From the general formula for the symplectic form, this is possible only if the surface has a boundary. The boundary of the future horizon $\ch$ is $\ch_+$,\footnote{For eternal black holes there is also a past boundary $\ch_-$. }  and $v$-independent gauge transformations are indeed nonvanishing there.

Note that classically, in an asymptotically stationary situation,
$\Q_\ve^{\ch}$ always vanishes, because the electromagnetic fields (except for the $\ell=0$ mode) go to zero  at late times.  We already saw this in our linearized analysis: equation \eqref{sxo} says that the hard and soft part of the charge are classically equal and opposite. However, it is not zero as an operator, as we have just seen its commutators create large gauge transformations. Moreover,  
when we pass to the quantum theory and expand in creation/annihilation operators, the associated charges will create and annihilate soft photons on the horizon. Acting on the horizon Hilbert space with the operator $\Q_\ve^{\ch}$ will not change the energy, since it commutes with the Hamiltonian (as we know, because we can write the charge as a boundary integral at infinity), but it nonetheless inserts a soft photon, which will change the angular momentum of the black hole by one unit. Thus, a black hole with a soft photon is different from a black hole without one. The action of this infinitesimal symmetry is nontrivial:
\begin{align}
	\Q_\ve^{\ch}\ket{M}=\pa{\frac{1}{e^2}\int_{\ch} \mathrm{d}\ve\wedge\ast F}\ket{M}\neq0~,
\end{align}
where we use $\ket{M}$ to denote the portion of the black hole quantum state on the null surface ${\ch}$. The classical vanishing of 
the linearized horizon charge holds at the level of expectation values: 
\be \bra{M}\Q_\ve^{\ch}\ket{M}=0~.\ee

A similar situation was already encountered in the electromagnetic vacuum in the absence of black holes. Since the field strength vanishes, the charges $\Q^+_\ve$ are all classically zero. However, they are not zero as operators. Instead, they create and annihilate photons, mapping one classical vacuum to another with different angular momenta.

In conclusion, black holes have an infinite electromagnetic degeneracy that can be described at both the classical and quantum levels. At the classical level, the degeneracy arises from nontrivially-acting gauge transformations. At the quantum level, there is an alternate perturbative description in terms of zero-energy  soft photons  on the horizon. 

\subsection{Discussion}\label{disc}

The no-hair theorem tells us that Einstein-Maxwell gravity has a three-parameter\footnote{We suppress possible magnetic charge.} family of black hole solutions labeled by the mass $M$, charge $Q$, and total angular momentum $J$,  up to diffeomorphisms and gauge transformations. It has long been accepted that a few diffeomorphisms (for example, boosts or translations) act nontrivially on the physical phase space and impart a sparse head of hair on black holes. We have shown that there are in fact  an infinite number of diffeomorphisms, including an antipodal subgroup of $\mathrm{BMS^+\times BMS^-}$, that act nontrivially and thereby impart a lush head of soft hair on black holes. The soft hairdo is classically measured by conserved  superrotation charges at infinity. More hair is supplied by large gauge transformations. The hairdo is rearranged whenever anything is thrown into the black hole, and it thereby stores partial information about how the black hole is formed. Exactly how much information is stored, and how it is transmitted to the outgoing Hawking radiation, is presently unknown.  At the very least, understanding black hole information requires understanding the properties of soft hair and more generally keeping careful track of how much information flows into or is stored in the deep infrared \cite{Kapec:2016aqd}. 

Our analysis reveals  two subtly flawed assumptions underlying the argument given in Hawking \cite{Hawking:1976ra} that there is an information paradox:
\begin{itemize}
\item The adage ``black holes have no hair" is misleading at best. In reality they carry an infinite number of conserved charges which generalize the ten Poincar\'{e} and one electric charges.  These provide some  memory of how the black hole was formed. Moreover, since these charges are exactly conserved, they must affect the quantum state of the outgoing Hawking radiation, correlating it with both the quantum state of the black hole and outgoing soft radiation \cite{Strominger:2017aeh}. 
The argument in Hawking \cite{Hawking:1976ra} incorrectly assumed such correlations do not exist. 
\item ``When the black hole evaporates completely, there is no place where the information can be stored" is a second incorrect assumption in Hawking \cite{Hawking:1976ra}. We have seen that the vacuum itself is infinitely degenerate and hence can store information. In particular, from what is understood so far, it is a logical possibility  that  the final vacuum state and or the outgoing soft quanta could be correlated with the outgoing Hawking radiation in such a way as to maintain quantum purity.
\end{itemize}
Of course, it is always possible that these assumptions are ultimately inessential and the paradox can be resolved without taking them into account.  This does not seem to be the case so far, but much remains to be understood. 

To phrase the issue more generally,  soft gravitons are produced 
in every scattering process. The IR pole in the soft theorem says that their production is more ubiquitous than might have been expected. In fact, infinitely many are produced in any physical process. The soft modes are correlated highly with the hard modes \cite{Strominger:2017aeh,Carney:2017jut} and can store information at little or no cost in energy.  Many are made in any  process of black hole formation/evaporation in a manner that is highly regulated by an infinite number of conservation laws. It strikes me as implausible that we could solve the  information paradox in asymptotically flat spacetime without a good understanding  of these modes. 

What would it actually mean to solve the black hole information paradox? First, a convincing solution must microscopically reproduce the Bekenstein-Hawking entropy $S_{BH}=\frac{\text{Area}}{4G}$ with the right factor of $1/4$ in front, not just a squiggly line. As we now understand them, 
soft modes on the horizon are both too many and too few for this purpose (although see Donnay et al. \cite{Donnay:2016ejv}). They are too many because there is no UV cutoff. Presumably, modes that are localized to shorter than the Planck length on the horizon should be thrown out, but we do not know how to do this systematically. They are too few because of a variant of the flavor problem: there are many different  ways of throwing things into the black hole that  produce the same soft hairdo. Second, a satisfactory understanding of black hole information should not be merely verbal  but should  enable explicit computations. In particular, one would like  an algorithm for computing the rate at which the information is returned to infinity \cite{Strominger:2009aj}.

The last 40 plus years of contemplating the wonders of quantum black holes have been truly amazing. 
It has led us to unexpected and fundamental new insights about the structure of classical spacetime, 
exotic quantum states of matter, quantum chaos, nonperturbative string theory, and a potentially revolutionary holographic reformulation of the laws of nature!  At the same time, clearly much more is to come,  as the basic paradox in Hawking \cite{Hawking:1976ra} remains unresolved, while the current literature is replete with promising ideas and directions. The adventure continues!

\section{Exercises {\it with Prahar Mitra and Monica Pate}}

This section contains exercises. We set $G=1$ in all exercises and solutions.
\begin{enumerate}

\item Exercise 1. (Section 2) 

Consider a scattering process in which an incoming null shockwave of the form
\begin{equation}
\begin{split}
j_v (v,r,z,\bz)  = \frac{1}{r^2} \delta (v) Y_{\ell m} (z,\bz) ~, \qquad j_r   = j_z   = 0  ~, 
\end{split}
\end{equation}
scatters at the origin $u=v=0$ into an outgoing null shockwave
\begin{equation}
\begin{split}
j_u ( u , r , z , \bz ) = \frac{1}{r^2} \delta (u) Y_{\ell'm'}  (z,\bz)  ~, \qquad j_r  = j_z = 0~, 
\end{split}
\end{equation}
where  $Y_{\ell m} \big(z,\bz\big) $ are the usual spherical harmonics on $S^2$. 

\begin{enumerate}
\item Find a solution for the field strength sourced by the null shockwave described above, assuming that
\begin{equation}
\begin{split}
F^{(2)}_{vr}  = \lim_{r\to\infty} \left[ r^2 F_{vr} \right] = 0 ~,
\end{split}
\end{equation}
and that $F_{vz}^\0 = \lim\limits_{r\to\infty} F_{vz} $ only has support at the source $v=0$.

\item Using the result of part (a), compute $A_z^{(0)}$ on $\ci^+$ and $\ci^-$ assuming $A^{-\0}_z \big|_{\ci^-_-} = 0$ and $A_z^\0 \big|_{\ci^+_-} = A_z^{-\0} \big|_{\ci^-_+}$. 

\end{enumerate}

\item Exercise 2. (Section 2)

Working in Lorenz gauge $\nabla^\mu A_\mu = 0$ and, using a large-$r$ expansion, show that in pure abelian gauge theory, the four-dimensional gauge field $A_\mu (u,r,z,\bz)$ is determined, up to $u$-independent integration constants by the equations of motion with boundary data (initial conditions)
\begin{equation}
\begin{split}
A_z^\0(u,z,\bz) = \lim_{r\to\infty} A_z(u,r,z,\bz) ~. 
\end{split}
\end{equation}

Hint: Use the residual gauge freedom to set $A_u^{(1)} = 0$.

\item Exercise 3. (Section 2)

\begin{enumerate}
\item  

Starting from the usual action for free abelian gauge theories 
\begin{equation}
\begin{split}
S = - \frac{1}{2e^2} \int F \wedge \ast F   ~, 
\end{split}
\end{equation}
use the formalism in  \cite{Wald:1999wa} to derive the symplectic form \eqref{smp}.

\item Work out the symplectic form explicitly on $\Sigma = \ci^+$, and  derive the commutators \eqref{dto} and \eqref{dtp1}.

\end{enumerate}

\item Exercise 4. (Section 2)

 Starting from the free field mode expansion
\begin{equation}
\begin{split}
A_\mu (x) =  e \sum_{\alpha=\pm} \int \frac{d^3q}{(2\pi)^3} \frac{1}{2\omega_q} \left[ \ve_\mu^{\alpha*} (\vec{q}) a_\alpha  (\vec{q}) e^{i q \cdot x }  + \ve_\mu^{\alpha} (\vec{q}) a_\alpha^\dagger  (\vec{q}) e^{- i q \cdot x }  \right] ,
\end{split}
\end{equation}
show that
\begin{equation}
\begin{split}\label{exercise3}
e^2 \p_z N(z,\bz) = -  \frac{1}{8\pi}  \frac{\sqrt{2} e }{1+ z \bz } \lim_{\omega \to 0^+ } \left[  \omega a_+ ( \omega {\hat x} )  + \omega a_-  ( \omega {\hat x} )^\dagger \right] ~ ,
\end{split}
\end{equation}
where ${\hat x}$ is the unit three-vector
\begin{equation}
\begin{split}
{\hat x} = \left( \frac{z+\bz}{1+z\bz} , \frac{-i(z-\bz)}{1+z\bz} , \frac{1-z\bz}{1+z\bz} \right) ~. 
\end{split}
\end{equation}


\item Exercise 5. (Section 2)

In this problem, you will show that the soft photon theorem for a single outgoing photon,
	\begin{equation}
\begin{split}\label{softth}
\bra{\text{out}} a^{\text{out}}_\pm ( \vec{q} ) {\cal S} \ket{\text{in}}    \to e \left( \sum_{k=1}^{m} Q^\text{out}_k \frac{ p^\text{out}_k \cdot \ve_\pm }{ p^\text{out}_k \cdot q }  - \sum_{k=1}^n Q^\text{in}_k \frac{ p^\text{in}_k \cdot \ve_\pm }{ p^\text{in}_k \cdot q }  \right) \bra{\text{out}}  {\cal S} \ket{\text{in}}    , 
\end{split}
\end{equation}
is equivalent to the Ward identity for large gauge transformations.

\begin{enumerate}

\item For massless momenta, use parameterizations of the type  \eqref{nullmomentum}, namely,
\begin{equation}
\begin{split}
p^\mu =  E \left( 1 ~, \frac{z+\bz}{1+z\bz} ~, \frac{-i(z-\bz)}{1+z\bz} ~, \frac{1-z\bz}{1+z\bz} \right) ~,
\end{split}
\end{equation}
and rewrite the soft theorem \eqref{softth} in terms of $z$, $z_k^{\text{in}}$, $z_k^{\text{out}}$, and $N_z$.

\item   For massive particles of mass $m$, use the momentum parameterization \eqref{pos},
	\be
		\vec{p} = m \rho \hat{x} ~, 
	\ee
	while keeping the parameterization $		q = \omega (1, \hat{q}) $ for the photon momentum, and rewrite the soft theorem for a scattering amplitude of massive particles in the $(\rho, \hat{x})$ coordinates.

\end{enumerate}

\item Exercise 6. (Section 2)

In this problem, we study large gauge transformations in Lorenz gauge, $\nabla^\mu A_\mu = 0$. These are gauge transformations that are nonzero on $\ci^\pm$ and satisfy $\nabla^\mu \nabla_\mu \ve = 0$. We will work in coordinates $(\nu,r,\theta,\phi)$, where $\nu = \frac{t}{r}$.

\begin{enumerate}

\item On a Penrose diagram, draw lines of constant $\nu$ and lines of constant $r$. 

\item Determine the scalar wave equation in these coordinates.

\item Find the most general non singular $r$-independent solution that is of the form $\ve_{\ell m} (\nu,\theta,\phi) = A_\ell(\nu) Y^\ell_m(\theta,\phi)$. Be sure to demand that the solution is continuous and finite at both $\nu = \pm1$ and the origin $\nu=\pm\infty$ (by choosing appropriate branch cuts at $\nu=\pm1$).

\item Determine the near-$\ci^+$ expansion of $\ve_{2m} (u,r,\theta,\phi)$ to $\CO(r^{-1})$. 

\item Verify from a large-$r$ expansion around $\ci^+$ (in retarded radial coordinates) that the form of the subleading behavior obtained in part (d) is required for any solution of the wave equation whose limit on $\ci^+$ is an arbitrary nonvanishing and nonsingular function on $S^2$.

\end{enumerate}

\item Exercise 7. (Section 2)

In this problem, you will derive the symplectic form for massive scalar fields near future timelike infinity.  To resolve $i^{+}$, work in  the coordinates \eqref{hyperboliccoord}
\cite{Campiglia:2015qka},
	\be	
		\tau  = \sqrt{t^2 -r^2}~, 
		~~~~~~~~~~~~ \rho = \frac{r}{\sqrt{t^2 -r^2}}~. \label{eq:CLcoord}
	\ee

\begin{enumerate}
	\item Start with the free field mode expansion for a real massive scalar 
		\be	
			\phi(x) = \int \frac{d^3 p}{(2 \pi)^3}  \frac{1}{2 \w_p} \left [a( \vec{p}) e^{ip \cdot x} + a^\dagger( \vec{p}) e^{-ip \cdot x}   \right]~,
		\ee
		and determine the leading order term in the asymptotic expansion of the field in the limit $\tau \rightarrow \infty$.
	\item Determine the symplectic form on a Cauchy surface $\Sigma$ for a free massive real scalar, governed by the action
		\be
			S = - \frac{1}{2} \int d^4x  \sqrt{-g} \left[ (\partial_\mu \phi)^2 + m^2 \phi^2\right]~.
		\ee
	\item Now take $\Sigma$ to be a constant-$\tau$ slice in the coordinates defined in (\ref{eq:CLcoord}), and determine the symplectic form in the limit $\tau \rightarrow \infty$.
	\item Verify that the  commutation relations determined by this symplectic form are the standard free field commutation relations:
		\be
			[a(\vec{p}), a^\dagger(\vec{p}')] = (2 \pi)^3 (2 \w_p) \delta^{3}( \vec{p} - \vec{p}')~.
		\ee
\end{enumerate}

\item Exercise 8. (Section 2)

Consider the hard part $\SF^h[\chi]$ of the fermionic charge that generates the soft-photino theorem,
\begin{equation}
\begin{split}
\SF^h [\chi] = \frac{1}{4} \int du d^2 z \g_{z\bz} \chi(z,\bz) u \p_u {\bar k}_- ~, 
\end{split}
\end{equation}
where ${\bar k}_-$ is the large-$r$ limit of the fermionic current. For minimally coupled $\CN=1$ supersymmetric QED, ${\bar k}_- = \sqrt{2} Q \phi {\bar \psi}_-$, where $\phi$ and ${\bar \psi}_-$ are the boundary values of the bulk fields $\Phi$ and ${\bar \Psi}_\da$, respectively:
\begin{equation}
\begin{split}
{ \Phi} (u,r,z,\bz) &= \frac{1}{r} { \phi} (u,z,\bz) + \CO(r^{-2}) ~, \\
\bar \Psi_\da (u,r,z,\bz) &= \frac{1}{r} \bar \psi_- (u,z,\bz)  \bar \xi^-_\da + \CO(r^{-2})~. 
\end{split}
\end{equation}
$\xi^\pm_\a$ and ${\bar \xi}^\mp_\da$ are commuting two-component helicity basis spinors (i.e., they are eigenspinors of $\sigma_z{}^z$)
\begin{equation}
\begin{split}
( \sigma_z{}^z \xi^\pm )_\a = \pm \frac{1}{2} \xi_\a^\pm~, \qquad \left( \xi_\a^\pm \right)^* = {\bar \xi}_\da^\mp ~. 
\end{split}
\end{equation}
Using the form above, derive the action of $\SF^h [\chi]$ on the radiative modes $\p_u {\bar \phi}$ and $\psi_+$. Note that the action of $\SF^h [\chi]$ mimics that of the superconformal generators $S_\a$ except that it acts only on charged matter.

Hint: You can either derive the symplectic form on $\ci^+$ for scalars and fermions or rewrite the fields on $\ci^+$ in terms of the usual creation and annihilation operators.

\item Exercise 9. (Section 4)

\begin{enumerate}
\item Consider an amplitude $\CA^{a_1,a_2}_{n+2}(q_1,\ve_1,q_2,\ve_2 ;  p_1,\cdots,p_n)$ containing two outgoing gluons with color, momentum, and polarization $(a_1,q_1,\ve_1)$ and $(a_2,q_2,\ve_2)$, and $n$ particles charged under the gauge group. Derive the following double-soft limit
\begin{equation}
\begin{split}
\left[ \lim_{q_1 \to 0} , \lim_{q_2 \to 0 } \right]  \CA^{a_1,a_2}_{n+2}(q_1,\ve_1,q_2,\ve_2 ;  p_1,\cdots,p_n) ~. 
\end{split}
\end{equation}

\item Show that when the two gluons have the same helicity the above quantity vanishes.

\end{enumerate}

\item Exercise 10. (Section 5)

Any four-dimensional Lorentzian metric can be written in Bondi gauge as
\begin{equation}
\begin{split}
ds^2 = - U du^2 - 2e^{2\beta} du dr + g_{AB} \left( dx^A + \frac{1}{2} U^A du \right) \left( dx^B + \frac{1}{2} U^B du \right)~. 
\end{split}
\end{equation}
Asymptotically flat spacetimes are defined by the following falloff conditions for the Weyl tensor:
\begin{equation}
\begin{split}\label{eq1}
C_{rzrz} \sim \CO(r^{-3}) ~, \qquad  C_{rurz}, C_{rur\bz} \sim \CO(r^{-3}) \qquad \text{at large $r$,}
\end{split}
\end{equation}
along with some other assumptions of uniform smoothness that will not be considered here. In \eqref{bondimetricangularmomentum}, asymptotically flat spacetimes are described by the large-$r$ expansion
\begin{equation}
\begin{split}
ds^2 = - du^2 - 2 du dr + 2r^2 \g_{z\bz} dz d\bz + \frac{2m_B}{r} du^2 + r C_{zz} dz^2   + U_z du dz   + c.c. + \cdots ~.
\end{split}
\end{equation}
Show that this metric satisfies the first equation of \eqref{eq1}. Use the second equation to derive a constraint on $U_z$ and $U_\bz$.

\item Exercise 11. (Section 5)

In retarded Bondi gauge, a four-dimensional Lorentzian spacetime can be written as 
\begin{equation}
\begin{split}
ds^2 =  - U du^2 - 2 e^{2\beta} du dr  + g_{AB} \left( dx^A + \frac{1}{2} U^A du \right) \left( dx^B + \frac{1}{2} U^B du \right) ~,
\end{split}
\end{equation}
where $\det g = r^4 \det \g$, and $\g_{AB}$ is the two-dimensional metric on the round sphere, $S^2$. As discussed in section \ref{sec:asymptoticallyflatspacetimes}, the asymptotic falloff conditions for the component fields are 
\begin{equation}
\begin{split}
U &= 1 - \frac{2m_B}{r} + \CO(r^{-2}) ~, \qquad \beta = \CO(r^{-2}) ~, \\
 U_A &= \frac{1}{r^2} D^B C_{BA} + \CO(r^{-3}) ~, \qquad g_{AB} = r^2 \g_{AB} + r C_{AB} + \CO(1)  ~,
\end{split}
\end{equation}
where $\g_{AB}$ is the round metric on $S^2$, $D_A$ is its covariant derivative, and all indices $(A,B,C,\cdots)$ are raised and lowered with respect to $\g_{AB}$. 

\begin{enumerate}

\item Show that the coordinate condition $\p_r \det \left( \frac{g_{AB}}{r^2} \right) = 0$ implies that $C_{AB}$ is traceless. 

\item Find the most general diffeomorphism $\xi$ that preserves the form of this metric and that satisfies the asymptotic falloffs $\xi^u,\xi^r \sim \CO(1)$ and $\xi^z,\xi^\bz \sim \CO(r^{-1})$ at large $r$. \emph{Be sure to keep track of the leading and the first subleading terms in the large-$r$ expansion.}

\item The diffeomorphism in part (b) is parametrized by a function $f(x^A)$ on the sphere. Show that, to leading order, the Lie algebra of supertranslations is abelian: 
\begin{equation}
\begin{split}
\left[ \xi(f_1) , \xi(f_2) \right] = \CO(r^{-1}) ~. 
\end{split}
\end{equation}

\item Show that the action of $\xi(f)$ on $m_B$, $C_{AB}$, and $N_{AB} = \p_u C_{AB}$ is
\begin{equation}
\begin{split}
\CL_f C_{AB}  &= f \p_u C_{AB} - 2 D_A D_B f + \g_{AB} D^2 f ~, \\
\CL_f N_{AB}  &= f \p_u N_{AB} ~, \\
\CL_f m_B &= f \p_u m_B + \frac{1}{4} \left( N^{AB} D_A D_B f + 2 D_A N^{AB} D_B f \right) ~. 
\end{split}
\end{equation}
\emph{Note that the traceless condition on $C_{AB}$ is preserved under supertranslations.}

\item The function $f(x^A)$ can be expanded in spherical harmonics on the sphere. Show that the $\ell = 0$ and $\ell = 1$ modes in this expansion correspond to the standard global translations
	in Minkowski spacetime. Find the action of these modes on $C_{AB}$, $N_{AB}$, and $m_B$.

\end{enumerate}

\item  Exercise 12. (Section 5)

In addition to the supertranslations, the four-dimensional BMS algebra also contains generators  of an (extended) Lorentz group. We will study these in this problem.
	\begin{enumerate}
		\item Determine the Killing vector $\zeta$ that generates Lorentz transformations in flat spacetime in the $(u,r,z,\bz)$ coordinates. Show that it can be parametrized in terms of a global CKV $Y^A$ on $S^2$, and find $\zeta(Y)$.\footnote{A conformal Killing vector (CKV) on $S^2$ is one that satisfies $D_A Y_B + D_B Y_A = \g_{AB} D_C Y^C$. In the $(z,\bz)$ coordinates, this equation reads $\p_\bz Y^z = \p_z Y^\bz = 0$, implying that any holomorphic vector field $Y^z(z)$ is locally a CKV. The ones that are defined globally on the sphere $S^2$ take the special form $Y^z(z) = a+bz+cz^2$ for $a,b,c\in\mcc$. }

		\item Show that $\left[ \zeta(Y_1) , \zeta(Y_2) \right] = \zeta \left( [ Y_1 , Y_2 ] \right)$. This shows that the Lorentz algebra is isomorphic to $SL(2,\mcc)$, the algebra of global CKVs of $S^2$.
		\item Show that $\zeta(Y)$ preserves the asymptotic form of the metric in asymptotically flat spacetimes even when $Y$ is a local CKV of $S^2$, except at possibly isolated analytic singularities. 
		Find the action of $\zeta(Y)$ on $C_{AB}$. These diffeomorphisms are called superrotations.

	\end{enumerate}

\item  Exercise 13. (Section 6)

In this problem, we study the gravitational memory effect experienced by inertial observers due to supertranslations.
	\begin{enumerate}
		\item Working in retarded Bondi coordinates, determine the first subleading correction to the component of 
			 the four-velocity of an inertial time-like observer that is stationary with respect to retarded time at leading order (i.e., $v^\mu = \delta^\mu_u+ \co(r^{-1})$).
			 Then determine the trajectory with this four-velocity that passes through the point $(u_0, r_0, z_0, \bz_0)$ and is valid for $|u-u_0| \ll r_0$.  Be sure to determine any
			 subleading corrections needed to reproduce the subleading corrections to the four-velocity.

		\item Now consider a nearby inertial observer, who at $u = u_0$ is located at the same radius $r_0$ but at a different 
			point $(z_1, \bz_1)$ on the asymptotic $S^2$, where $|z_1 - z_0| \sim \co (r_0^{-1})$.  Determine the proper distance 
			between the observers at retarded time $u>u_0$ to leading order in the  large-$r_0$ limit.  Assume $u-u_0 \ll r_0$, so you can use the trajectory determined in part (a).

		\item A  burst of radiation of the following form passes the observers:
			\be
				T^M_{uu}(u, r,  z ,\bz) = \frac{\mu}{4\pi r^2} \delta (u-u_{\text{rad}}) \gamma^{z \bz }\delta^2(z-z_{\text{rad}})~, 
			\ee
			where
			\be
			u_{\text{rad}}> u_0~,~~~~\frac{|u_{\text{rad}}- u_0|}{r_0} \ll 1 ~. 
			\ee
			Determine the change in proper distance between the observers as a result of the passage of this radiation.  You will need to work consistently to 
			an order at which this change is nonzero.  Also, assume that the Bondi mass $m_B$ is independent of $(z, \bz)$.
			
		\item Find the supertranslation $f$ that gives rise to the same change in proper distance as calculated in part (c). 
	\end{enumerate}

%

\item Exercise 14. (Section 7)

Consider a Schwarzschild black hole with mass $M$ perturbed by a null shockwave at $v = v_0$ described by the stress tensor
\begin{equation}
\begin{split}
T_{vv} &= \frac{1}{4\pi r^2} \left[   \mu + \frac{1}{4} D^2 \left( D^2 + 2 \right) f \right] \delta \left( v - v_0 \right)  - \frac{3 M }{ 8 \pi r^3 } D^2 f  \delta \left( v - v_0 \right)  ~, \\
T_{vA} &= - \frac{3 M }{ 8 \pi r^2 } D_A f \delta \left( v - v_0 \right) ~. 
\end{split}
\end{equation}

\begin{enumerate}
\item Solve the linearized Einstein's equations about the initial Schwarzschild solution in Bondi gauge. Assume that only the zero mode of the Bondi mass aspect $m_B$ on $\ci^-$ changes under this perturbation. 

\item Show that the final black hole (after $v > v_0$) can be obtained instead by sending a spherically symmetric null shockwave with stress tensor
\begin{equation}
\begin{split}
T_{vv} = \frac{\mu}{4\pi r^2} \delta \left( v - v_0 \right)  ~, 
\end{split}
\end{equation}
followed by an infinitesimal supertranslation.

\end{enumerate}

\item Exercise 15. (Section 7)

Consider the Vaidya black hole created by a spherically symmetric null shockwave at $v = 0$. We define the horizon large gauge charge as
\begin{equation}
\begin{split}
Q^{\ch}_\ve \equiv \frac{1}{e^2} \int_{\CH^+_+} \ve \ast F ~, 
\end{split}
\end{equation}
where $\CH^+_+$ is the future of the horizon of the Vaidya black hole. 

\begin{enumerate}
\item Using the constraints derived from Maxwell's equations on the horizon $\ch$, rewrite $Q^{\ch}_\ve$ as an integral over $\ch$, and determine the soft and hard parts of this charge. Write the soft charge explicitly in the $(v,r,z,\bz)$ coordinates. 

\item Determine the symplectic form for the gauge fields on $\ch$. (Hint: You can use the symplectic form $\Omega_\Sigma$ derived in exercise 3 and set $\Sigma = \ch$). Write it out explicitly in $(v,r,z,\bz)$ coordinates. 

\item By using the explicit forms of the charge and symplectic form derived in parts (a) and (b), show that $\hat{Q}^{\ch}_\ve$ generates large gauge transformations on the horizon, that is,
\begin{equation}
\begin{split}
\left[\hat{Q}_\ve^{\ch} , A_z  \right]_{\ch} = i \p_z \ve  \big|_{\ch} ~. 
\end{split}
\end{equation}

\end{enumerate}

\end{enumerate}

\section{Solutions {\it with Prahar Mitra and Monica Pate}}

\begin{enumerate}
\item[1(a).] First consider the electric field produced by an incoming null shockwave. We work in the $(v,r,z,\bz)$ coordinates. Maxwell's equations take the explicit form
\begin{equation}
\begin{split}
\left( \p_v + \p_r \right) \left(  r^2F_{rv}   \right) -   \g^{z\bz} \left(  \p_z F_{ v \bz}   +  \p_\bz  F_{ v z} \right)  &= e^2  \delta (v) Y_{\ell m}    ~,  \\
\p_r \left(  r^2  F_{rv} \right) + \g^{z\bz} \left( \p_z   F_{r\bz}  + \p_\bz  F_{rz}  \right)   &= 0 ~, \\
r^2 \left( \p_v + \p_r \right)  F_{rz}   + r^2 \p_r  F_{vz}   - \p_z \left( \gamma^{z\bz} F_{z\bz}  \right) &= 0  ~. 
\end{split}
\end{equation}
We also have the Bianchi identities, namely,
\begin{equation}
\begin{split}
\p_v F_{z\bz} + \p_\bz F_{vz} - \p_z F_{v\bz} &= 0 ~, \\
\p_r F_{z\bz} + \p_\bz  F_{rz}  - \p_z  F_{r\bz}  &= 0 ~, \\
\p_v  F_{rz} - \p_z  F_{rv} - \p_r F_{vz} &= 0 ~. 
\end{split}
\end{equation}
A simple solution satisfying the boundary conditions is
\begin{equation}
\begin{split}
F_{vz} = - e^2 \delta(v) \p_z \ve_{\ell m} ~, \qquad F_{vr} = F_{rz} = F_{z\bz} = 0 ~,
\end{split}
\end{equation}
where 
\begin{equation}
\begin{split}
 \ve_{\ell m} = \left\{ \begin{array}{cc} \log \big( 1 + z \bz  \big)    Y_{00} ~, & \ell = 0 ~, \\
- \frac{1}{\ell(\ell+1)} Y_{\ell m} ~, & \ell \neq 0 ~, 
\end{array} \right.
\end{split}
\end{equation}
and  $Y_{\ell m}$ satisfies
\begin{equation}
\begin{split}
2\g^{z\bz} \p_z \p_\bz Y_{\ell m} = - \ell ( \ell + 1 )Y_{\ell m}~. 
\end{split}
\end{equation}
This implies 
\begin{equation}
\begin{split}
2\g^{z\bz} \p_z \p_\bz \ve_{\ell m} = Y_{\ell m}~. 
\end{split}
\end{equation}
Any other solution differs from this one by a solution to the homogeneous Maxwell equation. 

Similarly, for the outgoing null shockwave, the solution is
\begin{equation}
\begin{split}
F_{uz} = - e^2 \delta(u) \p_z \ve_{\ell'm'} ~, \qquad F_{ur} = F_{rz} = F_{z\bz} = 0 ~. 
\end{split}
\end{equation}

\item[1(b).]  We have $F_{vz}^\0 = \p_v A_z^{-\0}$ and the boundary condition $A^{-\0}_z \big|_{\ci^-_-} = 0$, we find
\begin{equation}
\begin{split}
A^{-\0}_z (v,z,\bz)  = - e^2 \theta (v) \p_z \ve_{\ell m} (z,\bz) ~, 
\end{split}
\end{equation}
where $\theta$ is the Heaviside step function.
Then, using the antipodal matching condition, we find
\begin{equation}
\begin{split}
A_z^\0 \big|_{\ci^+_-} =  - e^2  \p_z \ve_{\ell m} (z,\bz)  ~. 
\end{split}
\end{equation}
Finally, integrating $F_{uz}^\0 = \p_u A_z^\0$ and using the above boundary condition, we find
\begin{equation}
\begin{split}
A_z^\0(u,z,\bz) =  - e^2  \p_z \ve_{\ell m} (z,\bz)   - e^2 \theta(u) \p_z \ve_{\ell'm'} (z,\bz)~. 
\end{split}
\end{equation}

\item[2(a).] In the $(u,r,z,\bz)$ coordinates, the Lorenz gauge condition and Maxwell equations take the form
\begin{equation}
\begin{split}\label{maxeq}
\left( \p_u - \p_r \right) \left(  r^2F_{ru}   \right) + \g^{z\bz} \left(  \p_z F_{ u \bz}   +  \p_\bz  F_{ u z} \right)  &= 0    ~,  \\
- \p_r \left(  r^2  F_{ru} \right) + \g^{z\bz} \left( \p_z   F_{r\bz}  + \p_\bz  F_{rz}  \right)   &= 0 ~,   \\
  r^2 \left( \p_u - \p_r \right)  F_{rz} +  r^2 \p_r  F_{uz} + \p_z \left( \gamma^{z\bz} F_{z\bz}  \right) &= 0  ~,  \\
- \p_u \big(  r^2 A_r \big) - \p_r \big( r^2   A_u  \big)  +   \p_r \big( r^2  A_r \big)   +  \g^{z\bz}  \left( \p_z A_\bz + \p_\bz A_z \right)    &= 0   ~. 
\end{split}
\end{equation}
To  impose the Bianchi identity, express the Maxwell equations in terms of the gauge field $A_\mu$, and in addition,  substitute the Lorenz gauge condition into each Maxwell equation:
\begin{equation}
\begin{split} \label{maxexp}
			- 2 r \p_r (r \p_u A_u) + \p_r(r^2 \p_r A_u) + 2 \gamma^{z \bz} \p_z \p_{\bz} A_u   &= 0~, \\
			 -2 \p_r (r A_u) - 2 \p_u \p_r (r^2 A_r) + \p_r^2 (r^2 A_r) +2 \gamma^{z \bz} \p_z \p_{\bz} A_r& = 0 ~, \\
			-2 r^2 \p_u \p_r A_z + r^2 \p_r^2 A_z+ 2r \p_z (A_r - A_u) + 2   \p_z ( \gamma^{z \bz}   \p_{\bz} A_{ z})&= 0~. 
\end{split}
\end{equation}

The gauge field can be expanded at large $r$ as
\begin{equation}
\begin{split}
A_u = \sum_{n=2}^\infty \frac{A_u^{(n)}}{r^n} ~, \qquad A_r = \sum_{n=2}^\infty \frac{A_r^{(n)}}{r^n}  ~, \qquad A_z = \sum_{n=0}^\infty \frac{A_z^{(n)}}{r^n}  ~. 
\end{split}
\end{equation}
Note that $A_u$ has a slightly different expansion than what was given in section \ref{qed}, namely, $A_u^{(1)}$ = 0.    The Lorenz gauge  condition leaves unfixed a set of residual gauge transformations satisfying 
$\square \ve = 0$.  Among these are a set of trivial gauge transformations with falloff behavior $\mathcal{O}(1/r)$ near $\mathcal{I}^+$.  This set of residual trivial gauge transformations  has precisely the 
 degrees of freedom of a massless scalar field with  the usual falloff near  $\mathcal{I}^+$, and thus, the residual trivial gauge symmetry can be used to set $A_{u}^{(1)} = 0$.
 Then the Lorenz gauge condition has the following expansion:
\begin{equation}
\begin{split}\label{lorexp}
			\p_u A_r^{(2)} &=   \gamma^{z \bz} \left ( \p_z A_{ \bz}^{(0)} +  \p_{\bz} A_{ z}^{(0)} \right)~,    \\
			\p_u A_r^{(3)}  & =  \gamma^{z \bz} \left ( \p_z A_{ \bz}^{(1)} +  \p_{\bz} A_{ z}^{(1)} \right)~,   \\
			\p_u A_r^{(n+2)} & = (n-1)(A_u^{(n+1)}-A_r^{(n+1)}) + \gamma^{z \bz} \left ( \p_z A_{ \bz}^{(n)} +  \p_{\bz} A_{ z}^{(n)} \right) ~,~~~~ n>1~.
\end{split}
\end{equation}
Using  the leading order equation, one can determine $A_{r}^{(2)}$ up to integration constants in terms of the boundary data $ A_{ z}^{(0)} $.
The first equation in \eqref{maxexp} has the following expansion:
\begin{equation}
\begin{split}\label{exer3}
		 	  \p_u A_u^{(2)} & = 0~, \\
			   \p_u A_u^{(n+1)}& = -\frac{1}{2n}\Big [n (n-1) +2 \gamma^{z \bz} \p_z \p_{ \bz}\Big]A_u^{(n)}    ~, ~~ ~~n>1~.
\end{split}
\end{equation}
  Hence,  we can determine all $A_u^{(n)}$ up to integration constants.
The second equation in \eqref{maxexp} has the following expansion:		 
\begin{equation}
\begin{split}
		 	  \p_u A_r^{(3)} &= -   A_u^{(2)}-   \gamma^{z \bz} \p_z \p_{ \bz}A_r^{(2)} ~, \\
			 \p_u A_r^{(n+1)} & =-  A_u^{(n)} - \left [ \frac{1}{2} (n-2) + \frac{1}{n-1}\gamma^{z \bz} \p_z \p_{ \bz} \right]A_r^{(n)}   ~, ~~~~n> 2~.
\end{split}
\end{equation}
 Hence,  with the previously determined set of $A_u^{(n)}$ and $A_r^{(2)}$, we can determine all $A_r^{(n)}$ up to integration constants.
 Finally, the third  equation in \eqref{maxexp} has the following expansion:	
\begin{equation}
\begin{split}
		 	   \p_u A_z^{(1)}& =    -   \p_z (\gamma^{z \bz} \p_{\bz} A_z^{(0)})~, \\
			  \p_u A_z^{(n+1)} & =   \frac{1}{n+1 } \left [-   \p_z A_r^{(n+1)} +   \p_z A_u^{(n+1)} - \frac{1}{2} n (n+1) A_z^{(n)}-   \p_z(\gamma^{z \bz} \p_{\bz} A_z^{(n)}) \right] , 
\end{split}
\end{equation}
for $n > 0$. With the previously determined sets of $A_u^{(n)}$ and $A_r^{(n)}$ plus the boundary data $A_z^{(0)}$, we can determine all $A_z^{(n)}$, up to integration constants.  Therefore, 
we have determined the full four-dimensional gauge field entirely in terms of $A_z^\0$ up to integration constants. 
Note the remaining (seemingly unused)
equations in \eqref{exer3} give rise to additional constraints on the integration constants.

\item[3(a).] Varying the action, we find
\begin{equation}
\begin{split}
\delta S = - \frac{1}{e^2} \int d \delta A \wedge \ast F = \frac{1}{e^2} \int  \left( d \ast F \right) \wedge \delta A  - \frac{1}{e^2} \int  d \left(   \ast F \wedge \delta A  \right) ~. 
\end{split}
\end{equation}
The equations of motion are $d \ast F = 0$. Following \cite{Wald:1999wa}, the (pre) symplectic potential current density can be read off as
\begin{equation}
\begin{split}
{\bf \Theta}\left( A , \delta A \right)  =   - \frac{1}{e^2}  \left( \ast F \right) \wedge \delta A ~. 
\end{split}
\end{equation}
The symplectic form is then
\begin{equation}
\begin{split}
 \Omega_\Sigma =  \int_\Sigma   {\boldsymbol \omega} \left( A , \delta_1 A , \delta_2 A \right) =   - \frac{1}{e^2}  \int_\Sigma \left[  \delta_1 \left( \ast F \right) \wedge \delta_2 A  -  \delta_2 \left( \ast F \right) \wedge \delta_1 A  \right]~. 
\end{split}
\end{equation}
In terms of a wedge product of differential forms on the symplectic manifold, this can be written as
\begin{equation}
\begin{split}
\Omega_\Sigma =  - \frac{1}{e^2} \int_\Sigma  \delta  \left( \ast F \right) \wedge \delta  A ~.  
\end{split}
\end{equation}

\item[3(b).]  For $\Sigma = \ci^+$, we are interested in the $uz\bz$ component of the above three-form. We find
\begin{equation}
\begin{split}
{\boldsymbol \omega}_{uz\bz} \big|_{\ci^+} &= - \frac{1}{e^2} \left[ \delta (\ast F)_{uz} \wedge \delta A_\bz + \delta (\ast F)_{\bz u} \wedge \delta A_z + \delta (\ast F)_{z\bz} \wedge \delta A_u \right]_{\mathcal{I}^+} \\
&= - \frac{i}{e^2} \left[ \delta F^\0_{uz} \wedge \delta A^\0_\bz + \delta F^\0_{u\bz} \wedge \delta A^\0_z   \right] ~,
\end{split}
\end{equation}
where
\begin{equation}
\begin{split}
(\ast F)_{uz} = \ve_{uzr\bz} F^{r\bz} = i \left( F_{uz} - F_{rz} \right) ~, \qquad (\ast F)_{z\bz} =  i r^2 \g_{z\bz} F_{ru} ~. 
\end{split}
\end{equation}
This implies
\begin{equation}
\begin{split}
\Omega_{\ci^+} =    \frac{1}{e^2} \int du d^2 z \left[ \delta F^\0_{uz} \wedge \delta A^\0_\bz + \delta F^\0_{u\bz} \wedge \delta A^\0_z   \right] ~.
\end{split}
\end{equation}
We now use this to write everything in terms of ${\hat A}_z$, $\phi$, and $N$. To do this, we write
\begin{equation}
\begin{split}
A_z^\0 = {\hat A}_z +  \p_z \phi ~, \qquad \int_{-\infty}^\infty du F^\0_{uz} = e^2 \p_z N ~, 
\end{split}
\end{equation}
which gives
\begin{equation}
\begin{split}
\Omega_{\ci^+} =  \frac{2}{e^2} \int du d^2 z  \p_u \delta {\hat A}_z \wedge \delta {\hat A}_\bz - 2   \int  d^2 z \delta \p_z \phi \wedge   \p_\bz  \delta N   ~. 
\end{split}
\end{equation}
We can then derive the brackets
\begin{equation}
\begin{split}
- \frac{2}{e^2} \left[ \p_u {\hat A}_z(u,z,\bz) , {\hat A}_\bw ( u',w,\bw) \right] &= i \delta(u-u') \delta^2(z-w)  ~, \\
2   \left[ \p_z \phi (z,\bz) , \p_\bw N(w,\bw) \right] &= i \delta^2 ( z - w )~. 
\end{split}
\end{equation}
Integrating the above, we find
\begin{equation}
\begin{split}
 \left[ {\hat A}_z(u,z,\bz) , {\hat A}_\bw ( u',w,\bw) \right] &= - \frac{ie^2}{4} \Theta(u-u') \delta^2(z-w)  ~, \\
\left[   \phi (z,\bz) ,  N(w,\bw) \right] &= - \frac{i}{4\pi  } \log | z - w |^2  + f(z,\bz) + g(w,\bw)  ~. 
\end{split}
\end{equation}

\item[4.] The free field mode expansion can be written as
\begin{equation}
\begin{split}
A_\mu (x) &=  e \sum_{\alpha=\pm} \int \frac{d^3q}{(2\pi)^3} \frac{1}{2\omega_q} \left[ \ve_\mu^{*\alpha} (\vec{q}) a_\alpha (\vec{q}) e^{- i \omega_q u - i \omega_q r \left( 1 - {\hat q} \cdot {\hat x} \right) } \right. \\
&\left. \qquad \qquad \qquad \qquad \qquad \qquad \qquad\qquad \quad  + \ve_\mu^{\alpha} (\vec{q}) a_\alpha^\dagger  (\vec{q}) e^{  i \omega_q u + i \omega_q r \left( 1 - {\hat q} \cdot {\hat x} \right) }  \right]  \\
&=  \frac{e}{ 8\pi^2 } \sum_{\alpha=\pm} \int_0^\infty d\omega_q \omega_q  \int_0^\pi d\theta \sin\theta \left[ \ve_\mu^{*\alpha} (\vec{q}) a_\alpha (\vec{q}) e^{- i \omega_q u - i \omega_q r \left( 1 - \cos\theta \right) }  \right. \\
&\left. \qquad \qquad \qquad \qquad \qquad \qquad \qquad\qquad  \quad + \ve_\mu^{\alpha} (\vec{q}) a_\alpha^\dagger  (\vec{q}) e^{  i \omega_q u + i \omega_q r \left( 1 -  \cos\theta \right) }  \right] . 
\end{split}
\end{equation}
At large $r$, we use the saddle point approximation. The exponent is stationary at $\theta =0,\pi$. The saddle point at $\theta = \pi$ does not contribute due to the Riemann-Lebesgue lemma (the assumptions of the lemma hold here as operator statements due to the cluster decomposition principle that all $\mathcal S$-matrices satisfy). Then, expanding the integrand around $\theta = 0$, we find
\begin{equation}
\begin{split}
A_\mu (x) &=  \frac{e}{ 8 \pi^2 } \sum_{\alpha=\pm} \int_0^\infty d\omega_q \omega_q \ve_\mu^{*\alpha} (\omega_q {\hat x}) a_\alpha (\omega_q {\hat x}) e^{- i \omega_q u}  \int_0^\pi d\theta  \theta  e^{ - i \omega_q r \theta^2/2  }  +  c.c. + \CO(r^{-2}) \\
&=  - \frac{i e}{8 \pi^2  r } \sum_{\alpha=\pm} \int_0^\infty d\omega_q  \left[   \ve_\mu^{*\alpha} (\omega_q {\hat x}) a_\alpha (\omega_q {\hat x}) e^{- i \omega_q u}  -   c.c. \right] + \CO(r^{-2}) ~. \\
\end{split}
\end{equation}
We are interested first in $A_z = \p_z x^\mu A_\mu$. To determine this, note that
\begin{equation}
\begin{split}
\p_z x^\mu \ve^+_\mu ( \omega_q {\hat x}) = 0 ~, \qquad \p_z x^\mu \ve^-_\mu ( \omega_q {\hat x}) = \frac{ \sqrt{2} r}{ 1 + z \bz } ~.  
\end{split}
\end{equation}
Using this, we find
\begin{equation}
\begin{split}
A_z  &=  - \frac{i }{ 8 \pi^2  }  \frac{ \sqrt{2} e  }{ 1 + z \bz }   \int_0^\infty d\omega_q  \left[ a_+ (\omega_q {\hat x}) e^{- i \omega_q u}  - a_-^\dagger (\omega_q {\hat x}) e^{  i \omega_q u}  \right] + \CO(r^{-1}) ~. \\
\end{split}
\end{equation}
Finally, we can immediately read off from this equation the large-$r$ mode:
\begin{equation}
\begin{split}
A^\0_z  &=  - \frac{i }{ 8 \pi^2  }  \frac{ \sqrt{2} e  }{ 1 + z \bz }   \int_0^\infty d\omega_q  \left[ a_+ (\omega_q {\hat x}) e^{- i \omega_q u}  - a_-^\dagger (\omega_q {\hat x}) e^{  i \omega_q u}  \right] ~. \\
\end{split}
\end{equation}
From this, we find
\begin{equation}
\begin{split}
e^2 \p_z N &= \frac{1}{2} \lim_{\omega \to 0^+} \int_{-\infty}^\infty du \left( e^{i \omega u} + e^{- i \omega u} \right) \p_u A_z^\0  \\
&=  - \frac{1}{ 8 \pi }  \frac{ \sqrt{2} e  }{ 1 + z \bz }   \lim_{\omega \to 0^+}  \left[  \omega a_+ (\omega  {\hat x})   +  \omega a_-^\dagger (\omega  {\hat x})   \right]    ~. 
\end{split}
\end{equation}

\item[5(a).] Rewriting expression (\ref{softth}) is a simple problem of algebra. We are required to substitute
\begin{equation}
\begin{split}
q^\mu &= \frac{\omega}{1+z\bz} \left( 1+z\bz , z + \bz , - i ( z - \bz ) , 1 - z \bz \right) ~, \\
p_k^\mu &= \frac{E_k}{1+z_k\bz_k} \left( 1 + z_k \bz_k , z_k + \bz_k , -i(z_k-\bz_k) , 1 - z_k \bz_k \right) ~, \\
\ve^\mu_+ (q) &= \frac{1}{\sqrt{2}} \left(   \bz , 1 , - i , - \bz \right)  ~, \\
 \ve^\mu_- (q) &= \frac{1}{\sqrt{2}} \left( z , 1 , i , - z \right)~. 
\end{split}
\end{equation}
To simplify the algebra a little, note that the polarizations satisfy the property
\begin{equation}
\begin{split}
\ve^\mu_+(q) = \frac{1}{\sqrt{2}\omega}  \p_z \left[ \left( 1 + z \bz \right) q^\mu  \right]  ~, \qquad \ve^\mu_-(q) = \frac{1}{\sqrt{2}\omega}  \p_\bz \left[ \left( 1 + z \bz \right) q^\mu  \right] ~. 
\end{split}
\end{equation}
Then the soft factor for the $k$th particle takes the form
\begin{equation}
\begin{split}
\frac{ p_k \cdot \ve_+ }{ p_k \cdot q } &= \frac{1}{\sqrt{2}\omega}   \frac{\p_z \left[ \left( 1 + z \bz \right) p_k \cdot   q  \right]    }{ p_k \cdot q }    \\
&=  \frac{1}{\sqrt{2}\omega}  \left( 1 + z \bz \right) \p_z \log   \left[ \left( 1 + z \bz \right) p_k \cdot   q  \right] ~. 
\end{split}
\end{equation}
A similar formula for negative helicity holds with $\p_z \to \p_\bz$. Finally, compute
\begin{equation}
\begin{split}
p_k \cdot q &= - \frac{2 \omega E_k}{  \left( 1 + z \bz \right) \left( 1 + z_k \bz_k \right) } |z-z_k|^2  ~. 
\end{split}
\end{equation}
Then the soft factor is
\begin{equation}
\begin{split}
\frac{ p_k \cdot \ve_+ }{ p_k \cdot q } &=  \frac{1}{\sqrt{2}\omega}  \left( 1 + z \bz \right) \p_z \log  |z-z_k|^2  =   \frac{1}{\sqrt{2}\omega} \frac{1+z\bz}{z-z_k} ~, \\
\frac{ p_k \cdot \ve_- }{ p_k \cdot q } &=  \frac{1}{\sqrt{2}\omega}  \left( 1 + z \bz \right) \p_\bz \log  |z-z_k|^2  =   \frac{1}{\sqrt{2}\omega} \frac{1+z\bz}{\bz - \bz_k} ~. 
\end{split}
\end{equation}
Thus, the soft-theorem can be written as
\begin{equation}
\begin{split}
& \lim_{\omega \to 0^+} \left[ \omega \bra{\text{out}} a^{\text{out}}_+ ( \omega {\hat x} ) {\cal S} \ket{\text{in}}   \right] \\
&\qquad \qquad = \frac{e}{\sqrt{2}} \left( 1 + z \bz \right) \left[ \sum_{k\in\text{out}} \frac{Q_k}{z-z_k}  - \sum_{k\in\text{in}} \frac{Q_k}{z-z_k}  \right]  \bra{\text{out}}  {\cal S} \ket{\text{in}}  ~.  
\end{split}
\end{equation}
A similar formula is true for the negative helicity photon. 
Finally,  note that we can write the limit in terms of an insertion of $N_z$, by using formula (\ref{exercise3}) derived in exercise  4, namely,
\begin{equation}
\begin{split}
e^2 \p_z N = - \frac{1}{8\pi^2} \frac{ \sqrt{2} e}{ 1 + z \bz } \lim_{\omega \to 0^+} \left[ \omega a_+ \left( \omega {\hat x} \right) + \omega a_-^\dagger \left( \omega {\hat x} \right) \right] ~. 
\end{split}
\end{equation}
This gives for an insertion of $N_z$,
\begin{equation}
\begin{split}
  \bra{\text{out}} \p_z N~ {\cal S} \ket{\text{in}} = - \frac{1}{8\pi^2}  \left[ \sum_{k\in\text{out}} \frac{Q_k}{z-z_k}  - \sum_{k\in\text{in}} \frac{Q_k}{z-z_k}  \right]  \bra{\text{out}}  {\cal S} \ket{\text{in}}  ~.  
\end{split}
\end{equation}

\item[5(b).] Again, it will be convenient to write the soft factor
\begin{equation}
\begin{split}
\frac{ p_k \cdot \ve_+ }{ p_k \cdot q } &=  - \frac{1}{\sqrt{2}\omega}  \left( 1 + z \bz \right) \p_z \log   \frac{\left( 1 + z \bz \right)^{-1} }{ p_k \cdot   q }~. 
\end{split}
\end{equation}
However, now we parametrize $p_k^\mu = m_k \left( \sqrt{ 1 +  \rho_k^2 } ~,   \rho_k {\hat x}_k \right) $ and $q^\mu = \omega \left( 1 ,  {\hat q} \right)$. Then, we find
\begin{equation}
\begin{split}\label{asdsadsacxzcxzc}
\frac{ p_k \cdot \ve_+ }{ p_k \cdot q } &=   - \frac{1}{2 \sqrt{2}\omega}  \left( 1 + z \bz \right) \p_z \log   \frac{ \g_{z\bz}  }{  \big( \sqrt{1+\rho_k^2} - \rho_k {\hat q} \cdot {\hat x}_k   \big)^2 } ~. 
\end{split}
\end{equation}
Note that the mass $m_k$ drops out of the formula. Now, recall the scalar propagator \eqref{wet}:
\begin{equation}
\begin{split}
G\left( \rho,{\hat x};{\hat q}\right) = \frac{1}{4\pi}  \frac{\g_{z\bz} }{ \big( \sqrt{1+\rho^2} - \rho {\hat q} \cdot {\hat x}   \big)^2}~. 
\end{split}
\end{equation}
This is precisely the term in the logarithm in the soft factor. Thus, we find
\begin{equation}
\begin{split}
\frac{ p_k \cdot \ve_+ }{ p_k \cdot q } &=   - \frac{1}{2 \sqrt{2}\omega}  \left( 1 + z \bz \right) \p_z \log  G\left( \rho_k,{\hat x}_k;{\hat q}\right)   ~. 
\end{split}
\end{equation}
Then the soft theorem can be written as
\begin{equation}
\begin{split} 
 \bra{\text{out}}   \p_z N ~ {\cal S} \ket{\text{in}} =    \bra{\text{out}}  {\cal S} \ket{\text{in}}  \left[ \sum_{k\in\text{out} }     -   \sum_{k\in\text{in} }   \, \, \right]   \frac{  Q_k   }{16\pi^2} \p_z \log   G\left( \rho_k,{\hat x}_k;{\hat q}\right)  ~.  
\end{split}
\end{equation}

\item[6(a).] The compactified coordinates on the Penrose diagram of Minkowski space are given by
\begin{equation}
\begin{split}
T =  \tan^{-1} \left( t+r \right)  + \tan^{-1} \left( t - r \right)  ~, \\
R =  \tan^{-1} \left( t+r \right)  - \tan^{-1} \left( t - r \right)  ~, \\
\end{split}
\end{equation}
with the coordinate range $R + |T|  < \pi$ and $R \geq 0$. Setting $t = \nu r$, we find
\begin{equation}
\begin{split}
T =  \tan^{-1} \left[ r \left( \nu + 1 \right) \right]   + \tan^{-1} \left[ r \left( \nu - 1 \right) \right]   ~, \\
R =  \tan^{-1} \left[ r \left( \nu + 1 \right) \right]   - \tan^{-1} \left[ r \left( \nu - 1 \right) \right]  ~. \\
\end{split}
\end{equation}
Lines of constant $\nu$ and constant $r$ are shown in figure \ref{compact} .
\begin{figure}[h!] 
\begin{center}
\includegraphics[scale=0.9]{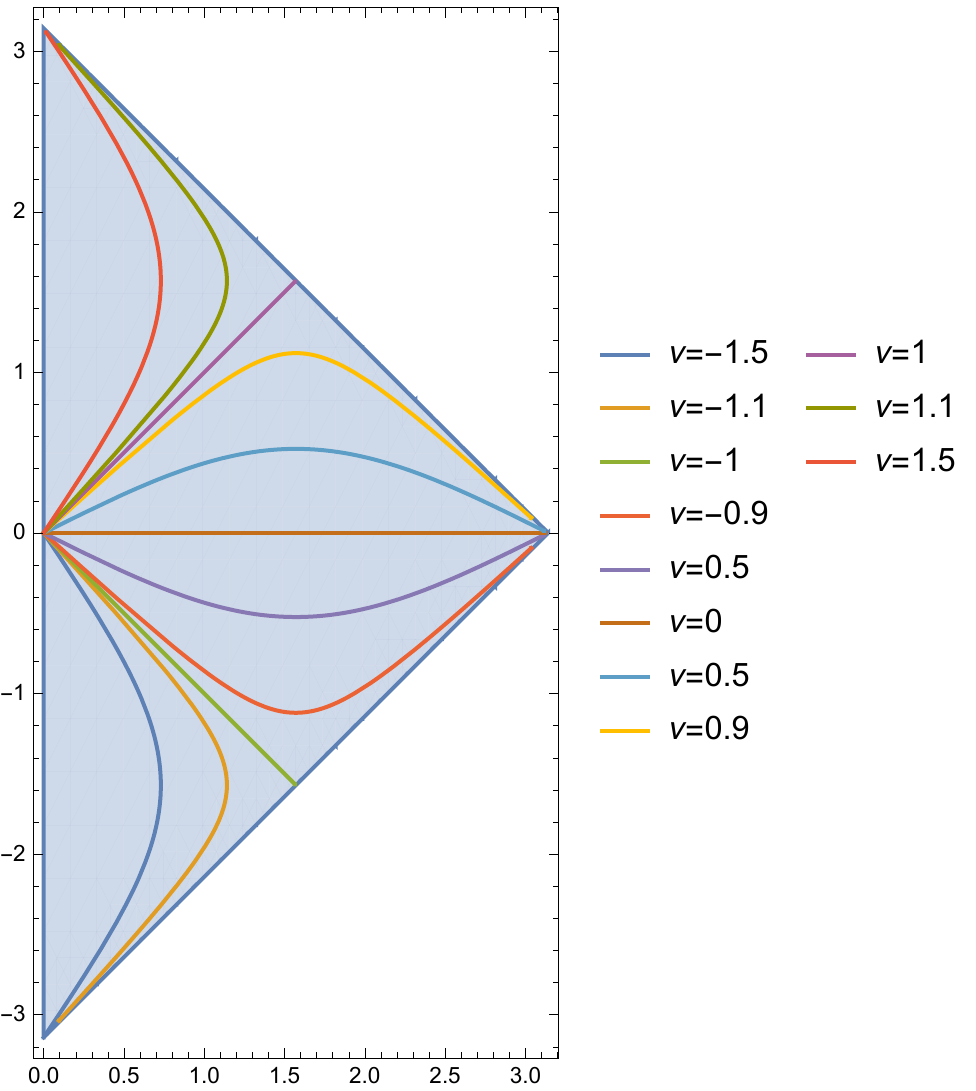}
\includegraphics[scale=0.9]{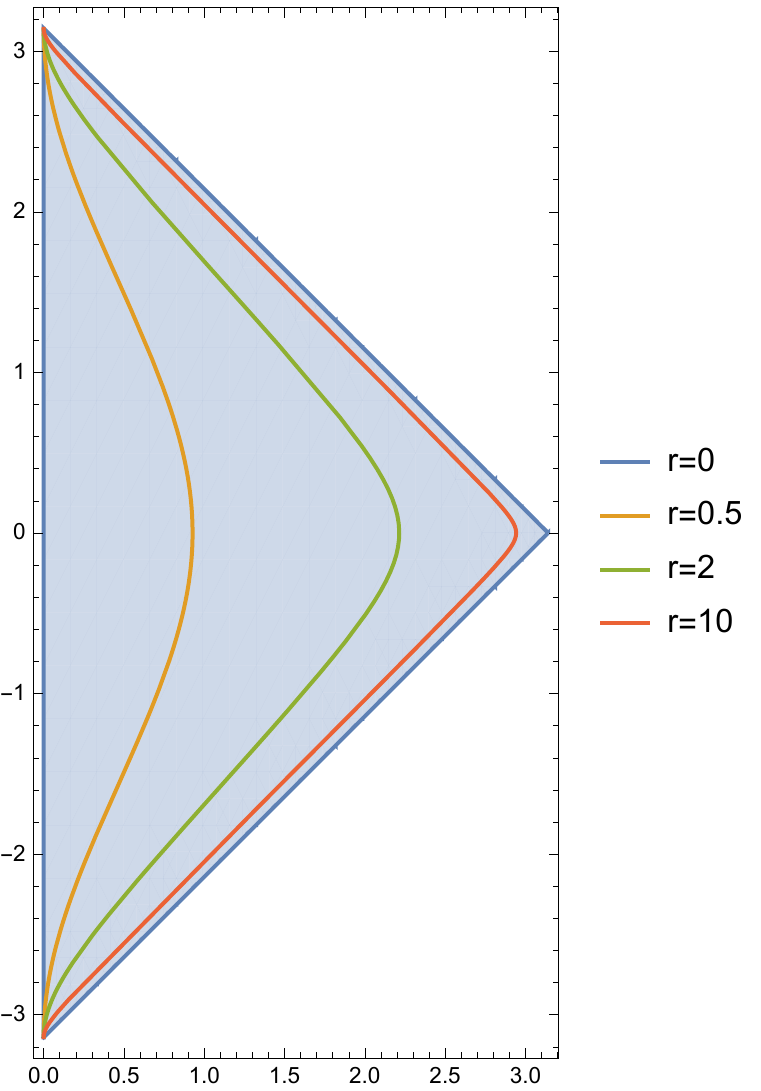}
\end{center}
\caption{Lines of constant $\nu$ (left) and constant $r$ (right).}\label{compact}
\end{figure}

\item[6(b).]  Let us first determine the metric in these coordinates. We start from the metric of Minkowski space in the standard spherical coordinates
\begin{equation}
\begin{split}
ds^2 = - dt^2 + dr^2 + r^2 d\Omega_2^2~. 
\end{split}
\end{equation}
Substituting in $t = \nu r$, we find
\begin{equation}
\begin{split}
ds^2 &= - (r d\nu + \nu dr )^2 + dr^2 + r^2 d\Omega_2^2  \\
&= - r^2 d\nu^2 - 2 \nu r d\nu dr + \left( 1 - \nu^2 \right) dr^2 + r^2 d\Omega_2^2 ~. 
\end{split}
\end{equation}
The inverse metric is
\begin{equation}
\begin{split}
g^{\mu \rho} = \frac{1}{r^2}  \begin{pmatrix}
- \left( 1 - \nu^2 \right) & - r \nu & 0 \\ - \nu r & r^2 & 0 \\ 0 & 0 & \g^{AB} 
\end{pmatrix} ~,
\end{split}
\end{equation}
where $x^A=(\theta,\phi)$, and $\g^{AB}$ is the inverse metric on the unit $S^2$. Note that $\sqrt{-g} = r^3 \sqrt{\g}$.
The scalar wave equation is then
\begin{equation}
\begin{split}
0 &= D^2 \ve = \frac{1}{\sqrt{-g}} \p_\mu \left( \sqrt{-g} g^{\mu \rho } \p_\rho \ve \right) \\
&= - \frac{1}{r^2} \left( 1 - \nu^2 \right) \p_\nu^2 \ve - \frac{2\nu}{r} \p_\nu \p_r \ve + \frac{2}{r} \p_r\ve + \p_r^2 \ve + \frac{1}{r^2} D^2 \ve ~,
\end{split}
\end{equation}
where $D^2$ is the Laplacian operator on $S^2$.

\item[6(c).]  For $r$-independent solutions, the scalar wave equation reduces to
\begin{equation}
\begin{split}
 \left( 1 - \nu^2 \right)  \p^2 _\nu \ve  =  D^2 \ve ~. 
\end{split}
\end{equation}
Now consider solutions of the form $\ve_{\ell m}(\nu,x^A) = A_\ell(\nu) Y^\ell_m(x^A)$. By noting that $Y^\ell_m(x^A)$ satisfies $D^2 Y^\ell_m = - \ell \left( \ell + 1 \right) Y^\ell_m$, we find the following equation for $A_\ell(\nu)$:
\begin{equation}\label{eqnAl}
\begin{split}
 \left( 1 - \nu^2 \right)  \p^2 _\nu A_\ell(\nu) = - \ell \left( \ell + 1 \right) A_\ell (\nu) ~. 
\end{split}
\end{equation}
When $\ell = 0$, the most general solution is $A_0(\nu) = d_1 \nu + d_2$. Requiring that the function be well defined at the origin, $\nu \to \pm\infty$ requires $d_1 = 0$. We also normalize the solution so that $A_\ell(1) = 1$. Thus, we must have $d_2 = 1$. So, for $\ell = 0$, the solution is
\begin{equation}
\begin{split}
A_0(\nu) = 1~. 
\end{split}
\end{equation}
We now come to the less trivial case $\ell > 0$, which will be our assumption for the rest of this problem. The first step is to note that by taking a derivative of the equation above we arrive at 
\begin{equation}
\begin{split}
 \left( 1 - \nu^2 \right)  A'''_\ell(\nu) - 2 \nu A''_\ell(\nu) + \ell \left( \ell + 1 \right) A_\ell' (\nu)  = 0 ~,
\end{split}
\end{equation}
which is the Legendre equation for $A'_\ell(\nu)$. Thus, we must have
\begin{equation}
\begin{split}
A'_\ell(\nu) = c_1  P_\ell(\nu) + c_2  Q_\ell(\nu)~. 
\end{split}
\end{equation}
Integrating this equation yields
\begin{equation}
\begin{split}\label{alansatz}
A_\ell(\nu) = \frac{c_1}{2\ell+1}   \left[ P_{\ell+1}(\nu) - P_{\ell-1} (\nu) \right] + \frac{c_2}{2\ell+1}   \left[ Q_{\ell+1}(\nu) - Q_{\ell-1} (\nu) \right] + c_3 ~, 
\end{split}
\end{equation}
where we have used the formula
\begin{equation}
\begin{split}
\int d\nu  P_\ell(\nu) = \frac{P_{\ell+1}(\nu) - P_{\ell-1} (\nu)}{2\ell+1}~, 
\end{split}
\end{equation}
and similarly for $Q_\ell$. Note that we originally started out with a second order differential equation, whereas our final answer for $A_\ell(\nu)$ contains three unknown constants. This is because we took a derivative of the original differential equation (\ref{eqnAl}) in the process. To fix one of the constants, let us plug in our ansatz for the solution into the original differential equation. To do this, we use the property
\begin{equation}
\begin{split}
\left[ \left( 1 - \nu^2 \right) \p_\nu^2 + \ell \left( \ell + 1 \right) \right] \left[ P_{\ell+1}(\nu) - P_{\ell-1}(\nu) \right]  = 0 ~, \\
\left[ \left( 1 - \nu^2 \right) \p_\nu^2 + \ell \left( \ell + 1 \right) \right] \left[ Q_{\ell+1}(\nu) - Q_{\ell-1}(\nu) \right]  = 0 ~. \\
\end{split}
\end{equation}
Thus, for \eqref{alansatz} to satisfy the differential equation, we must have $c_3 = 0$. So, we find the solution
\begin{equation}
 \begin{split}
A_\ell(\nu) = c_1  \left[ P_{\ell+1}(\nu) - P_{\ell-1} (\nu) \right] +  c_2  \left[ Q_{\ell+1}(\nu) - Q_{\ell-1} (\nu) \right]  ~. 
\end{split}  
\end{equation}
We are interested in studying the properties of this function at $\nu = \pm 1$ and $\nu = \pm \infty$. For this purpose, a more convenient representation of $Q_\ell(\nu)$ is
\begin{equation}
\begin{split}
Q_\ell(\nu) = \frac{1}{2}  P_\ell(\nu) \log \frac{1+\nu}{1-\nu} + F_\ell(\nu) ~, \qquad |\nu| < 1 ~,
\end{split}
\end{equation}
where
\begin{equation}
\begin{split}
F_\ell(\nu) = -  \sum_{k=0}^{\lfloor \frac{\ell-1}{2} \rfloor } \frac{2\ell - 4 k - 1 } { \left( 2 k + 1 \right) \left( \ell - k \right) }  P_{\ell - 2 k - 1 } (\nu)  ~. 
\end{split}
\end{equation}
Note that $F_\ell(\nu)$ is a polynomial in $\nu$. Then we have 
\begin{equation}
\begin{split}\label{adasddef}
A_\ell(\nu) &=  \left[ c_1 + \frac{c_2}{2}  \log \frac{1+\nu}{1-\nu}  \right] \left[ P_{\ell+1}(\nu) - P_{\ell-1} (\nu) \right]  + c_2 \left[ F_{\ell+1}(\nu) - F_{\ell-1} (\nu) \right]  , 
\end{split}
\end{equation}
for $ |\nu|<1$. Due to the logarithm in the parenthesis above, $A_\ell(\nu)$ has a branch point at $\nu = \pm1$. Thus, we must choose an appropriate branch cut. We would like to choose the branch cut so that $A_\ell(\nu)$ is defined for all $\nu \in \mrr$ and so that it does not diverge at both $\nu \to \pm\infty$. As a result, the branch cuts must be chosen as follows. First, we choose an arbitrary branch cut at $\nu = +1$. This determines a particular way of defining $A_\ell(\nu)$ for $\nu > 1$ and in the process also specifies the asymptotic behavior of $A_\ell(\nu)$ at $\nu \to \infty$. Requiring that this is finite will fix $c_1$ in terms of $c_2$. Once this has been done, the choice of branch at $\nu = - 1$ is fixed by requiring that the same solution (namely, for the same choice of $c_1$ and $c_2$) also be finite at $\nu \to -\infty$. 
We choose the branch cut at $\nu = 1$ so that it extends along $\nu = 1 + i t$ for $t > 0$. With this choice of the branch cut, the extension of $A_\ell(\nu)$ to $\nu > 1$ is fixed as
\begin{equation}
\begin{split}\label{eqsd}
A_\ell(\nu) &=  \left[ c_1 - \frac{ i c_2 \pi}{2} + \frac{c_2}{2}  \log  \frac{\nu+1}{\nu-1} \right] \left[ P_{\ell+1}(\nu) - P_{\ell-1} (\nu) \right]  + c_2 \left[ F_{\ell+1}(\nu) - F_{\ell-1} (\nu) \right] ~.  
\end{split}
\end{equation}
This extension can be seen by writing $\nu = 1 - r e^{i\theta}$ near the branch point. Note that as $\theta$ goes from $0$ to $\pi$, we move from $\nu < 1$ to $\nu > 1$ without crossing the branch cut. Plugging $\nu = 1 - r e^{i\theta}$ into \eqref{adasddef}, then setting $\theta = \pi$, and finally setting $r = \nu - 1$, we reproduce \eqref{eqsd}. 
We can now study the asymptotics of this function at large $\nu$. The leading term in the large $\nu$ expansion arises from $P_{\ell+1}(\nu)$, which goes like
\begin{equation}
\begin{split}
P_{\ell+1}(\nu) \quad \to \quad \frac{2^{-\ell-1} (2\ell+2)! }{  (\ell+1)!^2 } \nu^{\ell+1}     \qquad \text{as} \qquad \nu \to \infty ~. 
\end{split}
\end{equation}
Then we find
\begin{equation}
\begin{split}
A_\ell(\nu)  \to  \left( c_1 - \frac{ i c_2 \pi }{2 }  \right) \frac{2^{-\ell-1} (2\ell+2)! }{  (\ell+1)!^2 } \nu^{\ell+1} + \cdots ~. 
\end{split}
\end{equation}
The quantity above is divergent unless $c_1 = \frac{ic_2\pi}{2}$. A proper check requires that we also verify that the subleading divergent powers also do not contribute in this limit. The subleading divergent powers do not contribute because  
\begin{equation}
\begin{split}
\frac{1}{2}  \log \frac{\nu+1}{\nu-1}  P_{\ell}(\nu)   + F_{\ell}(\nu)    =  \CO\left( \nu^{-\ell-1} \right) \qquad \text{at large $\nu$} ~. 
\end{split}
\end{equation}
Thus, we have determined that for the choice $c_1 = \frac{ic_2\pi}{2}$, $A_\ell(\nu)$ is completely finite at large $\nu$.
We now consider the extension of $A_\ell(\nu)$ to $\nu < -1$. Since the asymptotics of the Legendre polynomials is the same at $\nu \to -\infty$ as it is for $\nu \to +\infty$, it is clear that a branch cut must be chosen so that $A_\ell(\nu)$ takes exactly the form \eqref{eqsd} for $\nu < -1$. This requires choosing a branch that starts at $\nu = -1$ and extends along $\nu = -1 + i t$ for $t>0$. The branch cuts for the function in the complex $\nu$ plane are shown in figure \ref{branch}.
\begin{figure}[ht!]
\begin{center}
\includegraphics{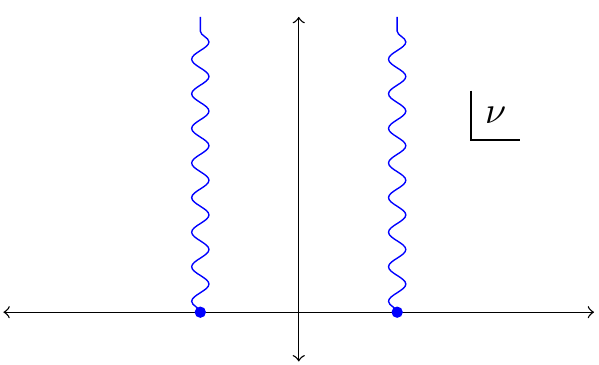}
\end{center}
\caption{Branch cuts for $A_\ell(\nu)$.} \label{branch}
\end{figure}
In summary, we have so far constructed branch cuts so that the function is well defined at $\nu = \pm\infty$. In doing so, we have defined the function everywhere except at $\nu = \pm 1$. Indeed, due to the presence of the branch point, it is not a priori evident that the function can be defined at these points. However, note that this is not the case. The only divergence at $\nu = \pm1$ arises from the logarithm. This is multiplied by $P_{\ell+1}(\nu) - P_{\ell-1}(\nu)$, which vanishes polynomially near $\nu = \pm1$.
Thus, $A_\ell(\pm1)$ is finite and well defined. In fact, the function is also continuous at this point, since the only nonvanishing term in this limit arises from $F_\ell(\nu)$, which is a polynomial and continuous everywhere. Explicitly, we have 
\begin{equation}
\begin{split}
A_\ell(\pm1) &= \lim_{h\to0^\pm} A_\ell(\pm1+h) = c_2 \left[ F_{\ell+1}(\pm1) - F_{\ell-1} (\pm1) \right]  = -  (\pm1)^{\ell}  \frac{ c_2\left( 2 \ell + 1 \right) }{ \ell \left( \ell + 1 \right) } ~. 
\end{split}
\end{equation}
As before, we normalize the function so that $A_\ell(1) = 1$. This fixes
\begin{equation}
\begin{split}
c_2 = - \frac{ \ell ( \ell + 1 ) }{ 2 \ell + 1 } ~. 
\end{split}
\end{equation}
In summary, for $|\nu|<1$, we have
\begin{equation}
\begin{split}
 A_\ell(\nu) =- \frac{ \ell ( \ell + 1 ) }{ 2 \ell + 1 } \left[  \frac{1}{2}  \log e^{i\pi} \frac{1+\nu}{1-\nu}  \left[ P_{\ell+1}(\nu) - P_{\ell-1} (\nu) \right]  +  F_{\ell+1}(\nu) - F_{\ell-1} (\nu)  \right]~, 
\end{split}
\end{equation}
whereas for $|\nu|>1$, we have
\begin{equation}
\begin{split}
 A_\ell(\nu) = -  \frac{ \ell ( \ell + 1 ) }{ 2 \ell + 1 } \left[ \frac{1 }{2} \left[ P_{\ell+1}(\nu) - P_{\ell-1} (\nu) \right]   \log  \frac{\nu+1}{\nu-1} +  F_{\ell+1}(\nu) - F_{\ell-1} (\nu) \right] ~. 
\end{split}
\end{equation}

\item[6(d).]  \label{partd}For $\ell = 2$, the solution in part (c) (when $\nu > 1$) takes the form
\begin{equation}
\begin{split}
A_2(\nu) = 3 \nu^2 - 2 + \frac{3}{2} \nu \left( 1 - \nu^2 \right) \log \frac{\nu+1}{\nu-1} ~. 
\end{split}
\end{equation}
Plugging in $\nu = \frac{t}{r} = 1 + \frac{u}{r}$ and expanding in powers of $r$, we find
\begin{equation}
\begin{split}
A_2\left( 1+\frac{u}{r}\right) = 1 + \frac{3u}{r} \left( 2 + \log \frac{u}{2r} \right) + \frac{3u^2}{2r^2} \left(1  + 3 \log  \frac{u}{2r}  \right) + \CO \big( r^{-3} \big)  ~. 
\end{split}
\end{equation}
This implies that the expansion of the full solution is 
\begin{equation}
\begin{split}\label{partd}
\ve_{2m}(u,r,\theta,\phi) &= Y^2_m(\theta,\phi) \left[ 1  + \frac{3u}{r} \left( 2 + \log \frac{u}{2r} \right) \right. \\
&\left. \qquad \qquad \qquad \qquad \qquad + \frac{3u^2}{2r^2} \left(1  + 3 \log  \frac{u}{2r}  \right) + \CO \big( r^{-3} \big) \right]  ~. 
\end{split}
\end{equation}

\item[6(e).]  We now solve the scalar wave equation at large $r$ in the retarded radial coordinates. The answer to part (d) of the problem shows that we must have a large-$r$ falloff of the form
\begin{equation}
\begin{split}\label{expansion}
\ve(u,r,z,\bz) &= \ve^\0(u,z,\bz) + \frac{1}{r} \ve^\1(u,z,\bz) + \frac{1}{r} f^\1(u,z,\bz)  \log\frac{u}{2 r}  \\
&\qquad \quad + \frac{1}{r^2 } \ve^\2(u,z,\bz) +  \frac{1}{r^2 } f^\2(u,z,\bz)  \log\frac{u}{2 r}  + \CO(r^{-3})~. 
\end{split}
\end{equation}
The scalar wave equation in the retarded radial coordinates is
\begin{equation}
\begin{split}
\left[ \p_r^2 - 2 \p_u \p_r + \frac{2}{r} \left( \p_r  - \p_u \right) + \frac{2}{r^2} \g^{z\bz} \p_z\p_\bz   \right] \ve (u,r,z,\bz) = 0  ~. 
\end{split}
\end{equation}
Plugging in the expansion \eqref{expansion} and expanding in large $r$, we find the following equations order-by-order in large $r$:
\begin{equation}
\begin{split}
\p_u \ve^\0 &= 0 ~, \qquad \p_u f^\1 +  \frac{1}{2} D^2 \ve^\0 = 0 ~, \qquad \p_u f^\2 + \frac{1}{2} D^2 f^\1 = 0 ~, \\
 \p_u  \ve^\2 &=  \frac{1}{2} D^2 f^\1 -  \frac{1}{2}  D^2 \ve^{(1)} - \frac{1}{2}  f^\1 - \frac{1}{u} f^\2  ~. \\
\end{split}
\end{equation}
The solutions to these equations are 
\begin{equation}
\begin{split}
f^\1 & = - \frac{u}{2} D^2 \ve^\0 + g^\1 ~, \\
f^\2 & =  \frac{u^2}{8}  \big( D^2 \big)^2  \ve^\0  -  \frac{u}{2} D^2  g^\1 + g^\2   ~, \\
 \ve^\2 &=   - \frac{3u^2}{16}  \big( D^2 \big)^2  \ve^\0 +  \frac{u^2}{8} D^2 \ve^\0  + u D^2  g^\1 - \frac{u}{2} g^\1   \\
 &\qquad \qquad \qquad \qquad \qquad \qquad   -   g^\2 \log u   -  \frac{1}{2}  D^2 \int du \ve^\1 + h^\2  ~, \\
\end{split}
\end{equation}
where all functions on the RHS are functions of $(z,\bz)$, except for $\ve^\1$, which is an arbitrary function of $(u,z,\bz)$. 
The solution is indeed the form of the solution $\ve_{2m}$ obtained in part (d). We can see this by choosing
\begin{equation}
\begin{split}
\ve^\0 = Y^2_m~, \qquad \ve^\1 = 6 uY^2_m ~, \qquad g^\1 = g^\2 = h^\2 = 0  ~. 
\end{split}
\end{equation}
In this case, we have
\begin{equation}
\begin{split}
f^\1   = 3 u  Y^2_m  ~, \qquad f^\2   =  \frac{9u^2}{2} Y^2_m~,  \qquad  \ve^\2 = \frac{3u^2}{2} Y^2_m  ~, 
\end{split}
\end{equation}
which is precisely the expansion \eqref{partd}.

\item[7(a).]  Inverting the coordinates, we have $r = \rho \tau$ and $t = \tau \sqrt{ 1 + \rho^2 } $. We parametrize the massive momentum as $p^\mu = \left( \w_p , \vec{p} \right)$. Then we have
\begin{equation}
\begin{split}
p \cdot x =  \tau  \left[ - \sqrt{p^2 + m^2 }  \sqrt{1+\rho^2}  +   \rho  \vec{p} \cdot {\hat x}  \right]  = -  \tau f(\vec{p})~. 
\end{split}
\end{equation}
In the mode expansion, we perform an integral over $\vec{p}_i$. In the limit of large $\tau$, the value of the integral is dominated by the integrand at the minimum of the function $f(\vec{p})$:
\begin{equation}
\begin{split}
\frac{\p f}{\p p_i} = 0 \qquad \implies \qquad   \frac{  \vec{p} }{  \sqrt{p^2+m^2}} = \frac{\rho}{ \sqrt{1+\rho^2}}  {\hat x}  ~. 
\end{split}
\end{equation}
Let us simplify this expression a little more. Squaring both sides, we find
\begin{equation}
\begin{split}
\frac{p^2}{p^2 + m^2 } = \frac{\rho^2}{1+\rho^2} \quad \implies \quad \rho = \frac{|\vec{p}|}{m}  \quad \implies \quad \w_p = m \sqrt{ 1 + \rho^2 } ~. 
\end{split}
\end{equation}
Then we find that the saddle point is at 
\begin{equation}
\begin{split}
\vec{p}  = m \rho {\hat x} ~. 
\end{split}
\end{equation}
Next, in the saddle point approximation, we must evaluate the function $f$ and its second derivative at this value. We find
\begin{equation}
\begin{split}
f(m\rho{\hat x}) =  m ~, \qquad \frac{\p}{\p p_i} \frac{\p}{\p p_j} f \bigg|_{\vec{p} = m\rho{\hat x}} =  \frac{1}{m} \left[ \delta_{ij} - \frac{ \rho^2}{1+\rho^2} {\hat x}_i {\hat x}_j  \right] =  \CP_{ij} ~. 
\end{split}
\end{equation}
Then around the saddle point,  expand
\begin{equation}
\begin{split}
p \cdot x = -m \tau -  \frac{\tau}{2} \CP_{ij} (p_i- m \rho \hat{x}_i) (p_j- m \rho \hat{x}_j) + \CO(p^3) ~. 
\end{split}
\end{equation}
At large $\tau$, we can then write the mode expansion as
\begin{equation}
\begin{split}
\phi(x) &=  \frac{1}{2m (2\pi)^3  \sqrt{1+\rho^2}}  \left [a(m\rho{\hat x}) e^{-im \tau } \int  d^3 p  e^{- i \frac{\tau}{2} \CP_{ij} p_i p_j}  \right. \\
&\left. \qquad \qquad \qquad \qquad  \qquad \qquad \qquad + a^\dagger(m\rho{\hat x}) e^{im \tau } \int  d^3 p  e^{  i \frac{\tau}{2} \CP_{ij} p_i p_j}   \right] \\
&=  \frac{ \sqrt{ \frac{(2\pi)^3}{ \tau^3 \det \CP } }  }{2m (2\pi)^3  \sqrt{1+\rho^2}}   \left [a( m\rho{\hat x}) e^{-im \tau } e^{ - 3 i \pi/4}  + a^\dagger(m\rho{\hat x}) e^{im \tau }   e^{   3 i \pi/4}  \right]~,
\end{split}
\end{equation}
where we used 
\begin{equation}
\begin{split}
\int_{-\infty}^\infty dx ~e^{i k x^2 } = \sqrt{ \frac{\pi}{k} } e^{i \pi/4}  ~. 
\end{split}
\end{equation}
Next we must evaluate $\det \CP$. An explicit computation shows
\begin{equation}\label{detP}
\begin{split}
\det \CP =  \frac{1}{m^3 \left( 1 + \rho^2 \right) } ~. 
\end{split}
\end{equation}
Using equation (\ref{detP}), we find
\begin{equation}
\begin{split}
 \phi(\tau,\rho,{\hat x})  =  \frac{\sqrt{m} }{2  (2\pi \tau )^{3/2} }  \left [a( m\rho{\hat x}) e^{-im \tau }  e^{ - 3 i \pi/4}  + a^\dagger(m\rho{\hat x}) e^{im \tau }    e^{   3 i \pi/4}  \right]~. 
\end{split}
\end{equation}

\item[7(b).] We start with the action
\begin{equation}\label{action}
\begin{split}
S = - \frac{1}{2} \int d^4 x \sqrt{-g} \left[ \left( \p_\mu \phi \right)^2 + m^2 \phi^2 \right] ~. 
\end{split}
\end{equation}
Varying (\ref{action})  with respect to $\phi$, we find
\begin{equation}
\begin{split}
\delta S =   \int d^4 x \sqrt{-g}   \left(  \Box   - m^2 \right) \phi    \delta \phi  -  \int d^4 x \sqrt{-g}  \nabla_\mu  \left(  \nabla^\mu  \phi  \delta \phi   \right)  ~. 
\end{split}
\end{equation}
The first term gives the equations of motion. The canonical one-form (symplectic potential current density) can be read off as $\Theta^\mu (\phi , \delta \phi ) = -   \nabla^\mu  \phi  \delta \phi    $. The symplectic form is then
\begin{equation}
\begin{split}
\Omega_\Sigma &= \int d\Sigma_\mu \omega^\mu = \int d\Sigma_\mu \left[ \delta_1 \Theta^\mu \left( \phi , \delta_2 \phi \right) - 1 \leftrightarrow 2 \right]  \\
& = -  \int d\Sigma_\mu \left[ \nabla^\mu  \delta_1  \phi  \delta_2  \phi     -  \nabla^\mu  \delta_2  \phi  \delta_1  \phi     \right]  ~. \\
\end{split}
\end{equation}

\item[7(c).] The Minkowskian metric in $(\rho,\tau,\theta,\phi)$ coordinates is
\begin{equation}
\begin{split}
ds^2 = - d\tau^2 + \tau^2 \left[ \frac{d\rho^2}{1+\rho^2} + \rho^2 \g_{AB} dx^A dx^B   \right] ~. 
\end{split}
\end{equation}
We consider the hypersurface $\Sigma$ to be $\tau = $ constant. Then the unit normal is
\begin{equation}
\begin{split}
n_\mu =  \p_\mu \tau =  \delta_\mu^\tau ~. 
\end{split}
\end{equation}
 The volume element is 
\begin{equation}
\begin{split}
d\Sigma_\mu =  d^3 V   \tau^{3} \delta_\mu^\tau~, 
\end{split}
\end{equation}
where $d^3V$ is the volume element on the unit hyperboloid. Using this volume element, we find
\begin{equation}
\begin{split}
\Omega_\tau  &  =  \int    d^3 V  \tau^{3}   \p_\tau  \delta   \phi  \wedge  \delta \phi     ~. \\
\end{split}
\end{equation}
To simplify this, we can now insert the large-$\tau$ expression for $\phi$. We then find at $\tau = \infty$ that
\begin{equation}\label{commRel}
\begin{split}
 \Omega_{\tau \to \infty}   =  -  \frac{  i m^2  }{2    (2\pi   )^{3} }    \int    d^3 V~  \delta a( m\rho{\hat x})   \wedge    \delta a^\dagger(m\rho{\hat x})    ~ .  
\end{split}
\end{equation}

\item[7(d).]The commutation relation obtained from equation (\ref{commRel}) is
\begin{equation}\label{commRel2}
\begin{split}
 \left[ a( m\rho{\hat x})   ,    a^\dagger(m\rho'{\hat x}')  \right] = \frac{2 \left( 2\pi \right)^3 }{ m^2 }   \delta \left( \rho - \rho' \right) \delta^2 \left( {\hat x} - {\hat x}' \right) ~, 
\end{split}
\end{equation}
where the $\delta$-functions are normalized so that
\begin{equation}
\begin{split}
\int d^3 V  \delta \left( \rho - \rho' \right) \delta^2 \left( {\hat x} - {\hat x}' \right)  =  1  ~. 
\end{split}
\end{equation}
To match (\ref{commRel2})  to the standard commutation relations in momentum space, we have to transform the $\delta$-functions above to the standard three-vector notation. We do this as follows. Recall that we are parameterizing
\begin{equation}
\begin{split}
\vec{p} = m \rho {\hat x} = \frac{ m \rho }{ 1 + z \bz } \left( z + \bz , - i (z - \bz ) , 1 - z \bz \right) ~. 
\end{split}
\end{equation}
Then the standard volume form in momentum space is $dp^1 \wedge dp^2 \wedge dp^3$. Using the parameterization above, we find
\begin{equation}\label{vol}
\begin{split}
dp^1 \wedge dp^2 \wedge dp^3 =  i m^3 \rho^2  \g_{z\bz}  d\rho \wedge dz \wedge d\bz ~. 
\end{split}
\end{equation}
From equation (\ref{vol}), we find
\begin{equation}\label{deltas}
\begin{split}
1 &= \int d^3 p \delta^3 \left( \vec{p} - \vec{p'} \right) \\
&= \int d\rho d^2 z  m^3 \rho^2  \g_{z\bz}  \delta^3 \left( \vec{p} - \vec{p'} \right)  \\
&=  \int d^3 V  m^2 \w_p   \delta^3 \left( \vec{p} - \vec{p'} \right) ~. 
\end{split}
\end{equation}
From equation (\ref{deltas}), we can read off
\begin{equation}
\begin{split}
 m^2 \w_p   \delta^3 \left( \vec{p} - \vec{p'} \right) =  \delta \left( \rho - \rho' \right) \delta^2 \left( {\hat x} - {\hat x}' \right)  ~. 
\end{split}
\end{equation}
Plugging this form into our commutation relations, we find
\begin{equation}
\begin{split}
 \left[ a( m\rho{\hat x})   ,    a^\dagger(m\rho'{\hat x}')  \right] =   \left( 2\pi \right)^3 \left( 2 \w_p \right)   \delta^3 \left( \vec{p} - \vec{p'} \right)  ~.  
\end{split}
\end{equation}

\item[8.] To determine the action of the charge, we first determine the symplectic form on $\ci^+$. We do so by determining the boundary fields $\phi$ and $\psi_+$ on $\ci^+$ using large-$r$ limits of the bulk mode expansions. For this, we will use the saddle point approximation. 
We start by determining the boundary field of a scalar. Bulk scalar fields have mode expansions
\begin{equation}
\begin{split}
\Phi(x) = \int \frac{d^3q}{ \left( 2 \pi \right)^3 } \frac{1}{2\omega_q} \left[ a ( \vec{q} ) e^{i q \cdot x } + {\tilde a}^\dagger ( \vec{q}) e^{- i q \cdot x } \right]  ~. 
\end{split}
\end{equation}
Here $a$ and ${\tilde a}$ satisfy the usual creation-annihilation commutation relations. Then, taking a large-$r$ limit and doing the usual saddle point procedure, we find
\begin{equation}
\begin{split}
\phi(u,z,\bz) = - \frac{i}{8\pi^2} \int_0^\infty d\omega \left[ a(\omega {\hat x}) e^{- i \omega u} - {\tilde a}^\dagger ( \omega {\hat x}) e^{i \omega u } \right]  ~. 
\end{split}
\end{equation}
Using the standard commutators, we find
\begin{equation}
\begin{split}
\left[ \phi (u,z,\bz) , {\bar \phi} (u',z',\bz') \right] = - \frac{i}{4} \g^{z\bz} \Theta(u-u') \delta^2 ( z - z' ) ~. 
\end{split}
\end{equation}
Bulk spinors\footnote{Spinor conventions of Wess and Bagger \cite{Wess:1992cp} are used here.} have the mode expansion
\begin{equation}
\begin{split}
\Psi_\a(x) = \int \frac{d^3q}{ \left( 2 \pi \right)^3 } \frac{1}{2\omega_q} \eta_\a(\vec{q}) \left[ b ( \vec{q} ) e^{i q \cdot x } + {\tilde b}^\dagger ( \vec{q}) e^{- i q \cdot x } \right]  ~, 
\end{split}
\end{equation}
where $\eta_\a(\vec{p}) {\bar \eta}_\da(\vec{p}) = p_\mu (\sigma^\mu)_{\a\da}$, and $b$ and ${\tilde b}$ satisfy the standard fermionic creation-annihilation commutation relations. 
To perform the large-$r$ expansion, we will first have to convert the conventions of Wess and Bagger \cite{Wess:1992cp} to the $(u,r,z,\bz)$ coordinate system. First, we determine the basis spinors. In the $(u,r,z,\bz)$ coordinates used here, we have
\begin{equation}
\begin{split}
\sigma_z{}^z  = \frac{1}{r^2} \g^{z\bz} \p_z x^\mu \p_\bz x^\nu \sigma_{\mu\nu} =  \begin{pmatrix} \frac{1-z\bz}{2(1+z\bz)} & \frac{\bz}{1+z\bz} \\ \frac{z}{1+z\bz} & - \frac{1-z\bz}{2(1+z\bz)}  \end{pmatrix} ~. 
\end{split}
\end{equation}
The eigenvalues of this matrix are $\pm \frac{1}{2}$ with eigenvectors
\begin{equation}
\begin{split}
\xi^{(+)}_\a = \sqrt{ \frac{1}{1+z\bz} } \begin{pmatrix} 1 \\ z  \end{pmatrix} ~, \qquad \xi^{(-)}_\a = \sqrt{ \frac{1}{1+z\bz} } \begin{pmatrix}  \bz \\ - 1  \end{pmatrix} ~. 
\end{split}
\end{equation}
The normalization of the eigenvectors is chosen so that $\xi^{(+)\a} \xi^{(-)}_\a  = 1$. 
Since $\eta_\a(\vec{p})$ is a two-component spinor, it can be expanded in this basis. If we parametrize the momentum as
\begin{equation}
\begin{split}
p^\mu = \frac{\omega}{1+z\bz} \left( 1 + z \bz , z + \bz , - i (z - \bz ) , 1 - z \bz \right) ~, 
\end{split}
\end{equation}
then $ \eta_\a(p) = \sqrt{2\omega} \xi^{(+)}_\a$. With all the explicit formulas determined, we can go ahead and take the large-$r$ expansion of $\Psi_\a$:
\begin{equation}
\begin{split}
\psi_{(+)}(u,z,\bz) = - \frac{i}{ 8 \pi^2   } \int_0^\infty d\omega_q   \sqrt{2\omega_q}   \left[ b ( \omega_q {\hat x} ) e^{- i \omega_q u   }  -  {\tilde b}^\dagger ( \omega_q {\hat x} ) e^{  i \omega_q u  } \right]  ~. 
\end{split}
\end{equation}
Using the commutation relations for $b$ and ${\tilde b}$, we find
\begin{equation}
\begin{split}
\left\{ \psi_{(+)}(u,z,\bz) , {\bar \psi}_{(-)}(u',z',\bz') \right\} &=  \g^{z\bz} \delta(u-u') \delta^2(z-z') ~. 
\end{split}
\end{equation}
Now we are ready to determine the action of the hard charge on the fields. The hard charge is
\begin{equation}
\begin{split}
\SF^h[\chi] = \frac{Q}{2\sqrt{2}} \int du d^2 z \g_{z\bz} \chi u \p_u \left(  \phi {\bar \psi}_\- \right) ~. 
\end{split}
\end{equation}
It is clear that this acts trivially on $\phi$ and ${\bar \psi}_\-$. However, we have
\begin{equation}
\begin{split}
\left[ \SF^h[\chi] , \p_u {\bar \phi} (u,z,\bz) \right] &= \frac{Q}{2\sqrt{2}} \int du' d^2 z' \g_{z'\bz'}\chi u' \p_{u'}  \p_u  \left( \left[  \phi (u',z',\bz')  , {\bar \phi} (u,z,\bz) \right]{\bar \psi}_\-  \right)  \\
&= - i  \frac{Q }{4\sqrt{2}} \chi (z,\bz) {\bar \psi}_\-  (z,\bz) ~, \\
\left\{ \SF^h[\chi] ,  \psi_\+ (u,z,\bz) \right\} &= \frac{Q}{2\sqrt{2}} \int du' d^2 z' \g_{z'\bz'} \chi u' \p_{u'} \left( \phi  \left\{  {\bar \psi}_\- ,  \psi_\+ (u,z,\bz) \right\} \right)  \\
&= -  \frac{Q}{2\sqrt{2}}  \chi (z,\bz)  \phi   (u,z,\bz) ~, 
\end{split}
\end{equation}
which determines the action of the hard charge on the fields.

\item[9(a).] We consider an $(n+2)$-point amplitude ${\cal A}_{n+2}^{a_1,a_2}(q_1,\ve_1 ; q_2 , \ve_2 )$. The amplitude also depends on the other external momenta, but we suppress the dependence for convenience. Taking first the limit $q_2 \to 0$ and using the soft gluon theorem, we find
\begin{equation}
\begin{split}
\lim_{q_2\to0}{\cal A}_{n+2}^{a_1,a_2}(q_1,\ve_1 ; q_2 , \ve_2 )  &\to g_{YM} \sum_{k=1}^n \frac{p_k \cdot \ve_2 }{ p_k \cdot q_2 } T_k^{a_2} {\cal A}^{a_1}_{n+1}(q_1,\ve_1)  \\
&\qquad \qquad\qquad   + g_{YM} \frac{q_1 \cdot \ve_2 }{ q_1 \cdot q_2 } \left( T_{\text{adj.}}^{a_2} \right)^{a_1 b} {\cal A}^b_{n+1}(q_1,\ve_1) ~, 
\end{split}
\end{equation}
where we have used the fact that the gluon transforms in the adjoint representation. The matrix elements of the adjoint representations are $\big( T_{\text{adj.}}^{a_2} \big)^{a_1 b} = - i f^{a_2a_1b} = i f^{a_1a_2b}$. Using this, we find
\begin{equation}
\begin{split}
\lim_{q_2\to0}{\cal A}_{n+2}^{a_1,a_2}(q_1,\ve_1 ; q_2 , \ve_2 )  & \to g_{YM} \sum_{k=1}^n \frac{p_k \cdot \ve_2 }{ p_k \cdot q_2 } T_k^{a_2} {\cal A}^{a_1}_{n+1}(q_1,\ve_1) \\
&\qquad \qquad \qquad\qquad   + i g_{YM} f^{a_1a_2b} \frac{q_1 \cdot \ve_2 }{ q_1 \cdot q_2 }   {\cal A}^b_{n+1}(q_1,\ve_1) ~. 
\end{split}
\end{equation}
Let us now take the second limit, $q_1 \to 0$: 
\begin{equation}
\begin{split}
\lim_{q_1 \to 0} \lim_{q_2\to0}{\cal A}_{n+2}^{a_1,a_2}(q_1,\ve_1 ; q_2 , \ve_2 )  & \to g^2_{YM} \sum_{k=1}^n \sum_{k'=1}^n \frac{p_k \cdot \ve_2 }{ p_k \cdot q_2 } \frac{ p_{k'} \cdot \ve_1 }{ p_{k'} \cdot q_1 } T_k^{a_2} T_{k'}^{a_1} {\cal A}_n \\
&\qquad\qquad   + i g^2_{YM} f^{a_1a_2b} \frac{q_1 \cdot \ve_2 }{ q_1 \cdot q_2 }  \sum_{k=1}^n \frac{p_k \cdot \ve_1 }{ p_k \cdot q_1} T^b_k {\cal A}_n ~. 
\end{split}
\end{equation}
When the limits are taken in the opposite order, we find
\begin{equation}
\begin{split}
\lim_{q_2 \to 0} \lim_{q_1\to0}{\cal A}_{n+2}^{a_1,a_2}(q_1,\ve_1 ; q_2 , \ve_2 )  & \to g^2_{YM} \sum_{k=1}^n \sum_{k'=1}^n \frac{ p_{k} \cdot \ve_2 }{ p_{k} \cdot q_2 }  \frac{p_{k'} \cdot \ve_1 }{ p_{k'} \cdot q_1 } T_{k'}^{a_1} T_{k}^{a_2} {\cal A}_n \\
&\qquad \qquad  - i g^2_{YM} f^{a_1a_2b} \frac{q_2 \cdot \ve_1 }{ q_2 \cdot q_1 }  \sum_{k=1}^n \frac{p_k \cdot \ve_2 }{ p_k \cdot q_2} T^b_k {\cal A}_n ~. 
\end{split}
\end{equation}
Now subtract the two. Note that in the double sum, when $k \neq k'$ the two quantities are equal, since the generators $T_k$ are acting on different sets of indices. The only nontrivial contribution to the difference occurs when $k = k'$. Thus, we find
\begin{equation}
\begin{split}
\left[ \lim_{q_1 \to 0}  \, , \,  \lim_{q_2\to0} \right] {\cal A}_{n+2}^{a_1,a_2}(q_1,\ve_1 ; q_2 , \ve_2 )  &\to g^2_{YM} \sum_{k=1}^n  \frac{p_k \cdot \ve_2 }{ p_k \cdot q_2 } \frac{ p_{k} \cdot \ve_1 }{ p_{k} \cdot q_1 } \left[ T_k^{a_2} , T_{k}^{a_1}  \right] {\cal A}_n   \\
&\qquad\qquad + i g^2_{YM} f^{a_1a_2b} \frac{q_1 \cdot \ve_2 }{ q_1 \cdot q_2 }  \sum_{k=1}^n \frac{p_k \cdot \ve_1 }{ p_k \cdot q_1} T^b_k {\cal A}_n  \\
&\qquad\qquad +  i g^2_{YM} f^{a_1a_2b} \frac{q_2 \cdot \ve_1 }{ q_2 \cdot q_1 }  \sum_{k=1}^n \frac{p_k \cdot \ve_2 }{ p_k \cdot q_2} T^b_k {\cal A}_n \\
&\to  i g^2_{YM} f^{a_1a_2b} \sum_{k=1}^n \left(   \frac{q_1 \cdot \ve_2 }{ q_1 \cdot q_2 }  \frac{p_k \cdot \ve_1 }{ p_k \cdot q_1}   \right. \\
&\left. \qquad\qquad+ \frac{q_2 \cdot \ve_1 }{ q_2 \cdot q_1 }  \frac{p_k \cdot \ve_2 }{ p_k \cdot q_2} -  \frac{p_k \cdot \ve_2 }{ p_k \cdot q_2 } \frac{ p_{k} \cdot \ve_1 }{ p_{k} \cdot q_1 } \right) T^b_k {\cal A}_n ~.  \\
\end{split}
\end{equation}
This can be written in a more concise way by adding to it a fourth term that is identically zero:
\begin{equation}
\begin{split}
-  i g^2_{YM} f^{a_1a_2b} \sum_{k=1}^n  \frac{q_1 \cdot \ve_2 }{ q_1 \cdot q_2 }   \frac{q_2 \cdot \ve_1 }{ q_2 \cdot q_1 }   T^b_k {\cal A}_n = -  i g^2_{YM} f^{a_1a_2b} \frac{q_1 \cdot \ve_2 }{ q_1 \cdot q_2 }   \frac{q_2 \cdot \ve_1 }{ q_2 \cdot q_1 }  \sum_{k=1}^n   T^b_k {\cal A}_n ~. 
\end{split}
\end{equation}
The last term vanishes due to global color charge conservation. 
Adding this, we can write the double-soft limit as
\begin{equation}
\begin{split}
& \left[ \lim_{q_1 \to 0} \, , \,   \lim_{q_2\to0} \right] {\cal A}_{n+2}^{a_1,a_2}(q_1,\ve_1 ; q_2 , \ve_2 ) \\
&\qquad \qquad \to  i g^2_{YM} f^{a_1a_2b} \sum_{k=1}^n   \left(   \frac{p_k \cdot \ve_2 }{ p_k \cdot q_2} -  \frac{q_1 \cdot \ve_2 }{ q_1 \cdot q_2 }  \right) \left(  \frac{q_2 \cdot \ve_1 }{ q_2 \cdot q_1 } - \frac{ p_{k} \cdot \ve_1 }{ p_{k} \cdot q_1 }    \right)   T^b_k {\cal A}_n ~.   \\
\end{split}
\end{equation}

\item[9(b).] All we have to do here is to plug in the explicit forms of the momenta in the $z$ parametrization. Recall that when the helicity is positive, 
\begin{equation}
\begin{split}
\frac{p_k \cdot \ve }{ p_k \cdot q} \propto  \frac{1}{z-z_k} ~. 
\end{split}
\end{equation}

Then the soft factor is
\begin{equation}
\begin{split}
 \left(   \frac{p_k \cdot \ve_2 }{ p_k \cdot q_2} -  \frac{q_1 \cdot \ve_2 }{ q_1 \cdot q_2 }  \right) \left(  \frac{q_2 \cdot \ve_1 }{ q_2 \cdot q_1 } - \frac{ p_{k} \cdot \ve_1 }{ p_{k} \cdot q_1 }    \right)  &\propto \left(  \frac{1}{w-z_k}  - \frac{1}{w - z }   \right) \left( \frac{1}{z-w}   - \frac{1}{z-z_k}   \right) \\
 &=  \frac{z_k-z}{(w-z_k)(w-z)}  \frac{w-z_k}{(z-w)(z-z_k)}  \\
 &= \frac{1}{(z-w)^2} ~.  \\
\end{split}
\end{equation}
This factor does not depend on $z_k$ at all. Thus, we can pull this factor out of the sum to get
\begin{equation}
\begin{split}
\left[ \lim_{q_1 \to 0} \, , \,  \lim_{q_2\to0} \right] {\cal A}_{n+2}^{a_1,a_2}(q_1,\ve_1 ; q_2 , \ve_2 ) &\propto  \frac{ i g^2_{YM} f^{a_1a_2b} }{ ( z - w )^2 }  \sum_{k=1}^n   T^b_k {\cal A}_n = 0 ~. 
\end{split}
\end{equation}
The argument follows through in the  same way when the helicity is negative (just take all $z$'s to $\bar z$'s).

\item[10.]We take the metric to be of the form
\begin{equation}
\begin{split}
ds^2 &= - du^2 - 2dudr+2r^2\g_{z\bz} dz d\bz + \frac{2m_B}{r} du^2 + r C_{zz} dz^2 + r C_{\bz\bz} d\bz^2 \\
	& \qquad \qquad \qquad \qquad \qquad  \qquad \qquad \qquad  \qquad  + U_z du dz + U_\bz du d\bz + \cdots ~. 
\end{split}
\end{equation}
The four-dimensional Weyl tensor of a spacetime is given by
\begin{equation}
\begin{split}
C_{\mu\nu\rho\sigma} = R_{\mu\nu\rho\sigma} + \frac{1}{2} \left( g_{\nu\rho} R_{\sigma\mu} + g_{\mu\sigma} R_{\rho\nu} -g_{\nu\sigma} R_{\rho\mu} - g_{\mu\rho} R_{\sigma\nu}  \right)  + \frac{1}{6} R \left( g_{\mu\rho} g_{\sigma\nu} - g_{\mu\sigma} g_{\rho\nu} \right) ~. 
\end{split}
\end{equation}
We are interested in computing the following components:
\begin{equation}
\begin{split}
C_{rzrz} &= R_{rzrz} - \frac{1}{2}  g_{zz} R_{rr}  ~, \qquad C_{rurz} = R_{rurz} + \frac{1}{2} \left( g_{ur} R_{rz}   -g_{uz} R_{rr}   \right)    ~. 
\end{split}
\end{equation}
For the metric of interest, we then find
\begin{equation}
\begin{split}
C_{rzrz} = \CO(r^{-3})~, \qquad C_{rurz} = - \frac{1}{4r^2} \left(  U_z - D^z C_{zz} \right)  + \CO(r^{-3}) ~, 
\end{split}
\end{equation}
where $D_z$ is the covariant derivative with respect to $\g_{z\bz}$. Thus, to have an asymptotically flat spacetime, we must have 
\begin{equation}
\begin{split}
U_z = D^z C_{zz} ~. 
\end{split}
\end{equation}

\item[11(a).] {We will show that $\p_r \det\left(\frac{g_{AB}}{r^2}\right) = 0$ implies that $C_{AB}$ is traceless.
		\be
			\begin{split}
				\det \left (\frac{g_{AB}}{r^2}\right) & = \det \left ( \gamma_{AB} + \frac{C_{AB}}{r} + \mathcal O(r^{-2})\right)  \\
				&= \det\gamma    \left(1+  \frac{C^{A}{}_{A}}{r} + \mathcal O(r^{-2})\right)~.
			\end{split}
		\ee
		
		Hence, 
		\be
			\begin{split}
				\p_r\det \left (\frac{g_{AB}}{r^2}\right)  & = -  \det\gamma  ~  \frac{C^{A}{}_{A}}{r^2} + \mathcal O(r^{-3}) ~.
			\end{split}
		\ee
		
		Therefore, 
		\be
			\p_r \det \left (\frac{g_{AB}}{r^2}\right) = 0 \qquad \implies \qquad  C^A{}_A = 0 ~.
		\ee

		}
		\item[11(b).] {  First, we must determine the variation of the metric under a diffeomorphism generated by the vector field $\xi$, where $\xi^u, \xi^r \sim \mathcal O (1)$ and $\xi^A \sim \mathcal O (r^{-1})$.
				The variation of the metric under a diffeomorphism is given by
				\be
					\mathcal L_\xi g_{\mu\nu} = \xi^\rho \p_\rho g_{\mu\nu} + g_{\mu\rho} \p_\nu \xi^\rho + g_{\nu\rho} \p_\mu \xi^\rho  ~.
				\ee
 		By definition, the asymptotic symmetries must preserve the falloff conditions for the fields: 
			\be
				\begin{split}
					g_{uu} &= -U+ \frac{1}{4}g_{AB}U^A U^B =- 1 + \frac{2m_B}{r} + \mathcal O (r^{-2})~,\\
					g_{ur} &= - e^{2 \beta} = -1 + \mathcal O ( r^{-2})~,\\
					g_{uA} & = \frac{1}{2}g_{AB}U^B = \frac{1}{2} D^B C_{BA}+ \mathcal O (r^{-1}) ~,\\
					g_{rA} &= g_{rr} = 0~,\\ 
					g_{AB}& =r^2 \gamma_{AB} + r C_{AB}+ h_{AB}^\0 +  {\mathcal O} (r^{-1} )~.
				\end{split}
			\ee
			In Bondi gauge, 
			\be
				\mathcal L_\xi g_{rr} =  2\p_r \xi^u g_{u r} ~,
			\ee
			which implies $\xi^u$ must be independent of $r$
			\be
				\xi^u = \xi^u(u, x^A)~.
			\ee
			To leading order in the asymptotic expansion, we have
			\be
				\mathcal L_\xi g_{ur} = \p_u \xi^u g_{ur} + \mathcal O (r^{-1})~.
			\ee
			Hence, $\xi^u$ must also be independent of $u$:
			\be
				\xi^u = f(x^A)~.
			\ee
			Then the remaining components have the following variations under $\xi$, with $\xi^u = f$:
 			\be
				\begin{split}
					\mathcal L_\xi g_{uu}& =    f\p_u g_{uu} + \xi^r \p_r g_{uu} +\xi^A \p_A g_{uu}  + 2\p_u \xi^r g_{r u}+ 2\p_u \xi^A g_{Au} 
						~,\\
					\mathcal L_\xi g_{rA}& =    \p_r \xi^C g_{CA} +g_{u r} \p_A f  
						~,\\
					\mathcal L_\xi g_{AB}& = f \p_u g_{AB} + \xi^r \p_r g_{AB} +\xi^C \p_C g_{AB} 
											+ g_{u B} \p_Af   + \p_A \xi^C g_{CB}
											\\& \qquad \qquad \qquad \qquad \qquad \qquad \qquad \qquad \quad + g_{u A} \p_Bf  + \p_B \xi^C g_{CA}
						~.
				\end{split}
			\ee
			Let us consider the asymptotic expansion of the vector field
				\be
					 \xi = f \p_u + \sum_{n = 0}^\infty \frac{\xi^{r(n)}}{r^n} \p_r+ \sum_{n = 1}^\infty \frac{\xi^{A(n)}}{r^n} \p_A~.
				\ee
				The Lie derivative of the metric is then
				\be \label{eq14}
				\begin{split}
					\mathcal L_\xi g_{uu}& =    -  2\p_u \xi^{r(0)}+   \frac{-  2\p_u \xi^{r(1)} +2 f  \p_u m_B+  \p_u \xi^{A(1)}  D^B C_{BA} }{r}  + \mathcal O (r^{-2})
						~,\\ 
					\mathcal L_\xi g_{rA}& =     - \gamma_{AB}  \left( \xi^{B(1)}  + D^B f\right) - \frac{2 \gamma_{AB}  \xi^{B(2)}+    C_{AB} \xi^{B(1)} }{r}     + \mathcal O ( r^{-2})	~,\\
					\mathcal L_\xi g_{AB}& =   r  \Big(f \p_u C_{AB} +2   \gamma_{AB}  \xi^{r(0)} + D_A \xi_B^{(1)}+ D_B \xi_A^{(1)}\Big)\\
					&\qquad\qquad \quad +  f \p_u h_{AB}^{(0)}+   C_{AB} \xi^{r(0)} +2   \gamma_{AB}  \xi^{r(1)} +  \frac{1}{2} D^C C_{CA} D_Bf  \\
					& \qquad \qquad\quad  + \frac{1}{2} D^C C_{CB} D_Af  + \xi^{C(1)}D_C C_{AB}+ D_A \xi_B^{(2)}\\
					&\qquad \qquad\quad  + C_{BC}D_A \xi^{C(1)} + D_B \xi_A^{(2)}+ C_{AC}D_B \xi^{C(1)}  + \mathcal O (r^{-1})  
						~.
				\end{split}
			\ee
			
			From $\mathcal L_\xi g_{rA}$ at $\mathcal O (1)$, we find
			\be
				\xi^{A(1)} = - D^A f~.
			\ee
			From $\mathcal L_\xi g_{rA}$ at $\mathcal O (r^{-1})$, we find
			\be
				\xi^{A(2)} 
				=  \frac{1}{2} C^{AB} D_B f~.
			\ee
			We require that the $\mathcal O (r)$ term of  $\mathcal L_\xi g_{AB}$ be traceless, which implies
			\be
				       \xi^{r(0)}  = \frac{1}{2} D_AD^A f~.
			\ee
			Finally, to determine the $\mathcal O(r^{-2})$ piece of $\xi^{r}$, we need to consider a higher order expansion of $g_{AB}$: 
			\be
				g_{AB} = r^2 \gamma_{AB} + r C_{AB} + h^{(0)}_{AB} + \mathcal O ( r^{-1}).
			\ee
			Then 
			\be
				\begin{split}
					\det \left ( \frac{g_{AB}}{r^2}\right) &= \det \gamma   \exp \tr{\log \left(\delta_A^B+ \frac{1}{r}C_{A}{}^B +\frac{1}{r^2}h^{(0)}_{A}{}^B+ \mathcal  O (r^{-3}) \right)}~\\
					  &= \det \gamma   
						  	 \left[ 1 +  \frac{1}{r}C_{A}{}^A  \right. \\
							 &\left. \qquad  +\frac{1}{r^2} \left(h^{(0)}_{A}{}^A- \frac{1}{2} C_A{}^B C_B{}^A + \frac{1}{2}  C_A{}^A C_B{}^B \right) + \mathcal  O (r^{-3}) \right] ~.\\
				\end{split}
			\ee
		Assuming that $C_{A}{}^A = 0$ to satisfy the $\mathcal O ( r^{-1})$ constraint, we additionally find that
			\be
				h^{(0)}_{A}{}^{A}  = \frac{1}{2}C_{AB}  C^{AB} ~.
			\ee
	This imposes the following constraint on the variations of the metric components
			\be
				\delta h^{(0)}_{A}{}^{A}  = C_{AB}\delta C^{AB} ~.
			\ee
	An explicit computation reveals
			\be
				\begin{split}
					\delta h^{(0)}_A{}^A &= f \p_u h^{(0)}_A{}^A + 4 \xi^{r (1)} + 2 D_A f D_B C^{AB} - C^{AB} D_A D_B f~,\\
					C_{AB}\delta C^{AB} &= \frac{f}{2} \p_u (C^{AB}C_{AB}) - 2 C^{AB}D_{A}D_B f~.
				\end{split}
			\ee
		Equating these terms, we find
			\be
				\xi^{r(1)}  = - \frac{1}{2}D_A f D_B C^{AB} - \frac{1}{4}C^{AB}D_A D_Bf ~.
			\ee
	Therefore, the most general diffeomorphism $\xi$ that preserves the metric to leading order is given by
			\be
				\begin{split} 
					\xi(f) &= f  \p_u + \left [-\frac{D^A f}{r} + \frac{\frac{1}{2} C^{AB} D_B f}{r^2} + \mathcal{O}\left( \frac{1}{r^3}\right)\right]\p_A \\
						 & \qquad\quad +\left[\frac{1}{2} D^2 f + \frac{ - \frac{1}{2}D_A f D_B C^{AB} - \frac{1}{4}C^{AB}D_A D_Bf }{r}+ \mathcal O \left(\frac{1}{r^2} \right) \right] \p_r~,
				\end{split}
			\ee
			where $f = f(x^A)$.

		}
		\item[11(c).] { We will show that the Lie algebra of the supertranslations at leading order is abelian, that is,
			\be
				[\xi({f_1}), \xi({f_2} )] =\mathcal{O} (r^{-1})~.
			\ee 
		Using the definition of the Lie bracket, we find	
			\be
				\begin{split}
					[\xi({f_1}), \xi({f_2} )]^u & =  \xi(f_1)^\mu  \p_\mu f_2   - \xi(f_2)^\mu  \p_\mu f_1 \\
									& =  \frac{-D^A f_1   D_A f_2   }{r} + \frac{\frac{1}{2}C^{AB}D_A f_1 D_B f_2}{r^2}- (f_1 \leftrightarrow f_2)  \\
									&= 0 ~, \\
					[\xi({f_1}), \xi({f_2} )]^r & =  \xi(f_1)^\mu  \p_\mu \left(\frac{1}{2}D^2 f_2\right)        - (f_1 \leftrightarrow f_2)+  \mathcal O (r^{-2}) \\ 
									& = \frac{1}{2r} D_A \left (- D^A f_1    D^2 f_2       + D^A f_2 D^2 f_1  \right) +  \mathcal O (r^{-2}) \\  
									& = \mathcal O (r^{-1})~, \\
					[\xi({f_1}), \xi({f_2} )]^A & =  \xi(f_1)^\mu  \p_\mu\left [-\frac{D^A f_2}{r}   \right]    - (f_1 \leftrightarrow f_2)+  \mathcal O (r^{-3})    \\
						& = \frac{1}{2}  \frac{ D^2 f_1 D^A f_2}{r^2}   +\frac{D^B f_1 D_BD^A f_2}{r^2} -   1 \leftrightarrow 2    +  \mathcal O (r^{-3}) \\
						& = \mathcal O(r^{-2})~.
				\end{split}
			\ee

		}
		\item[11(d).]  Recall in part (b), Eq.\eqref{eq14}, we found
			\be
				\begin{split}
					\mathcal L_\xi g_{uu}& =    -  2\p_u \xi^{r(0)}+   \frac{-  2\p_u \xi^{r(1)} +2 f  \p_u m_B+  \p_u \xi^{A(1)}  D^B C_{BA} }{r}  + \mathcal O (r^{-2})
						~,\\
					\mathcal L_\xi g_{AB}& =   r  \Big(f \p_u C_{AB} +2   \gamma_{AB}  \xi^{r(0)} + D_A \xi_B^{(1)}+ D_B \xi_A^{(1)}\Big) + \mathcal O (1)  
						~.
				\end{split}
			\ee
		Using the solution for part (b) and the variations above, we can calculate the action of $\xi(f)$ on $m_B$, $C_{AB}$ and $N_{AB}$: 
		\be
			\begin{split}
 						\mathcal L_f C_{AB} & = f \p_u C_{AB} +    \gamma_{AB}   D^2 f   - 2 D_AD_B f  
							~,\\
						\mathcal L_f N_{AB} & =  f \p_u N_{AB} 
							~,\\
						\mathcal L_f m_B & =  f  \p_u m_B   +   \frac{1}{4}   \Big(N^{AB}D_A D_Bf+ 2D_A f D_B N^{AB} \Big) 
						~.
			\end{split}
		\ee
		Note that $\gamma^{AB}\mathcal L_f C_{AB}  = 0$~.
\item[11(e).] 	 Let us now evaluate the vector field $\xi(f)$ on the $\ell =0$ and $\ell =1$ spherical harmonics.
			\be
				\begin{split}
					 \xi(Y^0_0 ) &=  Y^0_0  \p_u ~,\\
					 \xi(Y_1^m)& = Y_1^m  \p_u -\frac{\gamma^{AB} \p_B  Y_1^m}{r}  \p_A  + \frac{1}{2} D^2 Y_1^m  \p_r~.
				\end{split}
			\ee
		Consider the following normalization for the spherical harmonics:
			\be \label{sphharm}
				Y^0_0 = 1, ~~~~~~Y^{1}_1 = \frac{z}{1+ z \bz}~, 
				~~~~~~Y^{0}_1 = \frac{1-z \bz}{1+ z \bz}~,~~~~~~Y^{-1}_1 = \frac{\bz}{1+ z \bz}~.
			\ee
			Then we have 
			\be
				\begin{split}
					 \xi(Y^0_0 ) &=     \p_u ~,\\
					 \xi(Y_1^1)& = \frac{z}{1+ z \bz} (  \p_u -    \p_r )+ \frac{z^2}{2r}  \p_z - \frac{1}{2r} \p_\bz  ~,\\
					  \xi(Y_1^0)& = \frac{1- z\bz}{1+ z \bz} (  \p_u -    \p_r ) +\frac{z}{r}  \p_z+\frac{\bz}{r}  \p_\bz  ~,\\
					  \xi(Y_1^{-1})& = \frac{\bz}{1+ z \bz} (  \p_u -    \p_r )- \frac{1}{2r}  \p_z + \frac{\bz^2}{2r} \p_\bz ~. \\
				\end{split}
			\ee
		To compare with the standard global translations in Minkowski space, we begin with these translations in Cartesian coordinates:  
		\be
			T = \p_t~,~~~~~~~~~X_i = \p_i~.
		\ee
		In Bondi coordinates, the global translations take the following form:
		\be
			\begin{split}
				T &= \p_u~,\\
				X_1 & =  -\frac{z+\bz}{1+z \bz} \p_u + \frac{z+\bz}{1+z \bz } \p_r+ \frac{1-z^2}{2 r}\p_z +\frac{1-\bz^2}{2 r} \p_\bz ~,\\
				X_2 & =  \frac{i(z-\bz)}{1+z \bz} \p_u - \frac{i(z-\bz)}{1+z \bz} \p_r+ \frac{i(1+z^2)}{2 r}\p_z -\frac{i(1+\bz^2)}{2 r} \p_\bz ~,\\
				X_3 & =  -\frac{1-z\bz}{1+z \bz} \p_u + \frac{1-z\bz}{1+z \bz } \p_r- \frac{z}{ r}\p_z -\frac{ \bz}{ r} \p_\bz ~.
			\end{split}
		\ee
		We then find
		\be
		\begin{split}
			\xi (Y^0_0) &= T~, \\
			\xi(Y_1^1) &= -\frac{1}{2} ( X_1 +i X_2)~, \\
			\xi(Y_1^0) &= -X_3~, \\
			\xi(Y_1^{-1}) &= -\frac{1}{2} ( X_1 -i X_2) ~. 
			\end{split}
		\ee		
		The action of the $\ell = 0$ mode on the fields is given by 
		\be
			\begin{split}
				\mathcal L_{\xi(Y^0_0)} C_{AB}  &= Y^0_0 \p_u C_{AB}  
							~, \\ 
						\mathcal L_{\xi(Y^0_0)} N_{AB}  &= Y^0_0  \p_u N_{AB} 
							~,\\ 
						\mathcal L_{\xi(Y^0_0)} m_B  &=   Y^0_0 \p_u m_B    
						~.
			\end{split}
		\ee 		The action of the $\ell = 1$ modes on the fields is given by
		\be
			\begin{split} 
 						\mathcal L_{\xi(Y^m_1)} C_{AB}  &=Y^m_1 \p_u C_{AB}      
							~, \quad \quad 
						\mathcal L_{\xi(Y^m_1)}  N_{AB}  =   Y^m_1\p_u N_{AB}, \\
						\mathcal L_{\xi(Y^1_1)}  m_B & = Y^1_1   \p_u m_B   +   \frac{1}{4}   \Big( D^\bz N_{\bz \bz}- z^2 D^z N_{zz} \Big) ~,\\
%
						\mathcal L_{\xi(Y^0_1)}  m_B & = Y^0_1  \p_u m_B   -   \frac{1}{2}   \Big(  z D^{z} N_{zz} +\bz D^\bz N_{\bz\bz}\Big) ~,\\
%
						\mathcal L_{\xi(Y^{-1}_1)}  m_B & = Y^{-1}_1  \p_u m_B   +   \frac{1}{4}   \Big( D^z N_{zz}- \bz^2 D^\bz N_{\bz\bz} \Big)  ~,
			\end{split}
		\ee
		where we have used the fact that $D_zD_z Y^m_1 = 0$.

\item[12(a).] {  The Lorentz transformations in Cartesian coordinates are given by  
		\be
			K_i = x^0 \p_i + x^i \p_0, ~~~~~~~~~~~~  J_{ij} = x^i \p_j - x^j \p_i~, 
		\ee
		where $K_i$ are the boosts and $J_{ij}$ are the rotations.  The Cartesian coordinates are related to the $(u,r,z, \bz)$ coordinates by the coordinate transformations
		\be
			\begin{split}
				u = t- \sqrt{x^i x_i}~,~~~~~~~~
				r  =  \sqrt{x^i x_i}~,~~~~~~~~
				z  = \frac{x^1 + ix^2}{x^3+\sqrt{x^i x_i}}~.
			\end{split}
		\ee
		In these coordinates, the boosts and rotations take the following form:
		\be
			\begin{split}
				K_1 & =  -\frac{u (z+\bz)}{1+z \bz} \p_u+\frac{(r+u) (z+\bz)}{1+z \bz }\p_r \\
				&\qquad \qquad \qquad -\frac{\left(z^2-1\right) (r+u)}{2 r}\p_z-\frac{\left(\bz^2-1\right) (r+u)}{2 r}\p_\bz ~,\\
				K_2 & =  \frac{i u (z-\bz)}{1+z \bz}\p_u-\frac{i (r+u) (z-\bz)}{1+z \bz }\p_r \\
				&\qquad \qquad \qquad +\frac{i \left(z^2+1\right) (r+u)}{2 r}\p_z-\frac{i \left(\bz^2+1\right) (r+u)}{2 r}\p_\bz ~,\\
				K_3 & = -  \frac{u (1-z \bz )}{1+z \bz}\p_u+\frac{(r+u) (1-z \bz)}{1+z \bz }\p_r-\frac{z (r+u)}{r}\p_z-\frac{\bz (r+u)}{r}\p_\bz ~,\\
				J_{12} &= i z\p_z-i \bz\p_\bz~,\\
				J_{23} &= \frac{1}{2} i \left(z^2-1\right)\p_z-\frac{1}{2} i \left(\bz^2-1\right)\p_\bz ~,\\
				J_{31} & = \frac{1}{2} \left(z^2+1\right)\p_z+\frac{1}{2} \left(\bz^2+1\right) \p_\bz ~.
			\end{split}
		\ee
		Then in Bondi gauge, a generic Lorentz transformation is generated by the vector field $\zeta$, with components
		\be
			\begin{split}
				\zeta^u & =  \frac{u}{2}~ \frac{2 z (c-a^* )+2 \bz (c^* - a )+(b+ b^*) (1-z\bz  )}{1+z\bz  }~,\\
				\zeta^r & =  -\frac{u+r}{2}~ \frac{2 z (c-a^* )+2 \bz (c^* - a )+(b+ b^*) (1-z\bz  )}{1+z\bz  }~,\\
				\zeta^z & =  a + bz + c z^2 + \frac{u}{2r} \left( (a-c^*) + (b + b^*)z + (c-a^*) z^2 \right)~.
			\end{split}
		\ee 		In particular, the individual rotations and boosts are reproduced for specific choices of $a,b,$ and $c$:
		\be
			\begin{split}
				J_{12} &= \zeta|_{\substack {a=0\\ b = i\\ c = 0}} ~,~~~~~ ~~~~~~~~J_{23} = \zeta|_{\substack {a=- \frac{i}{2}\\ b = 0\\ c = \frac{i}{2}}}  ~,
						~~~~~ ~~~~~~~~J_{31}  = \zeta|_{\substack {a= \frac{1}{2}\\ b = 0\\ c = \frac{1}{2}}}  ~,\\
				K_{1}& = \zeta|_{\substack {a= \frac{1}{2}\\ b = 0\\ c = -\frac{1}{2}}} ~,~~~~~ ~~~~~~~~K_{2} =  \zeta|_{\substack {a= \frac{i}{2}\\ b = 0\\ c = \frac{i}{2}}} ~,
					~~~~~ ~~~~~~~~K_{3} = \zeta|_{\substack {a=0\\ b = -1\\ c = 0}} ~.
			\end{split}
		\ee
		We now express this vector in terms of the global CKV on the sphere $Y^z = a + bz + c z^2$.  Using the identities
		\be	\label{ident1}
			\begin{split}
				D_z Y^z + D_\bz Y^\bz&= \frac{2 z (c-a^* )+2 \bz (c^* - a )+(b+ b^*) (1-z\bz  )}{1+z\bz  }~,\\
				D^z \left (D_z Y^z + D_\bz Y^\bz \right)&=- \left( (a-c^*) + (b+b^*)z + (c-a^*) z^2 \right)~,
			\end{split}
		\ee
		we find
		\be
			\begin{split}
		 \zeta(Y) = \frac{u}{2} D_A Y^A \p_u  -\frac{u+r}{2}D_A Y^A   \p_r +\left [Y^A -\frac{u}{2r}D^A D_B Y^B \right]\p_A ~.
			\end{split}
		\ee
		Comparing \eqref{ident1} with \eqref{sphharm}, we have $D \cdot Y \sim Y_1^m$.		Noting that $D^z D_z Y^z = - Y^z$, we can write this vector field as presented in section \ref{sec:superrotation}
		\be
			\begin{split}
			\zeta(Y) &= \frac{u}{2} \left(D_z Y^z + D_\bz Y^\bz\right) \p_u  -\frac{u+r}{2} \left(D_z Y^z + D_\bz Y^\bz\right) \p_r ~\\&\qquad \qquad \qquad  +  \left [Y^z + \frac{u}{2r} \left(Y^z -D^zD_\bz Y^\bz \right) \right] \p_z
	  \\&\qquad \qquad \qquad + \left [Y^\bz + \frac{u}{2r} \left(Y^\bz -D^\bz D_z Y^z \right) \right] \p_\bz~.
			\end{split}
		\ee

		}
		\item[12(b).] { We show that the Lorentz algebra is isomorphic to the algebra of the global CKVs of $S^2$.  First note that, since the $r$ and $u$  components  of $\zeta$ are linear in $r$ and $u$, the 
		terms of the form $\zeta^\mu \p_\mu \zeta^\nu$ will always cancel when $\mu, \nu = u,r$: 
 
			\be
				\begin{split}
					[\zeta(Y_1), \zeta(Y_2) ]^u	& = \zeta(Y_1)^A D_A \zeta(Y_2)^u -  \zeta(Y_2)^A D_A \zeta(Y_1)^u ~\\
						 & =   \frac{u}{2}\left [Y_1^A   D_A D_BY_2^B-  Y_2^A  D_A D_BY_1^B \right]~\\
						  & =   \frac{u}{2}D_B\left [Y_1^A   D_A  Y_2^B-  Y_2^A  D_A  Y_1^B \right]~\\
						  & =   \frac{u}{2}D_A\left [Y_1,  Y_2 \right]^A~.
				\end{split}
			\ee
In a similar way, we have
			\be
				\begin{split}
					[\zeta(Y_1), \zeta(Y_2) ]^r	& = \zeta(Y_1)^A D_A \zeta(Y_2)^r -  \zeta(Y_2)^A D_A \zeta(Y_1)^r ~\\
						 & =   -\frac{u+r}{2}\left [Y_1^A   D_A D_BY_2^B-  Y_2^A  D_A D_BY_1^B \right]  
						 \\&=   -\frac{u+r}{2}D_A\left [Y_1,  Y_2 \right]^A~,\\
					[\zeta(Y_1), \zeta(Y_2) ]^A	& =[Y_1, Y_2]^A 	+ \frac{u^2}{4r^2} \bigg[ D\cdot Y_2  D^A D \cdot  Y_1 - D \cdot  Y_1 D^A D \cdot  Y_2    \\
						& \qquad  + D^C D \cdot Y_1  D_CD^A D \cdot Y_2   - D^C D \cdot  Y_2 D_CD^A D \cdot  Y_1  \bigg] \\
						&\qquad  -   \frac{u}{2r}\bigg [D^A D \cdot [Y_1,Y_2]  - (D^AY_1^C +D^CY_1^A )D_C D \cdot Y_2    \\& \qquad\quad +(D^A Y_2^C  +D^CY_2^A) D_C D \cdot  Y_1  +   D \cdot  Y_1 D^A D \cdot  Y_2 \\& \qquad\quad - D\cdot Y_2  D^A D \cdot  Y_1  \bigg] \\ 
						& =[Y_1, Y_2]^A   -\frac{u}{2r} D^AD \cdot [Y_1,Y_2]       ~ ,
				\end{split}
			\ee
			where we have used the identities $D_A Y_B + D_B Y_A = \gamma_{AB}D \cdot Y$ and $D_AD_B(D \cdot Y) = \frac{1}{2} \gamma_{AB} D^2 D \cdot Y = -  \gamma_{AB} D \cdot Y$.

		}
		\item[12(c).] {
			The diffeomorphism generated by $\zeta$ induces the following variations in the metric components:
			\be \label{12c}
				\begin{split}
					\mathcal L_\zeta g_{uu} &= \frac{  D \cdot  Y  (u  \p_u m_B+ 3 m_B)+2 Y^A\p_A m_B- \frac{1}{2} D^B C_{BA} D^AD \cdot Y}{r}  + \mathcal O(r^{-2})~,\\
					\mathcal L_\zeta g_{ur} &= 
					  		  \mathcal O (r^{-2}) ~,\\
					\mathcal L_\zeta g_{rr} &= 0~,\\
					\mathcal L_\zeta g_{uA} &=  \mathcal O(1) ~,\\
					\mathcal L_\zeta g_{rA} &=    \frac{u}{2} \left (   \frac{g_{AB}\gamma^{BC}}{r^2}     +g_{ur}   \delta_A^C   \right) D_C D \cdot  Y  ~,\\ 
					\mathcal L_\zeta g_{AB} &=  \frac{u}{2}  \left (D \cdot Y \p_u g_{AB} -D \cdot Y \p_r g_{AB}  +     g_{uB} D_AD \cdot Y+   g_{uA}   D_B D \cdot Y   \right)  \\
												&\qquad~  -\frac{u}{2r} \left [  D^C D \cdot Y D_C g_{AB} +g_{CB} D_A  D^C D \cdot Y   +g_{CA}  D_B  D^C D \cdot Y   \right] \\
												&\qquad~  -\frac{r}{2}D \cdot Y \p_r g_{AB}  + \mathcal L_Y g_{AB}         ~. 
				\end{split}
			\ee 
			The first four equations in (\ref{12c}) indicate that $\zeta$ preserves the asymptotic form of the components $g_{uu}$, $g_{rr}$, $g_{ur}$, and $g_{uA}$.   We now check the leading  and subleading  terms in the variations
			 of $g_{rA}$ and $g_{AB}$, using the asymptotic expansion $g_{AB} = r^2 \gamma_{AB} + r C_{AB} + \mathcal O (1)$:
			 \be
				\begin{split} 
					\mathcal L_\zeta g_{rA} &=    \frac{u}{2} \left ( \gamma_{AB}\gamma^{BC}     -  \delta_A^C + \mathcal O (r^{-1})   \right) D_C D \cdot  Y ~,\\
					\mathcal L_\zeta g_{AB}& = r^2 \left ( \mathcal L_Y \gamma_{AB} - \gamma_{AB} D\cdot Y\right)+   r \left [
						\frac{u}{2}  \left (D \cdot  Y   \p_u   C_{AB}  - 2   \gamma_{AB}  D \cdot  Y   \right. \right. \\
						&\left. \left. \qquad \qquad -2D_A D_B D \cdot Y   \right)      -\frac{1}{2}C_{AB}  D \cdot Y      +   \mathcal L_YC_{AB} \right] + \mathcal O (1)~ .
				\end{split}
			\ee 
 
		At order $\mathcal O (r^2)$, $\mathcal L_\zeta g_{AB}$ vanishes, due to the conformal Killing equation $\mathcal L_Y \gamma_{AB} = D_A Y_B + D_B Y_A = \gamma_{AB}D \cdot Y $. 
		The $\mathcal O (r)$ term gives the variation of $C_{AB}$:
		\be\begin{split}
			 \delta C_{AB} &= \frac{u}{2}  \left (D \cdot  Y   \p_u   C_{AB}  - 2   \gamma_{AB}  D \cdot  Y   -2D_A D_B D \cdot Y   \right)   \\
			 &\qquad \qquad -\frac{1}{2}C_{AB}  D \cdot Y   +   \mathcal L_YC_{AB} ~.
		\end{split}\ee		
		Note that
		\be 
				\g^{AB}\delta C_{AB}  = \frac{u}{2}  \left (  - 4  D \cdot  Y   -2D^2 D \cdot Y   \right)    +C^{AB} \left(D_A Y_B+ D_B Y_A\right)    = 0~,
		\ee
		where $D^2 D \cdot Y = -2 D \cdot Y$, and $C^{AB}\left(D_A Y_B+ D_B Y_A\right)  =C^{AB}\gamma_{AB} D \cdot Y $.  Hence, $C_{AB}$ remains traceless under the diffeomorphism generated by $\zeta$.
		
		}

\item[13(a).] In the limit $r \to \infty$, the metric in Bondi gauge  is flat, and $v^\mu = (1,0,0,0)$ is a solution for the four-velocity of a timelike observer.  
			  At finite but large $r$, an inertial observer's four-velocity must have subleading corrections, generically of the form \be
				\begin{split}
					v^u &= 1+ \frac{1}{r}v^{u(1)}+ \mathcal O (r^{-2}) ~,\\
					v^r & =  \frac{1}{r}v^{r(1)}+ \mathcal O (r^{-2})~ ,\\
 					v^z &=  \frac{1}{r}v^{z(1)}+  \frac{1}{r^2}v^{z(2)}+ \mathcal O (r^{-3})~,\\
 					v^\bz & = \frac{1}{r}v^{\bz(1)}+  \frac{1}{r^2}v^{\bz(2)}+ \mathcal O (r^{-3}) ~.
				\end{split}
			\ee			To compute the norm, we use the asymptotic form of the metric:
			\be
				\begin{split}
					ds^2 &= - du^2 - 2 du dr + 2 r^2 \gamma_{z \bz}d z d \bz+  \frac{2 m_B}{r } du^2  \\& \quad  \quad 
							+ D^zC_{zz}  dudz+  D^\bz C_{\bz \bz} dud\bz +  rC_{zz} dz^2+   rC_{\bz \bz} d\bz^2+ \cdots~.
				\end{split}
			\ee
			By requiring that $v^\mu v_\mu = -1 + \mathcal O (r^{-2})$, we find
			\be
				\begin{split}
					v^{z(1)} = v^{\bz(1)} = 0~,~~~~~~
					m_B = v^{u(1)}+v^{r(1)}~.
				\end{split}
			\ee			Hence, we will study a four-velocity of the asymptotic form
			\be
				\begin{split}
					v^u &= 1+ \frac{m_0 (u,z,\bz) }{r}+ \mathcal O (r^{-2}) ~,\\
					v^r & =  \frac{m_B (u,z, \bz) - m_0(u, z, \bz) }{r} + \mathcal O (r^{-2})~ ,\\
 					v^z &=    \frac{1}{r^2}v^{z(2)}+ \mathcal O (r^{-3})~,\\
 					v^\bz & =  \frac{1}{r^2}v^{\bz(2)}+ \mathcal O (r^{-3}) ~.
				\end{split}
			\ee
						We require that $v^\mu \nabla_\mu v^{u,r} = \mathcal O (r^{-2})$ and $v^\mu \nabla_\mu v^{z, \bz} = \mathcal O (r^{-3})$.  Explicitly, these conditions are given by
			\be
				\begin{split}
					\frac{\p_u m_0}{r}+ \Gamma^u_{\alpha \beta} v^\alpha v^\beta &=  \mathcal O (r^{-2})~,\\
					\frac{\p_u m_B- \p_u m_0}{r}+ \Gamma^r_{\alpha \beta} v^\alpha v^\beta &=  \mathcal O (r^{-2})~,\\
					\frac{\p_u v^{z (2)}}{r^2}+ \Gamma^z_{\alpha \beta} v^\alpha v^\beta &=  \mathcal O (r^{-3})~.
				\end{split}
			\ee
			Next, we need to compute the large-$r$ expansion of the Christoffel connection.  The only nonzero components at this order are 
			\be \label{187}
				\begin{split} 
					\Gamma^{r}_{uu} &=- \frac{\p_u m_B}{r}+ \mathcal O (r^{-2})~, ~~~~~~~~
					\Gamma^{z}_{uu} = \frac{D_z N^{zz}}{2r^2}+ \mathcal O (r^{-3})~.
				\end{split}
			\ee			Substituting (\ref{187}) into the geodesic equation, we find the following constraints:
			\be
				\begin{split}
					\p_u m_0 = 0~,~~~~~~~ \p_u v^{z(2)} = - \tfrac{1}{2}D_z N^{zz}~.
				\end{split}
			\ee
			Integrating the second condition, we find
			\be
				v^{z (2)}(u,z, \bz ) = - \tfrac{1}{2} \left(D_z C^{zz}(u,z, \bz) - D_z C^{zz}(u_0,z, \bz) \right)~.
			\ee
			Moreover, since $m_0$ is simply a function of $(z, \bz)$, we can also interpret it as a boundary condition.   Hence, we have 
			\be
				\begin{split}
				v^\mu  = \bigg (&1+ \tfrac{m_B(u_0, z, \bz) }{r}+ \mathcal O (r^{-2})~,~\tfrac{\Delta m_B(u, z, \bz) }{r}+ \mathcal O (r^{-2})~,\\&
							 - \tfrac{D_z   \Delta C^{zz}(u,z, \bz)  }{2r^2 } + \mathcal O (r^{-3}), 
							 - \tfrac{D_\bz  \Delta C^{\bz \bz}(u,z, \bz) }{2r^2 } + \mathcal O (r^{-3}) \bigg)~,
				\end{split}
			\ee
			where
			\begin{align}
			\begin{split}
	\Delta m_B(u, z ,\bz) &\equiv m_B(u, z, \bz) - m_B(u_0, z, \bz)~, \\
	 \Delta C^{zz}(u,z, \bz)  &\equiv C^{zz}(u,z, \bz) -   C^{zz}(u_0,z, \bz)~.
\end{split}
			\end{align}
			To leading order in $r_0$ and for small $\lambda$, the trajectory of an observer with this four-velocity is
			\be
				X^\mu (\lambda)  = (u_0 + \lambda, r_0, z_0, \bz_0)~.
			\ee			The first correction to the four-velocity may now be approximated as
			\be\begin{split}
				v^\mu(\lambda) &= \left (1+ \frac{m_B(u_0, z, \bz)}{r_0}~,~ \frac{\lambda \p_u m_B(u_0, z, \bz)}{r_0} ~, \right. \\ &\left. \qquad \qquad \qquad  
								- \frac{ \lambda D_z N^{zz}(u_0,z, \bz)}{2r_0^2}~,~ - \frac{ \lambda D_{\bz} N^{\bz \bz}(u_0,z, \bz)}{2r_0^2}\right)~.
			\end{split}\ee			
			The trajectory with these subleading corrections is
			\be
				\begin{split}
				X^\mu(\lambda) &= \bigg (u_0 + \lambda \left(1+ \frac{m_B(u_0, z, \bz)}{r_0} \right), 
									r_0 +\frac{\lambda^2 \p_u m_B(u_0, z, \bz)}{2r_0} ,\\ 
									& \quad \quad  
									 z_0- \frac{ \lambda^2 D_z N^{zz}(u_0,z, \bz)}{4r_0^2},
									  \bz_0 - \frac{ \lambda^2 D_{\bz} N^{\bz \bz}(u_0,z, \bz)}{4r_0^2} \bigg) +\cdots ~ \\
							& = \bigg (u, r_0  \left (1+  \frac{(u-u_0)^2 \p_u m_B(u_0, z, \bz)}{2r_0^2} \right) ,
									\\& \quad \quad  z_0- \frac{(u-u_0)^2 D_z N^{zz}(u_0,z, \bz)}{4r_0^2},  \bz_0 - \frac{ (u-u_0)^2 D_{\bz} N^{\bz \bz}(u_0,z, \bz)}{4r_0^2} \bigg)+ \cdots~.
				\end{split}
			\ee	
\item[13(b).]  Since both inertial observers pass through the point $(u_0, r_0)$, we can use the trajectory found in part (a), and simply modify the angular dependence.  To leading order
			in the large-$r_0$ limit, assuming $|u-u_0| \ll r_0$, we find 
			\be
				\Delta X^\mu \equiv X^\mu_1 - X^\mu_0 = \left(0, 0, z_1- z_0, \bz_1- \bz_0\right)+ \cdots~.
			\ee
			
			The leading order proper distance between the observers is then
			\be
				L \approx \sqrt{2 r_0^2\gamma_{z_0 \bz_0} \delta z \delta \bz} = \frac{2r_0|\delta z|}{1+ z_0 \bz_0}~, \quad \quad \quad \delta z \equiv z_1-z_0~.
			\ee
\item[13(c).]
			Due to the burst of radiation, the metric undergoes a transition.  Inverting the leading order constraint equation, we find that the leading order change in the $zz$-component is given by
		\be 
			\begin{split}
			\Delta C_{zz} (z , \bz) & \equiv  C_{zz} (z, \bz)\big|_{u = u_f} - C_{zz}(z, \bz)\big|_{u = u_0}~ \\
				&=2\int d^2z'~ \gamma_{z' \bz'} D_{z}^2 G(z, \bz; z', \bz')\left[\int_{u_0}^{u_f} du ~T_{uu}(u, z', \bz') + \Delta m_B(z', \bz')\right],
			\end{split}
		\ee  
		where $u_f>u_{\text{rad}}$,  $\frac{|u_f - u_0|}{r_0} \ll 1$, and 
		\be
			\begin{split}
				 G(z, \bz; z', \bz') &= \frac{1}{\pi} \frac{|z-z'|^2}{(1+ z \bz) (1+ z' \bz')} \log \left ( \frac{|z-z'|^2}{(1+ z \bz) (1+ z' \bz')} \right) ~, \\
			\Delta m_B& = m_B (u_f,z, \bz) - m_B (u_0,z, \bz) ~.
			\end{split}
		\ee 
		Henceforth we assume $m_B$ is independent of $(z,\bz)$.		To check this, note that
		\be
			D_{\bz}^2 D_z^2  G(z, \bz; z', \bz') = \gamma_{z \bz} \delta^2(z-z') + \cdots~.
		\ee
		To determine $\Delta m_B $, we used the constraint equation 
		\be
			\p_u m_B = \frac{1}{4} \left [D^2_z N^{zz} + D_{\bz}^2 N^{ \bz \bz}\right]- T_{uu}~.
		\ee
		We can extract the zero mode of this equation by integrating over the two-sphere
		\be
			 \p_u m_B =  - \frac{1}{4 \pi} \int d^2 z~ \gamma_{z \bz} T_{uu}~.
		\ee		Integrating over $u$, we find
		\be
			  \Delta m_B  =  - \frac{1}{4 \pi} \int_{u_0}^{u_f} \int d^2 z~ \gamma_{z \bz} ~T_{uu} ~.
		\ee
		Then the change in $C_{zz}$ takes the modified form 
		\be
			\begin{split}
				\Delta C_{zz} (z , \bz) &=2\int d^2z'~ \gamma_{z' \bz'} D_{z}^2 G(z, \bz; z', \bz') \\ &\times 
					\left [\int_{u_0}^{u_f} du ~ \left (T_{uu}(u, z', \bz') - \frac{1}{4 \pi }\int d^2 z'' \gamma_{z ''\bz''} T_{uu}(u, z'', \bz'')\right)\right]~.
			\end{split}
		\ee		$T_{uu}$ due to the burst of radiation is given by
		\be
			T_{uu}(u, z ,\bz)  = \mu  \delta (u -u_ \text{rad}) \frac{\delta ^2 (z -z_{\text{rad}})}{\gamma_{z \bz}}~.
		\ee		The change in $C_{z z}$ due to the burst in radiation is then as follows:
		\be \label{eq:Czzfinalsol}
			\begin{split}
				\Delta C_{zz} (z , \bz) &=2 \mu \int d^2z'~  D_{z}^2 G(z, \bz; z', \bz')  
					\left [ \delta ^2 (z' -z_{\text{rad}})- \frac{\gamma_{z' \bz'}}{4 \pi } \right]~ \\
					&=2 \mu   D_{z}^2 G(z, \bz; z_{\text{rad}}, \bz_{\text{rad}})  \\
					&\qquad \qquad \qquad -    \frac{\mu }{2 \pi }\int d^2z'~  \gamma_{z' \bz'} D_{z}^2 G(z, \bz; z', \bz')   ~.
			\end{split}
		\ee 
				Since this component of the metric never contributed to the proper distance between inertial observers at leading order, to leading order the separation between 
		the observers after the burst of radiation remains unchanged:
		\be
			L \approx  \frac{2r_0|\delta z|}{1+ z_0 \bz_0} \equiv L_0~.
		\ee		However, this component of the metric does play a role in subleading corrections.  Note that to calculate the first subleading correction to $L$, we can safely neglect any corrections
		to $\Delta X$, since they will not contribute at this order.  Hence, the first subleading correction comes from the metric used to compute the inner product of $\Delta X^\mu$.		At a given time
		$u$, the distance is given by
		\be
			\begin{split}
				L &\approx \sqrt{2 r_0^2\gamma_{z_0 \bz_0} \delta z \delta \bz+ r_0 C_{zz}(u, z_0, \bz_0)\delta z^2+r_0 C_{\bz \bz}(u, z_0, \bz_0)\delta \bz^2}~\\
					& = L_0\sqrt{1+ \frac{r_0}{L_0^2}
								  \left [ C_{zz}(u, z_0, \bz_0)  \delta z^2 +  C_{\bz \bz}(u, z_0, \bz_0) \delta \bz^2 \right]}~\\
					& \approx   L_0+ \frac{r_0}{2L_0}
								  \left [ C_{zz}(u, z_0, \bz_0)  \delta z^2 +  C_{\bz \bz}(u, z_0, \bz_0) \delta \bz^2 \right] ~.
			\end{split}
		\ee
				Hence, the change in distance between $u_0$ and $u_f$ is given by the difference in the subleading term at $u_f$ and $u_0$:
		\be \label{208}
			\Delta L = \frac{r_0}{2L_0}
								  \left [  \Delta C_{zz}( z_0, \bz_0)  \delta z^2 +   \Delta C_{\bz \bz}( z_0, \bz_0) \delta \bz^2 \right] ~.
		\ee 		
		\item[13(d).] To determine the supertranslation that would give rise to the same change in proper distance as equation (\ref{208}),  recall that under a supertranslation parametrized by a function on 
		 the sphere $f$, we have
		 \be
		 	\mathcal L_f C_{zz}  = f N_{zz} - 2 D^2_zf~.
		 \ee

		The change in the metric is related to the Lie derivative
		\be
			\Delta C_{zz}  = - \mathcal L _f C_{zz} = 2 D^2_zf~,
		\ee
		where $N_{zz}$ has been set to zero, since the region of $\ci^+$ before and after the radiation has passed is in a vacuum state.  	Then, using equation  \eqref{eq:Czzfinalsol}, we find
		\be 
			\begin{split}
				f (z , \bz) 
					&=  \mu    G(z, \bz; z_{\text{rad}}, \bz_{\text{rad}}) -    \frac{\mu }{4\pi }\int d^2z'~  \gamma_{z' \bz'}   G(z, \bz; z', \bz')   ~.
			\end{split}
		\ee

\item[14(a).]  To determine the linearized Einstein's equation, note that
\begin{equation}
\begin{split}
R_{\mu\nu}[g+h] &= R_{\mu\nu}[g] + \frac{1}{2} \left( \nabla_\rho \nabla_\mu h_{\nu}{}^\rho + \nabla_\rho \nabla_\nu h_\mu{}^\rho - \nabla^2 h_{\mu\nu} - \nabla_\mu \nabla_\nu h \right)  \\
&\qquad \qquad + \CO \left( h^2 \right) ~, \\
R[g+h] &= R[g] - R_{\mu\nu} [g] h^{\mu\nu} + \nabla_\mu \nabla_\nu h^{\mu\nu} - \nabla^2 h + \CO \left( h^2 \right)~, 
\end{split}
\end{equation}
where $\nabla_\mu$ is the covariant derivative w.r.t. $g_{\mu\nu}$. Using this, we find
\begin{equation}
\begin{split}
G_{\mu\nu} [g+h] &= G_{\mu\nu}[g] + \frac{1}{2} \left( \nabla_\rho \nabla_\mu h_{\nu}{}^\rho + \nabla_\rho \nabla_\nu h_\mu{}^\rho - \nabla^2 h_{\mu\nu} - \nabla_\mu \nabla_\nu h \right)  \\
&\qquad  +  \frac{1}{2} g_{\mu\nu} h^{\rho\sigma} R_{\rho\sigma} [g]   - \frac{1}{2} g_{\mu\nu} \nabla_\rho \nabla_\sigma h^{\rho\sigma} + \frac{1}{2} g_{\mu\nu}  \nabla^2 h   \\
&\qquad - \frac{1}{2} h_{\mu\nu} R[g]+ \CO \left( h^2 \right) ~. 
\end{split}
\end{equation}
We are interested in the linearized equations about the Schwarzschild background. In this case, $R_{\mu\nu}[g] = R[g] = 0$, which implies that the linearized Einstein equations are 
\begin{equation}
\begin{split}
 \nabla_\rho \nabla_\mu h_{\nu}{}^\rho + \nabla_\rho \nabla_\nu h_\mu{}^\rho - \nabla^2 h_{\mu\nu} - \nabla_\mu \nabla_\nu h     -  g_{\mu\nu} \nabla_\rho \nabla_\sigma h^{\rho\sigma} +  g_{\mu\nu}  \nabla^2 h = 0 ~. 
\end{split}
\end{equation} 
To solve the linearized Einstein's equations about the initial Schwarzschild solution in Bondi gauge, we consider a line element of the form
		\be
			ds^2 = - \left (1- \frac{2 M }{r} \right)dv^2 + 2 dvdr + 2r^2 \gamma_{z \bz} dz d \bz+   h_{\mu\nu} dx^\mu dx^\nu~,
		\ee
		where $h_{\mu\nu}$ is a small perturbation of the metric. 
		We require that $h_{\mu\nu}$ preserve the Bondi gauge condition and the Bondi mass only shifts by a constant.  In particular, since the metric describes a spacetime that is 
		perturbed by a null shockwave at $v = v_0$, we can take $h_{\mu\nu}$ to be of the following form:
		\be
			\begin{split}
				h_{vv}(v,r,z,\bz)  &= \theta (v-v_0) \left (\frac{2 \delta m}{r} + \frac{1}{r^2}\tdh_{vv}(r, z, \bz)  \right) ~,\\
				h_{vr}(v,r,z,\bz)  &=   \frac{\theta (v-v_0) }{r^2}\tdh_{vr}(r, z, \bz)   ~,\\
				h_{vz}(v,r,z,\bz)  &= \theta (v-v_0)    \tdh_{vz}(r, z, \bz)  ~,\\
				h_{zz}(v,r,z,\bz)  &= r\theta (v-v_0)    \tdh_{zz}(r, z, \bz)  ~,\\
				h_{rr} &= h_{rz}  = h_{z \bz} = 0~,
			\end{split}
		\ee
		where   $ \tdh_{\mu\nu}(r,z ,\bz)$ must approach a finite function on the two-sphere as $r \rightarrow \infty$.
		Note that at linear order, the Bondi condition $\p_r \det\frac{g_{AB}}{r^2}$ implies $\gamma^{AB}h_{AB} = 0$.
			Consider the $rr$-component of Einstein's equation,
		\be
			G_{rr} =\frac{2   \theta (v-v_0) \left(r \p_r \tdh_{vr}-2\tdh_{vr}\right)}{r^4} = 0
		\ee
			which implies that
		\be
			\tdh_{vr} = r^2 c_{1vr}(z,\bz)~.
		\ee
		However, to satisfy the boundary condition that $\tdh_{vr}$ is finite as $r\to\infty$, we must have $c_{1vr}( z ,\bz) = 0$.
				Using this result, next consider the $zz$-component of Einstein's equations:
		\be \label{zzconstraint}
			\begin{split}
				G_{zz} &= -r \delta(v-v_0) \p_r \tdh_{zz} + \frac{1}{2r^2} \theta(v-v_0)
					 \left (  2 r^2  \p_r D_z \tdh_{vz}   -2 M r \p_r \tdh_{zz} \right. \\
					 &\left. \qquad \qquad \qquad  \qquad \qquad  +r^2 (2 M-r)  \p_r ^2\tdh_{zz}  +2 M \tdh_{zz}  \right) \\\ &= 0~,
			\end{split}
		\ee
where $D_z$ is the covariant derivative w.r.t. the two-sphere.		From the first term, we find that $\tdh_{zz}$ must be $r$-independent. Using these results, we have
		\be
			G_{rz} = - \theta (v-v_0) \left (\frac{-2 r^2 \p_r^2  \tdh_{vz}  +4  \tdh_{vz} +2 D^z  \tdh_{zz} }{4 r^2}\right) = 0~. \label{rzconstraint}
		\ee
		If we take a derivative of the numerator with respect to $r$, we find a differential equations for only $\tdh_{vz}$ with the general solution
		\be
			\tdh_{vz}(r, z ,\bz)  =   \frac{c_{1vz}(z, \bz)}{r} + c_{2vz}(z ,\bz)+c_{3vz}(z, \bz)  r^2 ~.
		\ee
		The boundary conditions force $c_{3vz}(z, \bz)  = 0$.  Moreover, substituting back into \eqref{rzconstraint} implies that $c_{2vz} = - \frac{1}{2}D^z \tdh_{zz}$. From the second term 
		in \eqref{zzconstraint}, we learn that $\tdh_{zz} = \frac{1}{M}D_z c_{1vz}$.		Using these results, we have
		\be
			G_{z \bz}  = \theta (v-v_0) \left( \frac{  2 D^z c_{1 vz} + 2 D^{\bz} c_{1 v\bz} -2 r^2  \p_r^2 \tdh_{vv}+4  r  \p_r \tdh_{vv}  -4 \tdh_{vv} }{2r^2 (1+ z \bz)^2} \right) = 0~.
		\ee
		Again, taking an $r$-derivative of the numerator, we can solve for the $r$-dependence of $\tdh_{vv}$.  The most general solution consistent with the boundary conditions is 
		\be 
			\tdh_{vv}(r, z, \bz ) = \frac{1}{2} \left (   D^z c_{1 vz} +  D^{\bz} c_{1 v\bz}\right).
		\ee
		
		Using these results, the $vz$-component is of the form
		\be
			G_{vz} = \frac{3  \delta (v-v_0) c_{1vz}}{2r^2} + \theta (v-v_0) G'_{vz}(r, z, \bz)~,
		\ee
		where $ G'_{vz}(r, z, \bz)$ is defined to be the coefficient of the $\theta (v-v_0) $ term in $G_{vz}$.		Using the Einstein's equations, we find
		\be
			     c_{1vz} = - 2MD_z f  ~.
		\ee		Finally, we can compute
		\be \label{gvv}
			G_{vv} = \delta( v- v_0) \left(\frac{2 \delta m}{r^2}+   \frac{D^2 (D^2 +2) f}{2 r^2}-  \frac{3   M D^2 f   }{r^3} \right)~.
		\ee
		Using Einstein's equations, we find $\delta m  = \mu$.		At this point, we have completely determined the metric to be 
		\be
			\begin{split}
				g_{vv}& = -1+\frac{2 (M+\mu\theta(v - v_0)  )}{r}  -\frac{\theta(v - v_0)   MD^2 f  }{r^2} ~,\\
				g_{vr}& = 1~,\\
				g_{vz}& = \theta(v - v_0)D_z \left [\left(1- \frac{2M}{r} \right)  + \frac{1}{2} D^2  \right]f~, \\
				g_{zz} &=  -2r\theta(v - v_0)  D^2_z   f~,\\
				g_{z \bz}& = r^2 \gamma_{z \bz}~, \\
				g_{rr} &= g_{rz } = 0~.
			\end{split}
		\ee
One can check that this metric satisfies
 $G_{vr} = 0$, as desired. 
\item[14(b).] Note  that a spherically symmetric shockwave with stress tensor
			\be
				T_{vv} = \frac{\mu \delta(v-v_0)}{4 \pi r^2}~,
			\ee
		is a particular case of equation (\ref{gvv}) part  (a) with $f=0$, so that for $v>v_0$ the metric will take the form
		\be
			ds^2 = - \left (1- \frac{2( M+ \mu) }{r} \right)dv^2 + 2 dvdr + 2r^2 \gamma_{z \bz} dz d \bz~.
		\ee
		The exact form of the generator of supertranslations in a Schwarzschild background is given by
		\be
			\xi = f \p_v + \frac{1}{r} D^Af \p_A - \frac{1}{2}D^2 f \p_r~,
		\ee
		where $f = f(z, \bz)$.		Under the supertranslation, the components of the metric have the following variations:
		\be
			\begin{split}
				\mathcal L_f g_{vv} &=  \frac{D^2f M}{r^2}
					~, \quad \quad  
				\mathcal L_f g_{vr}  = \mathcal L_f g_{rr} =\mathcal L_f g_{rz}  =\mathcal L_f g_{z\bz} = 0~, \\
				\mathcal L_f g_{vz} &= - \frac{1}{2}D_z D^2 f  - D_z f +\frac{2D_z f M}{r} 
					~, \quad   \quad 
				\mathcal L_f g_{zz} = 2r D_z^2 f  ~.\\
			\end{split}
		\ee
				The change in the metric is related to the Lie derivative by $\delta g_{\mu\nu} = - \mathcal L_f g_{\mu\nu}$.  Hence, the resulting metric is given by
		\be
			\begin{split}
				g_{vv} &= -1+\frac{2 (M+\mu)}{r}- \frac{D^2f M}{r^2} ~, \\
				g_{vr} & = 1~,\\
				g_{vz} & =   \frac{1}{2}D_z D^2 f  + D_z f -\frac{2D_z f  M}{r}  ~,\\
				g_{zz} &= -2 r D_z^2 f~, \\
				g_{z \bz} & = r^2 \gamma_{z \bz}~, \\
				g_{rr}& = g_{rz}  = 0~, 
			\end{split} 
		\ee
		which matches the metric found in part (a) when $v>v_0$.

\item[15(a).] Maxwell's equations are
\begin{equation}
\begin{split}
d \ast F = e^2 \ast j~. 
\end{split}
\end{equation}
Then, using Stokes' law, we can write
\begin{equation}
\begin{split}
\hat{Q}^{\mathcal H}_\ve =  \frac{1}{e^2} \int_{{\mathcal H}} d \left[ \ve  \ast F  \right] = \frac{1}{e^2} \int_{\mathcal H} d\ve \wedge \ast F + \int_{\mathcal H} \ve \ast j ~. 
\end{split}
\end{equation}
The first term in this equation is the soft charge. To determine its explicit form, we need to determine the projection of the two-form $\ast F$ onto ${\mathcal H}$. To do this, note that the Vaidya metric takes the form
\begin{equation}
\begin{split}
ds^2 = - \left[ 1 - \frac{2M\theta(v)}{r} \right] dv^2 + 2 d v dr + 2 r^2 \g_{z\bz} dz d\bz  ~. 
\end{split}
\end{equation}
Using this form, we can determine $\ast F$ as
\begin{equation}
\begin{split}\label{eq6}
\ast F &= i F^z{}_z  dv \wedge dr - i r^2 \g_{z\bz} F_{vr} dz \wedge d\bz \\
&\quad \qquad + \left[ i F_{rz} dr \wedge dz - i \left[ F_{vz} + \left[ 1 - \frac{2M}{r} \theta(v) \right] F_{rz} \right] dv \wedge dz  + c.c. \right] . 
\end{split}
\end{equation}
This is derived using
\begin{equation}\label{237}
\begin{split}
(\ast F)_{\mu\nu} = \frac{1}{2} \ve_{\mu\nu \rho \sigma  } F^{\rho \sigma} ~. 
\end{split}
\end{equation}
We must now project equation (\ref{237}) onto ${\mathcal H}$. To do so, we note that the event horizon of the Vaidya shockwave metric is located at
\begin{equation}
\begin{split}
r_H(v) = \left\{ \begin{array}{cc}
\frac{v}{2} + 2 M ~,  & v \leq 0 ~, \\
2 M ~, & v > 0~. 
\end{array} \right.
\end{split}
\end{equation}
Then, setting $r=r_H(v)$ in \eqref{eq6}, we find
\begin{equation}
\begin{split}
\left( \ast F \right) \big|_{\mathcal H} &= - i r_H^2 \g_{z\bz} F^{(0)}_{vr} dz \wedge d\bz  + \left\{ - i \left[ F^{(0)}_{vz} + \frac{1}{2} \theta( -v ) F^{(0)}_{rz} \right] dv \wedge dz + c.c. \right\}  ~, 
\end{split}
\end{equation}
where here the superscript ${(0)}$ indicates the projection of the corresponding component of the field strength onto ${\mathcal H}$. Now we are finally ready to compute the charge. We take $\ve$ to be extended onto ${\mathcal H}$ so that it is constant along null generators (i.e., $\p_v \ve = 0$). Then we find
\begin{equation}
\begin{split}
\hat{Q}^{S,{\mathcal H}}_\ve = \frac{1}{e^2} \int d^2 z \ve \left( \p_z N_\bz + \p_\bz N_z \right)~, 
\end{split}
\end{equation}
where
\begin{equation}
\begin{split}
N_z = \int dv \left( F_{vz}^\0 + \frac{1}{2} \theta(-v) F_{rz}^\0 \right) ~. 
\end{split}
\end{equation}
Here $N_z$ is the analog of the zero mode on $\ci^+$. Just like on $\ci^+$, it is also a boundary term and only gets contributions from ${\mathcal H}^+$. To see this, we move to the advanced radial gauge:
\begin{equation}
\begin{split}
A_r = 0 ~, \qquad A_v \big|_{\mathcal H} = 0~. 
\end{split}
\end{equation}
In this gauge, we have
\begin{equation}
\begin{split}
F_{vz}^\0 + \frac{1}{2} \theta(-v) F_{rz}^\0 = \left(  \p_v  + \frac{1}{2} \theta(-v) \p_r \right) A_z \big|_{r=r_H} = \p_v A_z^\0 ~. 
\end{split}
\end{equation}
In the second equality, we first take the appropriate derivatives and \emph{then} set $r = r_H$ in $A_z$. By the chain rule, this is equal to first setting $r=r_H$ in $A_z$ and then taking a $v$-derivative. Thus, we have
\begin{equation}
\begin{split}
N_z =  \int dv \p_v A_z^\0 = A_z^+ - A_z^- ~. 
\end{split}
\end{equation}
The past boundary of ${\mathcal H}$ is really a point at $v=-4M, r = 0$, where there is a coordinate singularity. To regulate this, we take the boundary to be at $v = - 4M+ \e$ and then take $\e \to 0$. 

\item[15(b).] We recall the symplectic form determined in exercise 3:
\begin{equation}
\begin{split}\label{sympform}
\Omega_\Sigma = - \frac{1}{e^2} \int_\Sigma   \delta (  \ast F )  \wedge   \delta A  ~. 
\end{split}
\end{equation}
Plugging in the explicit forms of $\ast F$ and $A$ in the advanced radial gauge, we find
\begin{equation}
\begin{split}
\Omega_{\mathcal H}  &=  \frac{1}{e^2} \int_{\mathcal H}   dv d^2 z    \left(  \p_v \delta A^\0_z  \wedge  \delta A^\0_\bz    + \p_v \delta A^\0_\bz  \wedge   \delta A^\0_z     \right)  ~. \\
\end{split}
\end{equation}
Note that the symplectic form as well as the soft charge take exactly the same form as  on $\ci^+$. We can therefore proceed as we did there. Define 
\begin{equation}
\begin{split}
{\hat A}_z  = A_z^\0  - \frac{1}{2} \left( A^+_z + A^-_z \right) = A_z ^\0  - \partial_z \phi ~. 
\end{split}
\end{equation}
Using this, we find
\begin{equation}
\begin{split}\label{eq8}
\Omega_{{\mathcal H}} =   \frac{2}{e^2} \int dv d^2 z  \p_v \delta {\hat A}_z \wedge \delta {\hat A}_\bz -  \frac{1}{e^2} \int  d^2 z \left[ \delta \partial_z \phi \wedge \delta N_\bz + \delta \partial_\bz \phi \wedge \delta N_z \right] ~. 
\end{split}
\end{equation}
\item[15(c).]  Just as on $\ci^+$, we can write out the commutators from \eqref{eq8} as
\begin{equation}
\begin{split}
\left[ {\hat A}_z (v) , {\hat A}_\bw(v') \right]& = -   \frac{i e^2}{4} \Theta ( v - v' )  \delta^2 ( z - w ) ~, \\
\left[ \partial_z \phi  , N_\bw \right] &=  i e^2  \delta^2 ( z - w )  ~. 
\end{split}
\end{equation}
Using this, we find
\begin{equation}
\begin{split}
\left[ Q_\ve^{S,\cal H} ,  {\hat A}_z \right] = \left[ Q_\ve^{S,\cal H} ,  N_z \right]  = 0 ~, \qquad \left[ Q_\ve^{S,\cal H} ,  \partial_z \phi \right] = i \p_z \ve ~. 
\end{split}
\end{equation}

\end{enumerate}

\section{Acknowledgments}
I am grateful to Thomas Dumitrescu, Stephen Hawking, Temple He, Dan Kapec, Slava Lysov, Prahar Mitra, Monica Pate,  Sabrina Pasterski, Malcolm Perry, Achilleas Porfyriadis, Ana Raclariu, Shu-Heng Shao and Sasha Zhiboedov for invaluable discussion and collaboration on the topics in these lectures,
to Temple He, Dan Kapec,  Alex Lupsasca,  Prahar Mitra, Monica Pate, Abhishek Pathak   and Ana Raclariu  for help with the manuscript, and to Juan Maldacena for key observations at early stages of my research into IR structures.  Various components  of this work were supported by DOE grant DE-FG02-91ER40654, NSF grant 1205550, the John Templeton Foundation and the Simons Foundation.

\bibliography{soft_Arxiv_update}

\bibliographystyle{utphys}
%
\end{document}